\documentclass[11pt,a4paper]{article}
\pdfoutput=1

\usepackage[english]{babel} 
\usepackage[utf8]{inputenc}
\usepackage[left=01in,right=1in,top=1.0in,bottom=1.1in]{geometry}
\usepackage[labelfont=bf]{caption}
\usepackage[hyperfootnotes=false]{hyperref}
\usepackage[normalem]{ulem}
\usepackage{dsfont}
\usepackage{fontawesome5}
\usepackage{amsfonts}
\usepackage{amsmath}
\usepackage{amssymb}
\usepackage{amsthm}
\usepackage{bbold}
\usepackage{calc}
\usepackage{cancel}
\usepackage{float}
\usepackage{slashed}
\usepackage{mathrsfs} 
\usepackage{pdfpages}
\usepackage{multirow}
\usepackage{mathtools}
\usepackage{braket}
\usepackage{cite}
\usepackage{tikz}
\usepackage{tikzsymbols}
\usepackage{caption}
\usepackage{subcaption}
\usepackage{array}
\usepackage{colortbl} 	
\usepackage{fontawesome}

\definecolor{Gray}{gray}{0.95}
\definecolor{RGray}{gray}{0.90}
\definecolor{CGray}{gray}{0.92}
\usepackage{arydshln}
\usepackage{soul}
\usepackage{booktabs}

\definecolor{cadmiumgreen}{rgb}{0.0, 0.42, 0.24}

\usepackage[hang,flushmargin]{footmisc}
\usepackage{scalerel}
\usepackage{xcolor}

\definecolor{codegreen}{rgb}{0,0.6,0}
\definecolor{codegray}{rgb}{0.5,0.5,0.5}
\definecolor{codepurple}{rgb}{0.58,0,0.82}
\definecolor{backcolour}{rgb}{0.95,0.95,0.92}

\usepackage{listings}
\lstdefinestyle{mystyle}{
    backgroundcolor=\color{backcolour},   
    commentstyle=\color{codegreen},
    keywordstyle=\color{magenta},
    numberstyle=\tiny\color{codegray},
    stringstyle=\color{codepurple},
    basicstyle=\ttfamily\footnotesize,
    breakatwhitespace=false,         
    breaklines=true,                 
    captionpos=b,                    
    keepspaces=true,                 
    numbers=left,                    
    numbersep=5pt,                  
    showspaces=false,                
    showstringspaces=false,
    showtabs=false,                  
    tabsize=2
}
\hypersetup{
    colorlinks=true,
    linkcolor=cadmiumgreen,
    citecolor=cadmiumgreen,
    filecolor=cadmiumgreen,      
    urlcolor=cadmiumgreen,
    linktoc=all
}
\usepackage{lipsum}
\numberwithin{equation}{section}
\numberwithin{figure}{section}
\numberwithin{table}{section}

\renewcommand\thefootnote{\textcolor{cadmiumgreen}{\arabic{footnote}}}

\allowdisplaybreaks

\usetikzlibrary{decorations}

\pgfdeclaredecoration{complete sines}{initial}
{
    \state{initial}[
        width=+0pt,
        next state=sine,
        persistent precomputation={\pgfmathsetmacro\matchinglength{
            \pgfdecoratedinputsegmentlength / int(\pgfdecoratedinputsegmentlength/\pgfdecorationsegmentlength)}
            \setlength{\pgfdecorationsegmentlength}{\matchinglength pt}
        }] {}
    \state{sine}[width=\pgfdecorationsegmentlength]{
        \pgfpathsine{\pgfpoint{0.25\pgfdecorationsegmentlength}{0.5\pgfdecorationsegmentamplitude}}
        \pgfpathcosine{\pgfpoint{0.25\pgfdecorationsegmentlength}{-0.5\pgfdecorationsegmentamplitude}}
        \pgfpathsine{\pgfpoint{0.25\pgfdecorationsegmentlength}{-0.5\pgfdecorationsegmentamplitude}}
        \pgfpathcosine{\pgfpoint{0.25\pgfdecorationsegmentlength}{0.5\pgfdecorationsegmentamplitude}}
}
    \state{final}{}
}

\tikzset{vector/.style={decorate, decoration={complete sines, amplitude=8pt, segment length=10pt}}}

\tikzset{
wc/.style = {circle, fill, minimum size=#1,
              inner sep=0pt, outer sep=0pt},
wc/.default = 6pt % size of the circle diameter 
}

\graphicspath{{./Figures/}{./figures/}}

\newcommand{\be}{\begin{equation}}
\newcommand{\ee}{\end{equation}}
\newcommand{\bea}{\begin{eqnarray}}
\newcommand{\eea}{\end{eqnarray}}  

\newcommand{\gsim}{\lower.7ex\hbox{$\;\stackrel{\textstyle>}{\sim}\;$}}
\newcommand{\lsim}{\lower.7ex\hbox{$\;\stackrel{\textstyle<}{\sim}\;$}}
\newcommand{\cO}{{\mathcal O}}

\newcommand{\cC}{{\mathcal C}}
\newcommand{\cL}{{\mathcal L}}

\newcommand{\cA}{{\mathcal A}}
\newcommand{\cB}{{\mathcal B}}
\newcommand{\cF}{{\mathcal F}}

\newcommand{\cS}{{\mathcal S}}
\newcommand{\cT}{{\mathcal T}}
\newcommand{\cU}{{\mathcal U}}
\newcommand{\cN}{{\mathcal N}}

\newcommand{\cZ}{{\mathcal Z}}

\newcommand{\SMrep}[3]{({\bf #1},\,{\bf #2},\,#3)}
\newcommand{\SMrepbar}[3]{({\bf \bar#1},\,{\bf #2},\,#3)}
\newcommand{\C}[2]{C_{\underset{#2}{#1}}}
\newcommand{\cmd}[1]{{\texttt{#1}}\xspace}
\newcommand{\squarebrackets}[1]{\left[ #1 \right]} 
\newcommand{\brackets}[1]{\left( #1 \right)}
\newcommand{\braces}[1]{\left\lbrace #1 \right\rbrace}
\newcommand{\dd}{\mathop{}\!\mathrm{d}}

\newcommand{\HighPT}[0]{{\texttt{HighPT}}\xspace}
\newcommand{\Mathematica}[0]{{\texttt{Mathematica}}\xspace}

\definecolor{cadmiumgreen}{rgb}{0.0, 0.42, 0.24}

\definecolor{applegreen}{rgb}{0.55, 0.71, 0.0}

\newcommand\blfootnote[1]{%
  \begingroup
  \renewcommand\thefootnote{}\footnote{#1}%
  \addtocounter{footnote}{-1}%
  \endgroup
}
\makeatletter
\g@addto@macro\bfseries{\boldmath}
\makeatother

\begin{document}
\begin{flushright}
ZU-TH-28/22 \\
\end{flushright}

\begin{center}
\vspace{0.7cm}
{\Large\bf  Drell-Yan Tails Beyond the Standard Model}
\\[1.0cm]
{L.~Allwicher${}^{\,\bullet}$\blfootnote{\href{mailto:lukall@physik.uzh.ch}{lukall@physik.uzh.ch}},~D.~A.~Faroughy${}^{\,\bullet}$\blfootnote{\href{mailto:faroughy@physik.uzh.ch}{faroughy@physik.uzh.ch}}, F.~Jaffredo${}^{\,\bullet\bullet}$\blfootnote{\href{mailto:florentin.jaffredo@ijclab.in2p3.fr}{florentin.jaffredo@ijclab.in2p3.fr}}}, 
{O.~Sumensari${}^{\,\bullet\bullet}$\blfootnote{\href{mailto:olcyr.sumensari@ijclab.in2p3.fr}{olcyr.sumensari@ijclab.in2p3.fr}}, F.~Wilsch${}^{\,\bullet}$\blfootnote{\href{mailto:felix.wilsch@physik.uzh.ch}{felix.wilsch@physik.uzh.ch}}}\\
\vspace{0.7cm}

{\em\small ${}^{\ \bullet}$Physik-Institut, Universit\"at Z\"urich, CH-8057 Z\"urich, Switzerland\\[0.3em]
\small${}^{\bullet\bullet}$IJCLab, P\^ole Th\'eorie (Bat. 210), CNRS/IN2P3 et Universit\'e, Paris-Saclay, 91405 Orsay, France}
\end{center}
\vspace{0.5 cm}

\centerline{\large\bf Abstract}
\begin{quote}
We investigate the high-$p_T$ tails of the $pp\to \ell \nu$ and $pp \to \ell \ell$ Drell-Yan processes as probes of New Physics in semileptonic interactions with an arbitrary flavor structure. For this purpose, we provide a general decomposition of the $2\to2$ scattering amplitudes in terms of form-factors that we match to specific scenarios, such as the Standard Model Effective Field Theory (SMEFT), including all relevant operators up to dimension-$8$, as well as ultraviolet scenarios giving rise to tree-level exchange of new bosonic mediators with masses at the TeV scale. By using the latest LHC run-II data in the monolepton ($e\nu$, $\mu\nu$, $\tau\nu$) and dilepton ($ee$, $\mu\mu$, $\tau\tau$, $e\mu$, $e\tau$, $\mu\tau$) production channels, we derive constraints on the SMEFT Wilson coefficients for semileptonic four-fermion and dipole operators with the most general flavor structure, as well as on all possible leptoquark models. For the SMEFT, we discuss the range of validity of the EFT description, the relevance of $\cO(1/\Lambda^2)$ and $\cO(1/\Lambda^4)$ truncations, the impact of $d=8$ operators and the effects of different quark-flavor alignments. Finally, as a highlight, we extract for several New Physics scenarios the combined limits from high-$p_T$ processes, electroweak pole measurements and low-energy flavor data for the $b\to c\tau\nu$ transition, showing the complementarity between these different observables. Our results are compiled in {\tt HighPT}~\href{https://highpt.github.io}{\faicon{github}}, a package in {\tt Mathematica} which provides a simple way for users to extract the Drell-Yan tails likelihoods for semileptonic effective operators and for leptoquark models.
\blfootnote{\href{https://highpt.github.io}{\faicon{github} https://highpt.github.io}}
\end{quote}
\thispagestyle{empty}
\clearpage
\setcounter{page}{0}
\newpage
{
  \hypersetup{linkcolor=black}
  \tableofcontents
  \thispagestyle{empty}
}
\newpage
%%%%%%%%%%%%%%%%%%%%%%%%%%%%%%%%%%%%%%%%%%%%%%%%%%%%%%%%%%%%%%%%%%%%%%%%%%
\section{Introduction}
\label{sec:intro}
%%%%%%%%%%%%%%%%%%%%%%%%%%%%%%%%%%%%%%%%%%%%%%%%%%%%%%%%%%%%%%%%%%%%%%%%%%

Semileptonic transitions are powerful probes of physics beyond the Standard Model (SM) which can indirectly access new phenomena arising at scales well beyond the reach of the direct searches at particle colliders. The most sensitive observables are those suppressed in the SM, such as Flavor Changing Neutral Currents (FCNC), which allow us to indirectly probe New Physics scales ($\Lambda$) up to $\cO(10^5~\mathrm{TeV})$ with the current precision~\cite{Isidori:2010kg}. The physics of low-energy semileptonic transitions has recently attracted a lot of attention thanks to the rich program of experiments studying $K$- \cite{NA62:2017rwk,Yamanaka:2012yma}, $D$- \cite{BESIII:2020nme} and $B$-meson  \cite{LHCb:2018roe,Belle-II:2018jsg} decays, which will offer many opportunities to probe New Physics in the near future. In particular, current data with loop-level~\cite{LHCb:2014vgu,LHCb:2017avl,LHCb:2019efc,LHCb:2021trn,LHCb:2021lvy} and tree-level~\cite{BaBar:2013mob,Belle:2015qfa,LHCb:2015gmp,Belle:2016dyj,Belle:2017ilt,LHCb:2017rln,Belle:2019gij,Belle:2019rba} induced $B$-meson decays already shows intriguing patterns of deviations from the SM which are under scrutiny at LHCb and Belle-II. 

Although low-energy flavor observables provide the most stringent constraints on semileptonic transitions, their sensitivity depends fundamentally on the assumption regarding the (unknown) flavor structure of physics Beyond the SM (BSM). New Physics scenarios based on Minimal Flavor Violation~\cite{DAmbrosio:2002vsn} or the $U(2)$ symmetry~\cite{Barbieri:2011ci} can be fully compatible with current flavor data for much lower values of $\Lambda$, in the $\cO(1~\mathrm{TeV})$ range, due to the symmetry protection of quark-flavor violation. These scales are currently being directly probed at the LHC, which can provide useful complementary probes to flavor observables. In particular, this is the case for operators that are unconstrained or that can only be weakly constrained by low-energy processes, see e.g.~Ref.~\cite{Angelescu:2020uug,Fuentes-Martin:2020lea}.

Measurements in the tails of momentum-dependent distributions in $pp\to\ell\nu$ and $pp\to\ell\ell$ processes at the LHC have proven to be powerful probes of flavor physics at hadron colliders. Effective Field Theory (EFT) contributions to the Drell-Yan cross-sections can be energy enhanced, as long as the EFT approach is valid, being potentially larger than the SM background in the tails of the distributions~\cite{Farina:2016rws}. Furthermore, the parton content of the proton includes five different quark flavors that can be exploited to indirectly probe various semileptonic transitions in high-energy proton collisions. A notable example concerns the discrepancies observed in $b\to c \tau\nu$ \cite{BaBar:2013mob,Belle:2015qfa,LHCb:2015gmp,Belle:2016dyj,Belle:2017ilt,LHCb:2017rln,Belle:2019gij,Belle:2019rba}, for which LHC data from $pp\to\tau\tau$ at high-$p_T$ was used to discard several putative BSM explanations of these anomalies, see e.g.~Ref.~\cite{Faroughy:2016osc} and following works. Furthermore, Drell-Yan measurements at the LHC are important probes for leptoquark states with masses in the TeV range, and in particular, for those coupling to third generation fermions. Indeed, it was pointed out long ago in Ref.~\cite{Eboli:1987vb} that leptoquarks can contribute non-resonantly to dilepton production $pp\to\ell\ell$ at hadron colliders when exchanged in the $t$- or $u$-channels. These indirect processes provide an additional experimental handle that can complement the existing leptoquark pair-production searches, in some cases more efficiently than the leptoquark single-production processes, see e.g.~Ref.~\cite{CMS:2022zks,ATLAS:2022fho}.

There have been many studies that derive constraints on flavor-physics scenarios by using the processes $pp\to \ell\ell$ \cite{Faroughy:2016osc,Greljo:2017vvb,deBlas:2013qqa,Dawson:2018dxp,Cirigliano:2012ab,Fuentes-Martin:2020lea,Crivellin:2021rbf}, $pp\to \ell\nu$ \cite{Cirigliano:2012ab,Chang:2014iba,Cirigliano:2018dyk,Greljo:2018tzh,Fuentes-Martin:2020lea,Endo:2021lhi,Marzocca:2020ueu,Iguro:2020keo,Jaffredo:2021ymt,Bressler:2022zhv} and $pp\to \ell\ell^\prime$ (with $\ell\neq \ell^\prime$)~\cite{Angelescu:2020uug} at the LHC. However, these studies typically consider either specific types of processes, or they impose a given ansatz to the flavor pattern of the New Physics couplings. The complete combination of the LHC constraints on semileptonic interactions into a single framework was not available thus far. In this paper, we aim to amend this gap by combining the most recent LHC data from all possible monolepton and dilepton productions channels, without any assumption on the flavor of the colliding quarks. This combination will be done for the Standard Model EFT (SMEFT)~\cite{Buchmuller:1985jz,Grzadkowski:2010es}, with a consistent EFT expansion up to $\cO(1/\Lambda^{4})$ including $d\leq 8$ contributions at the cross-section level~(see Refs.~\cite{Boughezal:2021tih,Boughezal:2022nof,Kim:2022amu} for similar analyses), as well as for models with concrete mediators, which should be used if the experimental precision in a given channel is not sufficient to justify the EFT approach. In particular, implementing both EFT and concrete ultraviolet models will allow us to assess the range of EFT validity within a few selected examples. Furthermore, we will verify the EFT truncation by using a \emph{clipping} procedure, which imposes a maximal energy cut ($M_{\mathrm{cut}}$) on the data considered to extract the limits~\cite{Brivio:2022pyi}.

As an important by-product of our work, we introduce the {\tt Mathematica} package {\tt HighPT}~\cite{<HighPT:_A_Tool_for_High-pT_Collider_Studies_Beyond_the_Standard_Model><Allwicher;Lukas><Faroughy;Darius><Jaffredo;Florentin><Sumensari;Olcyr><Wilsch;Felix><2022>} that provides the complete SMEFT likelihood for semileptonic operators in Drell-Yan processes at the LHC.~\footnote{\HighPT is publicly available at \href{https://github.com/HighPT/HighPT}{\faicon{github} https://github.com/HighPT/HighPT} and further information is provided on the website \href{https://highpt.github.io}{https://highpt.github.io}.} This package will complement the ongoing effort to provide tools for the SMEFT phenomenology of low-energy flavor observables~\cite{Straub:2018kue,Aebischer:2018iyb,EOSAuthors:2021xpv,DeBlas:2019ehy}, as well as electroweak and Higgs data~\cite{DeBlas:2019ehy,Falkowski:2019hvp,Ethier:2021bye}. {\tt HighPT} also provides the likelihood for specific models of interest such as leptoquarks~\cite{Buchmuller:1986zs,Dorsner:2016wpm}, including their propagation effects at the LHC.~\footnote{Future releases of {\tt HighPT} will also include other mediators listed in Table~\ref{tab:mediators}.} The comparison of the constraints derived for both EFT and concrete models will allow the users to directly assess the validity of the EFT description for a given high-$p_T$ process.

The remainder of this paper is organized as follows. In Sec.~\ref{sec:dilepton-ffs}, we provide a  general description in term of form-factors of the neutral and charged Drell-Yan processes at hadron colliders. In Sec.~\ref{sec:ff-bsm}, we use this framework to describe two specific New Physics scenarios with arbitrary flavor structures: (i) the SMEFT with operators up to $d=8$, and (ii) the most relevant simplified models for new bosonic mediators exchanged at tree-level. In Sec.~\ref{sec:collider-limits}, we recast the most recent LHC searches in the monolepton and dilepton channels for all possible final states and use these to set the most stringent LHC limits on $d=6$ dipoles and semileptonic operators, as well as for all leptoquark mediators. We also discuss the validity of the $\cO(1/\Lambda^2)$ and $\cO(1/\Lambda^4)$ EFT truncations and assess the impact of dimension-$8$ operators for different initial quark flavors.
In Sec.~\ref{sec:example}, we apply our results to the New Physics interpretations of the ongoing discrepancies observed in $b\to c\tau \nu$ low-energy data, combining our LHC bounds with the relevant flavor and electroweak constraints. We summarize our findings and discuss the outlook for our study in Sec.~\ref{sec:outlook}.

%%%%%%%%%%%%%%%%%%%%%%%%%%%%%%%%%%%%%%%%%%%%%%%%%%%%%%%%%%%%%%%%%%%%%%%%%%
\section{Drell-Yan Production at Hadron Colliders}
\label{sec:dilepton-ffs}
%%%%%%%%%%%%%%%%%%%%%%%%%%%%%%%%%%%%%%%%%%%%%%%%%%%%%%%%%%%%%%%%%%%%%%%%%%

%%%%%%%%%%%%%%%%%%%
\begin{figure}[t!]
\begin{center}
    \includegraphics[width=.25\linewidth]{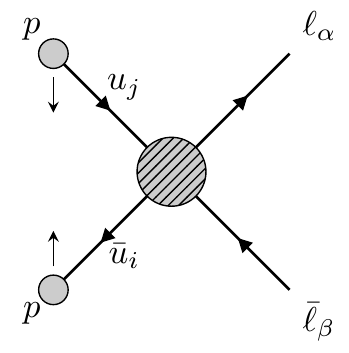}\hspace{1cm}
    \includegraphics[width=.25\linewidth]{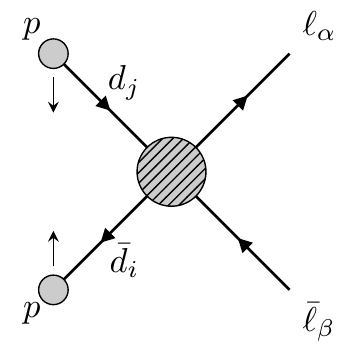}\hspace{1cm}
    \includegraphics[width=.25\linewidth]{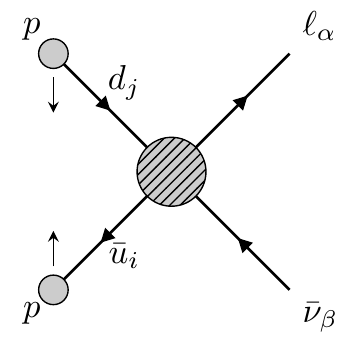}
    \caption{\sl\small Neutral and charged Drell-Yan production processes at proton-proton colliders.}
    \label{fig:DY_diags}
\end{center}
\end{figure}
%%%%%%%%%%%%%%%%%%%

In this Section, we provide a general description of the processes $pp \to \ell_\alpha^- \ell_\beta^+$ and $pp\to \ell_\alpha^\pm \nu_\beta$, in terms of form-factors, where $\alpha,\beta$ are generic lepton-flavor indices. This description has the advantage of covering both the EFT case, as well as scenarios containing new bosonic mediators that propagate at tree level.

%=====================================================
\subsection{Amplitude decomposition}
\label{subsec:amp_dec}
%=====================================================

First, we consider the scattering amplitude for the neutral Drell-Yan process $\bar{q}_i {q}_j \to \ell_\alpha^- \ell_\beta^+$ given by the first two diagrams in Fig.~\ref{fig:DY_diags}, with $q_i=\{u_i,d_i\}$, where quark and lepton flavor indices are denoted by Latin letters ($i,j=1,2,3$) and Greek letters ($\alpha,\beta=1,2,3$), respectively, unless stated otherwise.~\footnote{For up-type quarks the indices run as $i,j=1,2$ because of the negligible top-quark content of the proton at the LHC.} The most general decomposition of the four-point scattering amplitude that is Lorentz invariant and consistent with the $SU(3)_c\times U(1)_{\rm em}$ gauge symmetry reads  
\begin{align}
\begin{split}\label{eq:dilep-amp}
\mathcal{A}(\bar{q}_i q_j \to {\ell}_{\alpha}^-\ell^+_{\beta})\, =\, \frac{1}{v^2}\,\sum_{XY}&\,\Big\lbrace\,\,
\left(\bar \ell_\alpha\gamma^\mu   \mathbb{P}_X \ell_\beta\right)\left(\bar q_i\gamma_\mu  \mathbb{P}_Y q_j\right)\, [\mathcal{F}^{XY,\,qq}_{V}(\hat{s},\hat{t})]_{\alpha\beta ij}\\[0.15em]
& + \left(\bar \ell_\alpha  \mathbb{P}_X\ell_\beta\right)\left(\bar q_i  \mathbb{P}_Y q_j\right)\, [\mathcal{F}^{XY,\,qq}_{S}(\hat{s},\hat{t})]_{\alpha\beta ij}\\[0.25em]
& + \left(\bar \ell_\alpha\sigma_{\mu\nu} \mathbb{P}_X\ell_\beta\right)\left(\bar q_i \sigma^{\mu\nu}  \mathbb{P}_Y q_j\right) \, \delta^{XY} \, [\mathcal{F}^{XY,\,qq}_T(\hat{s},\hat{t})]_{\alpha\beta ij}\\[0.25em]
& +  \left(\bar \ell_\alpha\gamma_{\mu} \mathbb{P}_X\ell_\beta\right)\left(\bar q_i \sigma^{\mu\nu}  \mathbb{P}_Y q_j\right)\,\frac{ik_\nu}{v}  \, [\mathcal{F}^{XY,\,qq}_{D_q}(\hat{s},\hat{t})]_{\alpha\beta ij} \\[0.25em]
& + \left(\bar \ell_\alpha\sigma^{\mu\nu}   \mathbb{P}_X \ell_\beta\right) \left(\bar q_i \gamma_{\mu}   \mathbb{P}_Y q_j\right)\,\frac{ik_\nu}{v} \, [\mathcal{F}^{XY,\,qq}_{D_\ell}(\hat{s},\hat{t})]_{\alpha\beta ij}\,\Big\rbrace\,,
\end{split}
\end{align}
where $X,Y\!\in\!\{L,R\}$ are the chiralities of the anti-lepton and anti-quark fields, $\mathbb{P}_{R,L}=(1\pm\gamma^5)/2$ are the chirality projectors, $v=(\sqrt{2}G_F)^{-1/2}$ stands for the electroweak vacuum-expectation-value~(vev), and fermion masses have been neglected. Here, it is understood that $q$ ($\bar q$) and $\ell$ ($\bar\ell$) denote the Dirac spinors of the incoming quark (anti-quark) and outgoing anti-lepton (lepton) fields, respectively. The four-momentum of the dilepton system is defined by $k = p_q+p_{\bar{q}}$, and we take the Mandelstam variables to be $\hat s = k^2= (p_q+p_{\bar q})^2$, $\hat t = (p_{q}-p_{\ell^-})^2$ and $\hat u= (p_{q}-p_{\ell^+})^2=-\hat s-\hat t$. For each of the five components in Eq.~\eqref{eq:dilep-amp} we define the neutral current form-factor $\cF_{I}^{XY,\,qq}(\hat s,\hat t)$ where $I\in\{V,S,T,D_\ell,D_q\}$ labels the corresponding {\em vector}, {\em scalar}, {\em tensor}, {\em lepton-dipole} and {\em quark-dipole} Lorentz structures, respectively. These form-factors are dimensionless functions of the Mandelstam variables that describe the underlying local and non-local semileptonic interactions between fermions with fixed flavors and chiralities. Note, in particular, that the tensor form-factor is non-vanishing only for $X\!=\!Y$.~\footnote{This can be shown, e.g., using the identity $\sigma^{\mu\nu} \gamma_5 = i \varepsilon^{\mu\nu\alpha\beta}\sigma_{\alpha\beta}/2$\,.}

Similarly, the most general scattering amplitude for the charged current Drell-Yan process can be written as
\begin{align}
\begin{split}\label{eq:dilep-amp-charged}
\mathcal{A}(\bar{u}_i d_j \to {\ell}_{\alpha}^-\bar\nu_{\beta})\, =\, \frac{1}{v^2}\,\sum_{XY}&\,\Big\lbrace\,\,
\left(\bar \ell_\alpha\gamma^\mu   \mathbb{P}_X \nu_\beta\right)\left(\bar u_i\gamma_\mu  \mathbb{P}_Y d_j\right)\, [\mathcal{F}^{XY,\,ud}_{V}(\hat{s},\hat{t})]_{\alpha\beta ij}\\
& + \left(\bar \ell_\alpha  \mathbb{P}_X\nu_\beta\right)\left(\bar u_i  \mathbb{P}_Y d_j\right)\, [\mathcal{F}^{XY,\,ud}_{S}(\hat{s},\hat{t})]_{\alpha\beta ij}\\[0.25em]
& + \left(\bar \ell_\alpha\sigma_{\mu\nu} \mathbb{P}_X\nu_\beta\right)\left(\bar u_i \sigma^{\mu\nu}  \mathbb{P}_Y d_j\right) \, \delta^{XY} \, [\mathcal{F}^{XY,\,ud}_T(\hat{s},\hat{t})]_{\alpha\beta ij}\\[0.25em]
& +  \left(\bar \ell_\alpha\gamma_{\mu} \mathbb{P}_X\nu_\beta\right)\left(\bar u_i \sigma^{\mu\nu}  \mathbb{P}_Y d_j\right)\,\frac{ik_\nu}{v}  \, [\mathcal{F}^{XY,\,ud}_{D_q}(\hat{s},\hat{t})]_{\alpha\beta ij} \\[0.25em]
& + \left(\bar \ell_\alpha\sigma^{\mu\nu}   \mathbb{P}_X \nu_\beta\right)\left(\bar u_i \gamma_{\mu}   \mathbb{P}_Y d_j\right)\, \frac{ik_\nu}{v} \, [\mathcal{F}^{XY,\,ud}_{D_\ell}(\hat{s},\hat{t})]_{\alpha\beta ij}\,\Big\rbrace\,,
\end{split}
\end{align}
\noindent where the dilepton four-momentum is defined in a similar way by $k = p_d+p_{\bar{u}}$, and where we take the Mandelstam variables to be $\hat s = k^2= (p_d+p_{\bar u})^2$, $\hat t = (p_{d}-p_{\ell^-})^2$ and $\hat u= (p_{d}-p_{\nu})^2$\,. The charged current form-factors are denoted by $\smash{\cF^{XY,\,ud}_I(\hat s,\hat t)}$, with the same possible Lorentz structures as in the previous case.~\footnote{Note that the charge-conjugate process can be described by a similar expression to Eq.~\eqref{eq:dilep-amp-charged}. The relations between the $\bar{d}_j u_i \to {\ell}_{\alpha}^+\nu_{\beta}$ and the ${d}_j \bar{u}_i \to {\ell}_{\alpha}^-\bar{\nu}_{\beta}$ form-factors are spelled out in Appendix~\ref{sec:conventions}.} The above equation is also valid for $X=R$ in the presence of a light right-handed neutrino field $N\sim ({\bf 1},{\bf1},0)$ that is a singlet under the SM gauge group. The amplitudes in Eqs.~\eqref{eq:dilep-amp} and \eqref{eq:dilep-amp-charged} are written in the mass basis. Similar expressions in the weak interaction basis can be recovered by rotating the quark fields accordingly, as described in Appendix~\ref{sec:FFrotations}. From now on, we thus take all flavor indices in the basis of weak interactions if not mentioned otherwise.

%=====================================================
\subsubsection*{Related processes} 
%=====================================================

We briefly comment on two other types of semileptonic processes at hadron colliders that are closely related to Drell-Yan production. The first of these are the {\em quark-lepton fusion} processes $q_i \ell_\alpha \to q_j\ell_\beta$ and $d_i \ell_\alpha \to u_j\nu_\beta$. These probe the same semileptonic transitions entering Drell-Yan production. In this case, the initial lepton is taken as a partonic constituent of the proton with a parton distribution function~(PDF) that is suppressed by $\alpha_\mathrm{em}$~\cite{Buonocore:2021bsf}. By using crossing symmetry, it is straightforward to express the amplitudes in terms of the Drell-Yan from-factors described above,
\begin{align}
    \mathcal{A}(u_j \ell^+_\alpha \to u_i \ell^+_\beta)\ &=\ \mathcal{A}(\bar u_i u_j\to \ell^-_\alpha \ell^+_\beta)|_{\,s\to -t,\,t\to -s}\,,\\[0.3em]     \mathcal{A}(d_j \ell^+_\alpha \to d_i \ell^+_\beta)\ &=\ \mathcal{A}(\bar d_i d_j\to \ell^-_\alpha \ell^+_\beta)|_{\,s\to -t,\,t\to -s}\,,\\[0.3em]
    \mathcal{A}(d_j \ell^+_\alpha \to u_i \bar\nu_\beta)\ &=\ \mathcal{A}(\bar u_i d_j\to \ell^-_\alpha\bar\nu_\beta)|_{\,s\to -t,\,t\to -s}\,.
\end{align}
Another relevant probe of semileptonic transitions, also related to Drell-Yan production, are the {\it quark-gluon fusion} processes $q_jg\to q_i \ell^-_\alpha\ell^+_\beta$ and $q_jg\to q_i \ell_\alpha^{\mp}\bar \nu_\beta$. Since these are $2\to3$ scattering processes, they will suffer from an additional phase-space suppression when compared to the $2\to2$ Drell-Yan processes. In the following, given that both the quark-lepton and quark-gluon fusions are generically less efficient probes of New Physics, we will focus exclusively on the Drell-Yan production modes as they are currently the most relevant ones for phenomenology.~\footnote{A notable exception are the quark-gluon fusion processes $gb\to b\ell\ell$ and $gc\to b\ell\nu$. The enhancement of the gluon over the bottom PDF and the background reduction from the additional $b$-tagged jet make these processes important probes for New Physics entering third-generation semileptonic transitions~\cite{Altmannshofer:2017poe,Afik:2019htr,2008.07541}.} 

%=====================================================
\subsection{Form-factor parametrization}
\label{sec:FF_param}
%=====================================================

In the following, we introduce a general parametrization of the Drell-Yan form-factors that is useful for describing tree-level contributions from generic New Physics. We perform an analytic continuation of the scattering amplitudes to the complex $\hat s$ and $\hat t$ Mandelstam variables. Furthermore, we assume that the form-factors are analytic functions within some radius $\Lambda^2$ except for a finite set of simple poles in the $\hat s$,  $\hat t$ and $\hat u$ complex planes. This assumption captures all possible tree-level physics entering Drell-Yan production at collider energies below the scale $\Lambda^2\gg|\hat s|,|\hat t|$.~\footnote{This assumption leaves out scenarios with loop-level contributions from light degrees of freedom where branch cuts can appear.} We decompose each form-factor into a \textit{regular} term and a \textit{pole} term,
\begin{align}\label{eq:FF}
\cF_{I}(\hat  s,\hat{t})\ =\ \cF_{I,\, \rm Reg}(\hat s,\hat{t})\ +\ \cF_{I,\, \rm Poles}(\hat s,\hat{t})\,,  
\end{align}
encoding underlying local and non-local semileptonic interactions, respectively. To simplify the notation we drop the $XY$ and $qq^\prime$ superscripts wherever the equations hold true for any form-factor, and only keep the dependence on $I\in\{V,S,T,D_\ell,D_q\}$.

The regular form-factor $\cF_{I,\, \rm Reg}$ is an analytic function that describes local interactions, e.g.~four-point contact interactions, that arise from heavy unresolved degrees of freedom living at the scale $\Lambda$ beyond the characteristic energy of the scattering process. Within the radius $\Lambda^2$, this function admits a power series expansion of the form
\begin{align}\label{eq:regular}
    \cF_{I,\,\rm Reg}(\hat  s,\hat{t})\ =\ \sum_{n,m=0}^\infty \cF_{I \,(n,m)}\,\left(\frac{\hat s}{v^2}\right)^{\!n}\left(\frac{\hat t}{v^2}\right)^{\!m}\,,
\end{align}
\noindent where $\cF_{I\,(n,m)}$ are dimensionless expansion coefficients. 
This expression provides a convenient separation of contributions with different scaling on $\hat{s}$ and $\hat{t}$ and, in particular, of those that become dominant in the tails of the Drell-Yan distributions. The power series in Eq.~\eqref{eq:regular} is not to be confused with the EFT expansion in $1/\Lambda$, since each coefficient $\cF_{I \,(n,m)}$ receives contributions from an infinite tower of non-renormalizable operators, as will be discussed for the SMEFT in Sec.~\ref{sec:ff-bsm}. 

The pole form-factor $\cF_{I,\, \rm Poles}$ is a non-analytic function with a finite number of simple poles describing non-local tree-level interactions. We adopt the following parametrization,
\begin{align}\label{eq:poles}
    \cF_{I,\,\rm Poles}(\hat  s,\hat{t})\ &=\ \sum_a\frac{v^2\, \cS_{\,I\,(a)} }{\hat{s}-\Omega_a}
\ +\  \sum_b\frac{v^2\, \cT_{\,I\,(b)} }{\hat{t}-\Omega_b}
\ -\  \sum_c\frac{v^2\, \cU_{\,I\,(c)} }{\hat{s} + \hat {t} +\Omega_c}\,,
\end{align}
where the poles $\Omega_k=m_k^2-im_k\Gamma_k$ belong to each of the corresponding complex Mandelstam planes, with the last term representing the poles in the $u$-channel. The pole residues $\mathcal{S}_{I\,(a)}$, $\mathcal{T}_{I\,(b)}$ and $\mathcal{U}_{I\,(c)}$ are taken to be dimensionless parameters. Each term in Eq.~\eqref{eq:poles} describes the tree-level exchange of degrees of freedom in the $s$-, $t$- and $u$-channels, respectively, i.e.~these are the propagators for various bosons $a,b,c$ with masses $m_{a,b,c}$ and widths $\Gamma_{a,b,c}$ that can be resolved at the energy scales involved in the scattering. 

In principle, the simple-pole assumption for the form-factor singularities allows for the numerators in Eq.~\eqref{eq:poles} to be general analytic functions of the form $\mathcal{S}_{I\,(a)}(\hat s)$, $\mathcal{T}_{I\,(b)}(\hat t)$ and $\mathcal{U}_{I\, (c)}(\hat u)$, each describing the product of two local three-point interactions. However, the dependence of these residues on the Mandelstam variables can be completely removed from each pole by applying the identity, 
\begin{equation}\label{eq:trick}
\frac{\cZ_I(\hat z)}{\hat z-\Omega}\ =\ \frac{\cZ_I(\Omega)}{\hat z-\Omega}\ +\ h(\hat z,\Omega)\,,   
\end{equation}
\noindent where $h(\hat z,\Omega)$ is an analytic function of $\hat z=\{\hat s,\hat t,\hat u\}$ that can be reabsorbed into the regular form-factor by a redefinition of $\cF_{I,\, \rm Reg}$.~\footnote{The identity in Eq.~(\ref{eq:trick}) can be shown by power expanding the numerator $\cZ_I(\hat z)$ and decomposing into partial fractions each of the resulting terms as
\begin{align}\label{eq:partial_frac_decomp}
    \frac{\hat z^n}{\hat z-\Omega}\ =\ \frac{\Omega^n}{\hat z-\Omega}\ +\ \Omega^{n-1}\,\sum_{k=0}^{n-1}\left(\frac{\hat z}{\Omega}\right)^k\,.
\end{align}}
%

%=====================================================
\subsubsection*{Form-factors in the SM} 
%=====================================================

In the SM, the gauge bosons contribute to the Drell-Yan amplitudes in Eqs.~\eqref{eq:dilep-amp} and \eqref{eq:dilep-amp-charged} through the $s$-channel poles of the vector form-factors. It is therefore convenient to separate the effects of the SM from potential BSM effects by defining the $s$-channel vector residues in Eq.~\eqref{eq:poles} as
\begin{align}\label{eq:pole_decomp}
\mathcal{S}_{\,V\,(a)}\ =\ \mathcal{S}_{\,(a,\,\rm SM)}\ +\ \delta \mathcal{S}_{\,(a)} \,,
\end{align}
with $a\in\{\gamma,W,Z\}$, and where the $\delta \mathcal{S}_{(a)}$ parametrize potential modifications of the SM gauge couplings to fermions. The SM pole residues at leading-order read
\begin{align}\label{eq:SM_photon_pole}
\mathcal{S}_{(\gamma,\,\rm SM)}^{\,XY,\,qq}\ &=\ 4\pi\alpha_{\mathrm{em}}\,Q_\ell\, Q_q\,\mathbb{1}_\ell\mathbb{1}_q\,, \\[0.3em] \label{eq:SM_Z_pole}
\mathcal{S}_{(Z,\,\rm SM)}^{\,XY,\,qq}\ &=\ \frac{4\pi\alpha_\mathrm{em}}{c_W^2 s_W^2}\,g^X_\ell g_q^Y\,, \\[0.3em] \label{eq:SM_W_pole}
\mathcal{S}_{(W,\,\rm SM)}^{\,LL,\,ud}\ &=\ \frac{1}{2}g^2\, \mathbb{1}_\ell V \,,\end{align}
where $g_\psi^X\equiv (T^3_{\psi_X}-s_W^2Q_\psi)\,\mathbb{1}_\psi$ denote the $Z$-boson couplings to a fermion $\psi$ with electric charge $Q_\psi$ and weak isospin $T^3_{\psi_X}$, $g$ is the $SU(2)_L$ gauge coupling and $c_W \equiv \cos \theta_W$ and $s_W=\sin \theta_W$, where $\theta_W$ denotes the Weinberg angle. The $3\times3$ Cabibbo–Kobayashi–Maskawa~(CKM) matrix is labeled~$V$, and $\mathbb{1}_{\ell(q)}$ correspond to the $3\times 3$~unit matrices in lepton (quark) flavor-space with components $\delta_{\alpha\beta}$ ($\delta_{ij}$). See Appendix~\ref{sec:conventions} for our conventions. New Physics contributions to the form-factors $\mathcal{S}_{\,(a)}$, $\mathcal{T}_{\,(a)}$ and $\mathcal{U}_{\,(a)}$  will be discussed in Sec.~\ref{sec:ff-bsm}.

\subsection{Cross-sections}
The general amplitudes given in Eq.~\eqref{eq:dilep-amp} and \eqref{eq:dilep-amp-charged} can be used to compute the neutral- and charged-current cross-sections, respectively. After integrating over the azimuthal angle, the differential partonic cross-section for the Drell-Yan process is given by
\begin{align}\label{eq:dsigma}
 \resizebox{.91\hsize}{!}{ 	$\displaystyle\dfrac{{\rm d}\hat\sigma}{\rm{d}\hat t}(\bar{q}_i q^\prime_j \to {\ell}_{\alpha} \overline{\ell^{\prime}}_{\beta})  = \frac{1} {48\pi\,v^4}\,\sum_{XY} \sum_{IJ} M^{XY}_{IJ}(\hat s,\hat t)\,{\left[\mathcal{F}^{XY,\,qq^\prime}_I (\hat s,\hat t) \right]}_{\alpha\beta ij} {\left[\mathcal{F}^{XY,\,qq^\prime}_J(\hat s,\hat t)\right]}^{\ast}_{\alpha\beta ij}\,,$}
\end{align}
where neutral and charged currents are described by the same expression, where $q^{(\prime)}\in \{u,d\}$ can be either a down- or up-type quark, and $\ell^{\prime}\in \{\ell,\nu\}$ denotes both neutral and charged leptons, depending on the specific process. The indices $I,J\in\{V,S,T,D_\ell,D_q\}$ account for the different contributions and $M^{XY}$ are $5\times5$ symmetric  matrices that take the form 
\begin{align}
 M^{XY}(\hat s,\hat t)=\begin{pmatrix}
M_{VV}^{XY}(\hat t/\hat s) & 0 & 0 & 0 & 0 \\[0.3em]
0 & M_{SS}^{XY}(\hat t/\hat s) & M_{ST}^{XY}(\hat t/\hat s) & 0 & 0 \\[0.3em]
0 & M_{ST}^{XY}(\hat t/\hat s)  & M_{TT}^{XY}(\hat t/\hat s)  & 0 & 0 \\[0.3em]
0 & 0 & 0 & \frac{\hat s}{v^2} M_{DD}^{XY}(\hat t/\hat s)& 0 \\[0.3em]
0 & 0 & 0 & 0 & \frac{\hat s}{v^2}M_{DD}^{XY}(\hat t/\hat s)
\end{pmatrix} \,,
\end{align} 
\noindent where the different $M_{IJ}^{XY}$ entries are polynomials in the angular variable $\omega \equiv \hat{t}/\hat{s}$ defined by
\begin{align}
    M_{VV}^{XY}(\omega) &= (1+2\omega)\delta^{XY}+\omega^2\,, \\[0.3em]
    M_{SS}^{XY}(\omega)  &= 1/4\,, \\[0.3em]
    M_{TT}^{XY}(\omega) &= 4(1+2\omega)^2\delta^{XY}\,, \\[0.3em]
    M_{ST}^{XY}(\omega) &= -(1+2\omega)\delta^{XY} \,, \\[0.3em]
    M_{DD}^{XY}(\omega) &= -\omega(1+\omega)\,.
\end{align}
The quantity $\omega=-(1-\cos\theta_\ell)/2$ is a function of the emission angle $\theta_\ell$ of the lepton $\ell^-$ with respect to the incoming quark in the center-of-mass frame. At the differential level, there is only an interference term between the scalar and tensor structures that vanishes for any observable that is symmetric in $\theta_\ell$ with respect to $\pi/2$, including e.g.~the full cross-section.

%=====================================================
\subsubsection*{Hadronic cross-sections}
%=====================================================
The hadronic cross-section $\sigma$ at a proton-proton collider can be written, following the conventions of Ref.~\cite{Campbell:2006wx}, as the convolution of the partonic cross-section~$\hat{\sigma}(\bar q_i q_j \to \ell_\alpha \bar\ell'_\beta)$ with the PDFs $f_{\bar{q}_i}(x,\mu_F)$ and $f_{q_j}(x,\mu_F)$, summed over all possible incoming quark flavor combinations,
\begin{align}
\sigma (pp \to \ell_\alpha \bar\ell'_\beta) &= \sum_{ij} \int_0^1 \dd x_1 \dd x_2 \squarebrackets{f_{\bar{q}_i}(x_1,\mu)f_{q_j}(x_2,\mu)\,\hat{\sigma}(\bar q_i q_j \to \ell_\alpha \bar\ell'_\beta) + (\bar{q}_i \leftrightarrow q_j)}
 \,,
\end{align}
where $x_{1,2}$ are the fractions of momenta that the scattering quarks carry relative to the momenta of the corresponding protons. We set the factorization and renormalization scales equal to the scale of the hard scattering~$\mu=\sqrt{\hat{s}}$. The hadronic cross-section can be more conveniently expressed as
\begin{align}
\begin{split}
\sigma\brackets{pp \to {\ell}_\alpha \bar{\ell}_\beta^\prime} 
= \sum_{ij} \int \frac{\dd \hat{s}}{s} \,\cL_{ij}\brackets{\hat s} \hat{\sigma}(\bar q_i q_j \to \ell_\alpha \bar\ell'_\beta)  \,,
\end{split}
\end{align}
where $\sqrt{s}=13~\mathrm{TeV}$ is the proton-proton center-of-mass energy for the LHC searches considered in this paper, and where $\cL_{ij}(\hat s)$ are the dimensionless parton-parton luminosity functions \cite{hep-ph/0604120,Campbell:2006wx} defined as 
% %
\begin{align}\label{eq:parton-luminosities}
	\cL_{ij} (\hat s) \equiv \int_{{\hat s}/{s}}^1 \frac{\dd x}{x} \squarebrackets{f_{\bar{q}_i}\brackets{x,\mu} f_{q_j}\brackets{\frac{\hat s}{sx},\mu} + (\bar{q}_i \leftrightarrow q_j)} \, .
\end{align}
In Sec.~\ref{sec:collider-limits}, we will confront the predictions in Drell-Yan production from different BSM models with the LHC run-II measurements in the high-$p_T$ tails of various momentum-dependent distributions. For the neutral Drell-Yan process, we compute the particle-level distribution of the invariant mass~$m_{\ell\ell}$ of the dilepton system in terms of the form-factors introduced in Eqs.~\eqref{eq:dilep-amp}. Combining the previous results we find the expression for the hadronic cross-sections restricted to a specific invariant mass bin $B\equiv[m_{\ell\ell_\mathrm{0}}, m_{\ell\ell_\mathrm{1}}]$ to be given by

\begin{align}\label{eq:master-formula}
 \resizebox{.91\hsize}{!}{   
$\sigma_{B}(pp\to\ell^-_\alpha\ell^+_\beta)=\displaystyle\frac{1}{48\pi v^2}\sum_{XY,\,IJ}\sum_{ij}\,\int_{m^2_{\ell\ell_0}}^{m^2_{\ell\ell_1}}\frac{\text{d}\hat{s}}{s} \int_{-\hat s}^{0} \frac{\text{d}\hat  t}{v^2}\,M^{XY}_{IJ}\cL_{ij} \squarebrackets{\cF_I^{XY,qq}}_{\alpha\beta ij} \squarebrackets{\cF_J^{XY,qq}}_{\alpha\beta ij}^{\ast}\,,$}
\end{align}

\vspace{0.2cm}

\noindent where summing over up- and down-type quarks $q\in \{u,d\}$ is implied. Similarly, for the charged Drell-Yan process we compute the particle-level distribution of the transverse momentum of the charged lepton $p_T(\ell^\pm)$. In this case the cross-section $\sigma_B(pp\to\ell^\pm_\alpha\nu_\beta)$ restricted to a specific high-$p_T$ bin $B\equiv [p_{T_0},p_{T_1}]$ takes the same form as in Eq.~\eqref{eq:master-formula} but with the integration boundaries changed to~\footnote{Notice that for $\hat{s} < 4 p^2_{T_1}$ we find $\hat{t}_2^- = \hat{t}_2^+$, whereas for $\hat{s}<4 p^2_{T_0}$ the cross-section vanishes. Taking the limit $p_{T_0}\to 0$ and $p_{T_1}\to\infty$ yields again the integration boundaries for the full angular integration.}
\begin{align}\label{eq:t-integral_shift}
\int_{m^2_{\ell\ell_0}}^{m^2_{\ell\ell_1}}\! \dd\hat{s}\ \longrightarrow\ \int_{4p^2_{T_{0}}}^{s}\! \dd\hat{s}\ \ \ \ \text{and}  \ \ \ \ \int_{-\hat s}^{0} \dd\hat{t}\ \longrightarrow\ \left(\int_{\hat{t}_0^+}^{\hat{t}_1^+}\! \dd\hat{t}\ + \int^{\hat{t}_0^-}_{\hat{t}_1^-}\! \dd\hat{t}\right)\,,
\end{align}
where
\begin{small}
\begin{align}
\hat{t}_i^\pm (\hat{s}) &= - \frac{\hat{s}}{2} \brackets{1 \pm \sqrt{1- \min \braces{1,\, \frac{4p^2_{T_i}}{\hat{s}}}}}\,.
\label{eq:t-cuts}
\end{align}
\end{small}\newline
For the sake of presentation, we have not explicitly expanded the form-factors in Eqs.~\eqref{eq:dsigma} and~\eqref{eq:master-formula} in terms of the various regular and pole form-factors defined in~\eqref{eq:regular} and~\eqref{eq:poles}. Complete expressions for the hadronic cross-sections in terms of these parameters can be easily extracted for any bin using the \Mathematica package \HighPT.

%%%%%%%%%%%%%%%%%%%%%%%%%%%%%%%%%%%%%%%%%%%%%%%%%%%%%%%%%%%%%%%%%%%%%%%%%%
\subsection{High-$p_T$ Tails}
%%%%%%%%%%%%%%%%%%%%%%%%%%%%%%%%%%%%%%%%%%%%%%%%%%%%%%%%%%%%%%%%%%%%%%%%%%

As mentioned above, the high-energy regime of the dilepton invariant mass or the monolepton transverse mass are known to be very sensitive probes for a variety of New Physics models affecting semileptonic transitions. In the SM, the partonic cross-section scales as $\sim\!1/E^2$ at high-energies, leading to smoothly falling tail for the kinematic distributions of momentum-dependent observables. 
The presence of new semileptonic interactions can impact substantially the shape of these distributions at high energies, either via resonant or non-resonant effects. The most pronounced New Physics effect is the appearance of a resonant feature on top of the smoothly falling SM background, i.e.~a peak in the dilepton invariant mass spectrum, or an edge in the monolepton transverse mass spectrum. This observation would indicate that a heavy colorless particle was produced on-shell in the $s$-channel. Non-resonant effects from contact interactions, or leptoquark states exchanged in the $t/u$-channels, on the other hand, lead to more subtle non-localized features that appear in the tails of the distributions. Indeed, energy-enhanced interactions coming from non-renormalizable operators modify the energy-scaling of the cross-section, leading to an apparent violation of unitarity at high-energies. The effects from leptoquarks exchanged in the $t/u$-channels lead to a similar behavior \cite{Eboli:1987vb,Bessaa:2014jya,Davidson:2014lsa}. After convolution with the quark PDFs, the non-resonant features are more difficult to uncover than a resonance peak, but they are the only potentially observable effect if the collision energy is not sufficient to produce the New Physics particles on-shell.

Finally, we remark that for the quark-lepton fusion process $q_i \bar\ell_\alpha\to q_j\bar\ell_\beta^\prime$, leptoquarks are exchanged in the $s$-channel, leading to a resonance peak in the jet-lepton invariant mass distribution \cite{hep-ph/9406235,Buonocore:2020erb,Greljo:2020tgv,Dreiner:2021ext}, whereas the colorless mediators, now exchanged in the $t/u$-channels, will produce non-resonant effects in the tails of the distributions. The lepton PDFs have been recently computed in Ref.~\cite{Buonocore:2020erb} and could be used to give a robust estimation of the event yields. In this paper we will not provide limits from quark-lepton fusion, focusing only on the Drell-Yan processes. The reason for this is two-fold: (i) given that both processes are related through crossing symmetry, it would be enough to focus on the best performing of the two processes, i.e.~Drell-Yan, with a PDF enhancement of order $\cO(\alpha_s/\alpha)^2$ with respect to quark-lepton fusion; (ii) dedicated LHC resonance searches in jet-lepton final states that can be recasted for quark-lepton fusion are not yet available.~\footnote{\,Several searches for SUSY in $\ell+ nj$ final with $n$ large are available in the literature and could be recasted to give limits on semileptonic New Physics \cite{Dreiner:2021ext}. However, these are far from being optimized and give just a marginal improvement in some regions of parameter space when compared to Drell-Yan.} Note, however, that for leptoquark mediators that couple to valence quarks the resonant enhancement can compensate for the lepton PDF suppression in order to produce competitive limits.

\section{Semileptonic Scattering Beyond the SM}
\label{sec:ff-bsm}

%%%%%%%%%%%%%%%%%%%%%%%%%%%%%%%%%%%%%%%%%%%%%%%%%%%%%%%%%%%%%%%%%%%%%%%%%%
\subsection{SMEFT up to $\cO(1/\Lambda^4)$}
%%%%%%%%%%%%%%%%%%%%%%%%%%%%%%%%%%%%%%%%%%%%%%%%%%%%%%%%%%%%%%%%%%%%%%%%%%

If there is a large separation between the scale of New Physics $\Lambda$ and the electroweak scale~$v$, extending the SM to the SMEFT gives a general description of physical processes in the infrared without having to specify the details of the ultraviolet completion. Below the cutoff, these interactions can then be described in terms of $SU(3)_c\times SU(2)_L\times U(1)_Y$ invariant non-renormalizable operators built out of the SM degrees of freedom. The SMEFT Lagrangian is given by
\begin{equation}
\label{eq:SMEFT}
    \cL_\mathrm{SMEFT}\ =\ \cL_\mathrm{SM}\ +\ \sum_{d,k} \dfrac{\cC_k^{(d)}}{\Lambda^{d-4}}\,\cO^{(d)}_k\ +\ \sum_{d,k} \bigg[\dfrac{\widetilde{\cC}_k^{(d)}}{\Lambda^{d-4}}\,\widetilde{\cO}^{(d)}_k\,+\,\rm{h.c.} \bigg]\,,
\end{equation}
 where the first term corresponds to the SM Lagrangian, $\cO_i^{(d)}$ ($\smash{\widetilde\cO_k^{(d)}}$) denote Hermitian (non-Hermitian) operators of dimension $d>4$, and the ultraviolet physics is encoded in the Wilson coefficients $\smash{\cC^{(d)}_k}$ ($\smash{\widetilde{\cC}^{(d)}_k}$). The complete classification of SMEFT operators for $d=6$ and $d=8$ can be found in Refs.~\cite{Buchmuller:1985jz,Grzadkowski:2010es} and \cite{Li:2020gnx,Murphy:2020rsh}, respectively. In this paper, we consider the \emph{Warsaw} operator basis at $d\!=\!6$ from Ref.~\cite{Grzadkowski:2010es}, as well as its extension to $d\!=\!8$ from Ref.~\cite{Murphy:2020rsh}.  Our conventions for the SMEFT operators are given in Appendix~\ref{sec:conventions}. 

To consistently describe a given scattering cross-section at the LHC up to order $\cO(1/\Lambda^4)$ in the EFT expansion, it is necessary to include not only the contributions from dimension-6 operators, but also the interference terms between $d=8$ operators and the SM contributions since they appear at the same order,
\begin{align}
\label{eq:xsec}\nonumber
    \hat{\sigma}\, \sim\,  \int[\mathrm{d}\Phi]\,\Bigg{\lbrace} |\cA_{\rm SM}|^2&+ \frac{v^2}{\Lambda^2}\sum_i 2\,\mathrm{Re}\Big{(}\cA_i^{(6)}\,\cA_{\rm SM}^\ast\Big{)}\\
    \ &+ \frac{v^4}{\Lambda^4}\bigg{[}\sum_{ij}2\,\mathrm{Re}\Big{(} \cA_i^{(6)}\,\cA_j^{(6)\,\ast} \Big{)} + \sum_{i} 2\,\mathrm{Re}\Big{(}\cA_i^{(8)}\,\cA_{\rm SM}^\ast  \Big{)}\bigg{]}+ \ \dots\Bigg{\rbrace}\,,
\end{align}
\noindent where $[\mathrm{d}\Phi]$ denotes the corresponding Lorentz invariant phase-space measure, $\cA_{\rm SM}$ is the SM amplitude, and  $\cA_i^{(6)}$ and $\cA_i^{(8)}$ stand for the New Physics contributions from dimension-$6$ and dimension-$8$ operators, respectively. The dependence on the scale $\Lambda$ is explicitly factorized in each term to emphasize their order in the EFT expansion.

\begin{table}[b!]
    \centering
    {\renewcommand{\arraystretch}{1.5}
    \begin{tabular}{lc|c c c | c c c c}
        \multicolumn{2}{c|}{Dimension} &     \multicolumn{3}{c|}{$d=6$} & \multicolumn{4}{c}{$d=8$}  
        \\\hline\hline
        \multicolumn{2}{l|}{Operator classes}  &  $\psi^4$ & $\psi^2 H^2 D$ & $\psi^2 X H$ & $\psi^4 D^2$ & $\psi^4 H^2$ & $\psi^2 H^4 D$ & $\psi^2 H^2 D^3$
        \\\hline
        \multicolumn{2}{l|}{Amplitude  scaling}  & $E^2/\Lambda^2$ & $v^2/\Lambda^2$ & $v E/\Lambda^2$ & $E^4/\Lambda^4$ & $v^2 E^2/\Lambda^4$ & $v^4/\Lambda^4$ & $v^2 E^2/\Lambda^4$\\\hline
        \multirow{2}{*}{Parameters} & \# $\mathbb{Re}$  & 456 & 45 & 48 & 168 & 171 & 44 & 52 
        \\
        & \# $\mathbb{Im}$ & 399 & 25 & 48 & 54 & 63 & 12 & 12\\
    \end{tabular}
    }
    \caption{\sl\small Counting of SMEFT parameters relevant to the high-$p_T$ observables and the corresponding energy scaling of the amplitude for each class of operators. The number of real and imaginary free parameters that contribute to the Drell-Yan cross-sections at order $\cO(1/\Lambda^{4})$ are listed for each operator class. In total we find $549~(472)$ real (imaginary) parameters at $d=6$ and an additional $435~(141)$ real (imaginary) parameters at $d=8$, where for the latter we only consider those parameters that affect the interference of these operators with the~SM.}
    \label{tab:parameter_counting}
\end{table}

In this paper, we are interested in the high-energy tails of the momentum-dependent distributions at the LHC. In this regime, only the energy-enhanced terms in Eq.~\eqref{eq:xsec} that are proportional to $E/\Lambda$ will be relevant, where $\smash{E= \sqrt{\hat{s}}}$, while those scaling as powers of $v/\Lambda$ will be sub-dominant. There are three types of operators that directly contribute to the processes $\bar{q}_i q_j \to {\ell}_{\alpha} \bar\ell_{\beta}$  and $\bar{u}_i d_j \to {\ell}_{\alpha}\bar{\nu}_{\beta}$ at tree level 
%which are relevant 
up to order $\cO(1/\Lambda^4)$ in the EFT expansion: 
\begin{itemize}
    \item[$\bullet$] The semileptonic four-fermion operators in the classes {\bf$\psi^4$}, {\bf$\psi^4H^2$} and {\bf$\psi^4D^2$}\,;
    \item[$\bullet$] The Higgs-current operators in the classes {\bf$\psi^2 H^2 D$}, {\bf$\psi^2 H^4 D$} and {\bf$\psi^2 H^2 D^3$}\,;
    \item[$\bullet$] The dipole operators in the class {\bf$\psi^2XH$}\,.
\end{itemize} 

\noindent These operators are defined in Appendix~\ref{sec:SMEFT-ops}, with the $d=6$ ones listed in Table~\ref{tab:dim6_ops}, and the $d=8$ operators in Tables~\ref{tab:dim8_ops_1} and \ref{tab:dim8_ops_2}. The energy scaling of the New Physics amplitude for large~$E$ is shown in Table~\ref{tab:parameter_counting} for each class of operators listed above, which is to be compared to the SM one that becomes constant for~$E\gg v$. 

Up to dimension-6, the semileptonic four-fermion operators {\bf$\psi^4$} give the dominant contributions at large~$E$ since they scale quadratically with the energy ($\propto E^2/\Lambda^2$). In particular, the chirality-conserving semileptonic operators of this type can also interfere with the SM contributions, giving rise to sizable effects. Dipole operators {\bf$\psi^2 X H$} also induce energy-enhanced contributions at the amplitude level ($\propto  vE/\Lambda^2$), but these are suppressed compared to the previous ones since they only increase linearly with $E$ and since they do not interfere with the SM for massless fermions. Moreover, the contributions from Higgs-current operators {\bf$\psi^2 H^2 D$} do not increase with $E$ since they only modify the $W$- and $Z$-couplings, being mostly relevant at the $W$- and $Z$-poles~\cite{Breso-Pla:2021qoe,Almeida:2021asy}. The $d=8$ operators appear in Table~\ref{tab:parameter_counting} with an additional factor of either $v^2/\Lambda^2$ or $E^2/\Lambda^2$ with respect to the $d=6$ contributions described above. Since we are interested in the large $E$ region, we will only keep in our numerical analyses the $d=8$ operators that display an energy enhancement with respect to the SM contributions.

Besides the \emph{direct} contributions to the Drell-Yan cross-sections, there can also be \emph{indirect} contributions arising from the redefinition of the SM inputs by the SMEFT operators. This redefinition induce $\cO(v^2/\Lambda^2)$ shifts to the SM contributions in Eq.~\eqref{eq:SM_photon_pole}--\eqref{eq:SM_W_pole} depending on the chosen scheme for the electroweak parameters~\cite{Brivio:2021yjb}, which we assume to be $\lbrace \alpha_\mathrm{em},\,G_F,\,m_Z \rbrace$. Examples of such operators are the Higgs-current operators~$\smash{\cO_{Hl}^{(3)}}$ or the purely leptonic $\smash{\cO_{ll}}$ which can contribute to the muon decay, for specific flavor indices, inducing a finite renormalization of~$G_F$. Similar redefinitions are also needed in the flavor sector, since the Higgs-current and the semileptonic operators can induce finite shifts of the CKM parameters that are needed to compute the LHC processes~\cite{Descotes-Genon:2018foz}. However, these redefinitions of electroweak and flavor parameters do not lead to energy-enhanced effects at the LHC, thus being negligible in our present analysis.

\begin{figure}
    \centering
    %\hfill\hfill\hfill
    %\hspace{1cm}
    \begin{subfigure}[b]{0.3\textwidth}
    \centering
    ~\includegraphics[width=\textwidth]{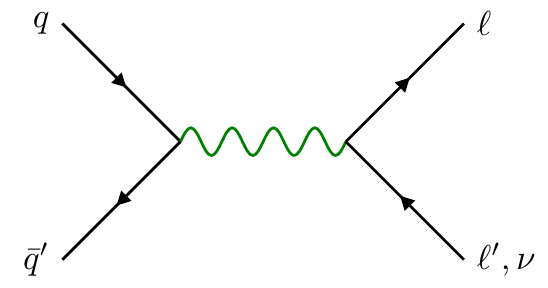}
    \caption{\sl}
    \end{subfigure}
    %\hfill
    \begin{subfigure}[b]{0.3\textwidth}
    \centering
    ~\includegraphics[width=0.8\textwidth]{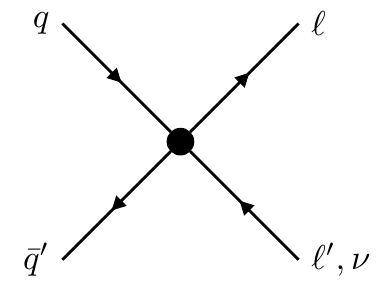}
    \caption{\sl}
    \end{subfigure}
    %\hfill\hfill\hfill
    \\
    \vspace{0.3cm}
    \begin{subfigure}[b]{0.3\textwidth}
    \centering
    ~\includegraphics[width=\textwidth]{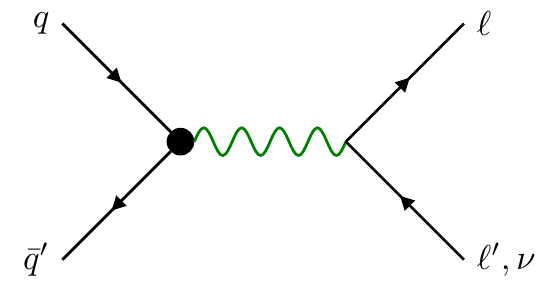}
    \caption{\sl}
    \end{subfigure}
    \begin{subfigure}[b]{0.3\textwidth}
    \centering
    ~\includegraphics[width=\textwidth]{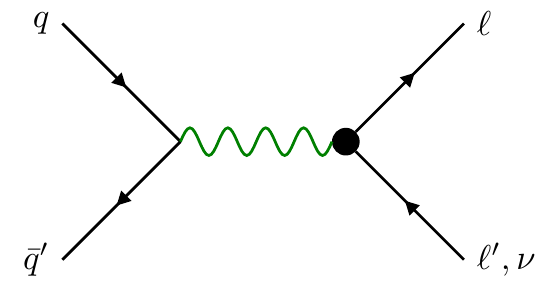}
    \caption{\sl}
    \end{subfigure}
    \begin{subfigure}[b]{0.3\textwidth}
    \centering
    ~\includegraphics[width=\textwidth]{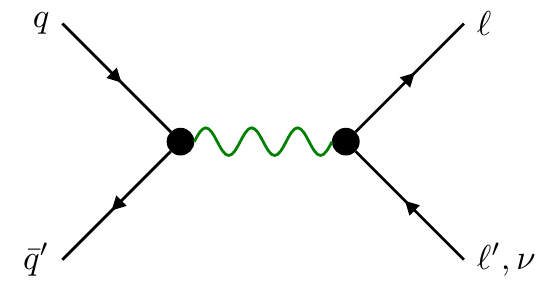}
    \caption{\sl}
    \end{subfigure}
 \caption{\sl\small Feynman diagrams of the leading contributions in the SM (a) and in the SMEFT (b)-(e) to the partonic processes $q \bar{q}^\prime \to \ell\ell^\prime$ and $u \bar{d}^\prime \to \ell\nu$ assuming an EFT expansion up to $\cO(1/\Lambda^{4})$. The green mediator represents the exchange of the SM gauge bosons $V \in \lbrace \gamma,Z,W\rbrace$ and the black vertices are insertions of the SMEFT effective interaction. }
    \label{fig:diagrams} 
\end{figure}

Lastly, we count the number of independent SMEFT parameters at mass dimension-6 and dimension-8 in Table~\ref{tab:parameter_counting}. For this counting it is necessary to separate operators that can contribute to LHC processes including all three quark generations and operators that can contribute only to processes involving the two light quark generation, i.e.~operators involving $SU(2)_L$ singlet up-type quarks~($u$), due to the negligible top quark~PDF. 
We find that there are 549 CP-even and 472 CP-odd parameters that can contribute at $d=6$ to the Drell-Yan processes. There are additional 435 CP-even and 141 CP-odd parameters that can contribute to these processes when $d=8$ operators are considered.~\footnote{Notice that we only count $d=8$~operators that interfere with the SM, and for which we consider a non-diagonal CKM matrix.} 

%=====================================================
\subsubsection*{Form-factors in the SMEFT}
%=====================================================
In the SMEFT the Drell-Yan amplitude can be written as a perturbative expansion in the small parameters $\hat s/\Lambda^2$, $\hat t/\Lambda^2$ and $v^2/\Lambda^2$. This EFT expansion can be matched to Eq.~\eqref{eq:regular} in order to determine the regular form-factor coefficients $\cF_{I\,(n,m)}$. These are given by an infinite perturbative series in the parameter $v^2/\Lambda^2$ of the form
\begin{align}\label{eq:Fnm_expansion}
\cF_{I\,(n,m)}\, &\ = \sum_{ d\,\ge\,2(n+m+3)}^{\infty}\, f_I^{(d)} \left(\frac{v}{\Lambda}\right)^{d-4}\,,
\end{align}
where $f_I^{(d)}$ correspond to linear combinations of $d$-dimensional Wilson coefficients. SMEFT operators of fixed dimension $d$ give rise to only a finite number of form-factor coefficients according to Eq.~\eqref{eq:Fnm_expansion}. For example, $d=6$ operators only contribute to the leading coefficient $\cF_{I\,(0,0)}$, while $d=8$ operators contribute to $\cF_{I\,(0,0)}$, as well as to the next order coefficients $\cF_{I\,(1,0)}$ and $\cF_{I\,(0,1)}$, and so on. In order to express the Drell-Yan form-factors in terms of the SMEFT Wilson coefficients we truncate the power expansion of the regular form-factors $\cF_{I,\,\rm Reg}$ in Eq.~\eqref{eq:regular} to order $n+m \le 1$. The form-factor parametrization given in Sec.~\ref{sec:FF_param} can be further simplified as
% %
\begin{align}
\cF_{S} &=\ \cF_{S\,(0,0)}\,, \label{eq:scalarFF}\\
\cF_{T} &=\ \cF_{T\,(0,0)}\,, \label{eq:tensorFF}\\
\cF_{V} &=\ \cF_{V\,(0,0)} + \cF_{V\,(1,0)}\,\frac{\hat s}{v^2} + \cF_{V\,(0,1)}\,\frac{\hat t}{v^2}+
\sum_{a} \frac{v^2\left[\cS_{(a,\,\rm SM)}+\delta \cS_{(a)}\right]}{\hat s-m^2_a+im_a\Gamma_a}\,, \label{eq:vectorFF}\\
\cF_{D_\ell} &= \sum_{a} \frac{v^2\, \cS_{D_\ell\,(a)}}{\hat s-m^2_a+im_a\Gamma_a}\,, \label{eq:dipole_l_FF}\\
\cF_{D_q} &= \sum_{a} \frac{v^2\, \cS_{D_q\,(a)}}{\hat s-m^2_a+im_a\Gamma_a}\,,\label{eq:dipole_q_FF}
\end{align}
where $a\in\{\gamma,Z\}$ when describing neutral Drell-Yan processes $\bar q_i q_j\to\ell^-_\alpha\ell^+_\beta$, and $a\in\{W^\pm\}$ when describing the charged Drell-Yan processes $\bar d_i u_j\to\ell^-_\alpha\bar\nu_\beta$ ($\bar u_i d_j\to\ell^+_\alpha\nu_\beta$), and $\cS_{(a,{\rm SM})}$ are defined in Eqs.~\eqref{eq:SM_photon_pole}--\eqref{eq:SM_W_pole}. This parametrization is enough to capture all possible effects to order $\cO(1/\Lambda^4)$ in semileptonic transitions.~\footnote{Note that the regular pieces of the scalar and tensor form-factors were truncated to order $n=m=0$ because when squaring the amplitude, the $d=8$ terms with $n+m=1$ do not interfere with the SM. For the dipoles we can set $\cF_{D_\ell,\,\rm Reg}$ and $\cF_{D_q,\,\rm Reg}$ to zero since these arise non-locally via SM gauge boson exchange. For the pole form-factors $\cF_{I,\,\rm Poles}$, we only need to consider the vector poles and dipoles arising from the $s$-channel exchange of the SM gauge bosons.} 

Given that the scalar and tensor form-factors are independent of $\hat s$ and $\hat t$, the coefficients $\cF_{S\,(0,0)}$ and $\cF_{T\, (0,0)}$ map directly to the Wilson coefficients of the $d=6$ scalar and tensor operators in the class {\bf$\psi^4$}, as shown in Appendix~\ref{sec:SMEFT_match_scalar_tensor}. The dipole residues $\cS_{D\,(a)}$ also match trivially to the $d=6$ SMEFT dipole operators in the class {\bf$\psi^2XH$}, as shown in Appendix~\ref{sec:SMEFT_match_dipole}. 

The regular coefficients and the pole residues of the vector form-factors, on the other hand, are generated at $d=6$ and $d=8$. The leading coefficient $\cF_{V\,(0,0)}$ receives contributions from contact operators in the classes {\bf$\psi^4$} and {\bf$\psi^4H^2$} at $d=6$ and $d=8$, respectively, as well as from modified interactions between fermions and the SM gauge bosons from $d=8$ operators in class {\bf $\psi^2H^2D^3$}. The higher-order coefficients $\cF_{V\,(1,0)}$ and $\cF_{V\,(0,1)}$ receive contributions from the $d=8$ operators in class {\bf$\psi^4D^2$}. The pole residues $\delta \cS_{(a)}$ receive contributions from modified fermion interactions due to dimension-$6$ operators in class {\bf$\psi^2H^2D$} and from dimension-$8$ operators in class {\bf$\psi^2H^2D^3$}. Schematically, the matching between SMEFT Wilson coefficients and the form-factors takes the following form: 
\begin{align}
\cF_{V\,(0,0)} \ &=\ \frac{v^2}{\Lambda^2}\,\cC^{\,(6)}_{\psi^4}\ +\ \frac{v^4}{\Lambda^4}\,\cC^{\,(8)}_{\psi^4H^2}\ +\ \frac{v^2 m_a^2}{\Lambda^4}\,\cC^{\,(8)}_{\psi^2H^2D^3}\ + \ \cdots \,, \label{eq:FV00}\\
\cF_{V\,(1,0)} \ &=\ \frac{v^4}{\Lambda^4}\,\cC^{\,(8)}_{\psi^4D^2}\ + \ \cdots \,, \label{eq:FV10}\\
\cF_{V\,(0,1)} \ &=\ \frac{v^4}{\Lambda^4}\,\cC^{\,(8)}_{\psi^4D^2}\ + \ \cdots \,, \label{eq:FV01}\\
\delta \cS_{(a)} \ &=\ \frac{m_a^2}{\Lambda^2}\,\cC^{\,(6)}_{\psi^2H^2D}\,+\frac{v^2m_a^2}{\Lambda^4}\,\left(\left[\cC_{\psi^2H^2D}^{\,(6)}\right]^2+\cC_{\psi^2H^4D}^{\,(8)}\right)\  +\ \frac{m_a^4}{\Lambda^4}\,\cC_{\psi^2H^2D^3}^{\,(8)}\ + \ \cdots\,, \label{eq:delF}
\end{align}
where the squared term $[\cC_{\psi^2H^2D}^{\,(6)}]^2$ in Eq.~\eqref{eq:delF} corresponds to double vertex insertions of the corresponding dimension-6 operators, as depicted in diagram (e) of Fig.~\ref{fig:diagrams}. The ellipsis
%ellipses $\cdots$ 
indicate contributions from the neglected 
%dimension-$10$ 
higher-dimensional operators. The precise matching of the SMEFT to the vector form-factors can be found in Appendix~\ref{sec:SMEFT_match_vector}. 

Notice that the operators in the class {\bf$\psi^2H^2D^3$} contribute to $\smash{\cF_{V\,(0,0)}}$ and $\smash{\delta \cS_{(a)}}$ simultaneously. This can be understood by analyzing one of the operators in this class. As an example we take $\smash{\cO_{q^2H^2D^3}^{(1)}} = \smash{(\bar q_i\gamma^\mu D^\nu q_j)D_{(\mu} D_{\nu)}H^\dagger H}$ which, after spontaneous symmetry breaking, gives rise to a modified coupling between the $Z$ boson and quarks that is proportional to $(\hat s\, m_Zv/\Lambda^4) \, Z_\mu(\bar q_i\gamma^\mu q_j)$. This interaction contributes to neutral Drell-Yan production with an amplitude that scales as $\mathcal{A}(\bar{q}_i q_j \to \ell^-_\alpha \ell^+_\beta)\propto \hat s/(\hat s - m_Z^2)$. This amplitude can be brought to the form in Eq.~\eqref{eq:vectorFF} by using the partial fraction decomposition in Eq.~\eqref{eq:partial_frac_decomp}, which in diagrammatic form reads:
\begin{center}
\vspace{0.3cm}
\begin{tikzpicture}[scale=1.0,>=stealth,thick]
  \node[wc=6pt, label=above:$\hat s$] (v1) at (-0.7,0) {};
  \coordinate (v2) at (0.7,0);
  \draw[] (v1) +(-0.7,0.7) -- (v1);
  \draw[] (v1) -- +(-0.7,-0.7);
  \draw[vector] (v1) -- (v2);
  \node[] at (0,-0.4) {$Z$}; 
  \draw[] (v2) -- +(0.7,0.7);
  \draw[] (v2) +(0.7,-0.7) -- (v2);
  \node[scale=1.2] at (2.4,0) {$=$};
  \begin{scope}[xshift=4cm]
    \node[wc=6pt] (v1) at (0,0) {};
    \draw[] (v1) +(-0.7,0.7) -- (v1);
    \draw[] (v1) -- +(-0.7,-0.7);
    \draw[] (v1) -- +(0.7,0.7);
    \draw[] (v1) +(0.7,-0.7) -- (v1);
  \end{scope}
  \node[scale=1.2] at (5.7,0) {$+$};
  \begin{scope}[xshift=8cm]
    \node[wc=6pt, label=above:$\ m_Z^2$] (v1) at (-0.7,0) {};
    \coordinate (v2) at (0.7,0);
    \draw[] (v1) +(-0.7,0.7) -- (v1);
    \draw[] (v1) -- +(-0.7,-0.7);
    \draw[vector] (v1) -- (v2);
    \node[] at (0,-0.4) {$Z$}; 
    \draw[] (v2) -- +(0.7,0.7);
    \draw[] (v2) +(0.7,-0.7) -- (v2);
  \end{scope}
\end{tikzpicture}
\vspace{0.3cm}
\end{center}
The first contact diagram appearing above on the right-hand side of the equality corresponds to the last term in \eqref{eq:FV00}, while the second diagram corresponds to the last term in \eqref{eq:delF}.

%%%%%%%%%%%%%%%%%%%%%%%%%%%%%%%%%%%%%%%%%%%%%%%%%%%%%%%%%%%%%%%%%%%%%%%%%%
\subsection{Concrete UV Mediators}
%%%%%%%%%%%%%%%%%%%%%%%%%%%%%%%%%%%%%%%%%%%%%%%%%%%%%%%%%%%%%%%%%%%%%%%%%%

%%%%%%%%%%%%%%%%
\begin{table}[t!]

  \begin{center}
  {\renewcommand{\arraystretch}{2}
  \resizebox{1\textwidth}{!}{
    \begin{tabular}{@{\hspace{1em}}c@{\hspace{1em}}c@{\hspace{1em}}c@{\hspace{1em}}l@{\hspace{1em}}}
    %\hline\hline
    \ &   SM rep.  & Spin  & \hspace{3cm} $\cL_{\rm int}$\\
\hline\hline
$Z^\prime$ & \SMrep{1}{1}{0} & 1 & $\cL_{Z^\prime}=\sum_\psi\, [g^{\psi}_1]_{ab} \, \bar \psi_a \slashed Z^\prime \psi_b$\, , \  $\psi\in\{u,d,e,q,l\}$\\
$\widetilde Z$ & \SMrep{1}{1}{1} & 1 & $\cL_{\widetilde Z}=[\widetilde g^{q}_1]_{ij}\, \bar u_i\slashed {\widetilde Z}d_j+[\widetilde g^{\ell}_1]_{\alpha\beta}\, \bar e_\alpha\slashed {\widetilde Z} N_\beta$\\
$\Phi_{1,2}$ & \SMrep{1}{2}{1/2}& 0 &$\cL_{\Phi}=\displaystyle\sum_{a=1,2}\Big{\lbrace} [y_{u}^{(a)}]_{ij}\, \bar q_i u_j\widetilde H_a+[y_{d}^{(a)}]_{ij}\, \bar q_i d_j H_a+[y_{e}^{(a)}]_{\alpha\beta}\, \bar l_\alpha e_\beta H_a\Big{\rbrace}+\mathrm{h.c.}$\\
$W^\prime$ & \SMrep{1}{3}{0} & 1 & $\cL_{W^\prime}=[g^{q}_3]_{ij} \, \bar q_i  (\tau^I \,\slashed W^{\prime\, I}) q_j+[g^{l}_3]_{\alpha\beta} \, \bar l_\alpha (\tau^I \,\slashed W^{\prime\, I}) l_\beta$ \\
\hline
$S_1$ & $\SMrepbar{3}{1}{1/3}$  & 0 & $\cL_{S_1}= [y_1^L]_{i\alpha}\, S_1 \bar  q^c_i\epsilon l_\alpha+[y_1^R]_{i\alpha}\,S_1\bar u^c_i e_\alpha + {\  [\bar y_1^R]_{i\alpha}\,S_1\bar d^c_i N_\alpha}+\mathrm{h.c.}$\\ 
$\widetilde S_1$ & $\SMrepbar{3}{1}{4/3}$ & 0 & $\cL_{\widetilde S_1}=[\widetilde y_1^R]_{i\alpha}\,  \widetilde S_1\bar d^c_i e_\alpha+\mathrm{h.c.}$\\ 
$U_1$ & $\SMrep{3}{1}{2/3}$ & 1 & $\cL_{U_1}=[x_1^L]_{i\alpha} \, \bar q_i \slashed  U_{\!1} l_\alpha + [x_1^R]_{i\alpha} \,\bar d_i \slashed U_{\!1} e_\alpha + { [\bar x_1^R]_{i\alpha} \,\bar u_i \slashed U_{\!1} N_\alpha}+\mathrm{h.c.}$\\ 
$\widetilde U_1$ & $$\SMrep{3}{1}{5/3}$$ & 1 & $\cL_{\widetilde U_1}= [\widetilde x_1^R]_{i\alpha} \,\bar u_i \slashed {\widetilde U}_{\!1} e_\alpha+\mathrm{h.c.}$\\ 
$R_2$ & \SMrep{3}{2}{7/6} & 0  & $\cL_{R_2}= -[y_2^L]_{i\alpha} \, \bar u_i R_2 \epsilon l_\alpha+[y_2^R]_{i\alpha} \, \bar q_i e_\alpha R_2+\mathrm{h.c.}$ \\ 
$\widetilde R_2$ & \SMrep{3}{2}{1/6} & 0 & $\cL_{\widetilde R_2}=- [\widetilde y_2^L]_{i\alpha} \, \bar d_i \widetilde R_2 \epsilon l_\alpha+ { [\widetilde y_2^R]_{i\alpha} \, \bar q_i N_\alpha \widetilde R_2} +\mathrm{h.c.}$ \\ 
$V_2$ & $\SMrepbar{3}{2}{5/6}$ & 1 & $\cL_{V_{\!2}}=[x_2^L]_{i\alpha} \,\bar d_i^c \slashed V_{\!\!2}\epsilon l_\alpha+[x_2^R]_{i\alpha}\, \bar q_i^c\epsilon \slashed V_{\!\!2} e_\alpha +\mathrm{h.c.}$\\ 
$\widetilde V_2$ & $\SMrepbar{3}{2}{-1/6}$  & 1 & $\cL_{\widetilde V_{\!2}}=[\widetilde x_2^L]_{i\alpha} \,\bar u_i^c \slashed {\widetilde V}_{\!\!2}\epsilon l_\alpha+ { [\widetilde x_2^R]_{i\alpha} \,\bar q_i^c \epsilon \slashed {\widetilde V}_{\!\!2} N_\alpha} +\mathrm{h.c.}$\\ 
$S_3$ & $\SMrepbar{3}{3}{1/3}$ & 0 & $\cL_{S_3}=[y_3^L]_{i\alpha} \, \bar q^c_i \epsilon (\tau^I \, S_3^I) l_{\alpha}+\mathrm{h.c.}$\\ 
$ U_3$ & $\SMrep{3}{3}{2/3}$ & 1 & $\cL_{U_3}=[x_3^L]_{i\alpha} \, \bar q_i (\tau^I \, \slashed {U}_{\!3}^I )l_\alpha+\mathrm{h.c.}$ \\
\hline\hline
\end{tabular}
}}
\caption{\sl\small Possible bosonic mediators contributing at tree level to Drell-Yan production classified by their SM quantum numbers and spin. In the last column, we provide the interaction Lagrangian where $\epsilon\equiv i\tau_2$, $\psi^c\equiv i\gamma_2\gamma_0\bar\psi^T$ and $\widetilde H=i\tau_2 H^\ast$ is the conjugate Higgs doublet, where $\tau_i$ ($i=1,2,3$) denote the Pauli matrices. The right-handed fermion fields are defined as $u\equiv u_R$, $d\equiv d_R$, $e\equiv \ell_R$ and $N \equiv \nu_R$, and the left-handed fermion fields as $q\equiv (V^\dagger u_L,d_L)^T$ and $l\equiv (\nu_L,\ell_L)^T$. We adopt the notation from Ref.~\cite{Buchmuller:1986zs,Dorsner:2016wpm} for the leptoquark states. }
\label{tab:mediators}
\end{center}
\end{table}
%%%%%%%%%%%%%%%%

In this Section, we discuss the effects of new bosonic states mediating Drell-Yan processes at tree level. These states can be classified in terms of their spin and SM quantum numbers, namely $(SU(3)_c,\, SU(2)_L,\, U(1)_Y)$ with $Q=Y+T_3$. The possible semileptonic mediators are displayed in Table~\ref{tab:mediators}, where we also provide the relevant interaction Lagrangians with generic couplings in the last column. For completeness, we also allow for three right-handed neutrinos, denoted as $N_\alpha\sim({\bf 1}, {\bf 1},0)$, with $\alpha=1,2,3$.~\footnote{For simplicity we assume exactly three right-handed neutrinos, but this need not be the case.} Furthermore, we assume that the masses of these SM singlets are negligible compared to the collider energies and, if produced, they can escape detection as missing energy. 

The possible mediators fall into two broad categories, each with different collider phenomenology: (i) color-singlets exchanged in the $s$-channel, and (ii) leptoquarks, i.e.~color-triplets, exchanged in the $t/u$-channels. If the masses of these states are at the $\cO(\mathrm{TeV})$ scale their propagators will contribute to the residues $\cS_{I\,(a)}$, $\cT_{I\,(b)}$, $\cU_{I\,(c)}$ of the pole form-factors~\eqref{eq:poles}. Leptoquarks can be further classified using fermion number \cite{Dorsner:2016wpm}, defined as $F\equiv3B+L$ where $B$ ($L$) stands for Baryon (Lepton) number. For Drell-Yan production, the leptoquarks with fermion number $F=0$, such as $U_1$, $\widetilde U_1$, $R_2$, $\widetilde R_2$ and $U_3$, are exchanged in the $t$-channel, while the remaining leptoquarks $S_1$, $\widetilde S_1$, $V_2$, $\widetilde V_2$ and $S_3$, carrying fermion number $F=-2$, are exchanged in the $u$-channel. Note that certain leptoquark representations can also couple to diquark bilinears (not listed in Table~\ref{tab:mediators}) that pose a potential threat to the stability of the proton~\cite{Assad:2017iib}, unless a stabilizing symmetry is further introduced, see e.g.~Ref.~\cite{2202.05275}.  

\begin{figure}[t!]
    \centering
    \includegraphics[width=0.27\textwidth]{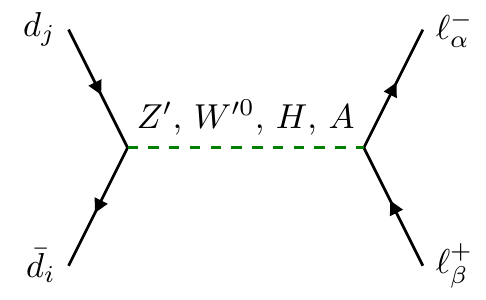}~\quad
    \includegraphics[width=0.27\textwidth]{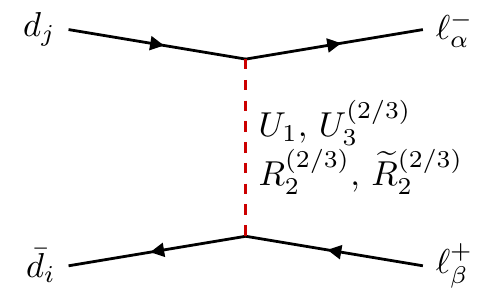}~\quad
    \includegraphics[width=0.27\textwidth]{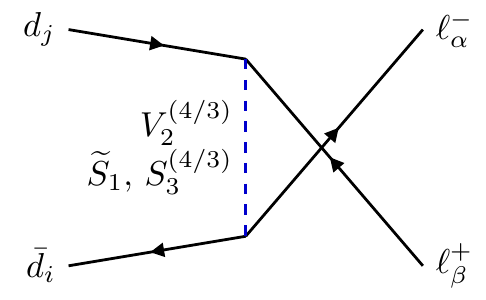}\\[0.7em]
    \includegraphics[width=0.27\textwidth]{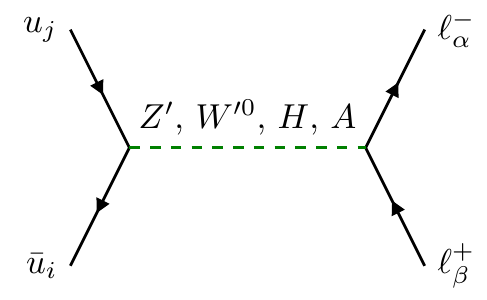}~\quad
    \includegraphics[width=0.27\textwidth]{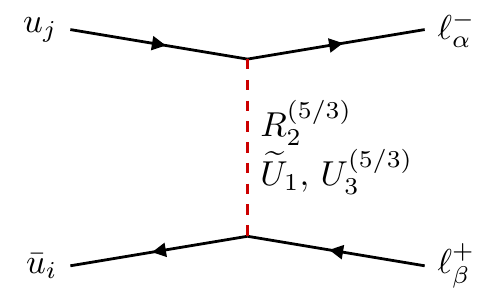}~\quad
     \includegraphics[width=0.27\textwidth]{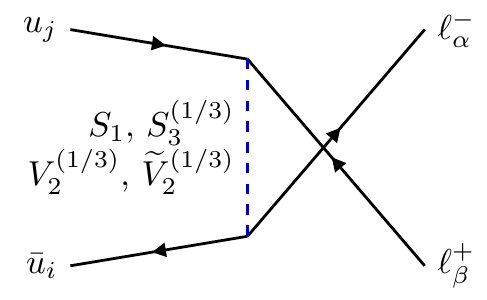}\\[0.7em]
    \includegraphics[width=0.27\textwidth]{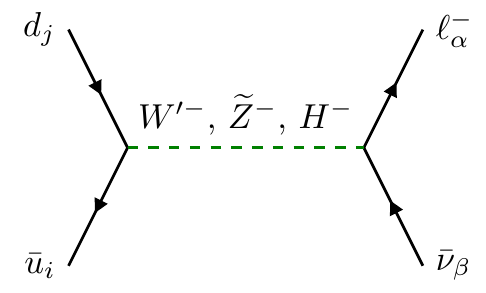}~\quad
    \includegraphics[width=0.27\textwidth]{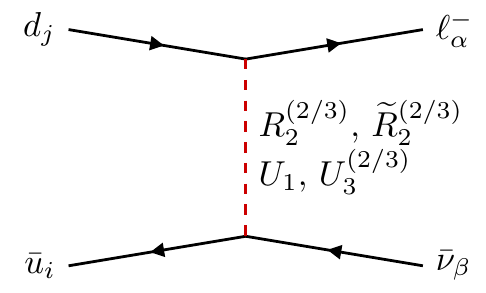}~\quad
    \includegraphics[width=0.27\textwidth]{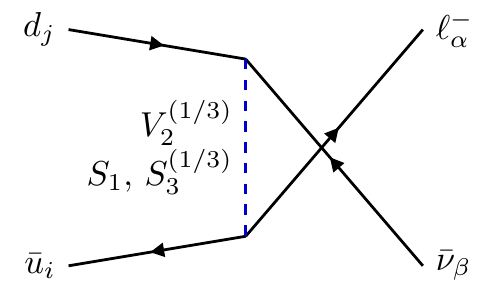}  
 \caption{\sl\small  Contributions to dilepton transitions $\bar d d\to \ell^-\ell^+$ (upper row), $\bar u u\to \ell^-\ell^+$ (middle row) and $\bar u d\to \ell^\pm\nu$ (lower row), from the tree-level exchange of the BSM mediators displayed in Table~\ref{tab:mediators} via the $s$-channel (left column), $t$-channel (middle column) and $u$-channel (right column).
 }
    \label{fig:mediator_diagrams}
\end{figure}

After electroweak symmetry breaking, each of the non-trivial $SU(2)_L$ multiplets decompose into physical eigenstates, 

\begin{align}
    W^{\prime\,} &= \frac{1}{\sqrt{2}}\begin{pmatrix}  W^{\prime0} & \sqrt{2}\,W^{\prime +}\\ \sqrt{2}\,W^{\prime -} & - W^{\prime0} \end{pmatrix}\,,\\[1.4em]
    R_2 &= \begin{pmatrix} R_2^{\,(+5/3)} \\ R_2^{\,(+2/3)}\end{pmatrix}\,,\ \ \ \widetilde{R}_2 =\begin{pmatrix} \widetilde R_2^{\,(+2/3)} \\ \widetilde R_2^{\,(-1/3)}\end{pmatrix}\,,\ \ \  V_2 = \begin{pmatrix} V_2^{\,(+4/3)} \\ V_2^{\,(+1/3)}\end{pmatrix}\,, \ \ \ \widetilde{V}_2 =\begin{pmatrix} \widetilde V_2^{\,(+1/3)} \\ \widetilde V_2^{\,(-2/3)}\end{pmatrix}\,, \\[1.4em]
    S_3 &= \frac{1}{\sqrt{2}} \begin{pmatrix} S_3^{\,(-1/3)} & \sqrt{2}\,S_3^{\,(+2/3)}\\ \sqrt{2}\,S_3^{\,(-4/3)} & - S_3^{\,(-1/3)} \end{pmatrix}\,, \ \ \
    U_3 =\frac{1}{\sqrt{2}}\begin{pmatrix}  U_3^{\,(+2/3)} & \sqrt{2}\,U_3^{\,(+5/3)}\\ \sqrt{2}\,U_3^{\,(-1/3)} & - U_3^{\,(+2/3)} \end{pmatrix}\,,
\end{align}

\noindent where the superscripts denote the electric charge of each component. The case of an additional Higgs doublet is to be treated separately as both doublets $H_{1,2}$ can acquire a vacuum expectation value, with the identification of the SM-like Higgs boson ($h$) depending on the parameters of the scalar potential. In this case, it is also fundamental to devise a mechanism that will suppress scalar mediated FCNC to make these models phenomenologically viable~\cite{Branco:2011iw}, such as a $\mathbb{Z}_2$ symmetry~\cite{Glashow:1976nt}, or assumptions regarding the alignment of the Yukawa and fermion-mass matrices~\cite{Pich:2009sp,Eberhardt:2020dat}. Here, we consider a parametrization of the Lagrangian in terms of the mass eigenstates,
%%%%%%%%%%%%%
\begin{align}
\mathcal{L}_\mathrm{\Phi} \supset &- \sum_{f=u,d,\ell} \sum_{i,j}  \bigg{\lbrace} \big{[}\xi_h^f\big{]}_{ij}\, \bar{f}_i f_j \, h  +\big{[}\xi_H^f\big{]}_{ij}\, \bar{f}_i f_j\,H - i \big{[}\xi_A^f\big{]}_{ij}\, \bar{f}_i \gamma_5 f_j\, A  \bigg{\rbrace} \\[0.3em]
&-\sum_{X=L,R}\sum_{i,j}\bigg{\lbrace} \Big{(} \big{[}\xi_{H^+}^{q_X}\big{]}_{ij} \bar{u}_{i} \mathbb{P}_X d_{j} + \big{[}\xi_{H^+}^{\ell_R}\big{]}_{ij} \bar{\nu}_{i} \mathbb{P}_R \ell_{j} + \big{[}\xi_{H^+}^{\ell_L}\big{]}_{ij} \bar{N}_{i} \mathbb{P}_L \ell_{j}\Big{)}H^+   +\mathrm{h.c.}\bigg{\rbrace}\,,\nonumber
\end{align}
%%%%%%%%%%%%%
with flavor indices $i,j$ and generic couplings $\xi_{\varphi}^f$ that can be easily matched to the models of interest~\cite{Branco:2011iw}. In the above equation, $H^\pm$ denotes the charged scalar, $h$ is the SM-like Higgs, $A$ the neutral CP-odd scalar and $H$ the heavy neutral CP-even scalar.  The contributions of each of these mediators to  Drell-Yan production can be found in the Feynman diagrams depicted in Fig.~\ref{fig:mediator_diagrams}. 
%

%=====================================================
\subsubsection*{Form-factors in concrete BSM models}
%=====================================================
The complete matching of the pole form-factors to concrete models is given in Appendix~\ref{sec:FF_concrete_UVmodel}, where the flavor structure of the residues of the pole form-factors take the following form:
 \begin{align}\label{eq:concrete_poles}
    [\cS_{I\,(a)}]_{\alpha\beta ij}\ &=\  [g^*_a]_{\alpha\beta}\,[g^*_a]_{ij}\,,\\[0.3em]
    [\cT_{I\,(b)}]_{\alpha\beta ij}\ &=\  [g^*_b]_{i\alpha}\, [g^*_b]_{j\beta}\,,\\[0.3em]
    [\cU_{I\,(c)}]_{\alpha\beta ij}\ &=\  [g^*_c]_{i\beta}\, [g^*_c]_{j\alpha} \,,
\end{align} 
\noindent for $I\in\{V,S,T\}$, where $g^*_{a,b,c}$ denote generic couplings of the mediators to fermions of a given chirality and each index $a,b,c$ labels the possible mediators contributing to the $s$-, $t$- and $u$-channels, respectively, as displayed in Fig.~\ref{fig:mediator_diagrams}. 

%=====================================================
\subsection{Other BSM scenarios}
\label{sec:BSMEFT}
%=====================================================
We have already discussed Drell-Yan production both in the SMEFT and in simplified models. These two cases cover the most common tree-level BSM scenarios but leave out some potentially interesting possibilities. For instance, one can extend the SM matter content with light right-handed fermion singlets  $N_{\alpha}\sim({\bf 1}, {\bf 1},0)$ and build an effective field theory known as $\nu$SMEFT~\cite{0806.0876,2001.07732}. At $d=6$, the resulting semileptonic four-fermion operators~\cite{1807.04753} are given by,
\begin{align}\label{eq:Ops_N_1}
\cO_{eNud}&=(\bar e \gamma^\mu N)(\bar u\gamma_\mu d)\,,\\[0.3em]
\cO_{l Nuq}&=(\bar l N)(\bar u q)\,,\\[0.3em]
\cO^{(1)}_{l Nqd}&=(\bar l N)\epsilon (\bar qd)\,,\\[0.3em]
\cO^{(3)}_{l N qd}&=(\bar l\sigma^{\mu\nu} N)\epsilon (\bar q\sigma_{\mu\nu} d)\,,\label{eq:Ops_N_2}
\end{align}
where flavor indices are omitted. These new local interactions will only contribute to the charged current Drell-Yan process $pp\to \ell^\pm_\alpha N_\beta$ without interfering with the SM. Higher order operators at dimension-$7$ or above can be dropped since these only contribute to the production cross-section starting at order $\cO(1/\Lambda^6)$. The mapping of the Wilson coefficients to the leading regular form-factors are provided in Appendix~\ref{sec:nuSMEFT}. Notice that the operators in Eqs.~\eqref{eq:Ops_N_1}-\eqref{eq:Ops_N_2} can be generated at tree level by integrating out at any of the heavy bosonic mediators coupling to $N_\alpha$ in Table~\ref{tab:mediators}.

%%%%%%%%%%%%%%%%%%%%%%%%%%%%%%%%%%%%%%%%%%%%%%%%%%%%%%%%%%%%%%%%%%%%%%%%%%
\section{Collider Limits}
\label{sec:collider-limits}
%%%%%%%%%%%%%%%%%%%%%%%%%%%%%%%%%%%%%%%%%%%%%%%%%%%%%%%%%%%%%%%%%%%%%%%%%%
 The detectors at hadron colliders are complex and imperfect environments with a finite experimental resolution and limited acceptance. When dealing with dilepton and monolepton searches at the LHC, differential distributions are measured from the reconstructed four-momenta of high-level objects such as isolated leptons, $\tau$-tagged jets, and missing transverse energy $\slashed{E}_{T}$. These objects are meant to approximate the underlying final-state leptons produced in the hard scattering. The theoretical predictions, on the other hand, are typically computed from the experimentally inaccessible final-state leptons. This mismatch between the predicted distribution of a particle-level observable $x$  and the measured distribution of the corresponding observable $x_{\rm obs}$ is described by the convolution
\begin{align}\label{eq:folding-kernel}
    \frac{\rm{d}\sigma}{{\rm d}x_{\rm obs}}\ =\ \int \mathrm{d}x\
K(x_{\rm obs}|x)\,  \frac{\rm{d}\sigma}{{\rm d}x} \,,
\end{align}
where $K(x_{\rm obs}|x)$ is a {\it kernel function} that parametrizes the detector response \cite{Blobel:1984ku}.~\footnote{For example, the particle-level observable relevant for resonance searches in ditau production at the LHC \cite{ATLAS:2020zms} is the invariant mass $x=m_{\tau\tau}$ of the ditau system. Given that $\tau$-leptons always decay into neutrinos, a precise experimental reconstruction of $m_{\tau\tau}$ is challenging. Therefore, what is actually measured is the quantity $x_{\rm obs}=m_T^{\rm tot}$ known as the {\it total transverse mass}, which serves as a proxy for $m_{\tau\tau}$. This observable is computed from the two leading $\tau$-tagged jets coming from the visible part of the hadronic decay of each underlying tau-lepton ($\tau_{h}$) and the missing transverse energy of the event which accounts for the undetected neutrinos.} In practice, for a given LHC search, both the measured and the particle-level distributions are binned into histograms leading to the discretization of Eq.~\eqref{eq:folding-kernel}. For a binning $\cA$ of $x_{\rm obs}$ and $\cB$ of $x$, the expected number of signal events $\cN_A$ in a bin $A\in\cA$ is given by

\begin{align}\label{eq:binned-folding}
    \cN_A\ =\ \sum_{B\,\in\,\cB}\,\cL_{\rm int} \cdot K_{AB}\cdot\sigma_B \,,
\end{align}
where $\cL_{\rm int}$ is the integrated luminosity used in the search, $\sigma_B$ is the particle-level cross-section restricted to a bin $B\in\cB$, and $K$ is a $N\!\times\!M$ {\em response matrix}, where $N$ and $M$ are the number of bins in $\cA$ and $\cB$, respectively. The response matrix represents the probability that an event produced in a bin $B$ of $x$ passes all event selections of the search and is measured in bin~$A$ of~$x_{\rm obs}$. 
When estimating the event yields $\cN_A$ of a BSM signal, each independent term contributing to the computation of the cross-section $\sigma_{B}$ (e.g. see Eq.~\eqref{eq:master-formula}) needs to be convoluted with a different $K_{AB}$ matrix since each term can respond differently to the selection cuts and the detector. Therefore, in full generality, the response matrices entering Eq.~\eqref{eq:binned-folding} are quantities depending on the chiralities $\lbrace X,Y\rbrace$ and flavors $\lbrace\alpha,\beta,i,j\rbrace$ of the external leptons and quarks, as well as the shape of the New Physics, i.e.~the regular and pole form-factors that are involved. It is clearly not possible to compute the entries of each response matrix from first principles. These must be estimated numerically for each LHC search using Monte Carlo event generators and detector simulators. 

%=====================================================
\subsection{Dilepton and monolepton searches at the LHC}
\label{sec:recasts}
%=====================================================

%%%%%%%%%%%%%%%%%%%%%
\begin{table}[t!]
    \renewcommand{\arraystretch}{1.3}
    \centering
    \begin{tabular}{l  c c c c c}
        Process  & Experiment & Luminosity & Ref. & $x_{\rm obs}$ & $x$
        \\\hline\hline
        $pp \to \tau\tau$ & ATLAS & $139\,\mathrm{fb}^{-1}$ & \cite{ATLAS:2020zms} & $m_T^{\rm tot}(\tau_h^1,\tau_h^2,\slashed{E}_T)$ & $m_{\tau\tau}$
        \\
        $pp \to \mu\mu$  & CMS & $140\,\mathrm{fb}^{-1}$ & \cite{CMS:2021ctt} & $m_{\mu\mu}$ & $m_{\mu\mu}$
        \\
        $pp \to ee$  & CMS & $137\,\mathrm{fb}^{-1}$ & \cite{CMS:2021ctt} & $m_{ee}$ & $m_{ee}$
        \\\hline
        $pp \to \tau\nu$ & ATLAS & $139\,\mathrm{fb}^{-1}$ & \cite{ATLAS:2021bjk} & $m_T(\tau_h,\slashed{E}_T)$ & $p_T(\tau)$
        \\
        $pp \to \mu\nu$  & ATLAS & $139\,\mathrm{fb}^{-1}$ & \cite{ATLAS:2019lsy} & $m_T(\mu,\slashed{E}_T)$ & $p_T(\mu)$
        \\
        $pp \to e\nu$  & ATLAS & $139\,\mathrm{fb}^{-1}$ & \cite{ATLAS:2019lsy} & $m_T(e,\slashed{E}_T)$ & $p_T(e)$
        \\\hline
        $pp \to \tau \mu$  & CMS & $138\,\mathrm{fb}^{-1}$ & \cite{CMS:2022fsw} & $m_{\tau_h\mu}^{\rm col}$ & $m_{\tau\mu}$
        \\
        $pp \to \tau e$  & CMS & $138\,\mathrm{fb}^{-1}$ & \cite{CMS:2022fsw} & $m_{\tau_he}^{\rm col}$ & $m_{\tau e}$
        \\
        $pp \to \mu e$ & CMS & $138\,\mathrm{fb}^{-1}$ & \cite{CMS:2022fsw} & $m_{\mu e}$ & $m_{\mu e}$\\
        \hline\hline
    \end{tabular}
    \caption{\sl\small Experimental searches by the ATLAS and CMS collaborations that have been recast in the \HighPT package~\cite{<HighPT:_A_Tool_for_High-pT_Collider_Studies_Beyond_the_Standard_Model><Allwicher;Lukas><Faroughy;Darius><Jaffredo;Florentin><Sumensari;Olcyr><Wilsch;Felix><2022>}. The last two columns refer to the measured and particle-level observables considered in our analyses. The data used in the present work corresponds to $x_\mathrm{obs} \geq 200\,\text{GeV}$, for all observables.}
    \label{tab:lhc-searches}
\end{table}
%%%%%%%%%%%%%%%%%%%%% 

The experimental searches considered in our analysis are collected in Table~\ref{tab:lhc-searches}. These correspond to data sets from the full run-II ATLAS and CMS searches for heavy resonances in dilepton and monolepton production at the LHC. In the last two columns we display the tail observables that are measured in each search~($x_{\rm obs}$) which serve as proxies for the particle-level observables~($x$) used to compute the signal cross-sections. Specific details concerning the definition of the measured observables, selection cuts and any other inputs used in these experimental analyses are available in the respective ATLAS and CMS papers listed in Table~\ref{tab:lhc-searches}.

Limits on the SMEFT and on mediator models are extracted with the \HighPT package~\cite{<HighPT:_A_Tool_for_High-pT_Collider_Studies_Beyond_the_Standard_Model><Allwicher;Lukas><Faroughy;Darius><Jaffredo;Florentin><Sumensari;Olcyr><Wilsch;Felix><2022>} where each Drell-Yan search has been repurposed for generic New Physics scenarios. For each signal hypothesis we compute the 95\%~confidence intervals using Pearson's $\chi^2$ test statistic. In order to obtain reliable limits, beforehand we made sure to combine the data between neighbouring experimental bins until $\cN^\mathrm{obs} \ge 10$ for any bin, where $\cN^\mathrm{obs}$ is the number of observed events (background errors are added in quadrature when combining). 

Internally, for each search in Table~\ref{tab:lhc-searches}, \HighPT extracts the number of signal events $\cN_A(\theta)$ in a bin $A\in \cA$ of $x_{\rm obs}$ by convoluting the relevant response matrix $K_{AB}$ with the analytical expressions for $\sigma_B$. These are computed with the PDF set {\tt PDF4LHC15\_nnlo\_mc}~\cite{1510.03865}. We denote by $\theta$ the parameters of the New Physics model that we wish to constrain, e.g.~form-factors or specific model parameters such as Wilson coefficients, or mediator masses and couplings. The $\chi^2$ is then built from the number of background events $\cN^b$, background uncertainties $\delta\cN^b$ and observed events $\cN^{\rm obs}$ provided by the experimental collaborations:
\begin{align}\label{eq:chi2}
    \chi^2(\theta)=\sum_{A\in\cA}\left(\frac{\cN_A(\theta)+\cN^{b}_A-\cN^{\rm obs}_A}{\Delta_A}\right)^2\,,
\end{align}
where the uncertainty $\Delta_A$ in bin $A$ is obtained by adding in quadrature the background and observed uncertainties, $\Delta_A^2=(\delta\cN^b_A)^2+ \cN_A^{\rm obs}$, where the last term corresponds to the Poissonian uncertainty of the data. 
Notice that we consider the pure SM contribution as a background included in~$\mathcal{N}_A^b$. Thus, $\mathcal{N}_A(\theta)$ contains only the New Physics contribution to the event yield, i.e. both the \textit{New Physics squared} contribution and the interference of the New Physics with the~SM.
The background predictions~$\mathcal{N}_A^b$ are taken from the experimental analyses, which are computed including higher-order corrections. Only the New Physics signal~$\mathcal{N}_A(\theta)$ is computed at tree-level using our form-factor decomposition presented in Sec.~\ref{sec:dilepton-ffs}.

The response matrices $K_{AB}$ have been provided in the \HighPT package for each LHC search. These were obtained from Monte Carlo simulations using the following pipeline: first all relevant operators in the SMEFT with $d\le8$ and all mediator Lagrangians in Table~\eqref{tab:mediators} were implemented with {\tt FeynRules}~\cite{Alloul:2013bka}. The resulting {\tt UFO} model files \cite{1108.2040} were then imported into {\tt MadGraph5}~\cite{Alwall:2014hca} and used to simulate statistically significant event samples for the dilepton and monolepton processes with all possible initial quark flavors. Samples were then showered and hadronized using {\tt Pythia8}~\cite{Sjostrand:2014zea}, and the final-state object reconstruction and detector simulation were performed using {\tt Delphes3}~\cite{deFavereau:2013fsa} tuned to match the experimental searches. After applying the same event selections as in each experiment, the events were binned into $x_{\rm obs}$ histograms. The simulation pipeline outlined above was used to produce a $x_{\rm obs}$ histogram from each bin $B\in \cB$ of $x$. The rows of the matrix $K_{AB}$ were then extracted from the resulting histograms.
The validation of our recast with the BSM benchmark scenarios provided by the experimental papers can be found in Appendix~\ref{app:recast-quality}.

The $\chi^2$ statistic described above is a reliable statistic whenever the number of observed, signal and background events in each bin are approximately Gaussian, which is usually the case for the bulk of the dilepton and monolepton differential distributions. Deep in the tails, however, the experimental bins will typically contain very low event yields, making these regions much more susceptible to data fluctuations. Indeed, if the tails are plagued with sizeable under-fluctuations, testing a signal hypothesis that relies on these regions of low-sensitivity can lead to spuriously strong exclusion limits for the New Physics parameters $\theta$. In order to remedy this, one can instead use the CL$_s$ method~\cite{Read:2000ru} based on the profiled Poissonian likelihood ratio~\cite{1007.1727}. We checked for the particular binnings used in our analyses that the two statistical methods give very good agreement. For an explicit comparison see Ref.~\cite{<HighPT:_A_Tool_for_High-pT_Collider_Studies_Beyond_the_Standard_Model><Allwicher;Lukas><Faroughy;Darius><Jaffredo;Florentin><Sumensari;Olcyr><Wilsch;Felix><2022>}. For this reason, in the following Sections we present all exclusion limits based on the $\chi^2$ statistic.

%%%%%%%%%%%%%%%%%%%
\begin{figure}[p!]
    \centering
    \resizebox{0.98\textwidth}{!}{
    \begin{tabular}{c c c}
        \includegraphics[]{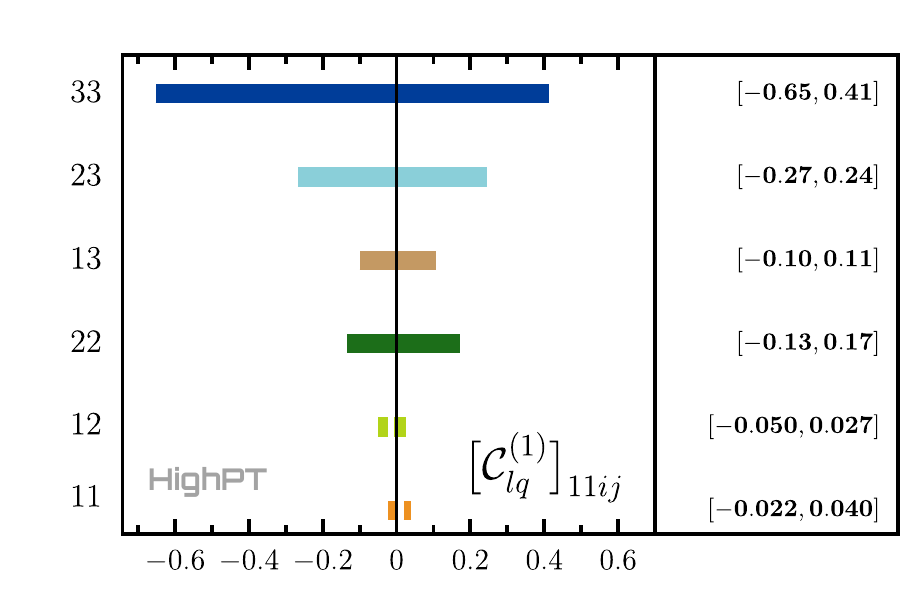} & \includegraphics[]{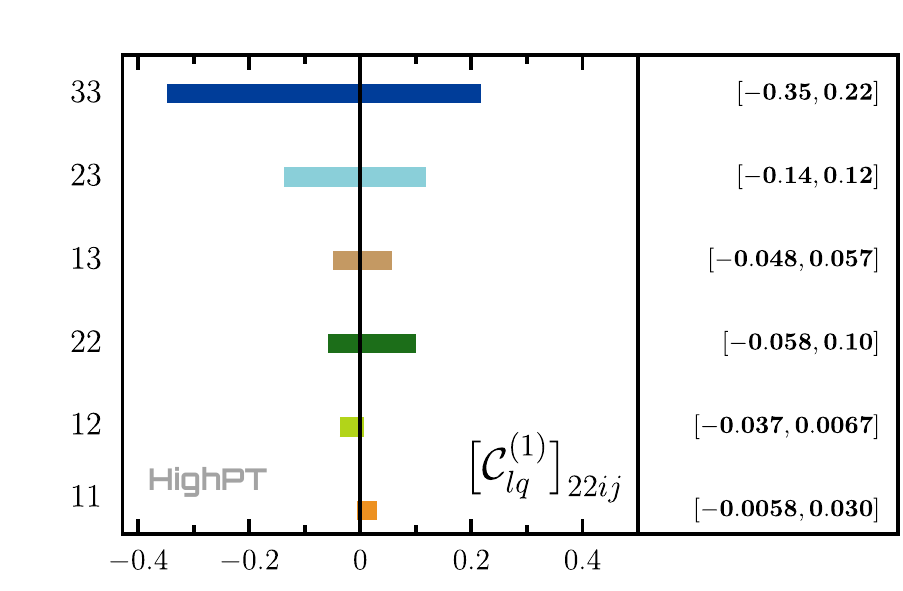} & \includegraphics[]{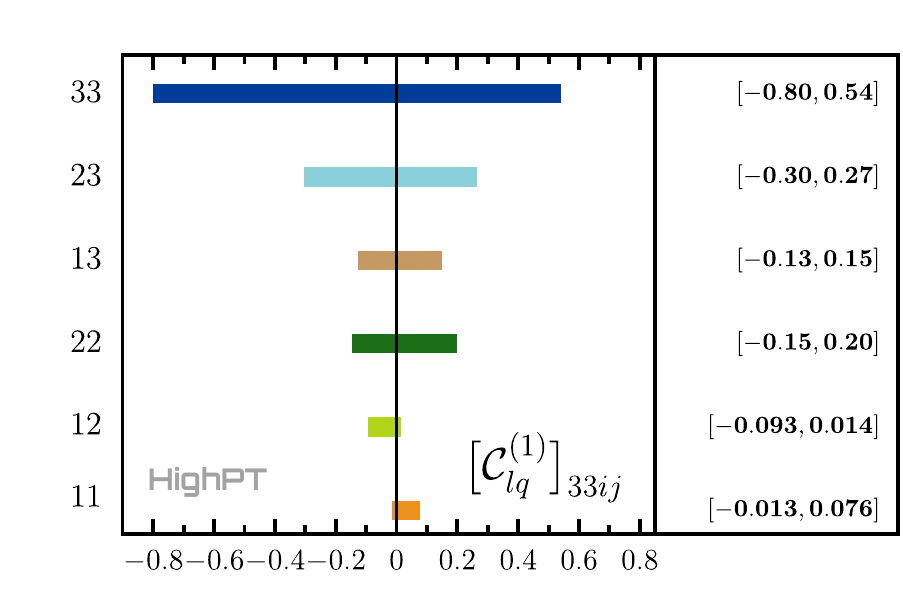} \\
        \includegraphics[]{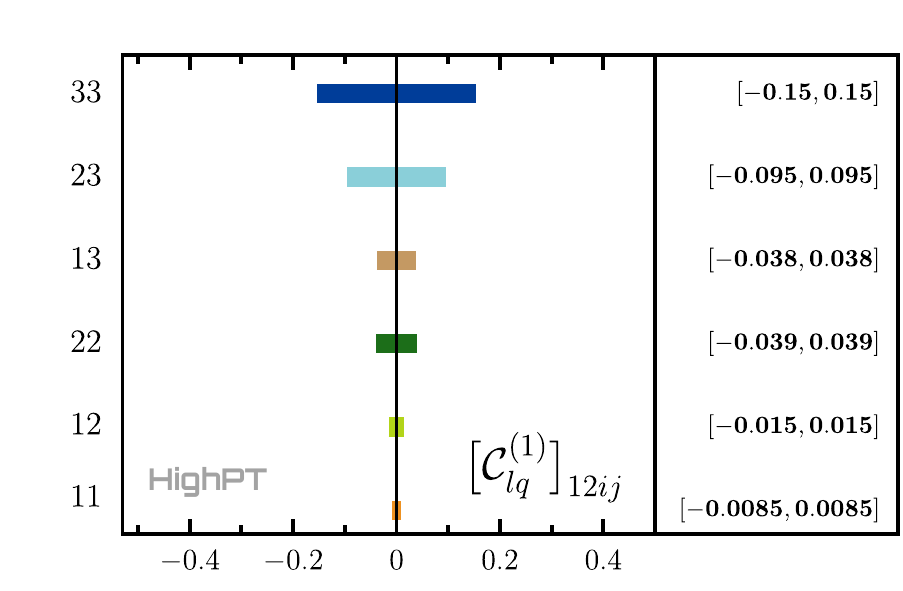} & \includegraphics[]{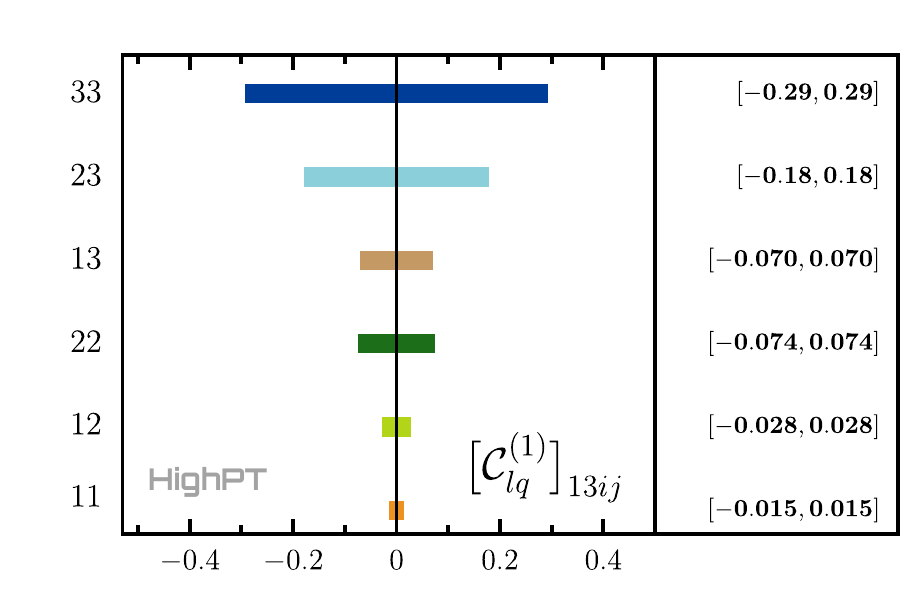} & \includegraphics[]{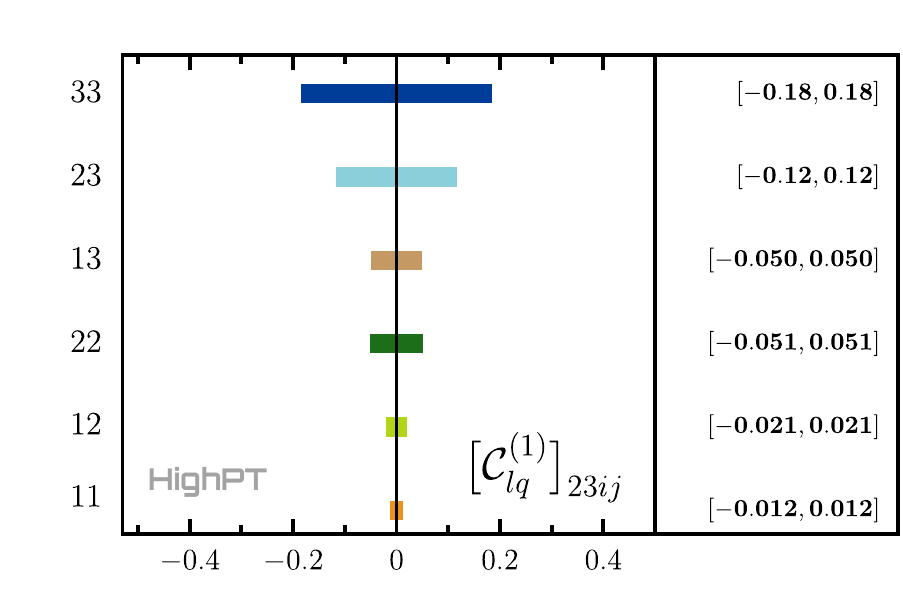} \\
        \includegraphics[]{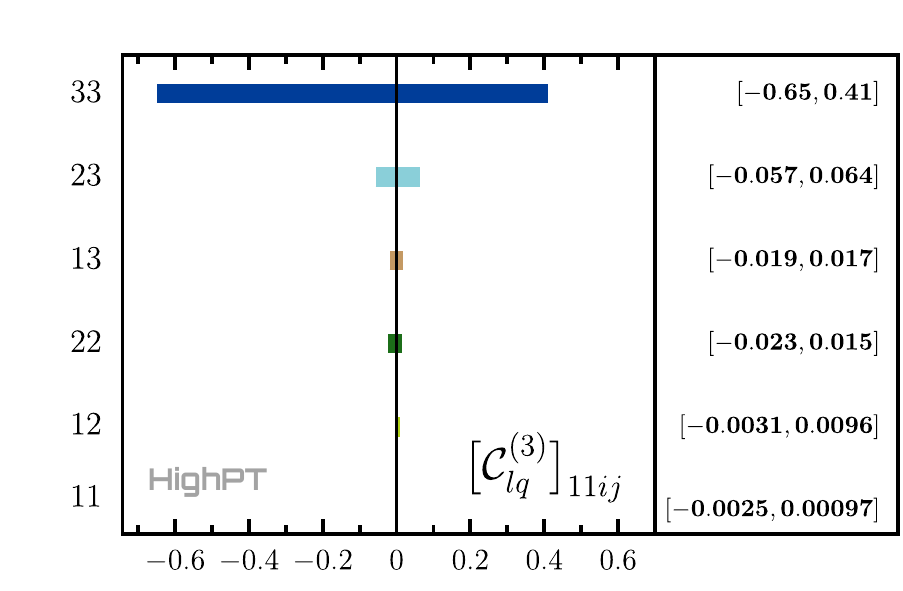} & \includegraphics[]{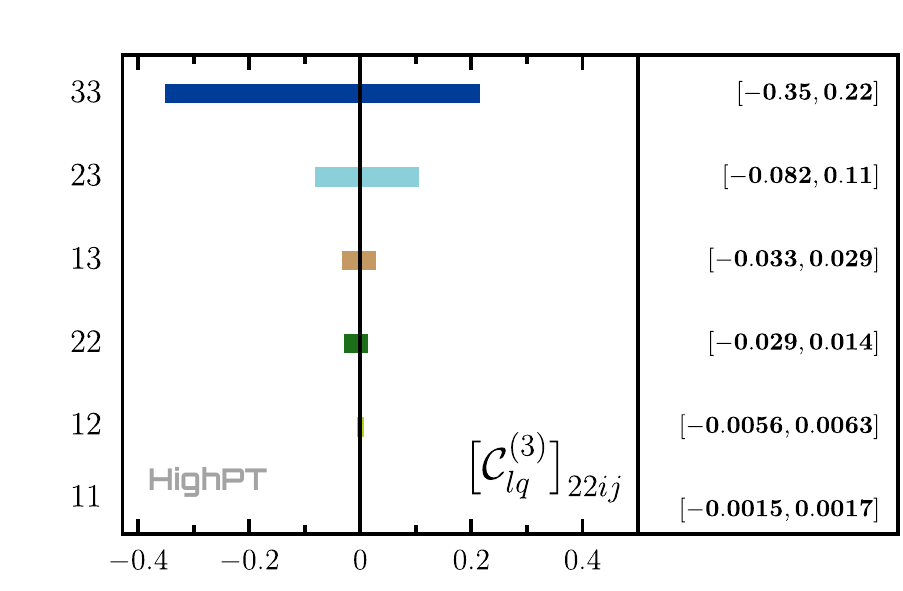} & \includegraphics[]{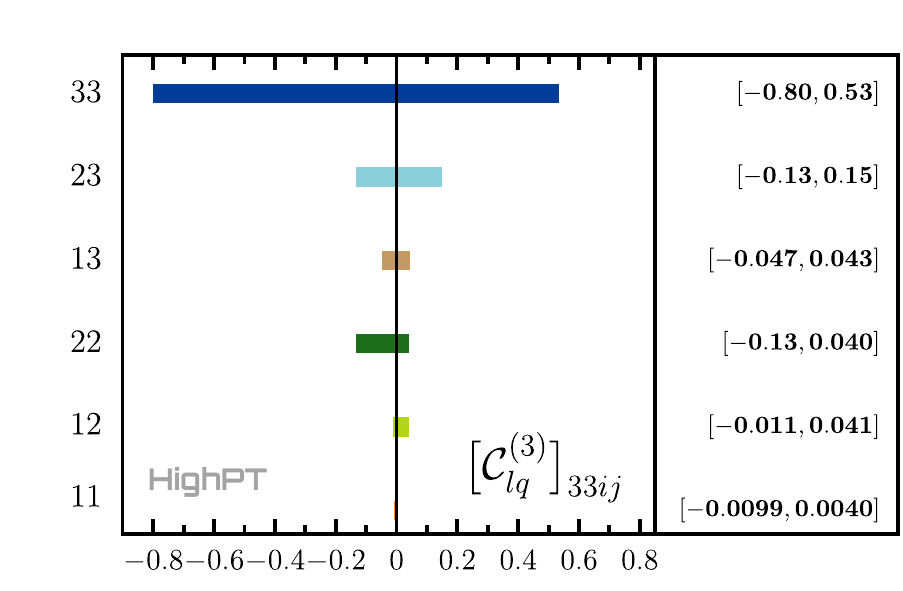} \\
        \includegraphics[]{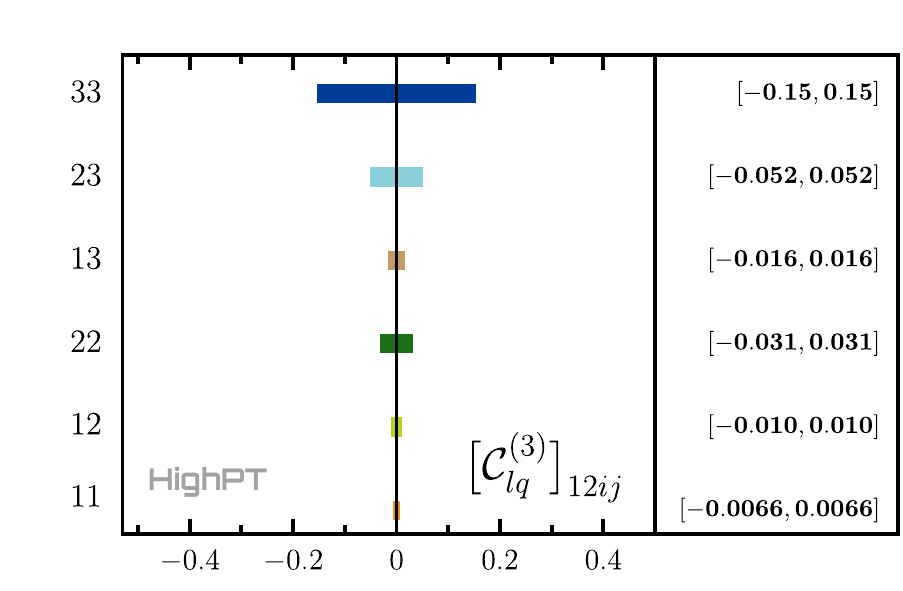} & \includegraphics[]{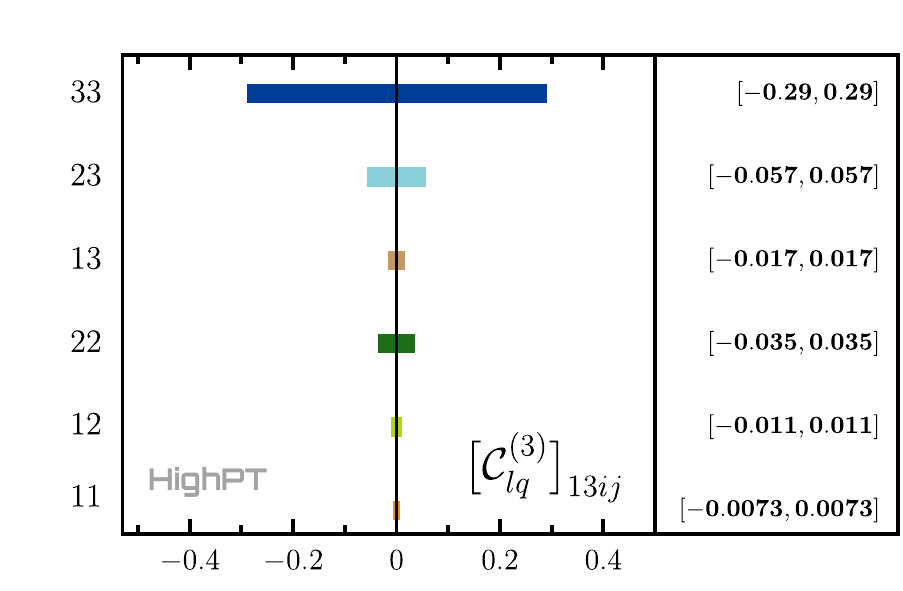} & \includegraphics[]{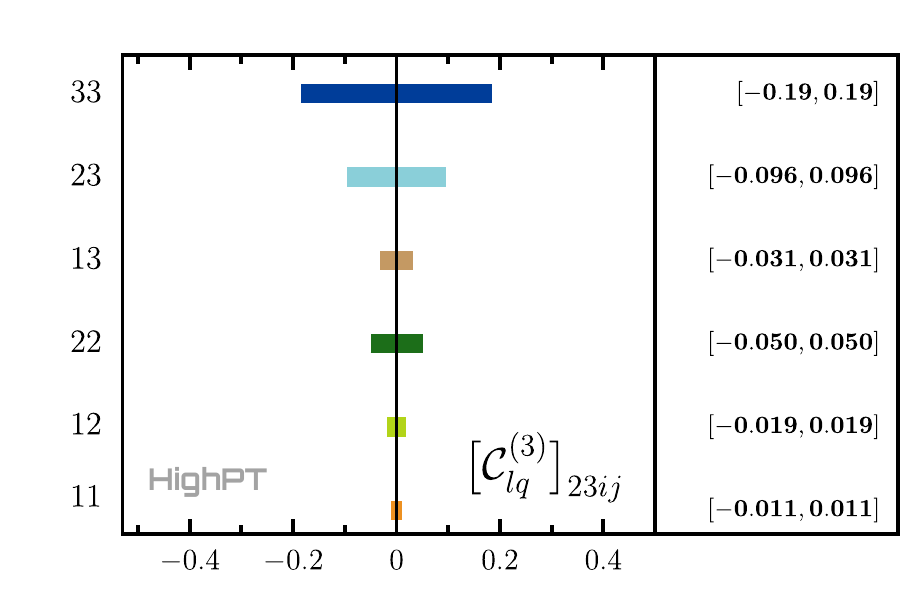}
    \end{tabular}
    }
    \caption{\sl\small LHC constraints on the SMEFT Wilson $\cC_{lq}^{(1,3)}$ coefficients with different flavor indices at $95\%$~CL, where a single coefficient is turned on at a time. Quark-flavor indices are denoted by~$ij$ and are specified on the left-hand side of each plot. All coefficients are assumed to be real and contributions to the cross-section up to and including $\cO(1/\Lambda^{4})$ are considered.  The scale $\Lambda$ is fixed to $1\,\mathrm{TeV}$.}
    \label{fig:single-WC-limits-lq}
\end{figure}
\begin{figure}[p!]
    \centering
    \resizebox{0.98\textwidth}{!}{
    \begin{tabular}{c c c}
        \includegraphics[]{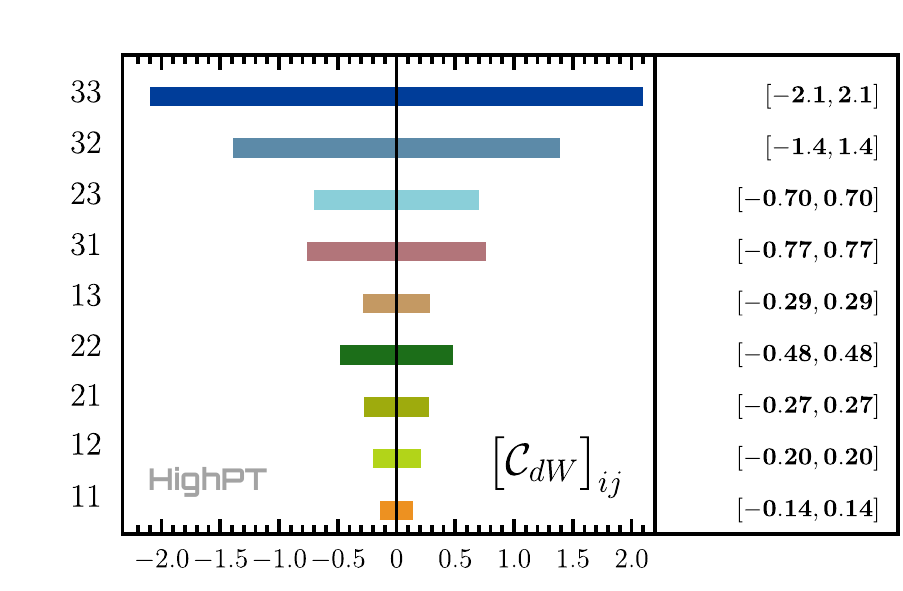} & \includegraphics[]{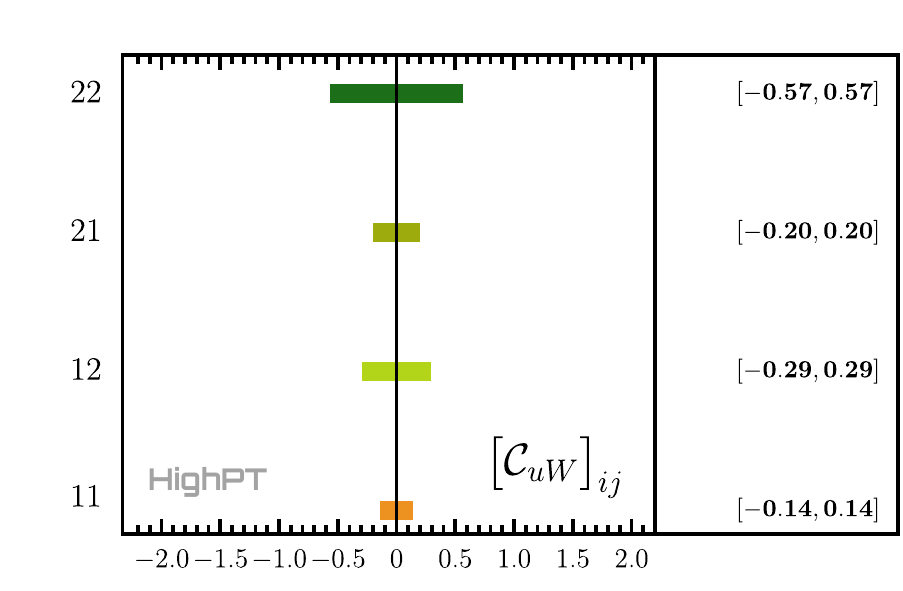} & \includegraphics[]{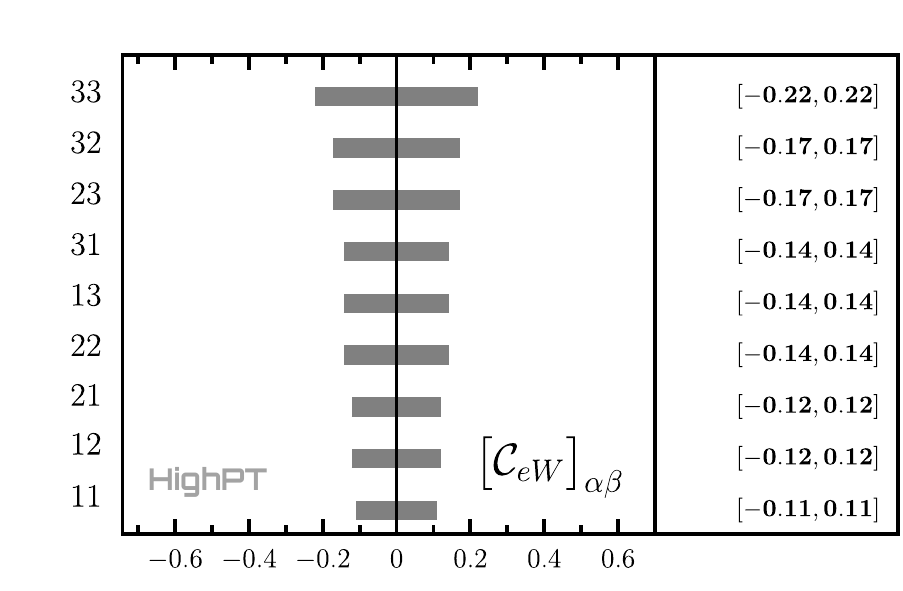} \\
        \includegraphics[]{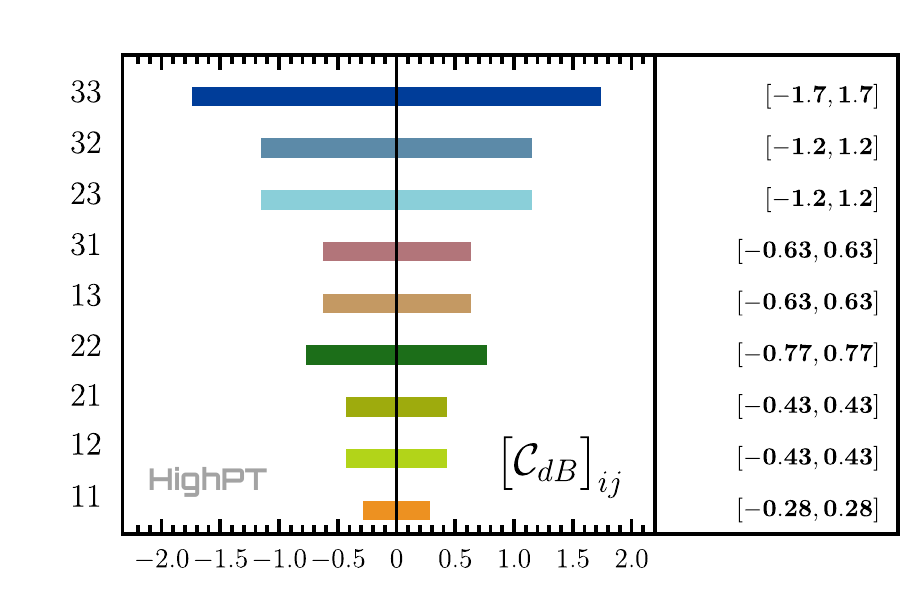} & \includegraphics[]{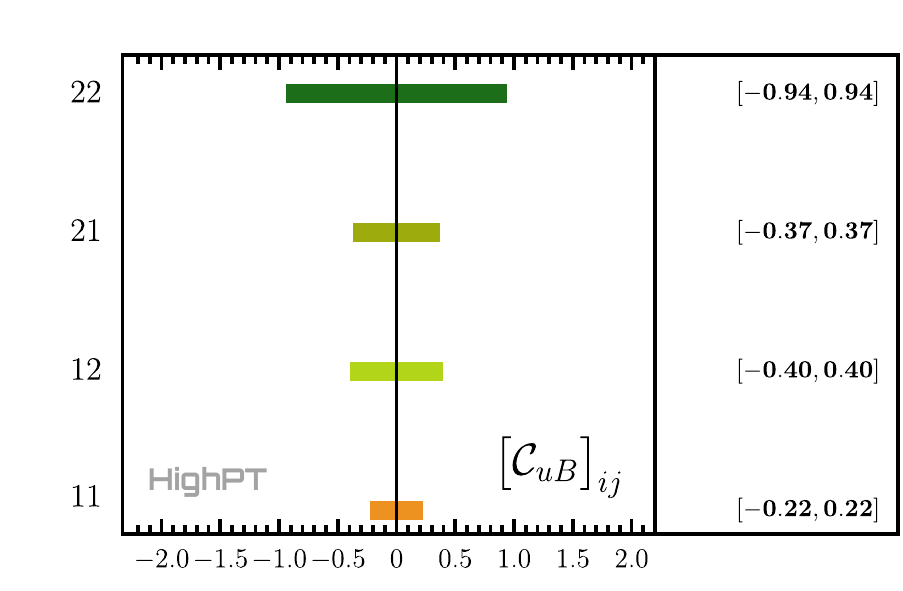} & \includegraphics[]{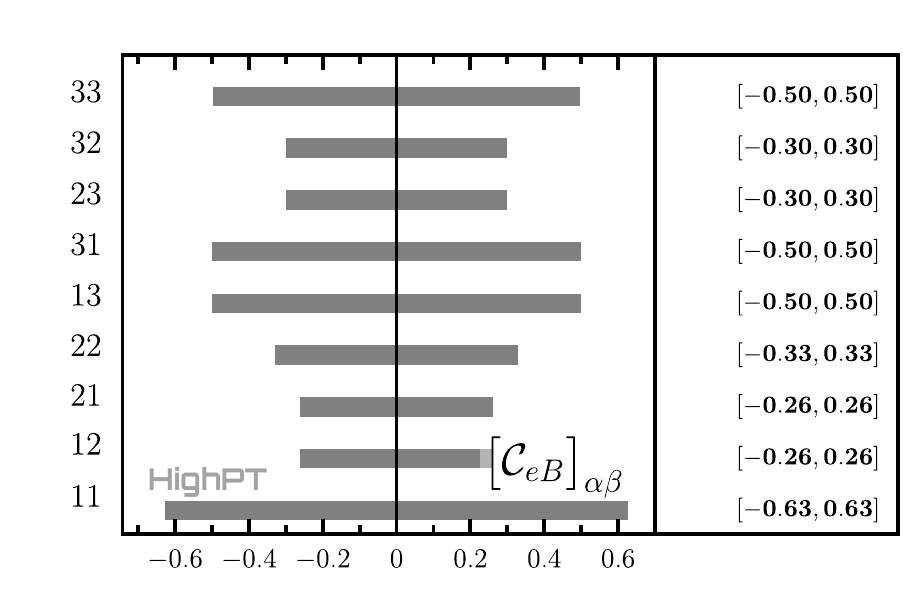}
    \end{tabular}
    }
    \caption{\sl\small LHC constraints on the SMEFT dipole Wilson coefficients with different flavor indices at $95\%$~CL, where a single coefficient is turned on at a time. Quark (lepton) flavor indices are denoted by~$ij$ $(\alpha\beta)$ and are specified on the left-hand side of each plot. All coefficients are assumed to be real and contributions to the cross-section at order $\cO(1/\Lambda^{4})$ are considered.  The scale $\Lambda$ is fixed to $1\,\mathrm{TeV}$.}
    \label{fig:single-WC-limits-dipoles}
\end{figure}
%%%%%%%%%%%%%%%%%%%

\subsection{Single-operator limits on the dimension-$6$ SMEFT}
\label{ssec:smeft-limits}

In this Section we present upper bounds on the dimension-$6$ SMEFT operators using LHC run-II data from the $pp\to\ell_\alpha^-\ell_\beta^+$ and $pp\to \ell_\alpha^{\pm}\nu_\beta$ Drell-Yan searches listed in Table~\ref{tab:lhc-searches}. Single-parameter limits are extracted for individual Wilson coefficients by assuming them, for simplicity, to be real parameters and by setting all other coefficients to zero. Since we are interested in tails of the Drell-Yan distributions, we only focus on four-fermion semileptonic operators, as well as the quark/lepton dipole operators, because these are the only ones that lead to an energy growth in the amplitude. Our limits are derived by keeping the $\cO(1/\Lambda^4)$ corrections from the dimension-$6$ squared pieces.~\footnote{The effects of dimension-$8$ operators will be discussed in the following Section.} Indeed, these are the lowest-order contributions driving the bounds for the dipole operators and for the four-fermion operators that are chirality-flipping or have a non-diagonal flavor structure, i.e.~any operator not interfering with the SM. For the flavor sector we assume flavor alignment in the down sector for the CKM matrix and a unit PMNS matrix. The results are presented in seven figures in Appendix~\ref{sec:SMEFT-limits} for all semileptonic operators with fixed leptonic indices, namely $ee$, $\mu\mu$, $\tau\tau$, $e\mu$, $e\tau$ and $\mu\tau$, and for all possible quark-flavor indices that can be probed at the LHC. All limits on the Wilson coefficients ($\cC$) are given at 95\%~CL at a fixed reference scale of $\Lambda=1$\,TeV with no loss of generality since the LHC processes probe the combination $\cC /\Lambda^2$ in the EFT limit.

To simplify the discussion of our results, we have replicated in Fig.~\ref{fig:single-WC-limits-lq} the LHC constraints for the following representative semileptonic operators,
\begin{align}
[\cO_{lq}^{(1)}]_{\alpha\beta ij}&=(\bar l_\alpha \gamma^\mu l_\beta)(\bar q_i \gamma_\mu q_j) \,,
&
[\cO_{lq}^{(3)}]_{\alpha\beta ij}&=(\bar l_\alpha \gamma^\mu\tau^I l_\beta)(\bar q_i \gamma_\mu\tau^I q_j) \,,
\end{align}

\noindent and, similarly, in Fig.~\ref{fig:single-WC-limits-dipoles} for the dipole operators,
\begin{align}
\left[\cO_{dB}\right]_{ij}&=(\bar q_i \sigma^{\mu\nu} d_j)\, H B_{\mu\nu} \,,
&
\left[\cO_{dW}\right]_{ij}&=(\bar q_i \sigma^{\mu\nu} d_j)\, \tau^I H W^I_{\mu\nu} \,,\nonumber
\\[0.3em]
\left[\cO_{uB}\right]_{ij}&=(\bar q_i \sigma^{\mu\nu} u_j)\, \widetilde{H} B_{\mu\nu} \,,
&
\left[\cO_{uW}\right]_{ij}&=(\bar q_i \sigma^{\mu\nu} u_j)\, \tau^I \widetilde{H} W^I_{\mu\nu} \,,
\\[0.3em]
\left[\cO_{eB}\right]_{\alpha\beta}&=(\bar l_\alpha \sigma^{\mu\nu} e_\beta)\, H B_{\mu\nu} \,,
&
\left[\cO_{eW}\right]_{\alpha\beta}&=(\bar l_\alpha \sigma^{\mu\nu} e_\beta)\, \tau^I H W_{\mu\nu}^I \,,\nonumber
\end{align}
where we keep the most general flavor structure. Note, in particular, that the singlet operator $\smash{\cO_{lq}^{(1)}}$ only contributes to neutral Drell-Yan processes, whereas the triplet $\smash{\cO_{lq}^{(3)}}$ contributes to both neutral- and charged-current processes, and similarly for the dipole operators. Neutral- and charged-current constraints have been combined to set the constraints on triplet operators depicted in Figs.~\ref{fig:single-WC-limits-lq} and \ref{fig:single-WC-limits-dipoles}. Note, in particular, that the interference of the SM with the contribution by $\smash{\cO_{lq}^{(3)}}$ has the same sign for up-type and down-type quarks. However, the interference of the SM with $\smash{\cO_{lq}^{(1)}}$ has opposite sign for up-type and down-type quarks, leading to an approximate cancellation of the interference term in this case. This explains why the constraints on the coefficient $\smash{\cC_{lq}^{(3)}}$ are considerably more stringent than for $\smash{\cC_{lq}^{(1)}}$, as can be seen by comparing the plots in the first row of Fig.~\ref{fig:single-WC-limits-lq} with the third row. For the Lepton Flavor Violating (LFV) couplings, the constraints for both operator types are of the same order of magnitude, as in this case the interference terms vanish and the constraints only derive from the New Physics squared contribution.

Before discussing the remaining results, it is worth mentioning that, while the single-parameter fits provide a useful benchmark leading to the most optimistic limits, these analyses completely neglect potential correlations between different Wilson coefficients that could relax substantially the constraints or even lead to flat directions in the parameter space. For this reason it would be preferable to perform a global fit, or at least a multi-parameter analysis involving the most relevant effective operators. While this lies outside the intended scope of this paper, in Sec.~\ref{sec:example} we carry out a two-parameter analysis fitting pairs of ultraviolet-motivated operators to Drell-Yan data, as well as combined fits to low-energy flavor and electroweak pole data. We leave for future work a more thorough multi-parameter analysis of the SMEFT with all three generations and different flavor structure hypotheses~\cite{Allwicher:2022mcg}. Note, however, that the full LHC likelihood for the $d=6$ SMEFT truncated at $\cO(1/\Lambda^4)$ including all $1021$ flavored Wilson coefficients (see Table~\ref{tab:parameter_counting}) is promptly available in {\tt HighPT}, which also allows to include dimension-8 coefficients.

\subsubsection*{SMEFT truncations at $\cO(1/\Lambda^2)$ and $\cO(1/\Lambda^4)$}
To assess the sensitivity of the LHC searches to the SMEFT truncation at $\cO(1/\Lambda^2)$ or $\cO(1/\Lambda^4)$, we investigate the impact on the upper limits on dimension-$6$ operators when specific portions of the Drell-Yan data are removed from the analysis. For definiteness, we focus on the dimuon searches and on single-parameter constraints for the vector operators $\smash{[\cO_{lq}^{(1)}]_{22ii}}$ and the quark dipole $\smash{[\cO_{dW}]_{ii}}$ for $i=1,2,3$. 

%%%%%%%%%%%%%%%%%%
\begin{figure}[t!]
\centering
        \includegraphics[width=0.92\linewidth]{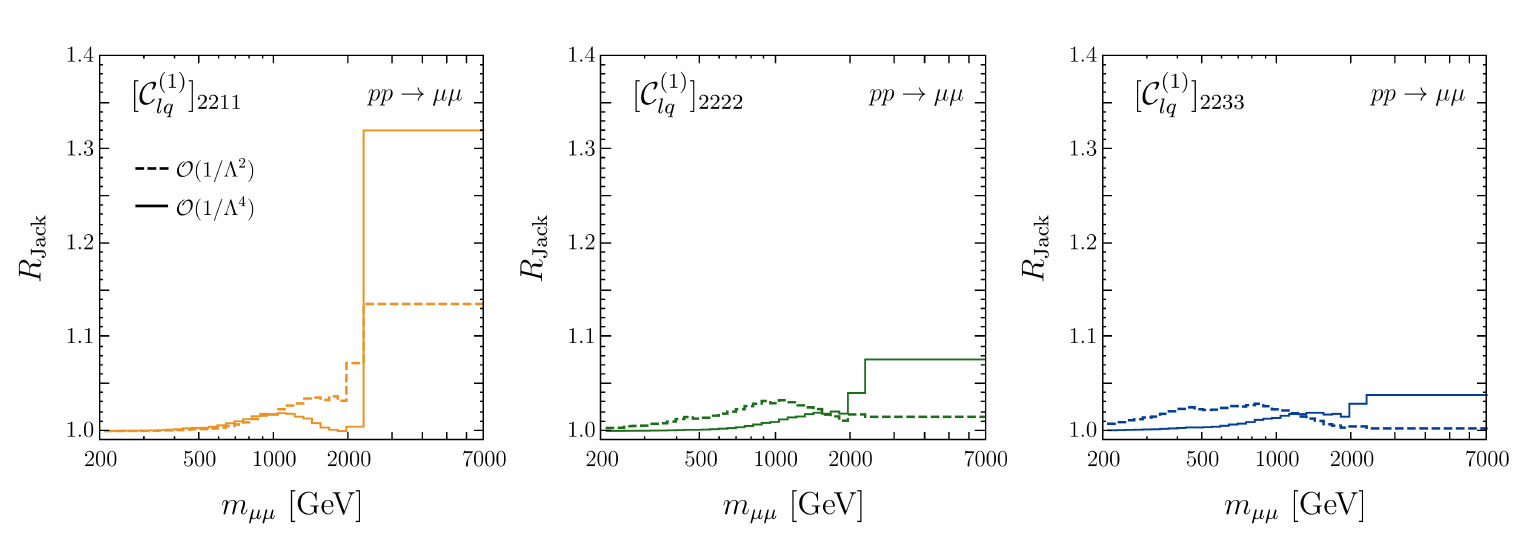} \\
        \vspace{-0.325cm}
        \includegraphics[width=0.92\linewidth]{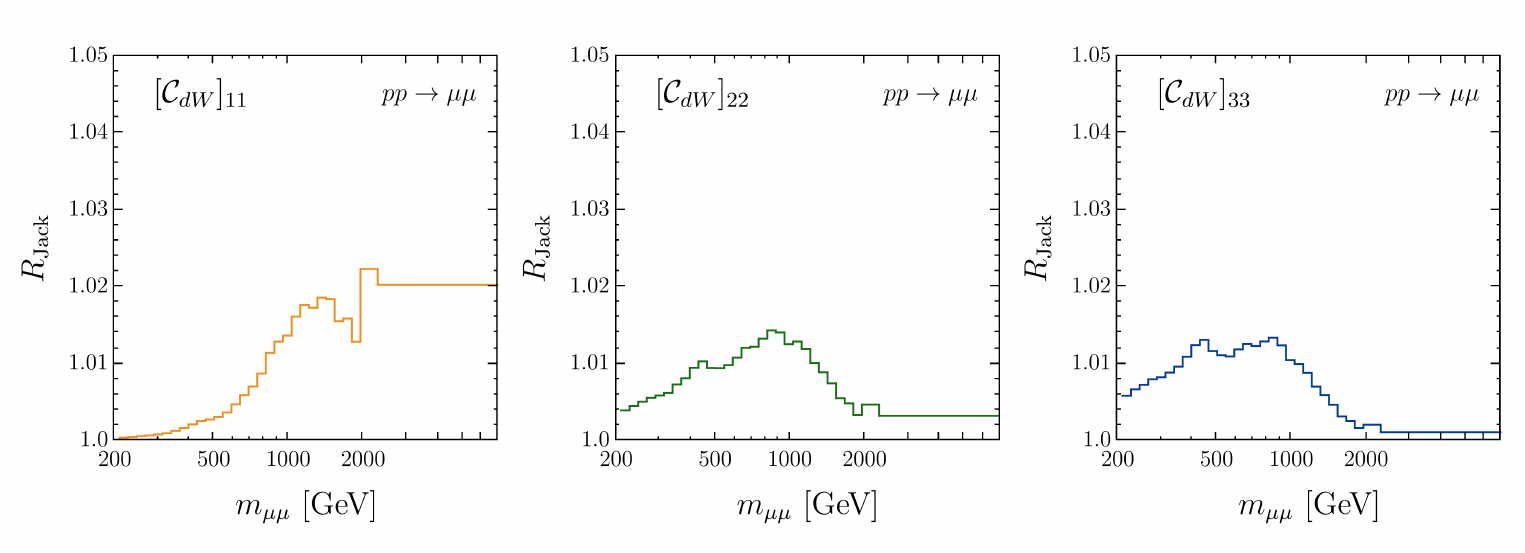} \\
    \caption{\sl\small The expected sensitivity of individual $m_{\mu\mu}$ bins to the dimension-$6$ flavor conserving  operators $\cO_{lq}^{(1)}$ and $\cO_{dW}$ using the Jack-knife ratio $R_{\rm Jack}$ (see text for details). The dashed and solid lines correspond to the EFT truncation at $\cO(1/\Lambda^2)$ and $\cO(1/\Lambda^4)$, respectively.}
    \label{fig:jacked_lims}
\end{figure}
%%%%%%%%%%%%%%%%%%

%%%%%%%%%%%%%%%%%%
\begin{figure}[t!]
\centering
        \includegraphics[width=0.3325\linewidth]{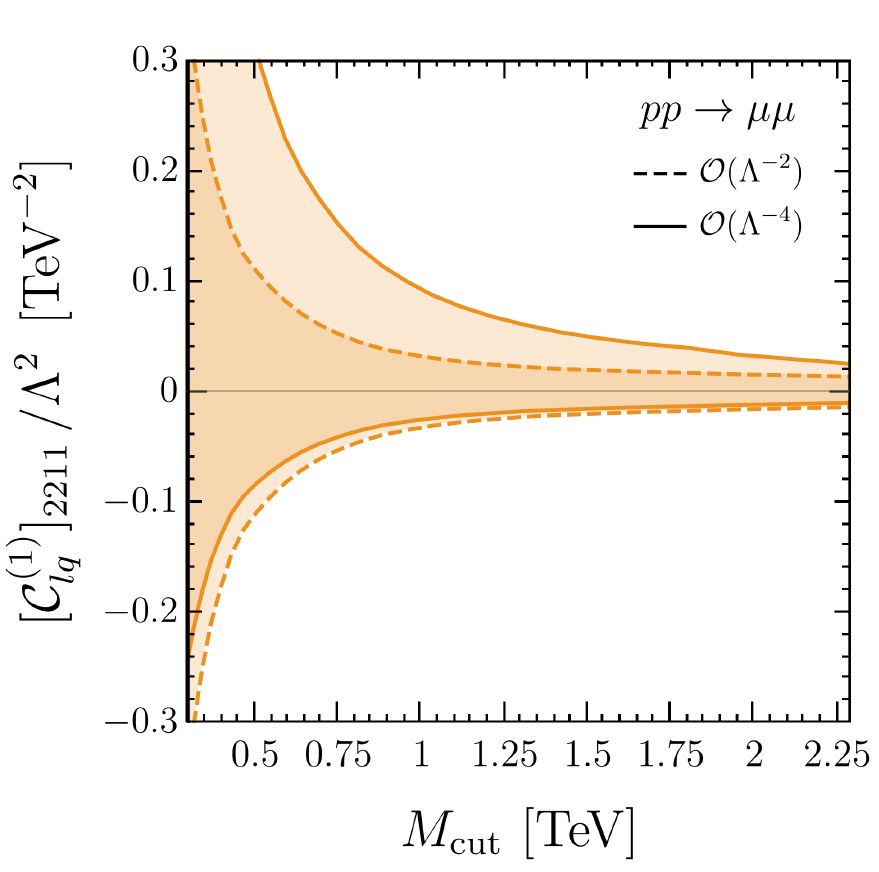}~~
        \includegraphics[width=0.32\linewidth]{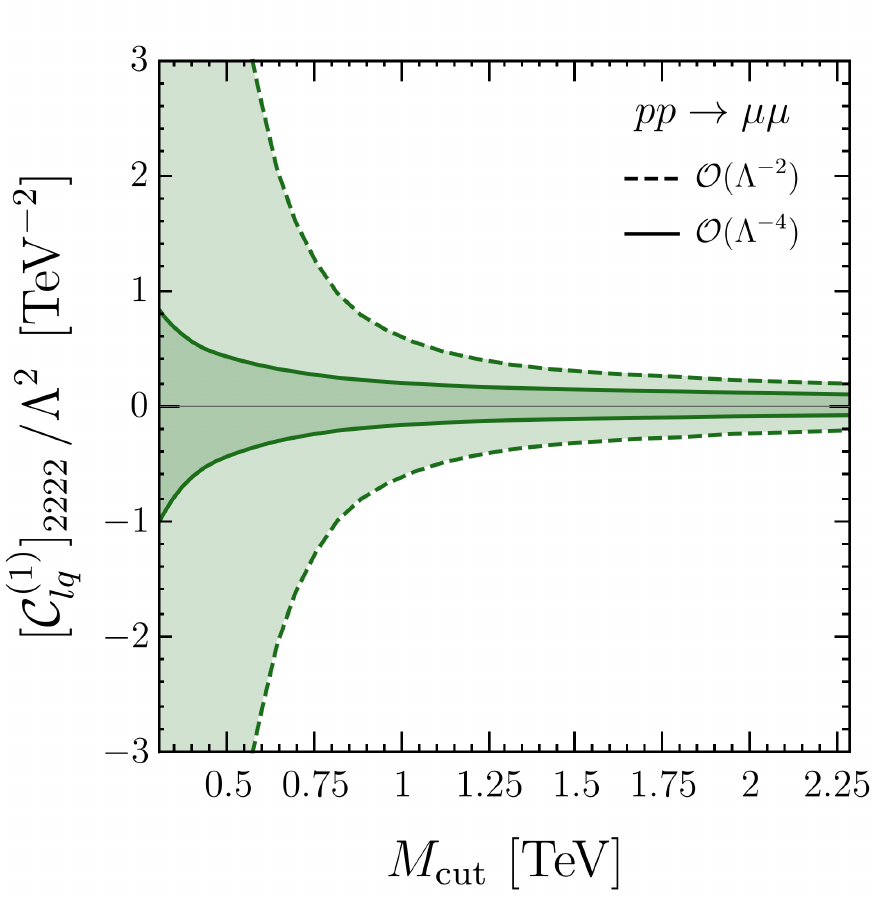}~~
        \includegraphics[width=0.32\linewidth]{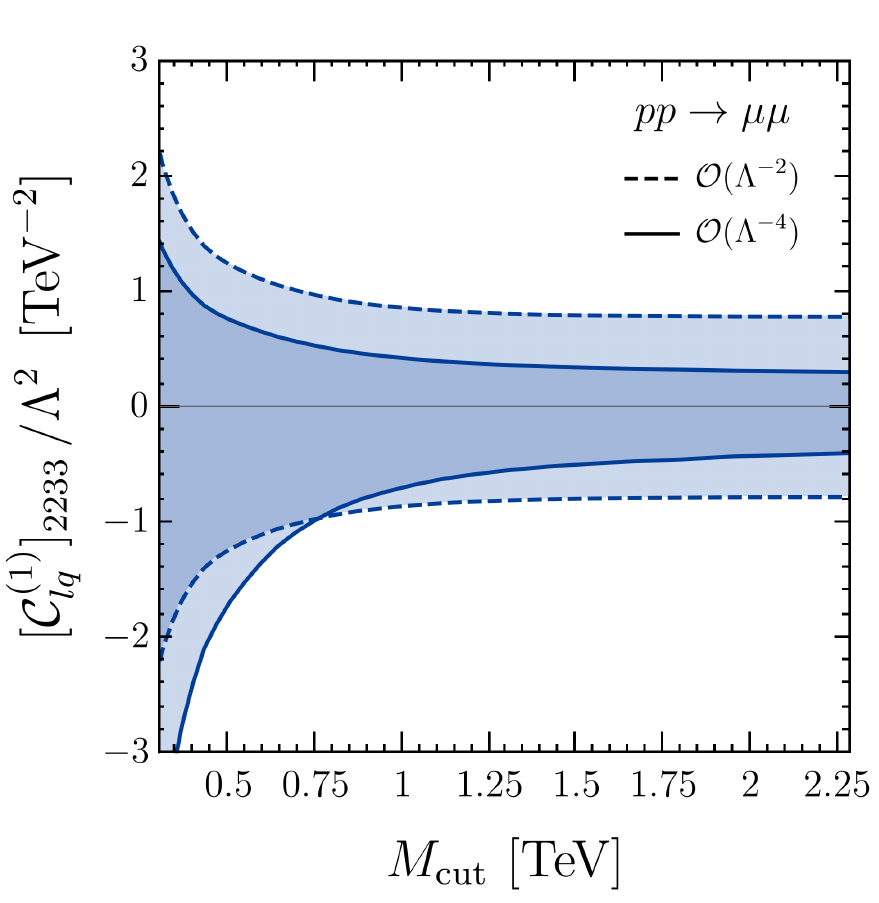}\\
        ~~\includegraphics[width=0.32\linewidth]{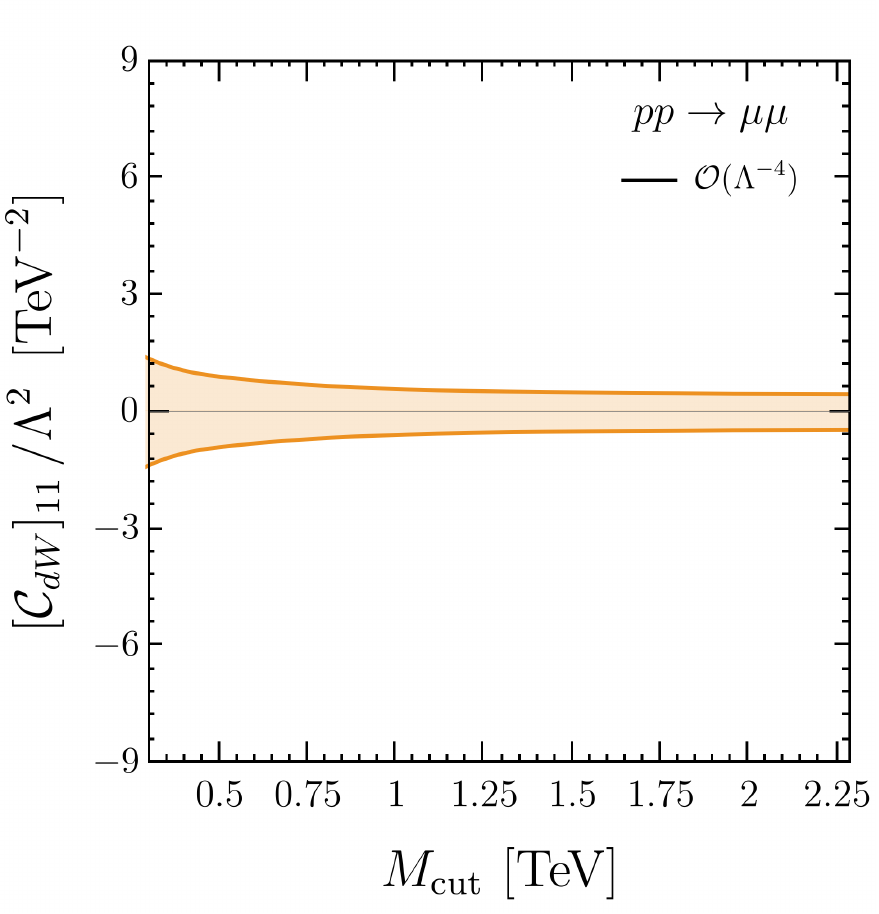}~~
        \includegraphics[width=0.32\linewidth]{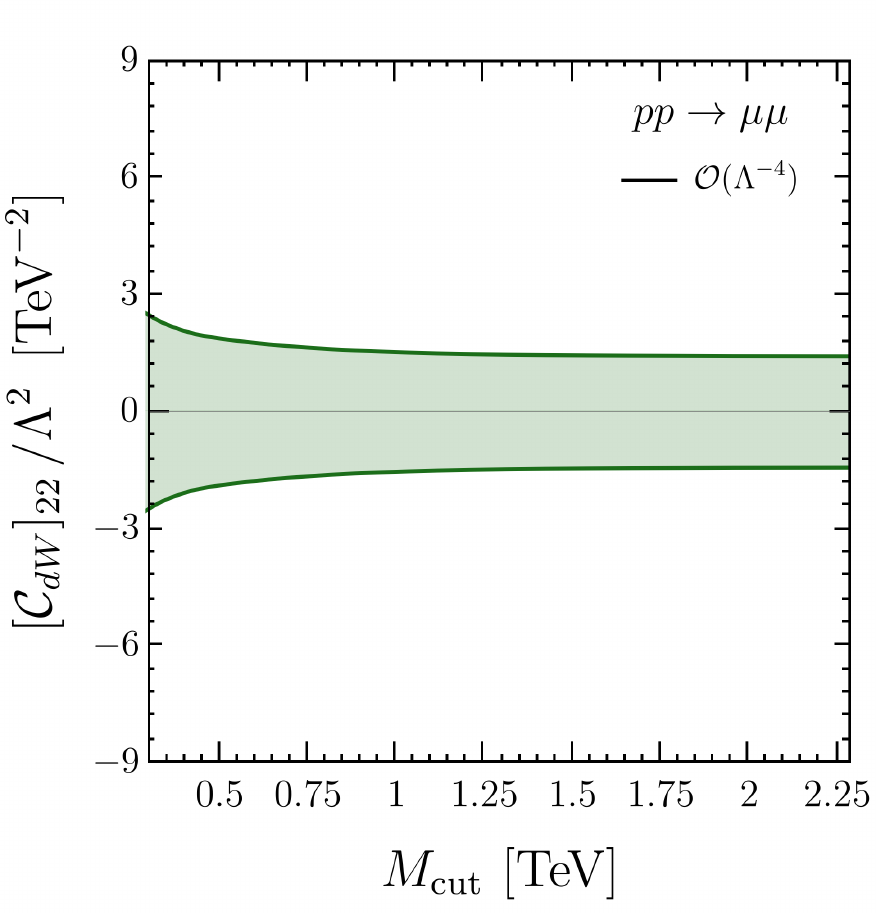}~~
        \includegraphics[width=0.32\linewidth]{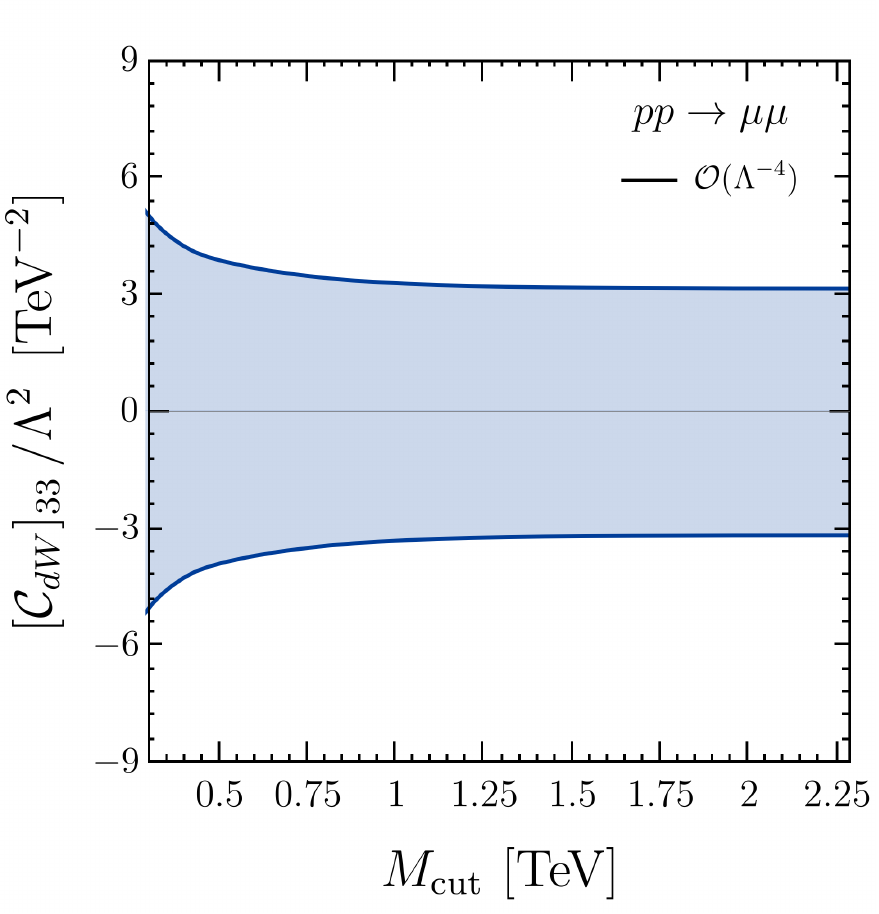}
    \caption{\sl\small Clipped expected limits from LHC dimuon searches for flavor conserving operators $\cO_{lq}^{(1)}$ and $\cO_{dW}$ as a function of the sliding maximal scale $M_{\rm cut}$. The dashed and solid contours correspond to the EFT truncation at $\cO(1/\Lambda^2)$ and $\cO(1/\Lambda^4)$, respectively.
    }
    \label{fig:clipped_lims}
\end{figure}
%%%%%%%%%%%%%%%%%%

First, we test the sensitivity of the constraints on the EFT operators to individual bins in the invariant-mass tails. For this purpose, we build the {\it jack-knife} function~\cite{10.2307/2334280, Greljo:2018tzh}, $\chi^2_{\rm Jack}(m_{\mu\mu})$, defined as the statistic that results from holding out a single invariant-mass bin at a time from the total $\chi^2$ function. In Fig.~\ref{fig:jacked_lims}, we show the quantity $R_{\rm Jack}$, defined as the ratio of the expected jack-knife limit over the expected limit extracted using the whole invariant-mass spectrum, for different quark flavors. For the semileptonic operators (upper row), when truncating the EFT at $\cO(1/\Lambda^2)$ (dashed lines), we see that for first generation valence quarks the last experimental bin, spanning $\{2300, 7000\}$\,GeV, is the most sensitive one in the search, while for second- and third-generation quarks the most sensitive bins are found at lower invariant masses, around $1000$~GeV and $800$~GeV, respectively. On the other hand, when truncating at $\cO(1/\Lambda^4)$ (solid lines) the most sensitive bin is always the highest mass bin of the search, irrespective of the quark flavor. Furthermore, we note that the last bin is less relevant for second and third generation quarks, whereas for valence quarks removing this bin can weaken the limits by as much as $\sim30$\%. From these results, we find that truncating the EFT at order $\cO(1/\Lambda^4)$ for semileptonic vector operators has a significant impact when setting limits on single operators and that the highest mass bins are the most sensitive ones in the dimuon search. This is not entirely surprising given that for high energies, the quadratic terms growing as $\hat s^2/\Lambda^4$ in the partonic cross-section can compete with the interfering term $\hat s/\Lambda^2$ that is typically more important at lower invariant masses. For the dipole operator, even though these are purely New Physics squared terms, the energy enhancement scales as $v^2\hat s/\Lambda^4$. The sensitivity of the search is therefore at lower invariant mass bins, similar to the four-fermion interference terms at $\cO(1/\Lambda^2)$, as shown in the second row of Fig.~\ref{fig:jacked_lims}.   

Next, we investigate how the limits on single operators are affected by restricting the Drell-Yan data in the tails with a maximal energy cut ($M_{\rm cut}$) \cite{1604.06444}. This procedure, known as {\it clipping}~\cite{2201.04974}, provides a useful way to obtain more robust EFT constraints. In Fig.~\ref{fig:clipped_lims} we show the expected limits for the vector operator (upper row) and the quark-dipole operator as a function of the sliding upper cut $M_{\rm cut}>m_{\mu\mu}$. The results are given for an EFT truncation at $\cO(1/\Lambda^2)$ (dashed lines) and $\cO(1/\Lambda^4)$ (solid lines). Here again we can appreciate the relevance of including $\cO(1/\Lambda^4)$ corrections for the vector operators. Irrespective of the truncation order and the operator, these results also indicate that the clipped limits for maximal values above $M_{\rm cut}\sim 1.5$\,TeV saturate, leading to upper limits that are comparable to those obtained when using the full kinematic spectrum.

\subsection{Flavor dependence}
\label{sec:flav_dependence}
We now discuss the constraints on the SMEFT coefficients shown in Fig.~\ref{fig:single-WC-limits-lq} from the perspective of flavor. For fixed leptonic flavors, as expected, we find that the most constrained coefficients are the ones involving valence quarks, but useful constraints are also obtained for operators involving the heavier $s$-, $c$- and $b$-quarks despite their PDF suppression. Overall, the upper limits for $\cO_{lq}^{(1,3)}$ and the quark dipoles for different $(i,j)$ indices follow approximately the expected hierarchies between the parton-parton luminosity functions $\cL_{ij}$ given in Eq.~\eqref{eq:parton-luminosities}.
For fixed quark flavor indices, we find comparable constraints for the $e$ and $\mu$ channels, with much weaker constraints for~$\tau$'s. Similar patterns are also observed for the other semileptonic and quark-dipole coefficients that are collected in Appendix~\ref{sec:SMEFT-limits}. For the leptonic dipoles $\cO_{eB}$, the limits for different lepton indices, which are all driven by valence-quark production, are determined by the experimental sensitivities of each LHC search. 

The constraints on the vector semileptonic operators with flavor-diagonal indices $i\!=\!j$ and $\alpha\!=\!\beta$ are not symmetric due to the interplay of interference and New Physics squared terms. The skewness of the limits towards a specific sign will depend on the relative size between these contributions, as well as on the size and sign of the fluctuations in the observed data in the most sensitive regions of the tails.~\footnote{The number of signal events, when turning on one (real) Wilson coefficient $\cC$ at a time, is given by $\cN(C)=\cN^{\rm int}\, C +\,\cN^{\rm NP} C^2$ where $C\equiv\cC/\Lambda^2$ is the New Physics (NP) parameter, $\cN^{\rm int}$ and $\cN^{\rm NP}$ are the SMEFT signals yields at $|\cC|/\Lambda^2=1\,{\rm TeV}^{-2}$ for the SM-NP interference and NP squared terms, respectively. It is convenient to write the $\chi^2$ function for the expected data in a specific bin as $\smash{\chi^2(C) = \chi^2_{\rm int}(C)\,\left[1+(\cN^{\rm NP}/\cN^{\rm int})\, C\right]^2}$ where $\chi^2_{\rm int}\propto C^2$ is the expected $\chi^2$ function when truncating the EFT expansion at $\cO(1/\Lambda^2)$ and the bracket contains the effects of $\cO(1/\Lambda^4)$ corrections from the dimension-$6$ squared terms. Since $\chi^2_{\rm int}$ is a symmetric function, the bracket above will {\it skew} the expected upper limits. For observed limits, there will be an additional asymmetry caused by the statistical fluctuations in the data.} As can be seen in the first three rows in Fig.~\ref{fig:single-WC-limits-lq}, the upper limits derived for the operators with first generation quarks are much more stringent for negative (positive) values of $\cC_{lq}^{(1)}$ ($\cC_{lq}^{(3)}$), while the asymmetry is much less pronounced for operators involving second or third generation quarks where the New Physics squared components dominate. 

For the operators with flavor-changing quark indices $i\!<\!j$ the limits are pretty much symmetric except for $(i,j)=(1,2)$. When turning on one of these operators, the cross-sections receives a positive definite contribution from the New Physics squared piece coming from the flavor-changing mode $\bar q_i q_j\to\ell^-_\alpha\ell^+_\alpha$, as well as subleading contributions from the CKM-suppressed flavor-conserving modes $\bar q_i q_i \to \ell^-_\alpha\ell^+_\alpha$. These last contributions can potentially skew the upper limits since they interfere with the SM. For $(i,j)=(1,2)$ the bounds are skewed because the $\bar u u\to\ell^-_\alpha\ell^+_\alpha$ modes can compete with the flavor-changing modes since these have a mild Cabibbo suppression of $\cO(\lambda)$ that can be compensated by the up-quark PDF. The effects of quark-flavor mixing through the CKM matrix will be discussed in more detail below. For $(i,j)=(1,3)$ and $(2,3)$, on the other hand, the limits appear symmetric because the flavor-conserving modes have a CKM-suppression of order $\cO(\lambda^2)$ and $\cO(\lambda^3)$, respectively, making these subleading with respect to the flavor-changing modes. Finally, the bounds on the LFV vector operators $\alpha<\beta$ and the dipole operators shown in the fourth and last row of Fig.~\ref{fig:single-WC-limits-lq} do not interfere with the SM, leading to perfectly symmetric bounds.

\subsubsection*{On the quark-flavor alignment}
While in the SM the only source of quark-flavor violation is the CKM matrix, this is generally no longer true in the presence of BSM Physics. In other words, the assumption concerning the flavor basis for quarks is fundamental, as a given operator can simultaneously contribute to several partonic processes depending on this choice, having a significant impact on the fits using LHC observables. The minimalistic approach is to consider right-handed fermions in the mass basis, and to assume the alignment between flavor and mass eigenstates for either left-handed up- or down-type quarks, with the CKM matrix appearing in the down- or up-type sectors, respectively. Scenarios with \emph{up-type alignment} turn out to be tightly constrained by $\Delta F= 1$ and $\Delta F= 2$ processes in the $K$- and $B$-meson sectors, which can be induced in this case via the CKM matrix~\cite{Isidori:2010kg}. The \emph{down-type alignment} is far less constrained by low-energy observable and it is typically a more convenient choice for phenomenology, as we have considered e.g.~in Fig.~\ref{fig:single-WC-limits-lq} and Appendix~\ref{sec:SMEFT-limits}.

We now explore the Drell-Yan processes induced via the CKM matrix for the operators that involve quark doublets, both for the \emph{up-} and \emph{down-type alignment}. We expect this effect to be prominent for SMEFT operators with second-generation quark indices.~\footnote{Similar effects are expected for fits in explicit New Physics models.} In these scenarios, the contribution of first generation valence quarks is only mildly Cabibbo suppressed $\cO(\lambda)$ and can thus compete with, and even dominate over, the direct contribution from the sea quarks. The effect of the alignment on fits of operators with first- or third-generation quarks is much less pronounced, because the leading contributions already comes from the valence quarks, or the Cabibbo suppression is stronger~($\cO(\lambda^3)$).

\begin{figure}[t!]
     \centering
    \begin{tabular}{r r r}
        \includegraphics[height=0.29\textwidth]{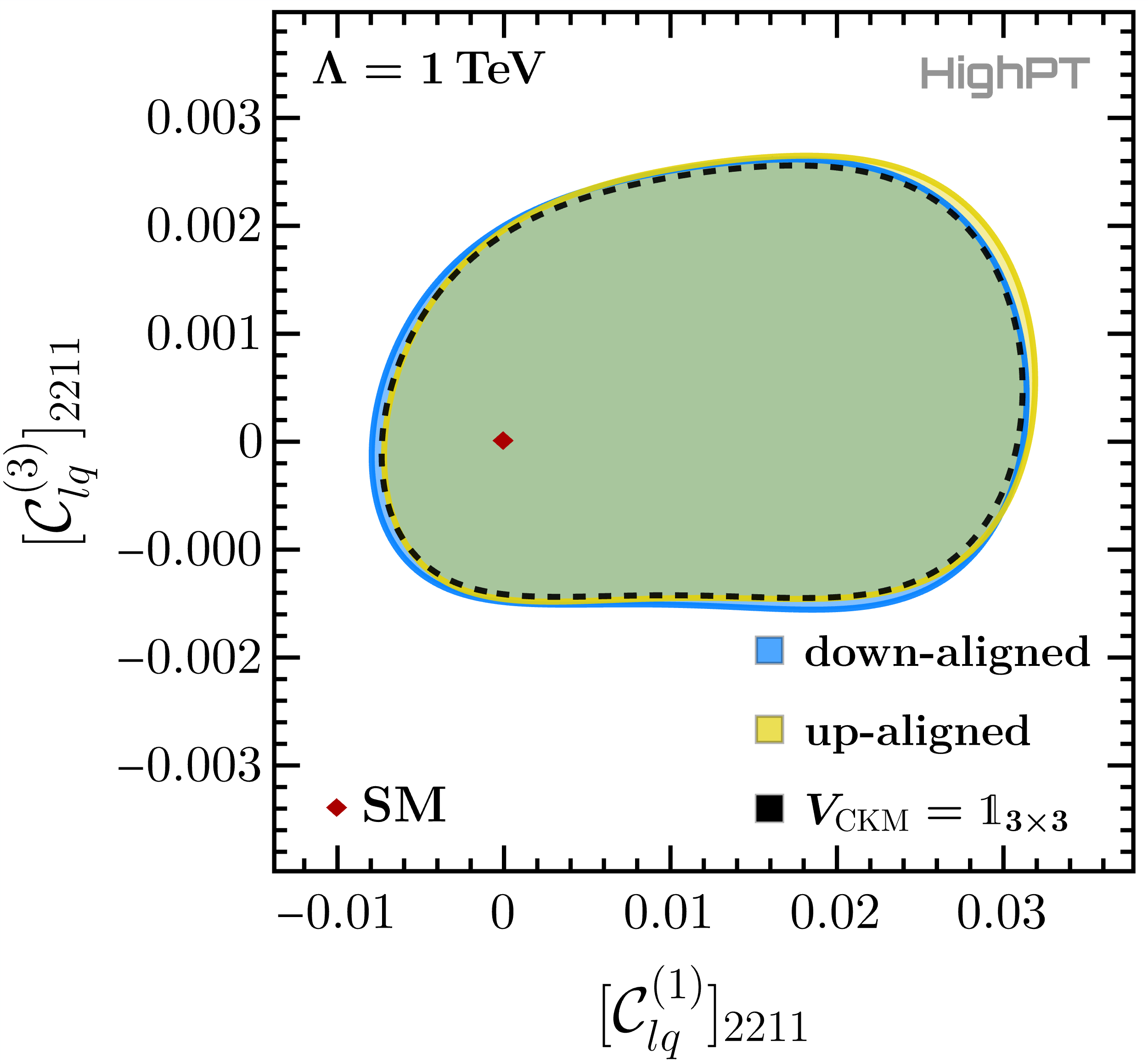} &
        \includegraphics[height=0.29\textwidth]{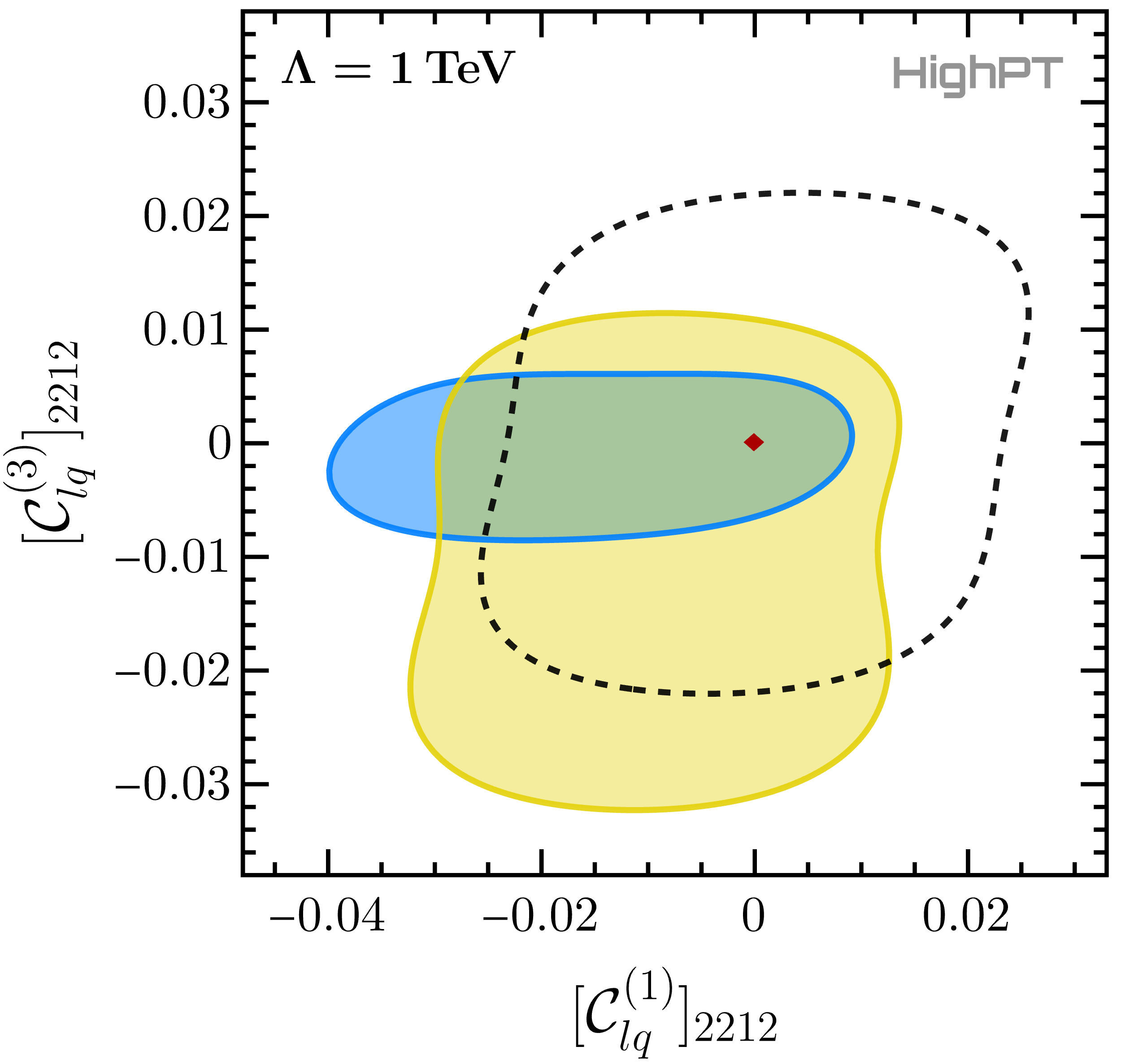} &
        \includegraphics[height=0.29\textwidth]{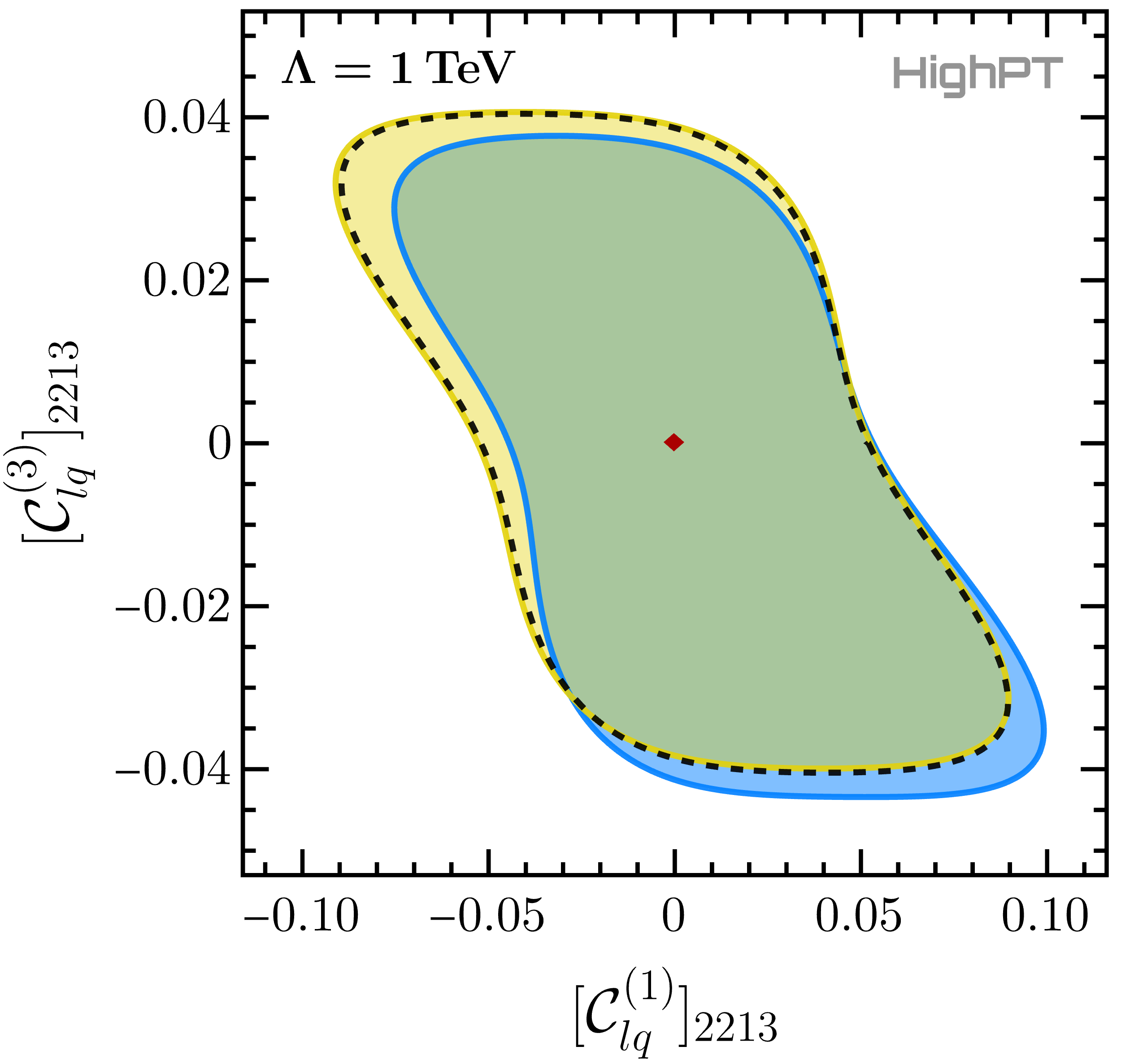} 
        \\
        \includegraphics[height=0.29\textwidth]{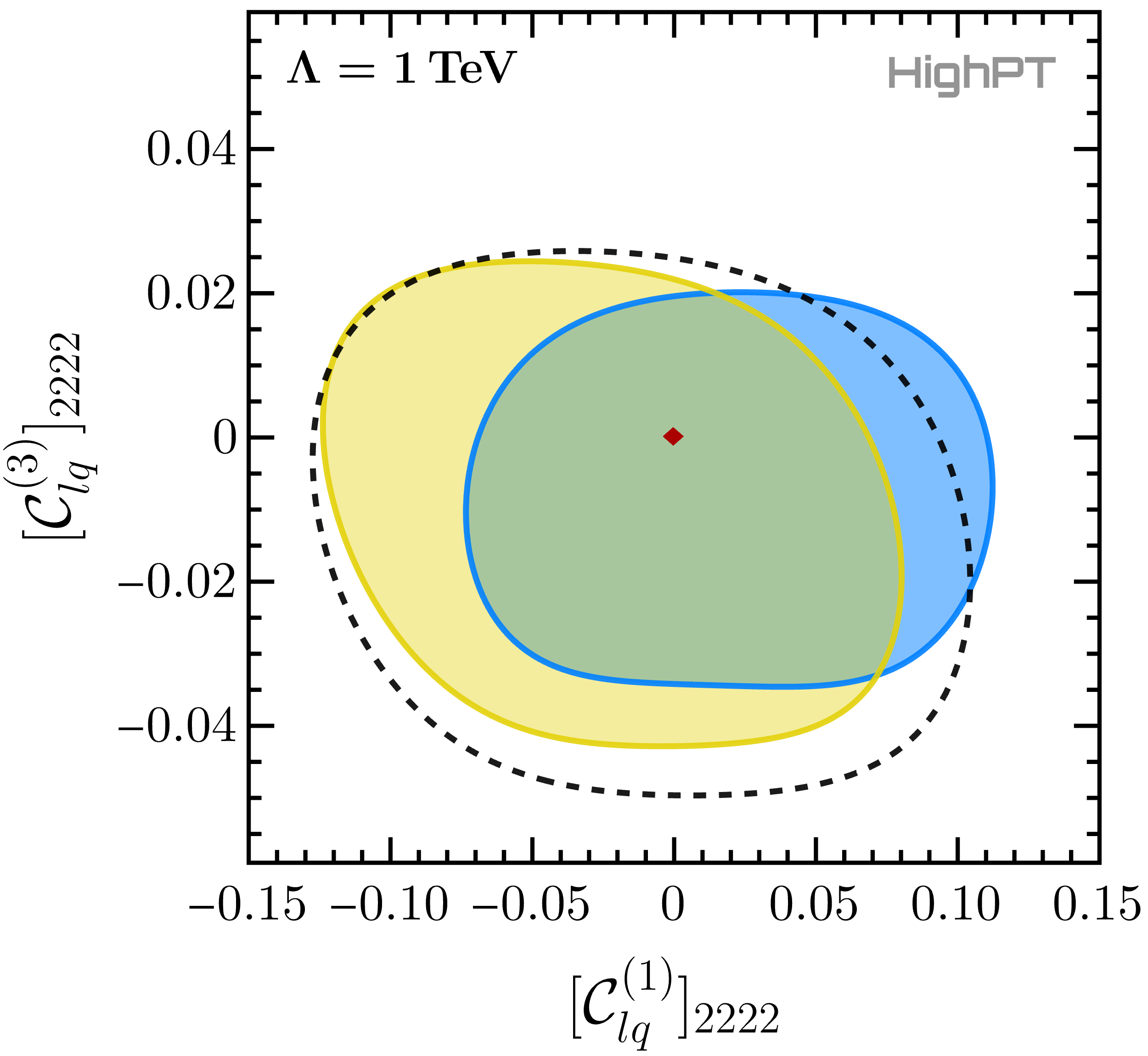} &
        \includegraphics[height=0.29\textwidth]{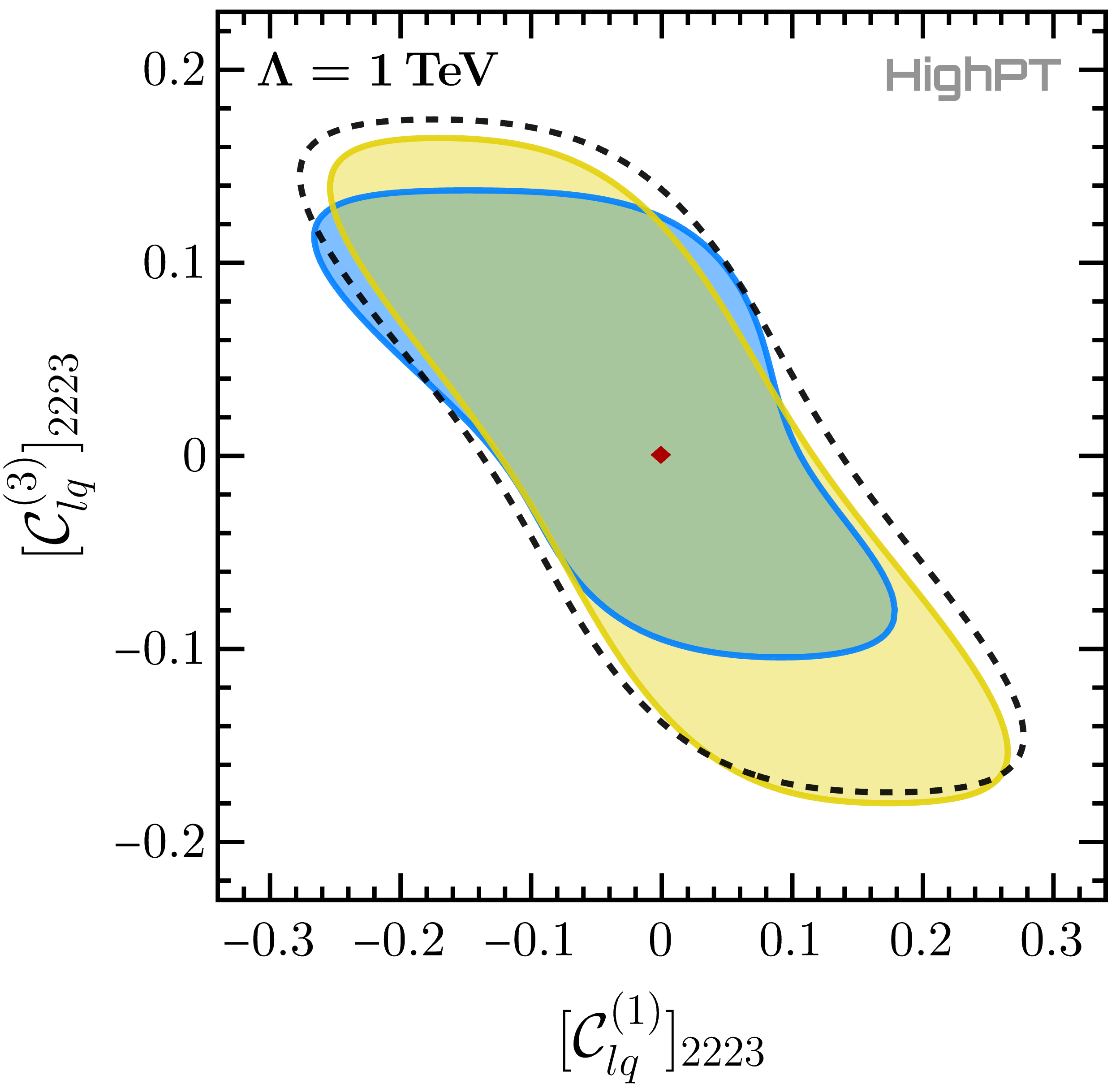} &
        \includegraphics[height=0.29\textwidth]{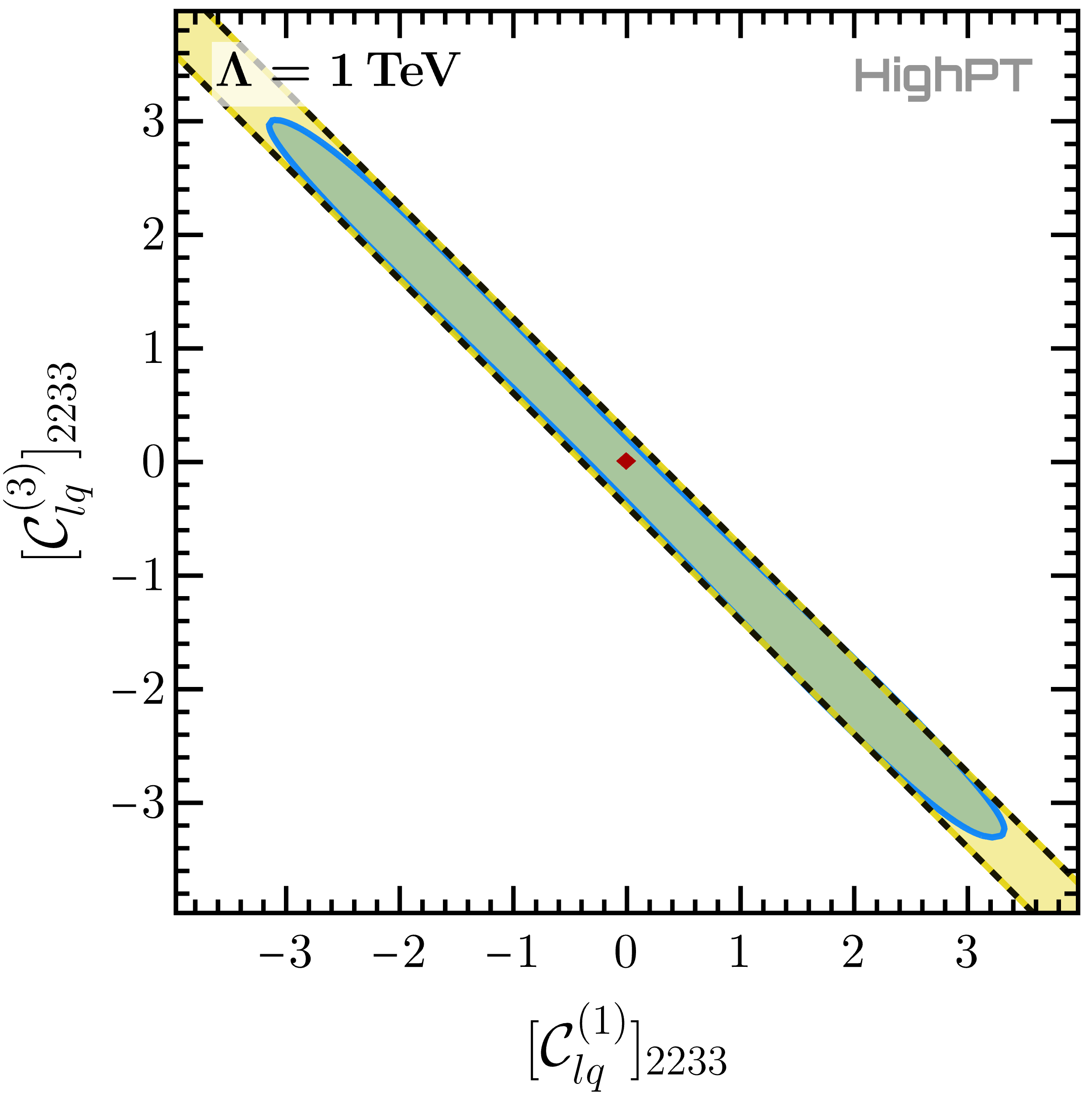}
    \end{tabular}
    \caption{\sl\small Two-parameter fit of the Wilson coefficients $\smash{[\cC_{lq}^{(1)}]_{22ij}}$ and $\smash{[\cC_{lq}^{(3)}]_{22ij}}$ for all allowed quark-flavor indices. We show the $95\%$ confidence regions for these coefficients assuming $V_\mathrm{CKM}=\mathbb{1}$ (black dashed line), a non-diagonal $V_\mathrm{CKM}$ with down alignment (blue region), and a non-diagonal $V_\mathrm{CKM}$ with up alignment (yellow region). Both dimuon and monomuon searches are considered for the constraints presented here. The coefficients are normalized by choosing~$\Lambda=1\,\mathrm{TeV}$.}
    \label{fig:basis_alignment}
\end{figure}

As an example, we investigate the Wilson coefficients $\smash{[\cC_{lq}^{(1)}]_{22ij}}$ and $\smash{[\cC_{lq}^{(3)}]_{22ij}}$ for all possible combinations of quark-flavor indices~$(i,j)$. We checked that similar results are obtained for other choices of lepton-flavor indices. Our results are presented in Fig.~\ref{fig:basis_alignment}, where we show the $95\%$~confidence regions for three different scenarios:
\begin{itemize}
    \item Assuming a (fictitious) diagonal CKM~matrix $V_\mathrm{CKM}=\mathbb{1}_{3 \times 3}$ (black dashed lines) only sea-quarks can contribute to operators with second and third generation quarks. The constraints obtained in this scenario are weaker than those obtained considering a non-diagonal~$V_\mathrm{CKM}$ due to the missing valence quark contributions. 
    \item Considering the CKM-matrix and assuming \emph{down alignment} (blue region), processes with up quarks can also contribute to the constraints for operators with charm quarks. The contribution of the former is suppressed by $\cO(\lambda)$, but this effect can be compensated by the enhancement of the up PDF with respect to the charm PDF. The constraints thus obtained are therefore stronger than the constraints obtained with a diagonal~$V_\mathrm{CKM}$. Furthermore, the confidence regions can be shifted with respect to the previous scenario due to the additional interference term of the New Physics contribution and the SM contribution with valence quarks.
    \item Considering the  CKM-matrix and assuming \emph{up alignment} (yellow region), operators with strange quarks receive contributions induced by the down-quark PDF, leading to a behaviour analogous to the down-aligned scenario. However, since the ratio of down to strange PDF is smaller than the ratio of up and charm PDF, the constraints obtained in this scenario are weaker than in the down-aligned case.
\end{itemize}
For operators with only first generation quarks the alignment has a negligible effect since the leading contribution to the constraints is already obtained by the valence quarks, see e.g. top left plot in Fig.~\ref{fig:basis_alignment}. As discussed before, the dominant effect of the alignment is obtained for operators with second generation quarks (center top, left bottom, and center bottom). For third generation quarks (right top, and right bottom) only a small effect of the alignment is found as the PDF enhancement cannot overcome the stronger Cabibbo suppression~$\cO(\lambda^3)$ in this case. The plot in the bottom right shows a flat direction due to the missing contribution from top quarks. 
Thus, the constraints in this case all stem from a single form-factor, which is related to both $\smash{[\cC_{lq}^{(1)}]_{2233}}$ and~$\smash{[\cC_{lq}^{(3)}]_{2233}}$, leading to the flat direction (see Eq.~\eqref{eq:FF_V_00_LL_dd}).
In the case of down alignment there is a non-vanishing top-quark contribution due to CKM rotations, breaking the flat direction. The bounds obtained this way are however quite weak.

In the analysis presented in this Section we restricted ourselves to the choices of up and down alignment or a diagonal CKM. However, our results are derived with \HighPT which also allows to define any arbitrary alignment of mass and flavor basis.
%

%%%%%%%%%%%%%%%%%%%
\begin{figure}[p!]
    \centering
    \includegraphics[width=.31\linewidth]{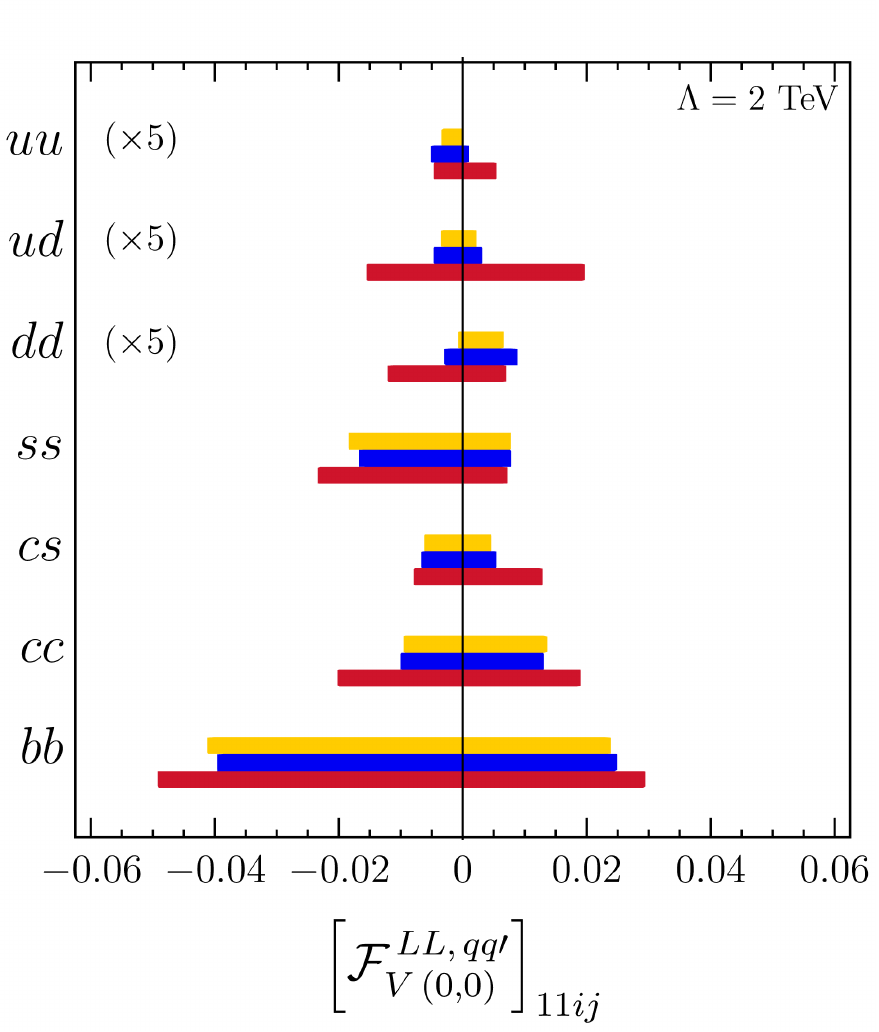} ~~
    \includegraphics[width=.31\linewidth]{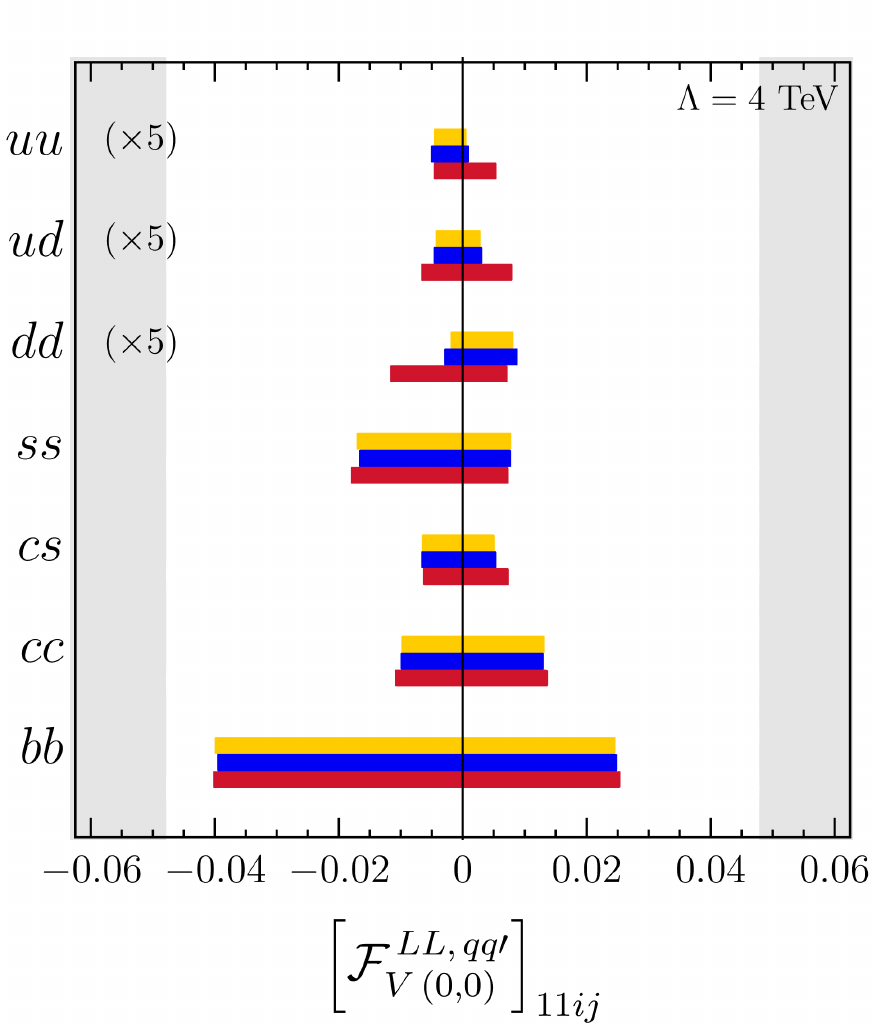} ~~    
    \includegraphics[width=.31\linewidth]{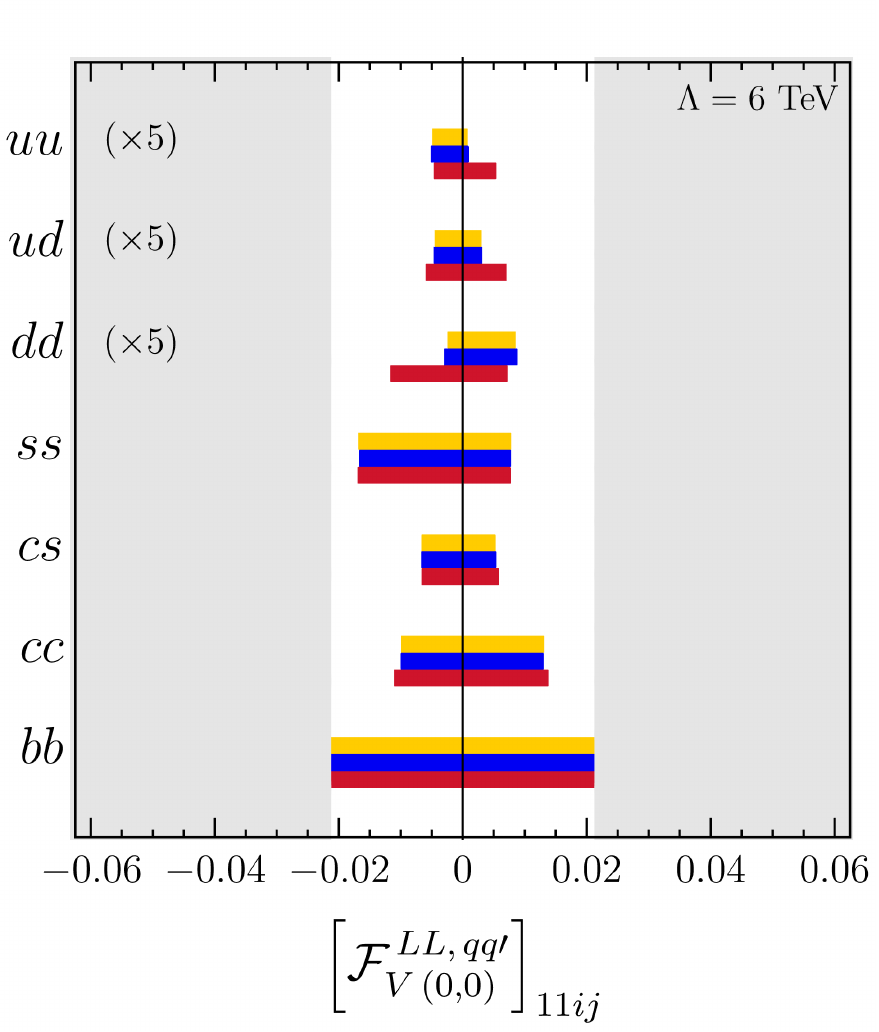}\\
    \vspace{0.5cm}
    \includegraphics[width=.31\linewidth]{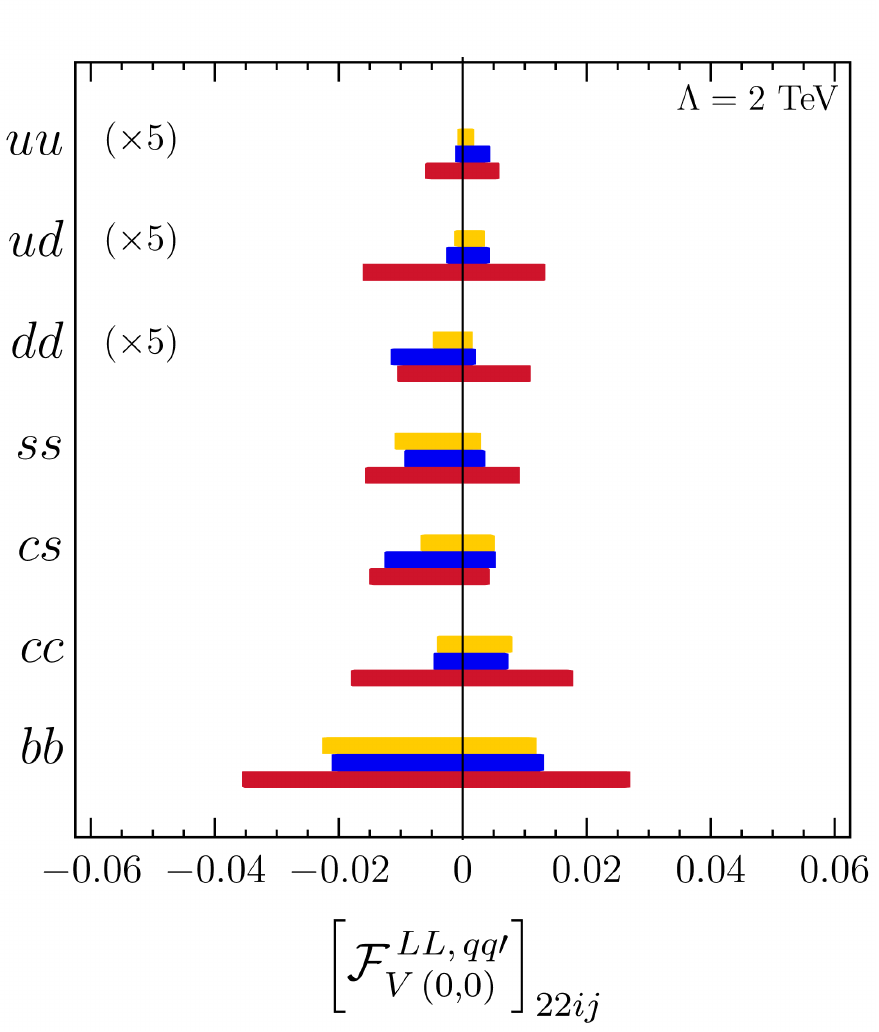} ~~
    \includegraphics[width=.31\linewidth]{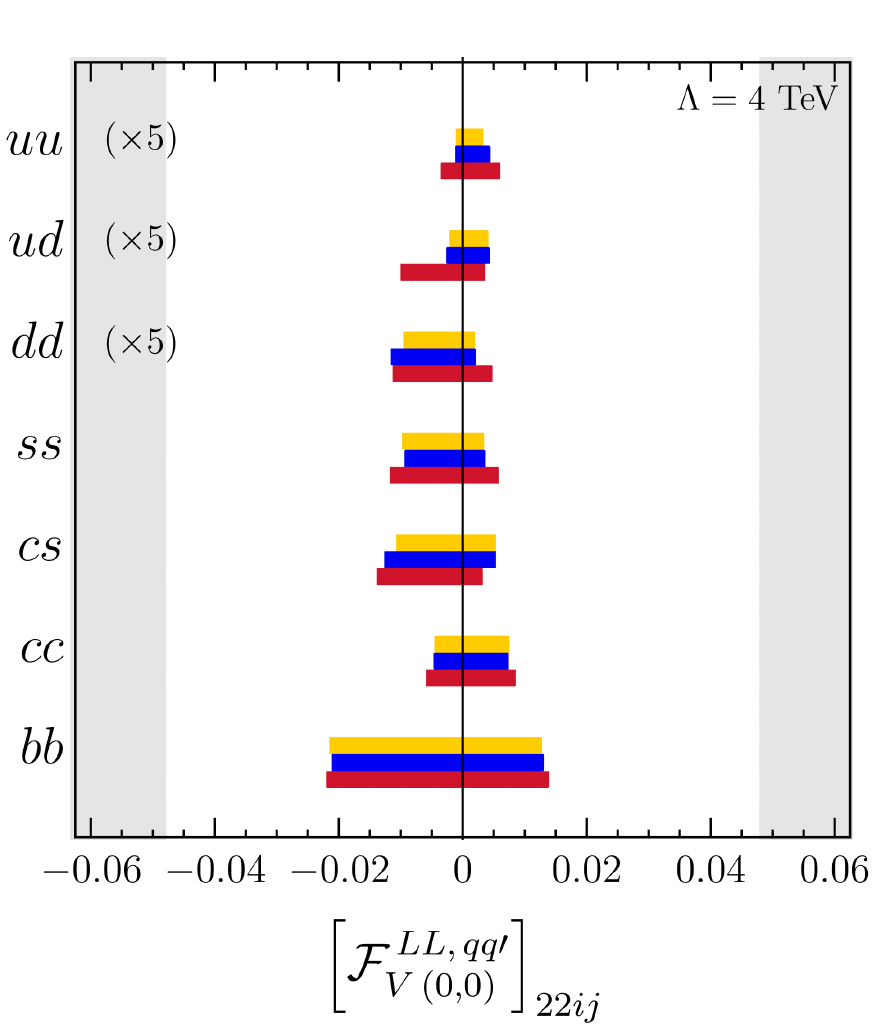} ~~  
    \includegraphics[width=.31\linewidth]{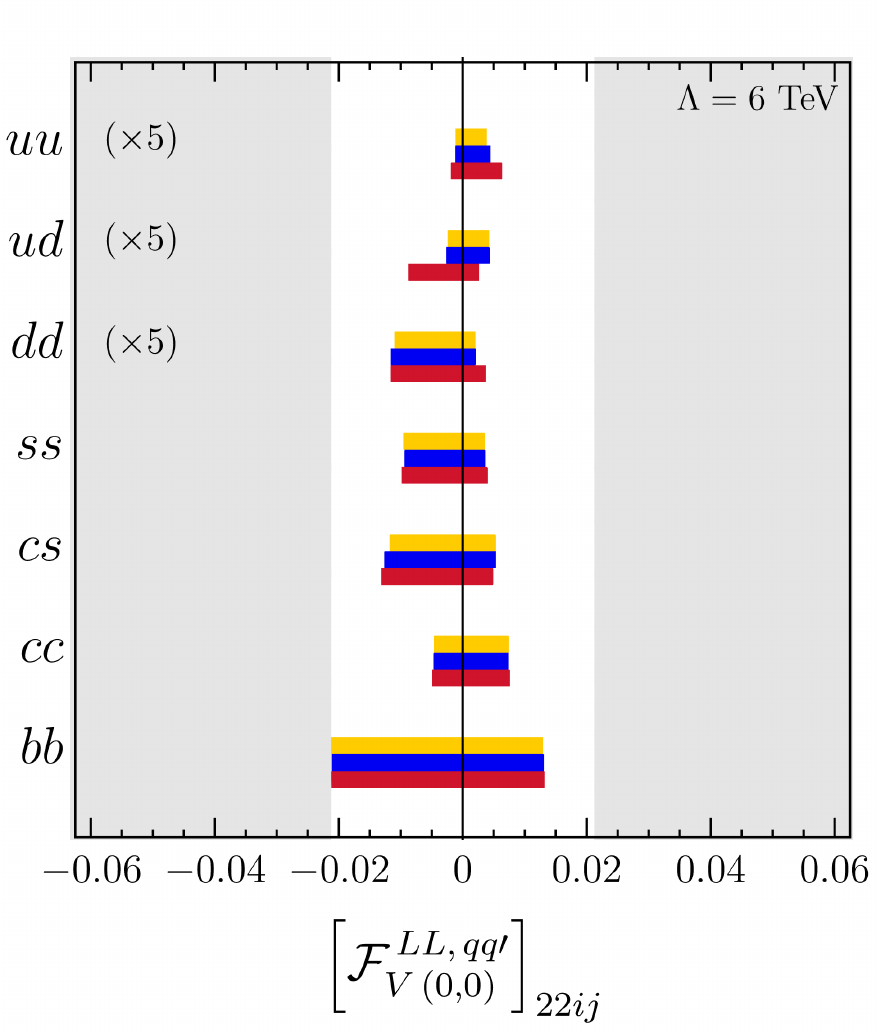}\\
    \vspace{0.5cm}
    \includegraphics[width=.31\linewidth]{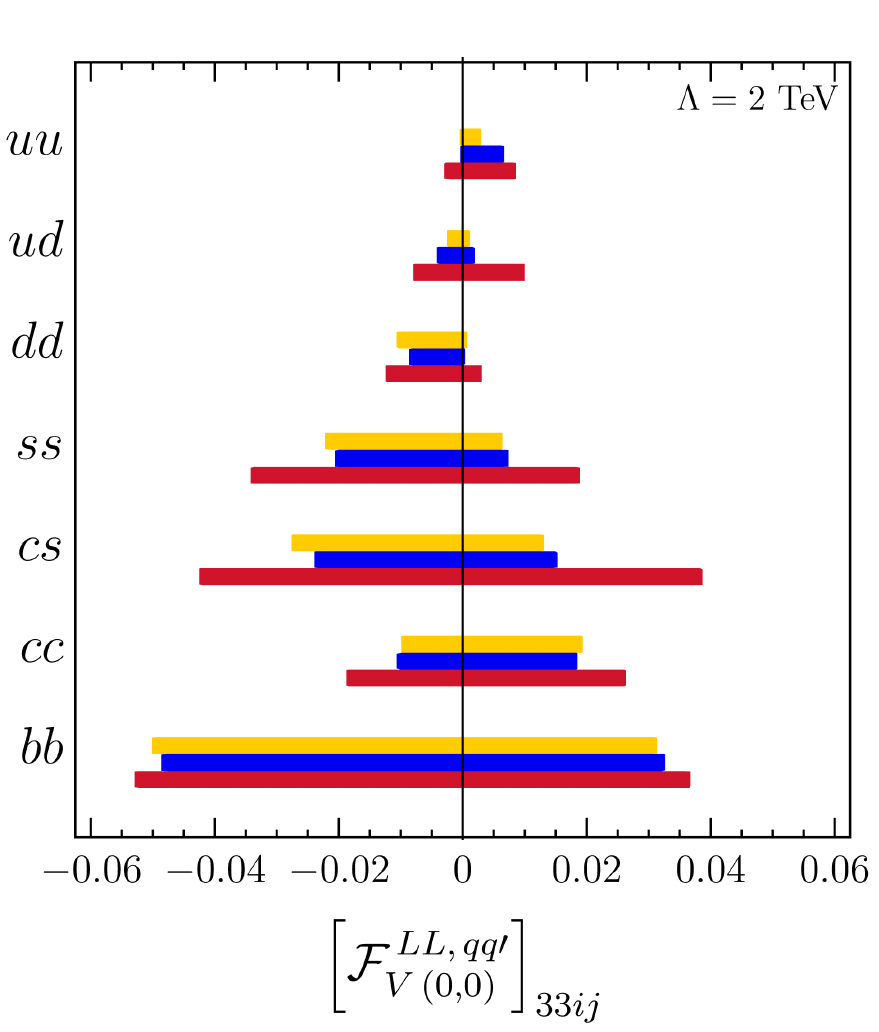} ~~
    \includegraphics[width=.31\linewidth]{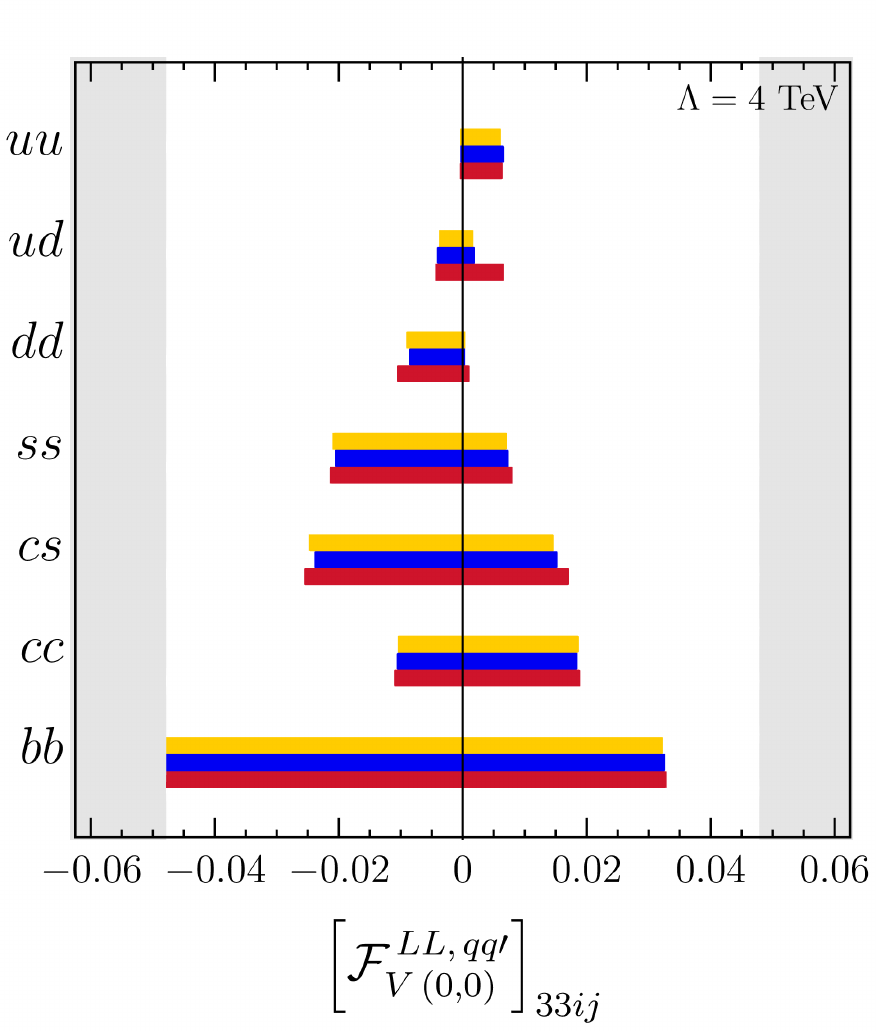} ~~  
    \includegraphics[width=.31\linewidth]{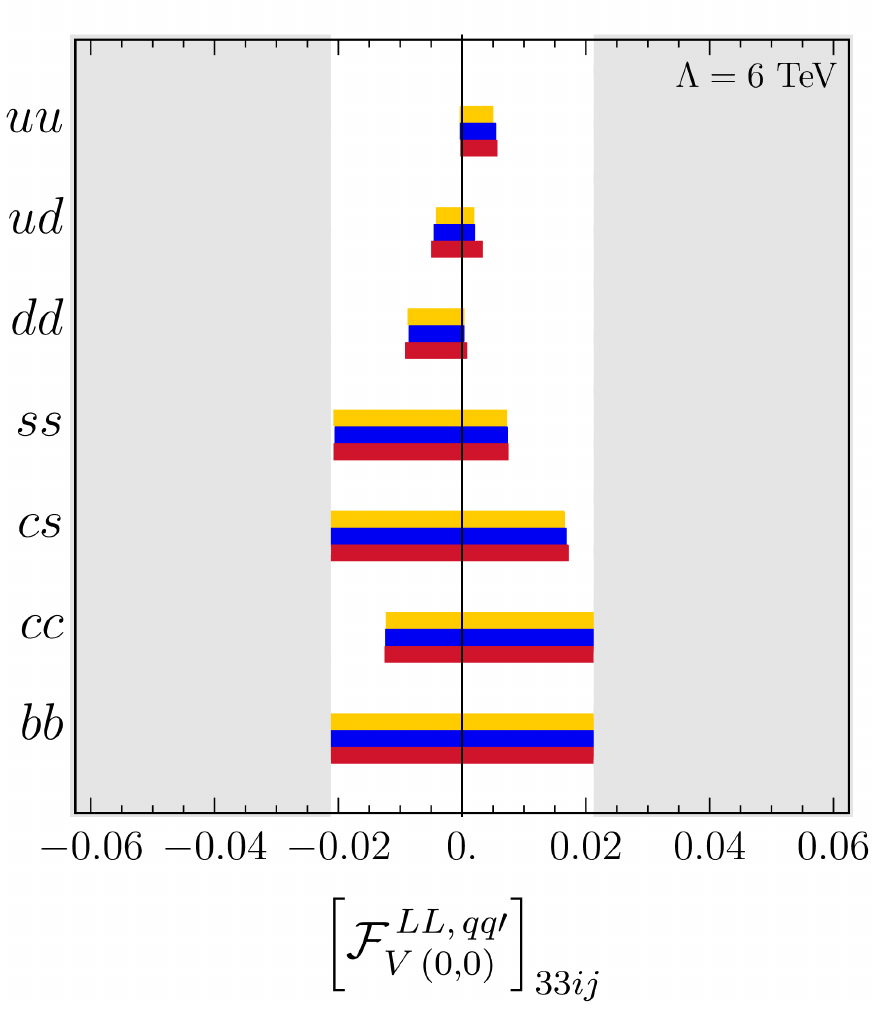} 
    \caption{\sl\small 95\% CL limits on the form-factor coefficients $\cF^{LL}_{V\,(0,0)}$ for first generation (top panel), second generation (middle panel) and third generation (bottom panel) leptons using LHC run-II searches in the dilepton and monolepton channels. Blue intervals are single-parameter limits, the red ones are limits marginalizing over momentum-dependent effects from $d=8$ operators, and the yellow ones are the expected limits from $d=6$ and $d=8$ operators assuming a specific UV scenario. For the first two rows, the limits for $uu,\, dd,\,ud$ have been rescaled by a factor of $5$ for visibility. The gray bands, given by $\smash{|\cF^{LL,\,qq^\prime}_{I\,(0,0)}|\ge 4\pi\,v^2/\Lambda^2}$, correspond to the region where perturbative unitarity is expected to break down. See text for more details.}
    \label{fig:FFs_limits_6vs8}
\end{figure}
%%%%%%%%%%%%%%%%%%%

\subsection{Impact of dimension-$8$ effects}

Up to now, we have neglected the effect of dimension-$8$ operators that contribute to the EFT truncation at $\cO(1/\Lambda^4)$. The impact of these higher-order corrections in Drell-Yan has been recently discussed in Refs.~\cite{Boughezal:2021tih,Boughezal:2022nof,Kim:2022amu} for searches with light leptons without including detector effects. These works have shown that including the $d=8$ contributions can have a significant impact when fitting SMEFT operators to Drell-Yan data at the LHC. In this Section we extend these analyses for generic flavor structures using the latest run-II LHC dilepton and monolepton searches. We present exclusion limits on the leading regular vector form-factor coefficient $\smash{\cF_{V\,(0,0)}}$ in the presence/absence of dimension-$8$ effects. 

For concreteness, we focus on the left-handed form-factor coefficients $\smash{\cF^{LL,qq^\prime}_{V\,(0,0)}}$. In the SMEFT, these are generated to lowest order by the $d=6$ operators $\smash{\cO^{(1,3)}_{lq}}$: 
\begin{align}\label{eq:ffLL00}
\cF_{V\,(0,0)}^{LL,\,uu}&\ \simeq\  \frac{v^2}{\Lambda^2} \,\left(\cC_{lq}^{(1)}-\cC_{lq}^{(3)}\right)\,,\\[0.3em]
\cF_{V\,(0,0)}^{LL,\,dd}&\ \simeq\  \frac{v^2}{\Lambda^2} \,\left(\cC_{lq}^{(1)}+\cC_{lq}^{(3)}\right)\,,\label{eq:FF_V_00_LL_dd}\\[0.3em]
\cF_{V\,(0,0)}^{LL,\,ud} &\ \simeq\  2\frac{v^2}{\Lambda^2}\,\cC^{\,(3)}_{lq}\,,
\end{align}
where we have dropped the subleading contributions of $\smash{\cO(v^4/\Lambda^4)}$ arising from the 12 dimension-$8$ operators $\smash{\cO^{(k)}_{l^2q^2H^2}}$, $\smash{\cO^{(k)}_{l^2H^2D^3}}$,  $\smash{\cO^{(k)}_{q^2H^2D^3}}$ with $k=1,2,3,4$ in the classes {\bf$\psi^4H^2$} and {\bf $\psi^2H^2D^3$} defined in Tables~\ref{tab:dim8_ops_1} and \ref{tab:dim8_ops_2} (for the exact expressions see Eqs.~\eqref{eq:FVLL00uu}, \eqref{eq:FVLL00dd} and \eqref{eq:FVLL00ud}). In general, these higher-order corrections will have a small impact when constraining the SMEFT and can be safely ignored in the remainder of this discussion \cite{Boughezal:2021tih}.

The contributions from momentum-dependent dimension-$8$ operators in class {\bf$\psi^4D^2$}, however, can be relevant when fitting the SMEFT to LHC data. The four operators $\smash{\cO_{l^2q^2D^2}^{(1,2,3,4)}}$, defined in Table~\ref{tab:dim8_ops_1}, match to the form-factor coefficients $\smash{\cF^{LL}_{V\,(1,0)}}$ and $\smash{\cF^{LL}_{V\,(0,1)}}$. Explicitly, one finds at leading order
\begin{align}
\cF_{V\,(1,0)}^{LL,\,uu} &\ \simeq\ \frac{v^4}{\Lambda^4} \,\cC_{l^2q^2D^2}^{\,(1+2-3-4)}\ \,, 
&
\cF_{V\,(0,1)}^{LL,\,uu}   &\ \simeq\    2\frac{v^4}{\Lambda^4}\, \cC_{l^2q^2D^2}^{\,(2-4)}\,, \\[0.3em]
\cF_{V\,(1,0)}^{LL,\,dd} &\ \simeq\  \frac{v^4}{\Lambda^4} \,\cC_{l^2q^2D^2}^{\,(1+2+3+4)}\ \,, 
&
\cF_{V\,(0,1)}^{LL,\,dd}   &\ \simeq\   2\frac{v^4}{\Lambda^4}\, \cC_{l^2q^2D^2}^{\,(2+4)}\,,\\[0.3em]
\cF_{V\,(1,0)}^{LL,\,ud} &\ \simeq\  2\frac{v^4}{\Lambda^4}\,\cC^{\,(3+4)}_{l^2q^2D^2} \,,
&
\cF_{V\,(0,1)}^{LL,\,ud} &\ \simeq\  4\frac{v^4}{\Lambda^4}\,\cC^{\,(4)}_{l^2q^2D^2}\,,\label{eq:ffLL01}
\end{align}
where the notation $\cC^{\,(1\,\pm\, 2\,\pm\cdots)}$ corresponds to the signed sum of Wilson coefficients defined in~Eq.~\eqref{eq:signed_sum_WC} in Appendix~\ref{sec:conventions}. In the amplitude, the energy-enhancement associated with these form-factor coefficients can potentially overcome the relative suppression of $\cO(v^2/\Lambda^2)$ with respect to $\cF^{LL,\,qq\prime}_{V\,(0,0)}$, specially in the high-$p_T$ tails of the differential distributions. 

In order to examine the importance of these dimension-$8$ effects, we set bounds on the leading dimension-$6$ form-factors $\cF^{LL}_{V\,(0,0)}$ with different flavor structures for three scenarios: 

\begin{itemize}
    \item[\bf{i)}]
    Single-parameter limits on $\smash{\cF^{LL}_{V\,(0,0)}}$ while completely neglecting the dimension-$8$ coefficients $\smash{\cF^{LL}_{V\,(1,0)}}$ and $\smash{\cF^{LL}_{V\,(0,1)}}$.
    
    \item[\bf{ii)}]
    Single-parameter limits on the $\smash{\cF^{LL}_{V\,(0,0)}}$ form-factor while turning on $\smash{\cF^{LL}_{V\,(1,0)}}$ such that $\smash{\cF^{LL}_{V\,(1,0)}=(v^2/\Lambda^2)\,\cF^{LL}_{V\,(0,0)}}$. This specific correlation between different order coefficients can arise from a ultraviolet setting where a heavy $s$-channel vector mediator has been integrated out at the scale $\Lambda$.
    
    \item[\bf{iii)}] 
    Marginalized limits on $\cF^{LL}_{V\,(0,0)}$ while profiling over $\cF^{LL}_{V\,(1,0)}$ and $\cF^{LL}_{V\,(0,1)}$. To enforce the correlations between dimension-$8$ form-factors that arise in the SMEFT from $SU(2)_L$ invariance, we marginalize over the quantities $\smash{2\,\cF^{LL}_{V\,(1,0)}-\cF^{LL}_{V\,(0,1)}}$ and $\smash{\cF^{LL}_{V\,(0,1)}}$. Notice that these two parameters map at $\cO(1/\Lambda^4)$ to independent Wilson coefficient combinations, each proportional to $\smash{\cC_{l^2q^2D^2}^{(1\pm3)}}$ ($\smash{\cC_{l^2q^2D^2}^{(3)}}$) and $\smash{\cC_{l^2q^2D^2}^{(2\pm4)}}$ ($\smash{\cC_{l^2q^2D^2}^{(4)}}$) for neutral (charged) currents, respectively.
    
\end{itemize}

\noindent While here we have focused on left-handed currents, a similar analysis can be performed with other form-factor helicities leading to similar conclusions. The marginalized limits in item {\bf iii)} are extracted using the {\it profiled} chi-square statistic $\chi^2(\theta,\widehat{\widehat{\nu}}(\theta))$ where $\theta$ represent the parameters of interest and $\nu$ the marginalized parameters. The double-hat notation \cite{ParticleDataGroup:2020ssz} is defined as the values of $\nu$ that minimize the $\chi^2(\theta,\nu)$ function for a given $\theta$. 

Furthermore, when minimizing the $\chi^2$ functions we constrain the form-factors to take values in the ranges $|\cF_{I\,(0,0)}|\le v^2\cC_*/\Lambda^2$, $|\cF_{I\,(1,0)}|\le v^4\cC_*/\Lambda^4$ and  $|\cF_{I\,(0,1)}|\le v^4\cC_*/\Lambda^4$, where we set $\cC_*=4\pi$, i.e.~the value for which the Wilson coefficients in Eqs.~\eqref{eq:ffLL00}--\eqref{eq:ffLL01} would enter the non-perturbative regime, assuming these come from a tree-level matching in the ultraviolet. Notice that this particular choice for $\cC_*$ has minimal impact when extracting the single-parameter limits on $\cF_{V\,(0,0)}$. In contrast, when including the effects of dimension-$8$ corrections via $\cF_{V\,(1,0)}$ and/or $\cF_{V\,(0,1)}$, the choice of $\cC_*$, which depends on the details of the UV completion, can affect substantially the fits. In this work we use the strong-coupling limit $\cC_*=4\pi$ as a benchmark since it is the choice that maximizes the effects of the dimension-$8$ operators. The LHC limits at 95\% CL for the form-factors can be found in Fig.~\ref{fig:FFs_limits_6vs8} for first generation (first row), second generation (second row) and third generation (third row) leptons at three different EFT cutoff choices: $\Lambda=2$\,TeV (left column), $\Lambda=4$\,TeV (middle column) and $\Lambda=6$\,TeV (right column). The limits for valence quarks in the first two rows have been rescaled by a factor of five for visibility and the gray regions correspond to the strong-coupling regime $\cC_*\geq4\pi$ discussed above. 

We show in blue the upper limits for $\cF_{V\,(0,0)}$ for scenario {\bf{i)}} i.e.~when all $d=8$ corrections are neglected. As expected, these are  mostly independent of the cutoff choice. In yellow we display the limits on $\cF_{V\,(0,0)}$ for scenario {\bf{ii)}} where the $d=6$ and $d=8$ Wilson coefficients are maximally correlated, e.g.~in $Z^\prime$ ($W^\prime$) models for neutral (charged) currents. One can see that these limits are most of the time similar to the blue ones for any quark flavor at $\Lambda=4~\mathrm{TeV}$ and $6~\mathrm{TeV}$, and for sea-quarks already at $2$\,TeV. This shows that for these cases the convergence of the EFT is good since the $d=8$ corrections have a small effect. This is not entirely surprising for these values of the cutoff given that the Wilson coefficients arise from expanding the same massive propagator ($s$-channel in this case) where the $d=8$ terms will always be parametrically suppressed with respect to the $d=6$ ones. However, for $uu$, $dd$ and $ud$ initial quarks at $\Lambda=2$\,TeV, the inclusion of $d=8$ corrections can strengthen the limits substantially and also lead to a slightly better fit to the observed data. This indicates that the EFT expansion is not to be trusted for valence quarks at such low cutoff scales. In red, we provide the limits for scenario {\bf{iii)}}, i.e.~marginalizing over $d=8$ contributions that are completely uncorrelated from the $d=6$ ones. Here we find that for valence quarks, marginalizing over the $\smash{\cF^{LL}_{V\,(1,0)}}$ and $\smash{\cF^{LL}_{V\,(0,1)}}$ lead to substantial corrections to $\smash{\cF^{LL}_{V\,(0,0)}}$ even at the large cutoff scales $\Lambda=4~\mathrm{TeV}$ and $6~\mathrm{TeV}$. Notice that the main effect is to both weaken and symmetrize the skewed $d=6$ limits in blue. The symmetrizing effect arises since the sign of the $d=8$ interference term is decorrelated from the sign of the $d=6$ term. For sea-quarks these effects are completely mitigated due to the limited energy reach from the PDF suppression. 

In summary, these results show that the effects of $d=8$ operators are only relevant for first generation (valence) quarks and only if the dimension-$6$ and $d=8$ Wilson coefficients are uncorrelated. This last condition likely requires a non-trivial cancellation of different contributions to $d=6$ operators from the ultraviolet. For example, in New Physics scenarios where a single heavy tree-level mediator is integrated out the dimension-$6$ and dimension-$8$ Wilson coefficients are expected to be maximally correlated.

\subsection{Constraints on leptoquark models}

In this Section we provide the LHC constraints on the couplings of the leptoquark states listed in Table~\ref{tab:mediators}. As already discussed, leptoquarks can be non-resonantly exchanged in the $t/u$-channels, as shown in Fig.~\ref{fig:mediator_diagrams}. We begin by setting upper limits on individual couplings at a time for a fixed leptoquark mass of $2$~TeV and a negligible width. For each leptoquark state, we take into account all components contributing to the relevant production channels and assume these to be mass-degenerate.~\footnote{For example, the $S_3\sim({\bf \bar3},{\bf 3},1/3)$ state being a $SU(2)_L$ triplet will contribute to $\bar d_i d_j\to\ell_\alpha\ell_\beta$ via the $S_3^{(4/3)}$ component and to $\bar u_i u_j\to\ell_\alpha\ell_\beta$ and $\bar d_i u_j\to\ell_\alpha\nu_\beta$ via the $S_3^{(1/3)}$ component.} We also include the interference with the SM background if present. Even though we explicitly include in our analysis the full effect of the propagators, it is worth noting that since leptoquarks are non-resonant it is possible to set fairly reliable limits on their couplings by using the constraints on the corresponding SMEFT operators for masses above a few $\mathrm{TeV}$, see e.g.~Refs.~\cite{Iguro:2020keo,Jaffredo:2021ymt}.

\begin{figure}[p!]
    \centering
    \resizebox{0.98\textwidth}{!}{
    \begin{tabular}{c c c}
        \includegraphics[]{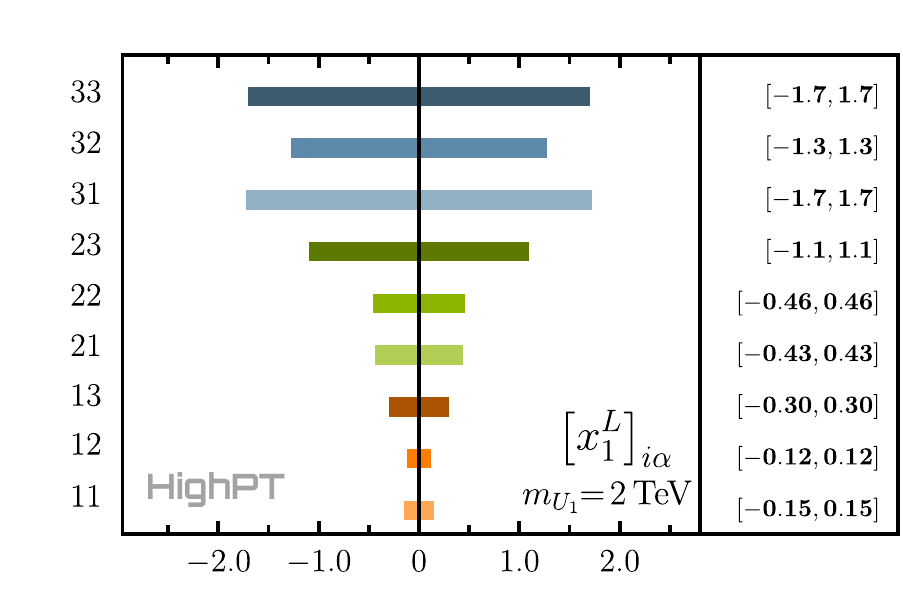} & \includegraphics[]{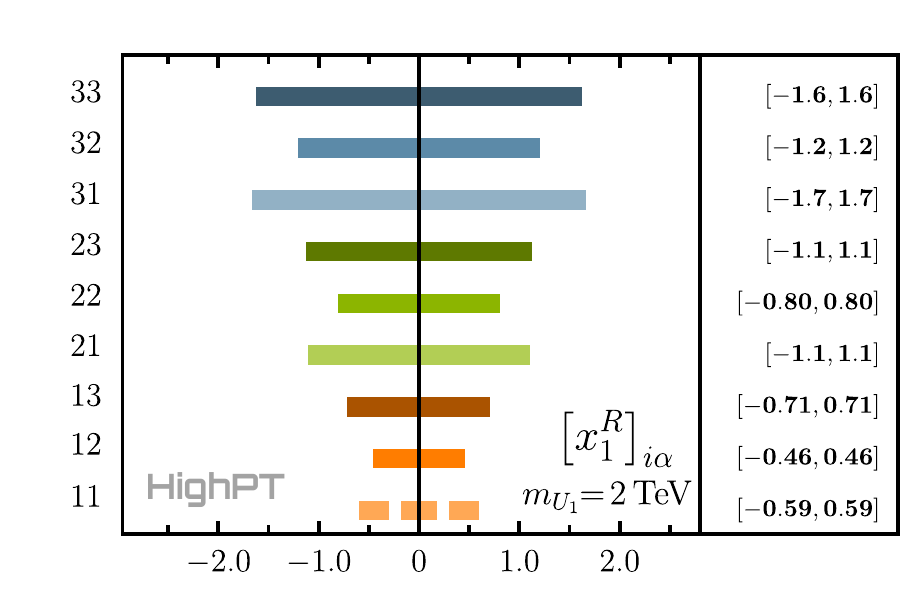} & \includegraphics[]{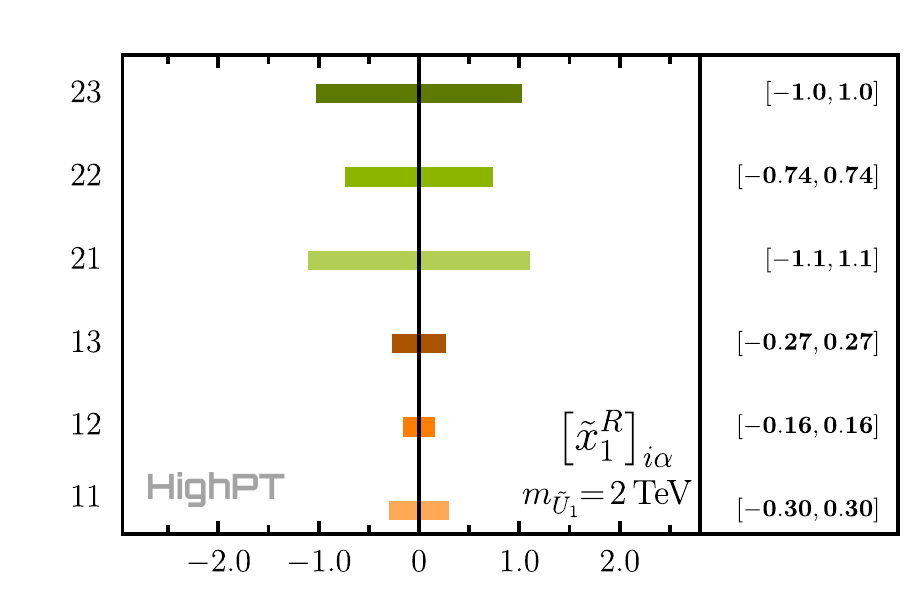}
        \\
        \includegraphics[]{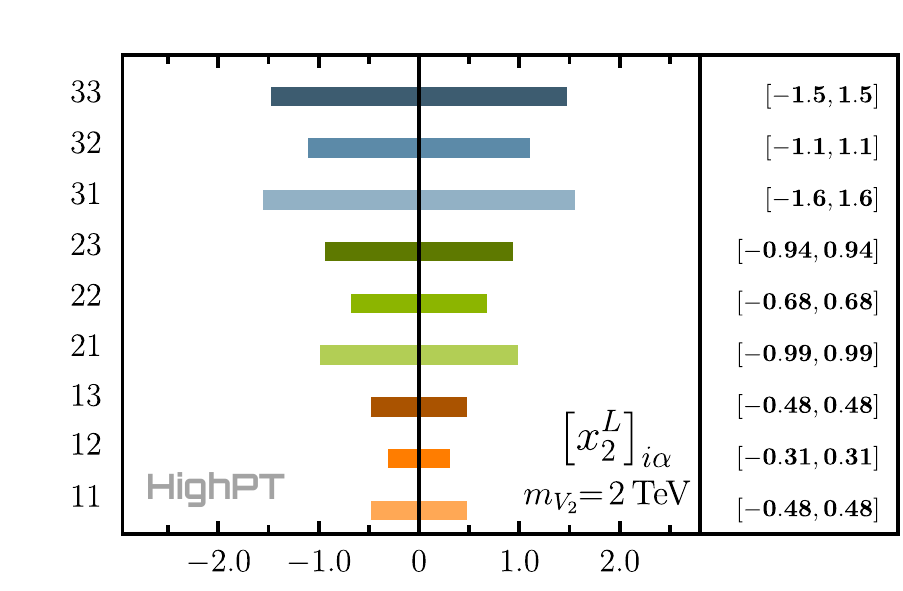} & \includegraphics[]{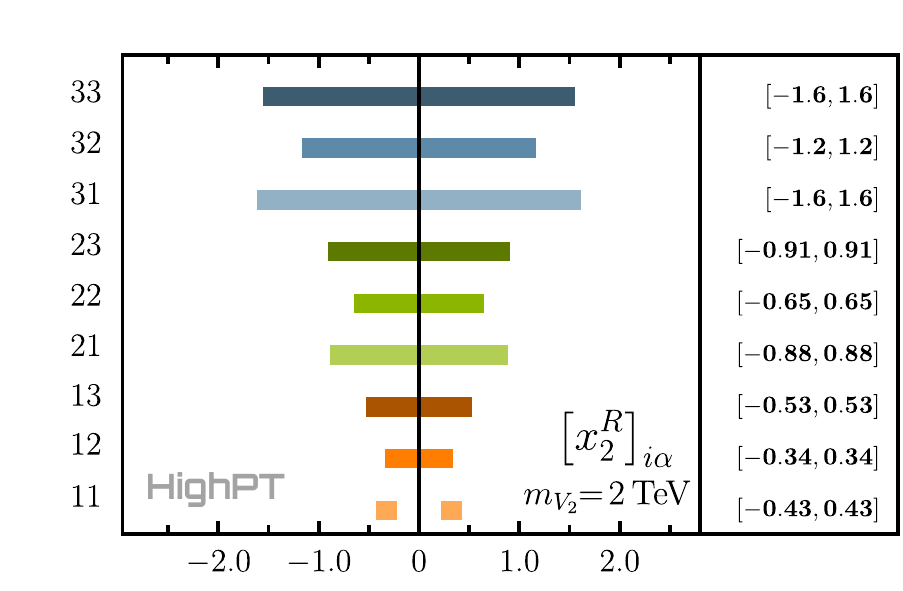} & \includegraphics[]{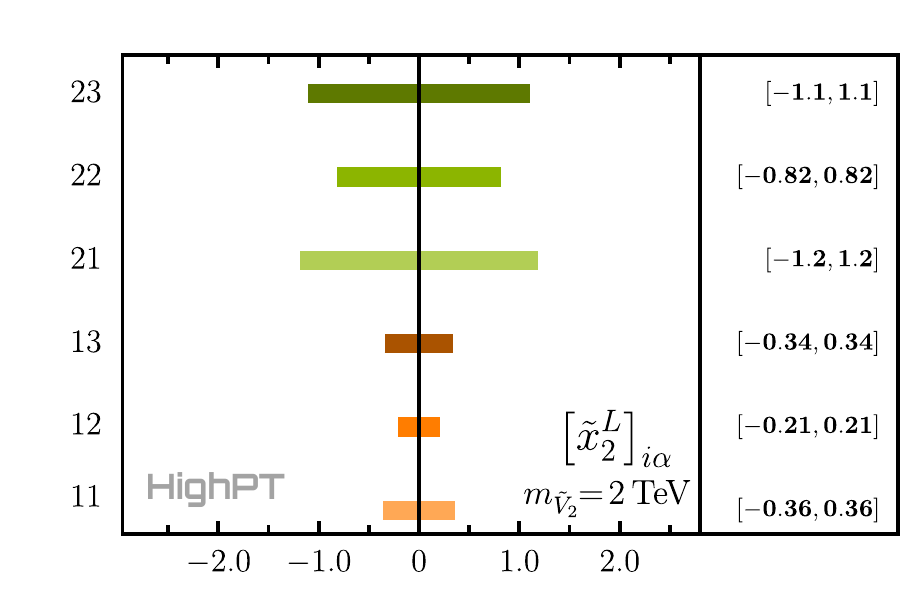}
        \\
        \includegraphics[]{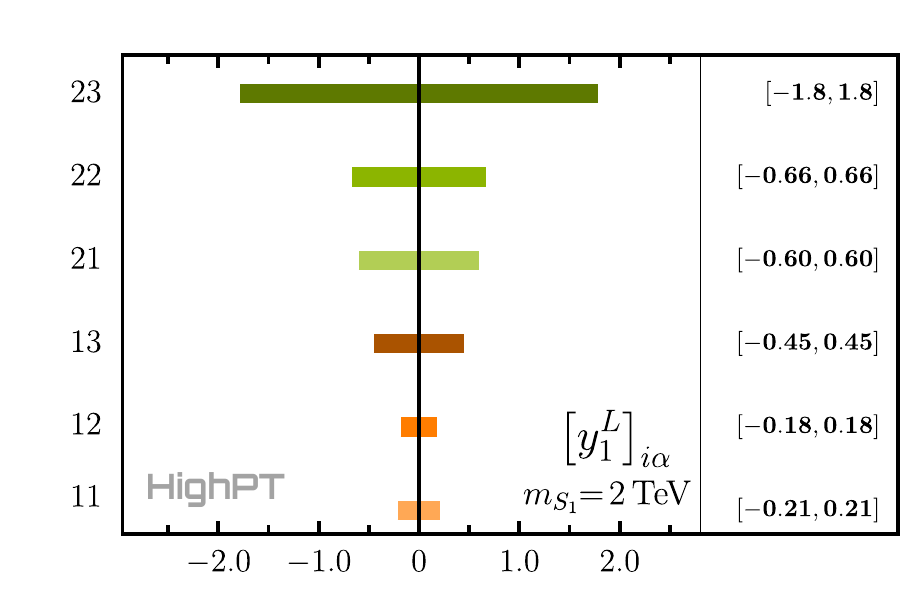} & \includegraphics[]{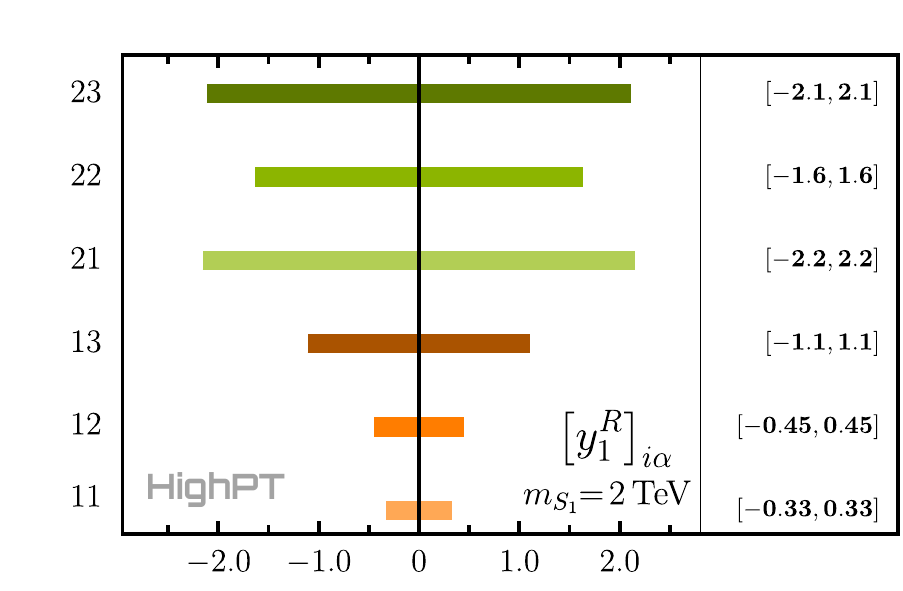} & \includegraphics[]{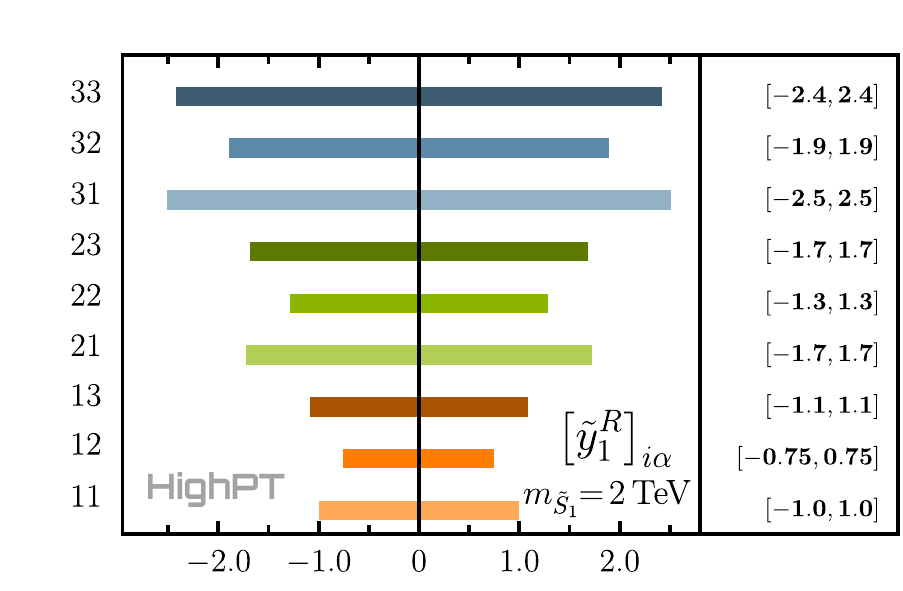}
        \\
        \includegraphics[]{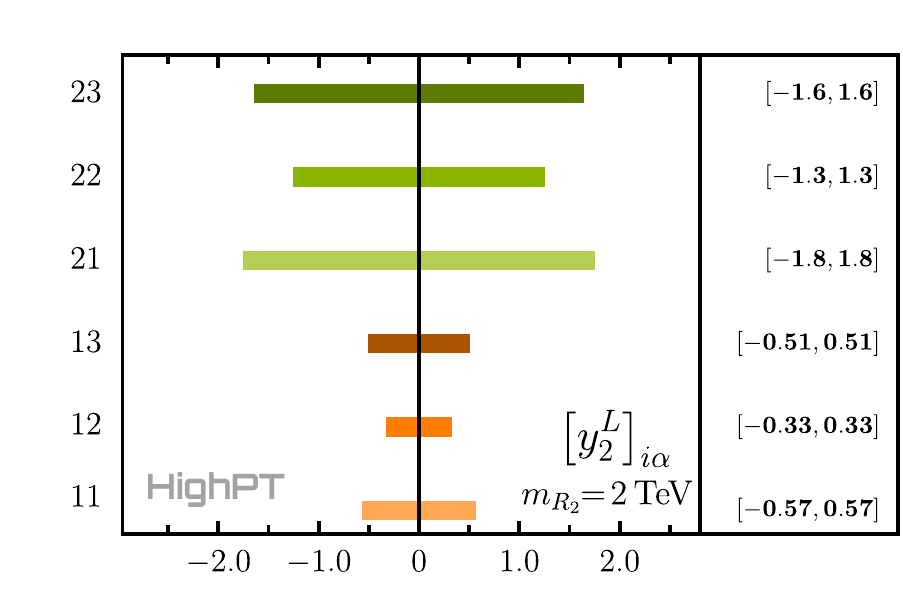} & \includegraphics[]{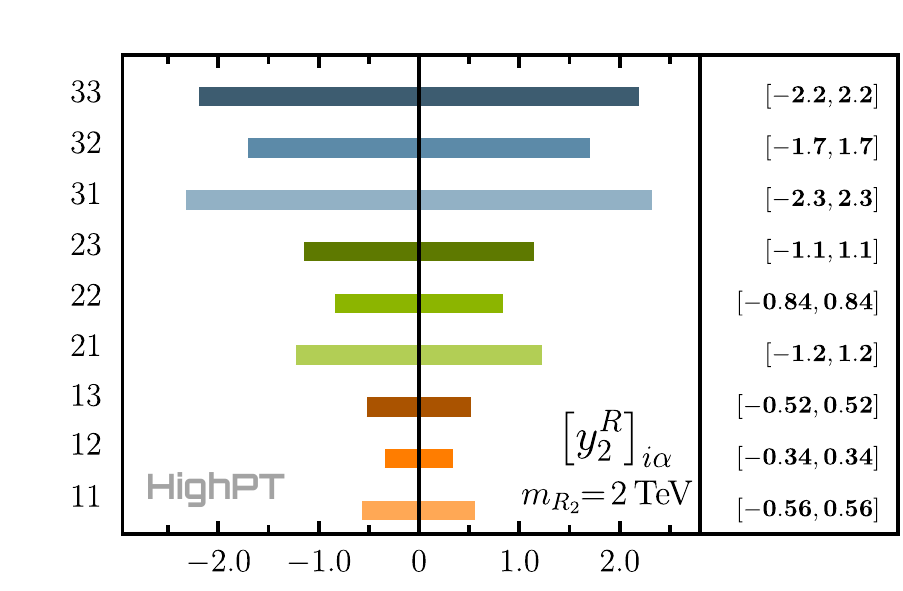} & \includegraphics[]{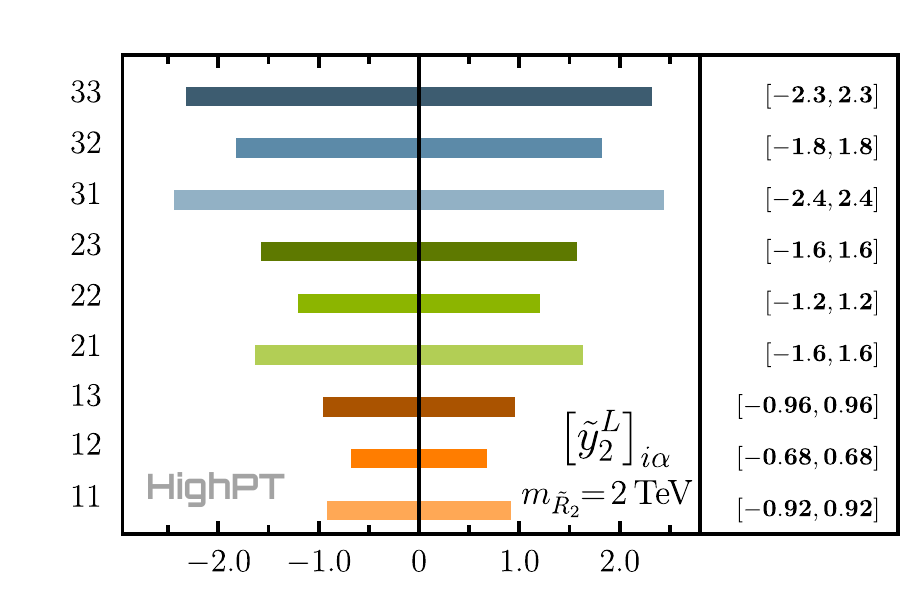} 
        \\
        \includegraphics[]{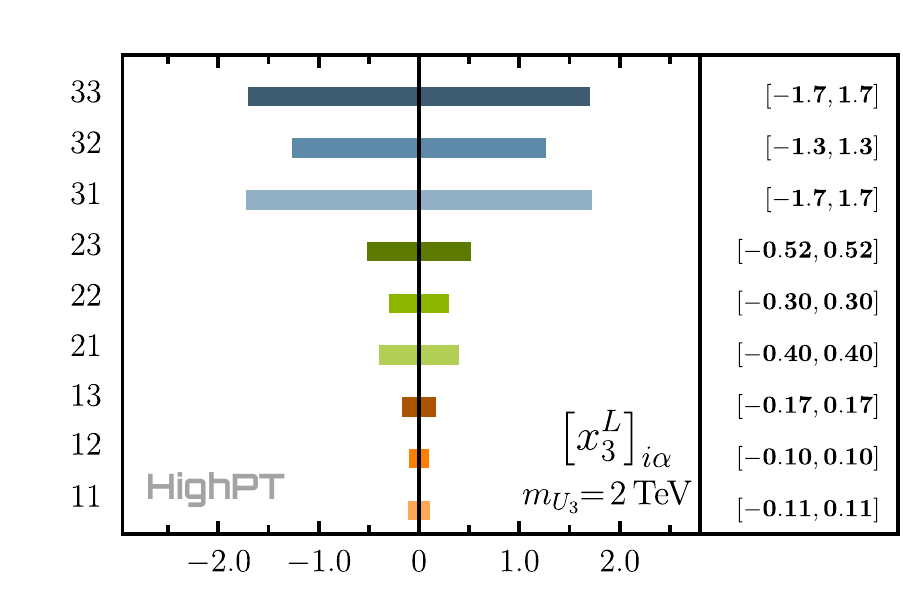} & \includegraphics[]{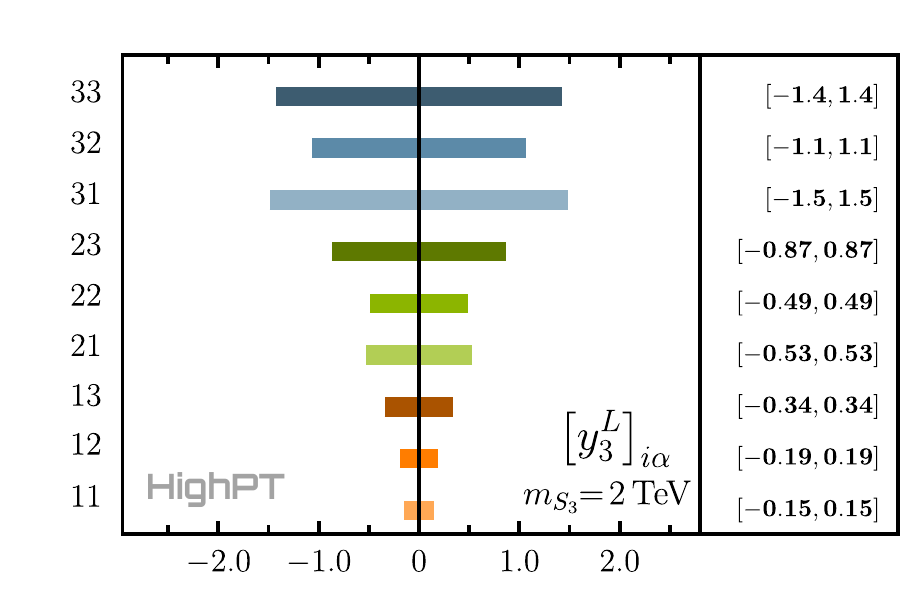}
    \end{tabular}
    }
    \caption{\sl\small LHC constraints at $95$\% CL on the coupling constants of all leptoquarks, where a single coupling is turned on at a time. The masses of all leptoquarks are fixed to~$2\,\text{TeV}$. The numbers on the left-hand side of each plot correspond to the respective quark and lepton flavor indices~$i\alpha$. See Table~\ref{tab:mediators} for the definition of the couplings. All constraints are compatible with the SM except for the bound on $\smash{[x_2^R]_{11}}$ for which we find a~$\sim 2 \sigma$ deviation.}
    \label{fig:LQ-couplings}
\end{figure}

The 95\% confidence intervals for each individual coupling are collected in Fig.~\ref{fig:LQ-couplings} for all possible scalar and vector leptoquarks coupling to all three leptons and first generation quarks (red intervals), second generation quarks (green intervals) and third generation quarks (blue intervals).~\footnote{For the right-handed coupling $\smash{[x_2^R]_{11}}$ of the $V_2$ vector leptoquark, we observe a mild discrepancy with the SM prediction with a low significance of~$\approx 2\,\sigma$.} Since the amplitude scales as the square of the probed couplings, the limits are therefore fully symmetric. For fixed quark-flavors, these results follow a similar pattern as in the SMEFT results presented in Sec.~\ref{ssec:smeft-limits}, where the strongest bounds correspond to the lightest quark flavors, as they have the largest PDFs. The only couplings that are not constrained by our observables are the ones that involve right-handed top-quarks. Interestingly, we also find that in most cases the bounds on the leptoquark couplings to $\mu$'s are considerably more constraining than the ones coupling to $e$'s or $\tau$'s. Surprisingly, we find that the bounds to the electron couplings turn out to be comparable to the tauonic ones, which is caused by a poor background description of the dielectron data over several high-mass $m_{ee}$ bins~\cite{CMS:2021ctt}.

The complete LHC likelihood for all leptoquarks with all possible flavor couplings are available in {\tt HighPT} for several benchmark masses.

\subsubsection*{Leptoquarks and LFV at the LHC}

Considering a single leptoquark coupling at a time will only lead to lepton flavor conserving Drell-Yan processes. However, if the leptoquark states couple to two different leptons, then they will also induce LFV modes in Drell-Yan production. In other words, the lepton flavor conserving modes $pp\to \ell_\alpha\ell_\alpha$ and $pp\to \ell_\beta\ell_\beta$, and the LFV process $pp\to \ell_\alpha\ell_\beta$ with $\alpha\ne\beta$ would be correlated. These three production modes are perfectly complementary since the scattering amplitudes are proportional to different combinations of the leptoquark couplings. 

\begin{figure}[!t]
    \centering
    \includegraphics[width=0.31\textwidth]{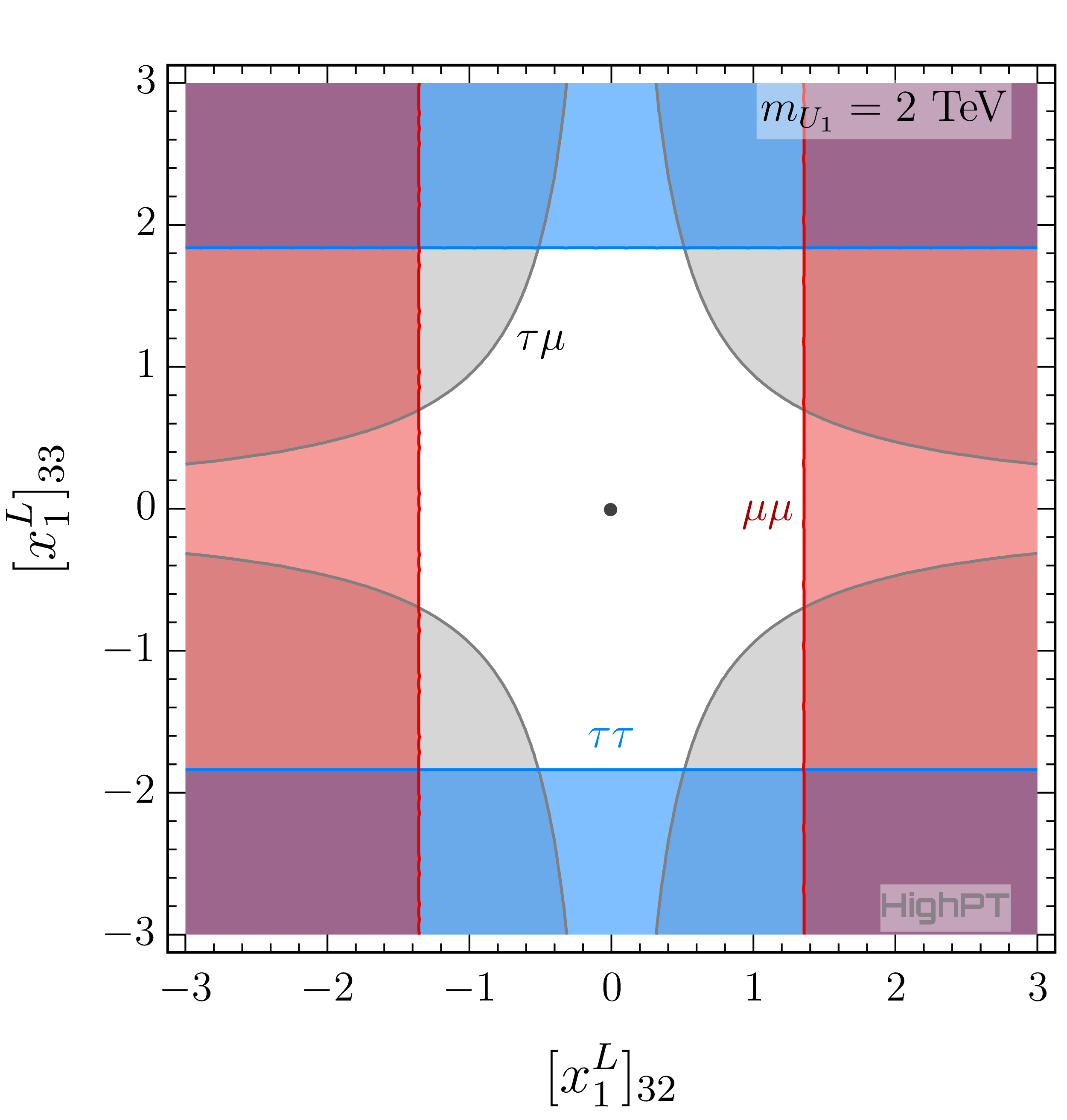}~~
    \includegraphics[width=0.31\textwidth]{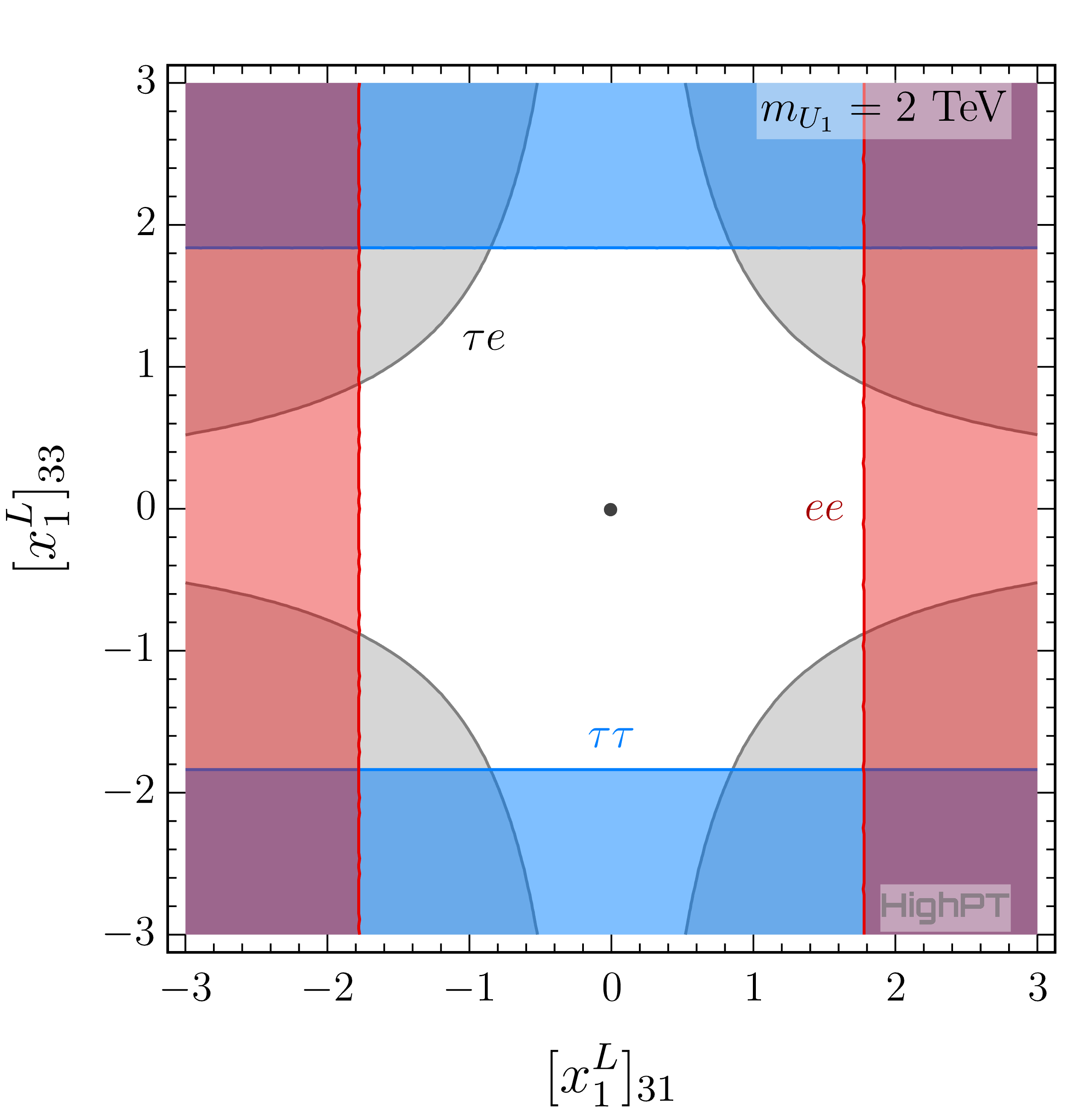}~~
    \includegraphics[width=0.31\textwidth]{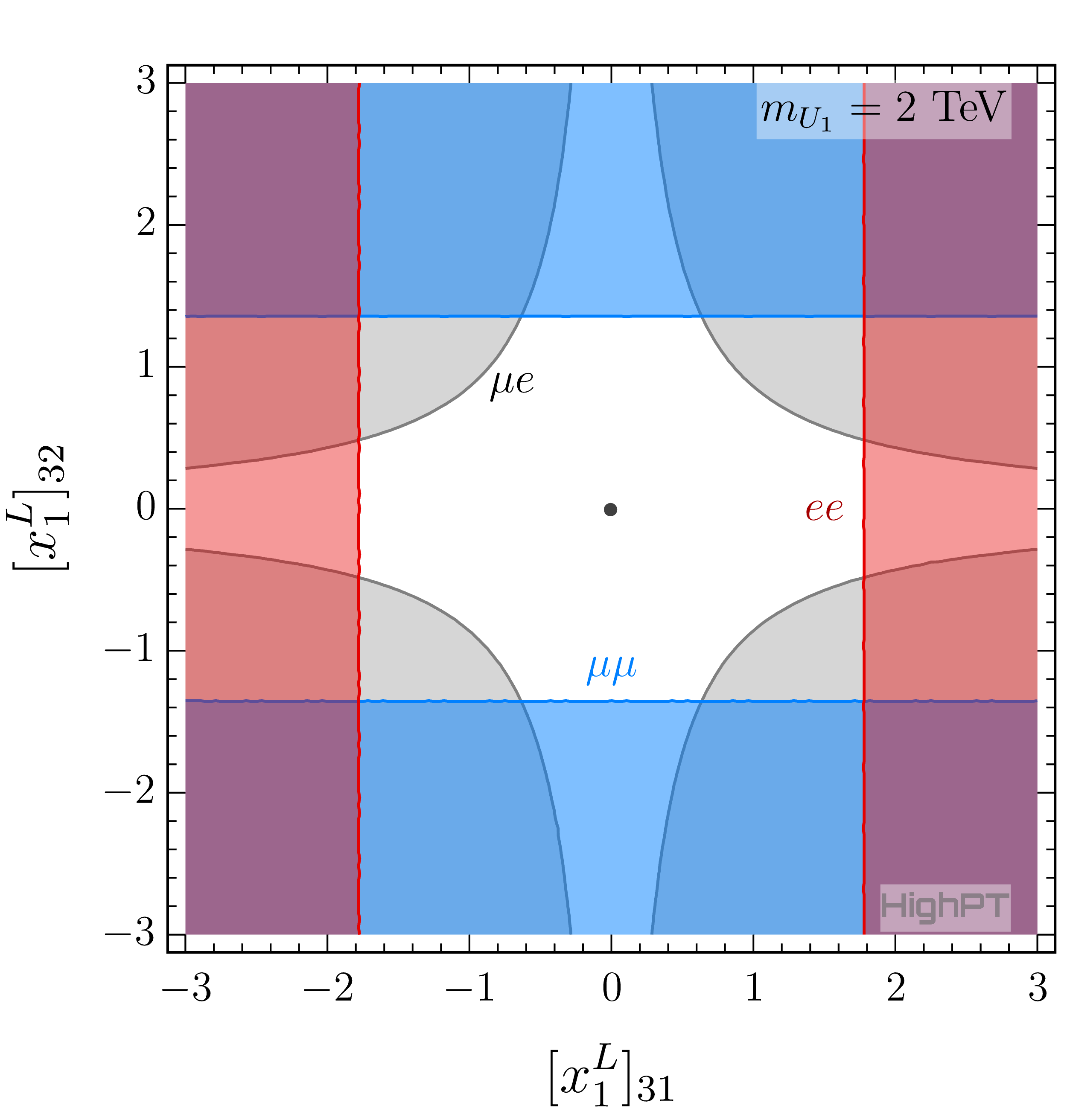}
    \caption{\sl\small  Two-dimensional exclusion regions on the $U_1\sim ({\bf 3},{\bf 1},2/3)$ leptoquark left-handed couplings derived from $pp\to\ell\ell$ and $pp\to\ell^\prime\ell^\prime$ (red and blue contours), and  $pp\to\ell\ell^\prime$ (gray contours), with $\ell\neq \ell^\prime$. All contours are depicted at $2\sigma$.}
    \label{fig:LFV}
\end{figure}

In the following, we set $95\%$ exclusion limits on leptoquark couplings using data from both LFV and lepton flavor conserving tails to highlight their complementarity. For concreteness, we consider the $U_1\sim({\bf 3},{\bf 1},2/3)$ vector leptoquark with mass $m_{U_1}=2$\,TeV and couplings to left-handed currents, 

\begin{equation}
    \cL_{U_1} \supset [x_1^L]_{i\alpha} \, U_{\!1}^\mu\,\bar q_i  \gamma_\mu l_\alpha+\mathrm{h.c.}\,.
\end{equation}

\noindent We assume that $U_1$ couples exclusively to the third-generation quarks, with nonzero couplings $[x_1^L]_{3\alpha}$ and $[x_1^L]_{3\beta}$ and $\alpha<\beta$. The results for the two-parameter limits are shown in the coupling planes given in Fig.~\ref{fig:LFV}. There, the blue and red regions corresponds to the excluded region from lepton flavor conserving searches, while the region in gray corresponds to the excluded region from the LFV searches. 
One can see that for this particular example the LFV searches can probe regions of parameter space that are not covered by the lepton flavor conserving modes. This complementarity can be understood from the inequality,
\begin{equation}
2\,\left|[x_1^L]_{i\alpha}\,[x_1^L]_{i\beta}^\ast\right| \leq \left|[x_1^L]_{i\alpha}\right|^2 + \left|[x_1^L]_{i\beta}\right|^2\,,
\end{equation}
where the left-hand side is the combination of couplings entering $pp\to \ell_\alpha\ell_\beta$, whereas the right-hand side can be bounded by lepton flavor conserving searches.

%%%%%%%%%%%%%%%%%%%%%%%%%%%%%%%%%%%%%%%%%%%%%%%%%%%%%%%%%%%%%%%%%%%%%%%%%%
\section{Combining Flavor and LHC Constraints: a Case Study}
\label{sec:example}
%%%%%%%%%%%%%%%%%%%%%%%%%%%%%%%%%%%%%%%%%%%%%%%%%%%%%%%%%%%%%%%%%%%%%%%%%%

In this Section, we illustrate the relevance of our results by combining the high-$p_T$ constraints derived in Sec.~\ref{sec:collider-limits} with the ones obtained from flavor and electroweak observables. The New Physics scenarios that can accommodate the hints of Lepton Flavor Universality (LFU) violation in charged-current $B$-meson decays will be considered as an example. This study case will be illustrative of the potential of high-$p_T$ physics to probe flavor physics operators, since the explanations of these discrepancies require relatively low values of the New Physics scale, in the few TeV range, where LHC constraints are known to be useful~\cite{Faroughy:2016osc,Greljo:2018tzh,Marzocca:2020ueu,Endo:2021lhi,Iguro:2020keo,Jaffredo:2021ymt}. 

\subsection{LFU violation in $b\to c \ell \nu$}

We start by reminding the reader of the status of LFU tests in the $b\to c\ell\nu$ transition and the EFT description of these observables. The deviations from LFU in charged currents $B$-decays have been observed in the ratios defined by~\cite{BaBar:2013mob,Belle:2015qfa,LHCb:2015gmp,Belle:2016dyj,Belle:2017ilt,LHCb:2017rln,Belle:2019gij,Belle:2019rba},~\footnote{Similar LFU tests have been performed by LHCb in the $B_c\to J/\psi \ell \nu$~\cite{LHCb:2017vlu} and $\Lambda_b\to \Lambda_c \ell \nu$~\cite{LHCb:2022piu} channels, but with sizable experimental uncertainties. Currently, these observables have a minor impact in the $b\to c\tau\nu$ fit.}
%%%%%%%%%%%%%%%%%
\begin{equation}
R_{D^{(\ast)}} = \left.\dfrac{\mathcal{B}(B\to D^{(\ast)} \tau \nu)}{\mathcal{B}(B\to D^{(\ast)} \ell \nu)} \right\vert_{\ell\in\lbrace e\,\mu\rbrace}\,,
\end{equation}
%%%%%%%%%%%%%%%%%

\noindent where light leptons $\ell\in \lbrace e,\mu \rbrace$ are averaged in the denominator. The current experimental averages reported by HFLAV~\cite{HFLAV:2019otj},
%%%%%%%%%%%%%%%%%
\begin{equation}
R_{D}^\mathrm{exp} = 0.34(3) \,,\qquad\qquad R_{D^ \ast}^\mathrm{exp} = 0.295(14)\,,
\end{equation}
%%%%%%%%%%%%%%%%%

\noindent appear to be systematically larger than the SM predictions~\cite{Na:2015kha,MILC:2015uhg,HFLAV:2019otj,<In_preparation><Allwicher;Lukas><Faroughy;Darius><Jaffredo;Florentin><Sumensari;Olcyr><Wilsch;Felix><2022>},
%%%%%%%%%%%%%%%%%
\begin{equation}
R_{D}^\mathrm{SM} = 0.294(4) \,,\qquad\qquad R_{D^ \ast}^\mathrm{SM} = 0.246(9)\,,
\end{equation}
%%%%%%%%%%%%%%%%%

\noindent amounting to a combined discrepancy at the $\approx 3 \sigma$ level~\cite{HFLAV:2019otj}.

%=====================================================
\subsubsection*{Low-energy EFT}
%=====================================================
We assume that New Physics only contributes to $b\to c\tau {\nu}$, i.e.~leaving the channels with electrons and muons unaffected, and we write the most general low-energy effective Lagrangian for this transition with operators up to $d=6$,
%%%%%%%%%%%%%%%%%
\begin{align}
\begin{split}
\label{eq:RD-Left}
    \mathcal{L}_\mathrm{eff}^{b\to c\tau \nu} =&-2 \sqrt{2}  G_F  V_{cb} \Big{[} (1+ C_{V_L}) \big{(} \bar{c}_L \gamma_\mu  b_L \big{)} \big{(} \bar{\tau}_L \gamma_\mu  \nu_L \big{)} + C_{V_R} \big{(} \bar{c}_R \gamma_\mu  b_R \big{)} \big{(} \bar{\tau}_L \gamma_\mu  \nu_L \big{)} \\
    &+C_{S_L} \big{(} \bar{c}_R  b_L \big{)} \big{(} \bar{\tau}_R \nu_L \big{)}+C_{S_R} \big{(} \bar{c}_L  b_R \big{)} \big{(} \bar{\tau}_R \nu_L \big{)}+C_{T} \big{(} \bar{c}_R  \sigma_{\mu\nu} b_L \big{)} \big{(} \bar{\tau}_R \sigma^{\mu\nu} \nu_L \big{)}
    \Big{]}+\mathrm{h.c.}\,,
\end{split}
\end{align}
%%%%%%%%%%%%%%%%%
where $C_i \equiv C_i(\mu)$ denotes the effective coefficients, defined at the scale $\mu=m_b$, and flavor indices are omitted. Many studies have determined the allowed values of the couplings $C_i \equiv C_i(\mu)$, at the scale $\mu=m_b$, by using the ratios $R_D$ and $R_{D^\ast}$, see e.g.~\cite{Blanke:2019qrx,Shi:2019gxi,Murgui:2019czp,Angelescu:2021lln} and references therein. The results from Ref.~\cite{Angelescu:2021lln} will be considered in this paper.

In addition to the ratios $R_{D^{(\ast)}}$, an important constraint on $C_P \equiv C_{S_R}-C_{S_L}$ can be derived from the $B_c$-meson lifetime~\cite{Alonso:2016oyd,Li:2016vvp}. Even though the $B_c\to\tau\nu$ branching fraction has not yet been measured, it is clear that the corresponding partial decay width should not saturate the value of $\Gamma_{B_c}$ determined experimentally~\cite{ParticleDataGroup:2020ssz}. In the following, we conservatively require that $\mathcal{B}(B_c\to \tau\nu)\lesssim 30\%$, which forbids the possibility of addressing both $R_D$ and $R_{D^\ast}$ with only the $C_{S_{L(R)}}$ coefficients~\cite{Alonso:2016oyd}.

%=====================================================
\subsubsection*{SMEFT description}
%=====================================================
In order to consistently explore the implications of the $R_{D^{(\ast)}}$ anomalies at high-$p_T$, the low-energy effective Lagrangian~\eqref{eq:RD-Left} must be replaced by the SMEFT Lagrangian. This comes with many important features, such as the correlations among different transitions that can arise from $SU(2)_L$ gauge invariance, which relates e.g.~modes with neutrinos and charged leptons~\cite{Aebischer:2015fzz}. Moreover, the Yukawa and electroweak running effects can induce sizable mixing among operators~\cite{Jenkins:2013zja,Jenkins:2013wua,Alonso:2013hga}. 

Firstly, the low-energy effective coefficients defined in Eq.~\eqref{eq:RD-Left} must be evolved from the scale $\mu=m_b$ up to $\mu = \mu_ \mathrm{ew}$ by using the renormalization group equations which are given e.g.~in Ref.~\cite{Gonzalez-Alonso:2017iyc}. The tree-level matching to the SMEFT Lagrangian at $\mu_ \mathrm{ew}$ then reads,
%%%%%%%%%%%%%%%%
\begin{align}
%\begin{split}
    C_{V_L}  &= -\dfrac{v^2}{\Lambda^2}\sum_i \dfrac{V_{2i}}{V_{23}} \left(\big{[}\mathcal{C}_{{lq}}^{(3)}\big{]}_{33 i3}+\big{[}\mathcal{C}_{{Hq}}^{(3)}\big{]}_{33}-\delta_{i3}\,\big{[}\mathcal{C}_{{Hl}}^{(3)}\big{]}_{33}\right)\,, \\[0.3em]
    C_{V_R}  &= \dfrac{v^2}{2\Lambda^2}\dfrac{1}{V_{23}}\big{[}\mathcal{C}_{{Hud}}^{(3)}\big{]}_{23}\,,  \\[0.3em]
    \label{eq:smeft-matching}
    C_{S_L}  &= -\dfrac{v^2}{2\Lambda^2} \dfrac{1}{V_{23}} \big{[}\mathcal{C}_{{lequ}}^{(1)}\big{]}_{3332}^\ast\,, \\[0.3em]
    C_{S_R}  &= -\dfrac{v^2}{2\Lambda^2} \sum_{i=1}^3\dfrac{V_{2i}^{\ast}}{V_{23}}\big{[}\mathcal{C}_{{ledq}}\big{]}^{\ast}_{333i}\,,\\[0.3em]
    C_T  &= -\dfrac{v^2}{2\Lambda^2} \dfrac{1}{V_{23}} \big{[}\mathcal{C}_{lequ}^{(3)}\big{]}^\ast_{3332}\,, 
%\end{split}
\end{align}
%%%%%%%%%%%%%%%%
where the scale $\mu_\mathrm{ew}$ is implicit in both sides of the above equations. The operator $\cO_{Hud}$ gives rise to lepton-flavor universal contributions, thus being irrelevant for $R_{D^{(\ast)}}$. Moreover, the operator $\smash{\cO_{Hl}^{(3)}}$ is tightly constrained at tree-level by the $Z$- and $W$-pole observables~\cite{Breso-Pla:2021qoe}.

From Eq.~\eqref{eq:smeft-matching}, we conclude that the possible New Physics explanations of $R_{D^{(\ast)}}$ must involve a combination of the operators $\smash{\lbrace \cO_{lq}^{(3)},\cO_{lequ}^{(1)},\cO_{lequ}^{(3)},\cO_{ledq}\rbrace}$ with appropriate flavor indices. The mediators that can induce these operators via tree-level exchange are known~\cite{deBlas:2017xtg}. They can be a scalar~$\Phi \sim (\mathbf{1},\mathbf{2},1/2)$ or vector $W^\prime \sim (\mathbf{1},\mathbf{3},0)$ color-singlet, or a scalar/vector leptoquark state~\cite{Buchmuller:1986zs,Dorsner:2016wpm}. The scalar doublet $\Phi$ is excluded from the constraints derived from the $B_c$~lifetime~\cite{Alonso:2016oyd}, whereas the vector triplet $W^\prime$ is tightly constrained by $pp\to \tau\tau$ at high-$p_T$ and by $\Delta F=2$ processes~\cite{Greljo:2015mma,Buttazzo:2017ixm}. Therefore, if we assume that a single mediator is behind the $R_{D^{(\ast)}}$ anomalies, this mediator must necessarily be a leptoquark. Among these mediators, only three are capable of explaining $R_{D^{(\ast)}}^\mathrm{exp}>R_{D^{(\ast)}}^\mathrm{SM}$ while being consistent with other existing bounds: (i) the vector $U_1 \sim (\mathbf{3},\mathbf{1},2/3)$, and the scalars (ii) $S_1 \sim (\mathbf{\bar{3}},\mathbf{1},1/3)$ and (iii) $R_2 \sim (\mathbf{3},\mathbf{2},7/6)$, see~\cite{Buttazzo:2017ixm,Angelescu:2018tyl,Angelescu:2021lln} and references therein.

\subsection{From EFT to concrete scenarios}
\label{ssec:lq-pheno} 

The viable leptoquark scenarios mentioned above predict specific combinations of effective semileptonic operators, as shown in Table~\ref{tab:LQmatching}. In order to successfully explain the $b\to c\tau \nu$ anomalies, the flavor pattern of the effective coefficients must couple exclusively, or predominantly to the second and third generation of quarks and leptons. The most relevant operators, at the matching scale $\Lambda$, in each of these scenarios read
%%%%%%%%%%%%%
\begin{align}
\label{eq:eft-lq-1}
    U_1\,: &~ \big{[}\mathcal{C}_{lq}^{(1)}\big{]}_{3323}=  \big{[}\mathcal{C}_{lq}^{(3)}\big{]}_{3323}\,, \qquad\;\;\;\,  \big{[}\mathcal{C}_{lq}^{(1)}\big{]}_{3333}= \big{[}\mathcal{C}_{lq}^{(3)}\big{]}_{3333}\,,  \qquad\;\;\;\;\;  \big{[}\mathcal{C}_{ledq}\big{]}_{3333} \,. \\[0.65em]
\label{eq:eft-lq-2}
    S_1\,: &~\big{[}\mathcal{C}_{lq}^{(1)}\big{]}_{3333}=- \big{[}\mathcal{C}_{lq}^{(3)}\big{]}_{3333}\,,  \quad \quad\;  \big{[}\mathcal{C}_{lequ}^{(1)}\big{]}_{3332} = - 4\, \big{[}\mathcal{C}_{lequ}^{(3)}\big{]}_{3332} \,. \\[0.65em]
\label{eq:eft-lq-3}
    R_2\,: &~\big{[}\mathcal{C}_{lequ}^{(1)}\big{]}_{3332} = 4\, \big{[}\mathcal{C}_{lequ}^{(3)}\big{]}_{3332} \,.
\end{align}
%%%%%%%%%%%%%
\noindent These operators contribute not only to the $b\to c\tau \nu$ transition, but also to many other precision observables that we briefly describe below:

\begin{itemize}
    \item[$\bullet$] $B\to K^{(\ast)} \nu \bar{\nu}$\,: The $b\to s\nu\nu$ transition provides stringent constraints on operators with left-handed leptons~\cite{Buras:2014fpa}. The observables based on this transition are particularly relevant to probe couplings to $\tau$-leptons, which are difficult to assess otherwise. The low-energy effective Lagrangian describing the $b\to s\nu\nu$ transition can be written as,
%%%%%%%%%%%%%%%%
\begin{align}
\label{eq:ope-bsnunu}
\mathcal{L}_\mathrm{eff}^{b\to s\nu\nu} =  \dfrac{4 G_F}{\sqrt{2}} V_{tb} V_{ts}^\ast \dfrac{\alpha_\mathrm{em}}{4\pi}\sum_{\alpha\beta}  \Big{\lbrace} [C_L]_{\alpha\beta} \, [O_L]_{\alpha\beta} + [C_R]_{\alpha\beta} \, [O_R]_{\alpha\beta} \Big{\rbrace}+\mathrm{h.c.}\,,
\end{align}
%%%%%%%%%%%%%%%
with
%%%%%%%%%%%%%%%%
\begin{align}
\label{eq:ope-bsnunu-bis}
[O_{L}]_{\alpha\beta} = (\bar{s}_{L}\gamma^\mu b_{L})(\bar{\nu}_{L\alpha}\gamma_\mu \nu_{L\beta})\,, \qquad 
[O_{R}]_{\alpha\beta}  = (\bar{s}_{R}\gamma^\mu b_{R})(\bar{\nu}_{L\alpha}\gamma_\mu \nu_{L\beta})\,.
\end{align}
%%%%%%%%%%%%%%%
The SM contributions are lepton-flavor conserving and they are given by the coefficient $C_L^\mathrm{SM}=-13.6(1.2)$, which includes NLO QCD corrections~\cite{Buchalla:1993bv,Buchalla:1998ba,Misiak:1999yg} and two-loop electroweak contributions~\cite{Brod:2010hi}. The low-energy Wilson coefficients can be matched to the semileptonic SMEFT operators at $\mu=\mu_\mathrm{ew}$,
%%%%%%%%%%%%%%%%
\begin{align}
[C_L]_{\alpha\beta} &=\delta_{\alpha\beta}\,C_L^\mathrm{SM}+\dfrac{2\pi}{\alpha_\mathrm{em} V_{tb}V_{ts}^\ast} \dfrac{v^2}{\Lambda^2} 
\left(\big{[}\mathcal{C}_{lq}^{(1-3)}\big{]}_{\alpha\beta 23}-\delta_{\alpha\beta}\big{[}\mathcal{C}_{Hq}^{(1-3)}\big{]}_{23}\right)\,,\\[0.4em]
[C_R]_{\alpha\beta} &=\dfrac{2\pi}{\alpha_\mathrm{em}  V_{tb}V_{ts}^\ast } \dfrac{v^2}{\Lambda^2} \left( \big{[}\mathcal{C}_{ld}\big{]}_{\alpha\beta 23} - \delta_{\alpha\beta}\big{[}\mathcal{C}_{Hd}\big{]}_{23}
\right)\,.
\end{align}
%%%%%%%%%%%%%%%
These effective coefficients can be evolved up to the scale $\Lambda$ by using the one-loop anomalous dimensions computed in Ref.~\cite{Jenkins:2013zja,Jenkins:2013wua,Alonso:2013hga}. The $B\to K^{(\ast)}\nu\bar{\nu}$ branching fractions can be easily computed in terms of the coefficients defined in Eq.~\eqref{eq:ope-bsnunu}~\cite{Buras:2014fpa}. The most stringent experimental limits are given by $\mathcal{B}(B^+\to K^+\nu\bar{\nu})<1.6\times 10^{-5}$ and $\mathcal{B}(B^0\to K^{\ast 0}\nu\bar{\nu})<2.7\times 10^{-5}$~\cite{BaBar:2013npw,Belle:2017oht,Belle-II:2021rof}, which lie just above the SM predictions, namely $\mathcal{B}(B^+\to K^+\nu\bar{\nu})^\mathrm{SM}=4.9(4)\times 10^{-6}$ and $\mathcal{B}(B^0\to K^{\ast 0}\nu\bar{\nu})^\mathrm{SM}=1.00(9)\times 10^{-6}$~\cite{Allwicher:2022mcg}. 

%%%%%%%%%%%%%%%%%%%%
%%%%%%%%%%%%%%%%%%%%
\begin{table}[tbp!]
    \renewcommand{\arraystretch}{1.6}
    \centering
    \begin{tabular}{cccc}
         Field & $S_1$ & $R_2$ & $U_1$ \\ \hline\hline
         %Spin & 0 & 0 & 1 \\ \hline
         Quantum Numbers & $(\mathbf{\bar{3}},\mathbf{1},1/3)$ & $(\mathbf{3},\mathbf{2},7/6)$ & $(\mathbf{3},\mathbf{1},2/3)$ \\ \hline
         $\left[\cC_{ledq}\right]_{\alpha\beta ij}$ & -- & -- & $2 [x_1^L]_{i\alpha}^* [x_1^R]_{j\beta}$ \\[0.3em]
         $\left[\cC_{lequ}^{(1)}\right]_{\alpha\beta ij}$ & $\frac{1}{2}[y_1^L]_{i\alpha}^*[y_1^R]_{j\beta}$ & $-\frac{1}{2}[y_2^R]_{i\beta}[y_2^L]_{j\alpha}^*$ & -- \\[0.3em]
         $\left[\cC_{lequ}^{(3)}\right]_{\alpha\beta ij}$ & $-\frac{1}{8}[y_1^L]_{i\alpha}^*[y_1^R]_{j\beta}$ & $-\frac{1}{8}[y_2^R]_{i\beta}[y_2^L]_{j\alpha}^*$ & -- \\[0.3em]
         $\left[\cC_{eu}\right]_{\alpha\beta ij}$ & $\frac{1}{2}[y_1^R]_{j\beta}[y_1^R]_{i\alpha}^*$ & -- & -- \\[0.3em]
         $\left[\cC_{ed}\right]_{\alpha\beta ij}$ & -- & -- & $-[x_1^R]_{i\beta}[x_1^R]_{j\alpha}^*$ \\[0.3em]
         $\left[\cC_{\ell u}\right]_{\alpha\beta ij}$ & -- & $-\frac{1}{2}[y_2^L]_{i\beta}[y_2^L]_{j\alpha}^*$ & -- \\[0.3em]
         $\left[\cC_{qe}\right]_{ij\alpha\beta}$ & -- & $-\frac{1}{2}[y_2^R]_{i\beta} [y_2^R]_{j\alpha}^*$ & -- \\[0.3em]
         $\left[\cC_{lq}^{(1)}\right]_{\alpha\beta ij}$ & $\frac{1}{4}[y_1^L]_{i\alpha}^*[y_1^L]_{j\beta}$ & -- & $-\frac{1}{2} [x_1^L]_{i\beta} [x_1^L]_{j\alpha}^*$ \\[0.3em]
         $\left[\cC_{lq}^{(3)}\right]_{\alpha\beta ij}$ & $-\frac{1}{4}[y_1^L]_{i\alpha}^*[y_1^L]_{j\beta}$ & -- & $-\frac{1}{2} [x_1^L]_{i\beta} [x_1^L]_{j\alpha}^*$ \\\hline\hline
    \end{tabular}
    \caption{\sl\sl \small
Matching of the leptoquarks to the semileptonic operators in the Warsaw basis~\cite{Grzadkowski:2010es}. In the matching conditions we have set $\Lambda = m_{LQ}$.}
    \label{tab:LQmatching}
\end{table}
%%%%%%%%%%%%%%%%%%%%
%%%%%%%%%%%%%%%%%%%%
 
    \item[$\bullet$] $W$- and $Z$-pole observables\,: The precise determinations of the $W$- and $Z$-couplings at LEP and the LHC can be used to constrain semileptonic interactions at loop level~\cite{Feruglio:2016gvd,Feruglio:2017rjo,Cornella:2018tfd}. The SMEFT operators describing modifications of the $Z$ and $W$ leptonic couplings up to $d=6$ read
%%%%%%%%%%%%%%%%
\begin{align}
\label{eq:ZWpoles}
\big{[}\cO_{Hl}^{(1)}\big{]}_{\alpha\beta} &= \big{(}H^\dagger  \overleftrightarrow{D}_\mu H \big{)} \bar{l}_\alpha \gamma^\mu l_\beta\,,\qquad \big{[}\cO_{He}\big{]}_{\alpha\beta} =\big{(}H^\dagger  \overleftrightarrow{D}_\mu H \big{)} \bar{e}_\alpha \gamma^\mu e_\beta\,,\\[0.35em]
\big{[}\cO_{Hl}^{(3)}\big{]}_{\alpha\beta} &= \big{(}H^\dagger  \overleftrightarrow{D}_\mu^I H \big{)} \bar{l}_\alpha \gamma^\mu \tau^I l_\beta\,.
\end{align}
%%%%%%%%%%%%%%%%
Chirality-conserving semileptonic operators such as $\smash{\cO_{lq}^{(1)}}$ and $\smash{\cO_{lq}^{(3)}}$ mix into these operators at one loop~\cite{Jenkins:2013zja,Jenkins:2013wua,Alonso:2013hga}. In particular, these effects can be sizable for semileptonic couplings to the top quark. We account for these contributions by using a leading-logarithmic approximation and we consider the recent fit to the $W$- and $Z$-couplings from Ref.~\cite{Breso-Pla:2021qoe}.
     
    \item[$\bullet$] $H\to\tau\tau$\,: Measurements of the Higgs Yukawa coupling to $\tau$-leptons at the LHC can also provide a useful constraint on specific semileptonic operators at one loop. This is the case for the chirality-breaking operators $\cO_{lequ}^{(1)}$ and $\cO_{ledq}$, since they mix at one loop with the following operator,
    %%%%%%%%%%%%%%%%
    \begin{align}
    \big{[}\cO_{eH}\big{]}_{\alpha\beta} = \big{(}H^\dagger H \big{)} \overline{l}_\alpha H e_\beta\,,
    \end{align}
    %%%%%%%%%%%%%%%%
    
    \noindent which induces a shift in the SM value of the $\tau$-lepton Yukawa after the electroweak-symmetry breaking. This contribution is particularly relevant if the semileptonic operators couple with third-generation quarks, due to the chirality-enhancement induced via the Yukawa (i.e.~$\propto m_t/m_\tau$)~\cite{Feruglio:2018fxo}. The latest PDG average for the $H\to \tau\tau$ signal strength reads~\cite{ParticleDataGroup:2020ssz},
    %%%%%%%%%%%%%%%%%
    \begin{equation}
        \mu_{\tau\tau}^\mathrm{exp} = \dfrac{\sigma(pp\to h) \cdot \mathcal{B}(H\to \tau\tau)}{\sigma(pp\to h)_\mathrm{SM} \cdot \mathcal{B}(H\to \tau\tau)_\mathrm{SM}}=1.15_{-0.15}^{+0.16}\,,
    \end{equation}
    %%%%%%%%%%%%%%%%%
    which is used to constrain the relevant operators at one loop, with a leading-logarithm approximation, and assuming that the Higgs production cross-section at the LHC is unaffected by New Physics.

\end{itemize}

 \noindent The observables discussed above will be incorporated in a future release of {\tt HighPT}, along with numerous other low-energy observables, in order to provide a complete likelihood for flavor physics~\cite{Allwicher:2022mcg}. 

\subsection{Numerical results}
\label{ssec:flavor-lq-numerical}

In this Section we combine the LHC constraints derived in Sec.~\ref{sec:collider-limits} with the flavor and electroweak precision observables discussed above. To illustrate the main features of the {\tt HighPT} package, which provides constraints for both EFT and concrete model scenarios, we perform a two-step analysis. In a first step, we will consider the minimal set of SMEFT operators that can accommodate $R_{D^{(\ast)}}$ in the viable scenarios described in Eqs.~\eqref{eq:eft-lq-1}--\eqref{eq:eft-lq-3}. In a second step, we directly consider the leptoquark models that predict these Wilson coefficients, including their propagation effects in the LHC observables. The comparison of the results obtained for the EFTs and the concrete models will allow us to directly assess the validity of the EFT approach for the high-$p_T$ observables that we have considered. Using the leptoquark models will also allow us to correlate the effective coefficients entering flavor processes in different sectors, as shown in Table~\ref{tab:LQmatching}. 

%%%%%%%%%%%%%%%%%%
\begin{figure}[p!]
    \centering
    \begin{subfigure}[b]{0.49\textwidth}
    \centering
    \includegraphics[width=\textwidth]{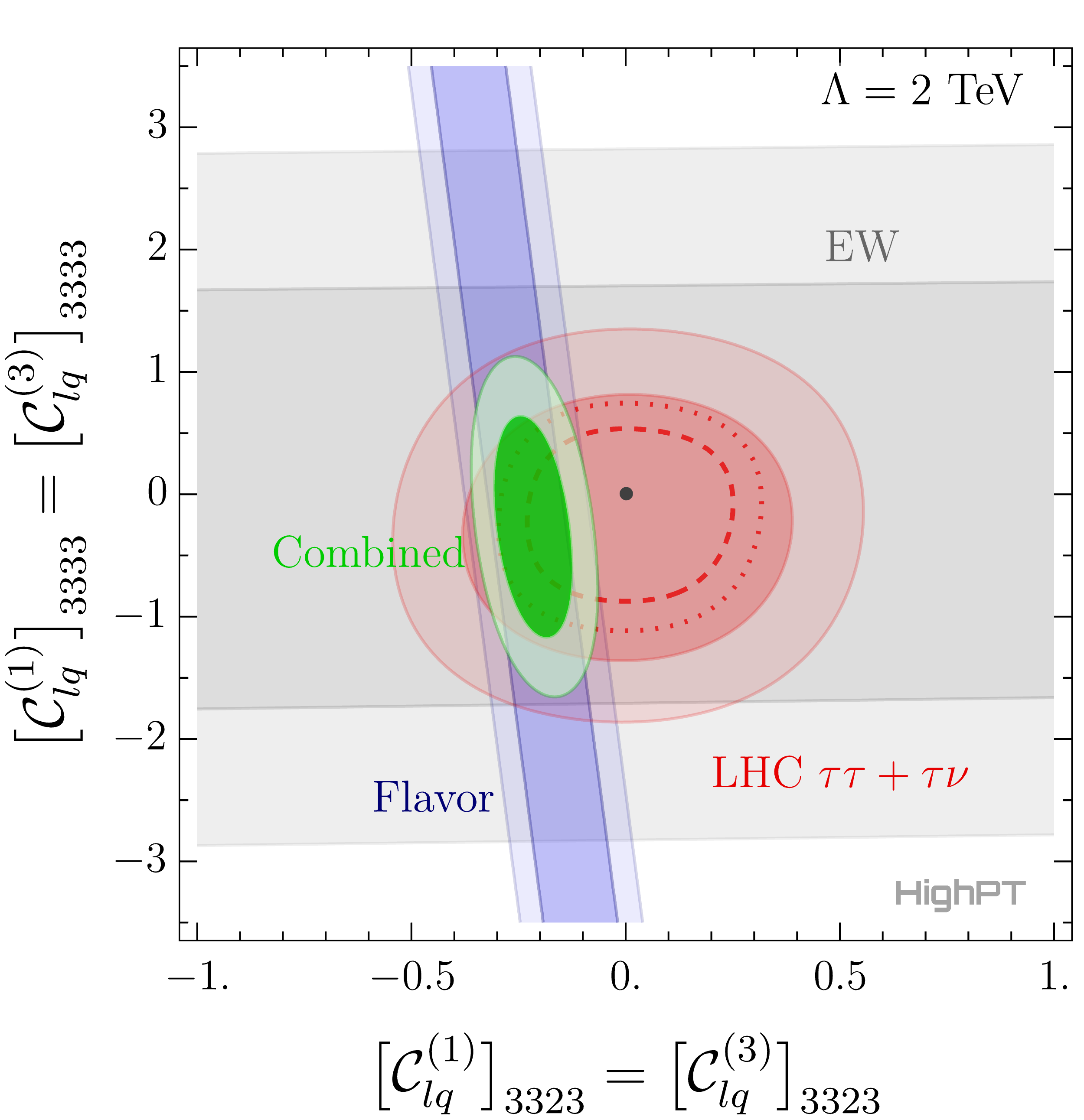}
    \caption{\sl}
    \end{subfigure}
    ~\begin{subfigure}[b]{0.49\textwidth}
    \centering
    \includegraphics[width=\textwidth]{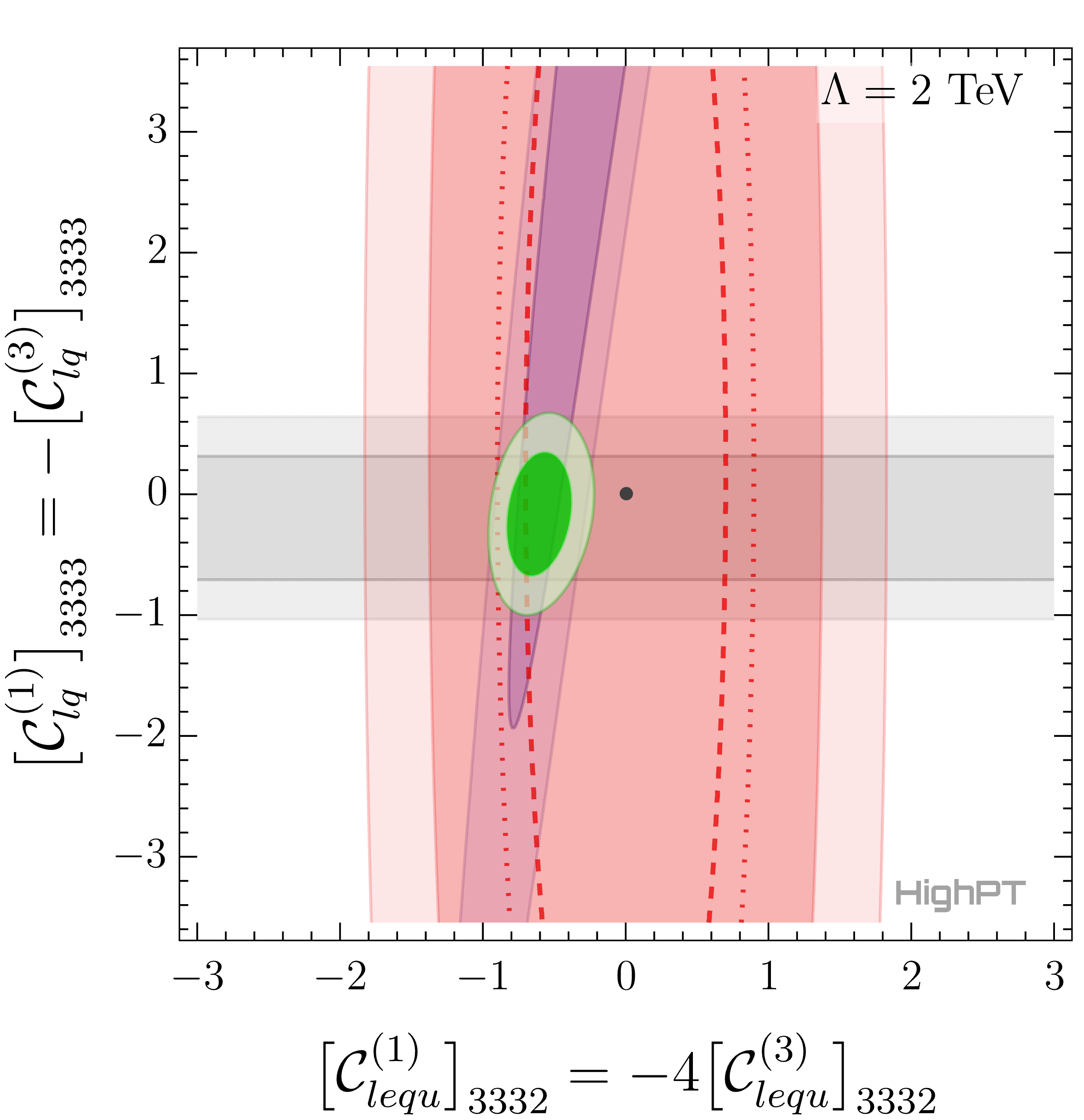}
    \caption{\sl}
    \end{subfigure}
    \\ \vspace{0.3cm}
    \begin{subfigure}[b]{0.49\textwidth}
    \centering
    \includegraphics[width=\textwidth]{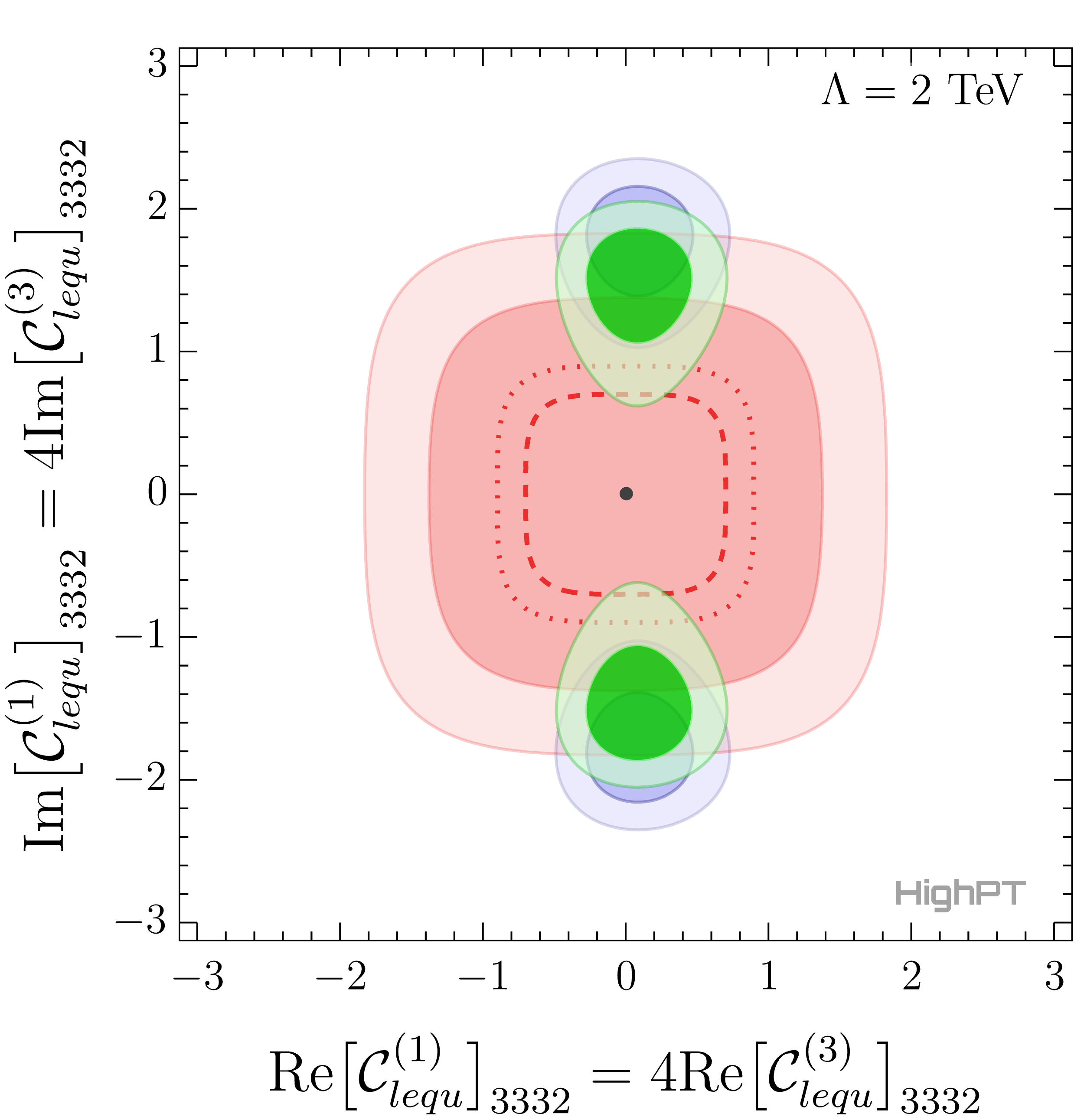}
    \caption{\sl}
    \end{subfigure}
    \caption{\sl\small Constraints on the SMEFT coefficients from flavor-physics (blue region), electroweak-precision (gray) and high-$p_T$ LHC observables (red). The combined fit is shown in green. For each type of observables, we show the $1\sigma$ ($2\sigma$) regions with darker (lighter) colors. The dashed (dotted) red line indicates the projection of the $1\sigma$ ($2\sigma$) region for an integrated luminosity of 3 ab$^{-1}$. Three effective scenarios are considered which are motivated by different leptoquark models, as explained in the text. The EFT cutoff is set to $\Lambda = 2$ TeV and the minimum values for the combined $\chi^2$ in our fits are given by $\chi^2_{\text{min}} = 28.3 \,(a),\, 29.5 \,(b),\, 78.0 \,(c)$.}
    \label{fig:SMEFTcombined}
\end{figure}
%%%%%%%%%%%%%%%%%%

\subsubsection*{EFT approach}

Starting with the EFT scenarios inspired by the viable leptoquarks, we consider the effective coefficients $\smash{\mathcal{C}_{lq}^{(1)}=\mathcal{C}_{lq}^{(3)}}$, which are predicted at tree level by the vector leptoquark $U_1$ with purely left-handed couplings, see Table~\ref{tab:LQmatching}.~\footnote{Notice that the presence of right-handed $U_1$ couplings is also allowed by current constraints, which would predict a different pattern of low- and high-energy observables~\cite{Cornella:2019hct,Cornella:2021sby}.} In the top left panel of Fig.~\ref{fig:SMEFTcombined}, we show the allowed Wilson coefficients with flavor indices that contribute directly to the $b\to c\tau{\nu}$ transition. The flavor constraints in this case are dominated by $R_{D^{(\ast)}}$ (blue region), which are combined with electroweak (gray) and LHC constraints (red). In this case, the LHC constraints are dominated by $pp\to\tau\tau$, whereas $pp\to\tau\nu$ gives weaker bounds. From Fig.~\ref{fig:SMEFTcombined}, we see that low- and high-energy observables are complementary, and the synergy of the different searches is fundamental to restrict the allowed region of the effective coefficients.

In a similar way, the scenario with $\mathcal{C}_{lequ}^{(1)}=-4\,\mathcal{C}_{lequ}^{(3)}$ and $\mathcal{C}_{lq}^{(1)}=-\mathcal{C}_{lq}^{(3)}$ is considered in Fig.~\ref{fig:SMEFTcombined}(b). This pattern of effective coefficients is predicted by the $S_1$ leptoquark at tree level~\cite{Sakaki:2013bfa,Angelescu:2018tyl,Bauer:2015knc,Becirevic:2016oho,Crivellin:2019dwb,Gherardi:2020qhc}. For simplicity, we assume real couplings and focus on the flavor indices $3332$ and $3333$ for the scalar/tensor and vector operators, respectively.~\footnote{Note that the effective coefficients $\mathcal{C}_{lq}^{(1)}=-\mathcal{C}_{lq}^{(3)}$ with flavor indices $3323$ also contribute to $R_{D^{(\ast)}}$, but these effective coefficients are subject to stringent constraints from $B\to K\nu\bar{\nu}$.} In this case, we find that the most relevant constraints arise from flavor observables, which are once again dominated by $R_{D^{(\ast)}}$, and from electroweak observables. In particular, the latter prevent an explanation of the $b\to c\tau{\nu}$ anomalies via only left-handed couplings in this scenario. Note, also, that LHC constraints turn out to be practically irrelevant, at the EFT level, since the contributions to $pp\to\tau\tau$ are CKM suppressed for the scalar/tensor operators, and absent for the particular combination of $\smash{\cC_{lq}^{(1)}}$ and $\smash{\cC_{lq}^{(3)}}$ under examination.

%%%%%%%%%%%%%%%%%%
\begin{figure}[p!]
    \centering
    \begin{subfigure}[b]{0.49\textwidth}
    \centering
    \includegraphics[width=\textwidth]{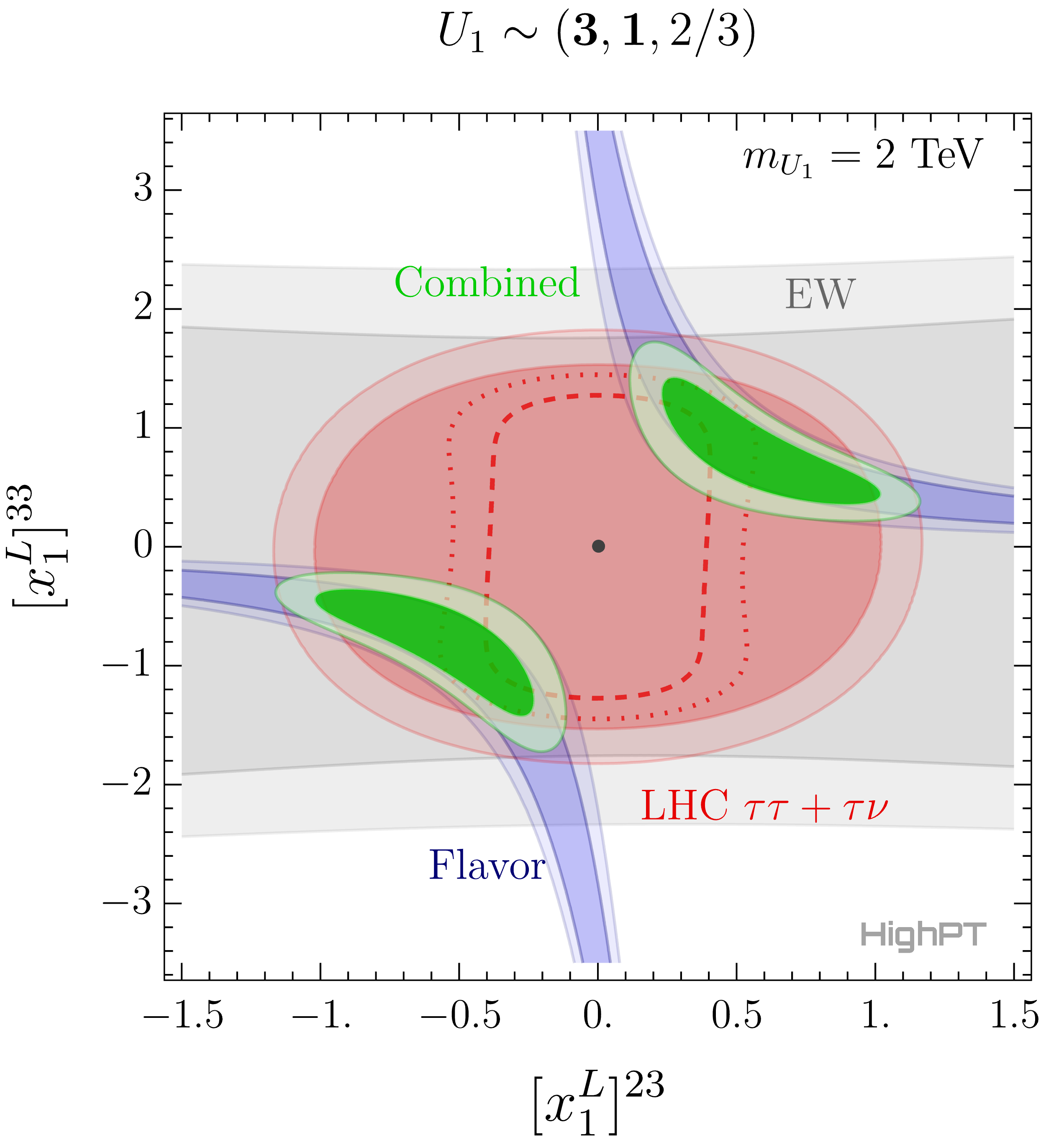}
    \caption{\sl}
    \end{subfigure}
    \begin{subfigure}[b]{0.49\textwidth}
    \centering
    \includegraphics[width=\textwidth]{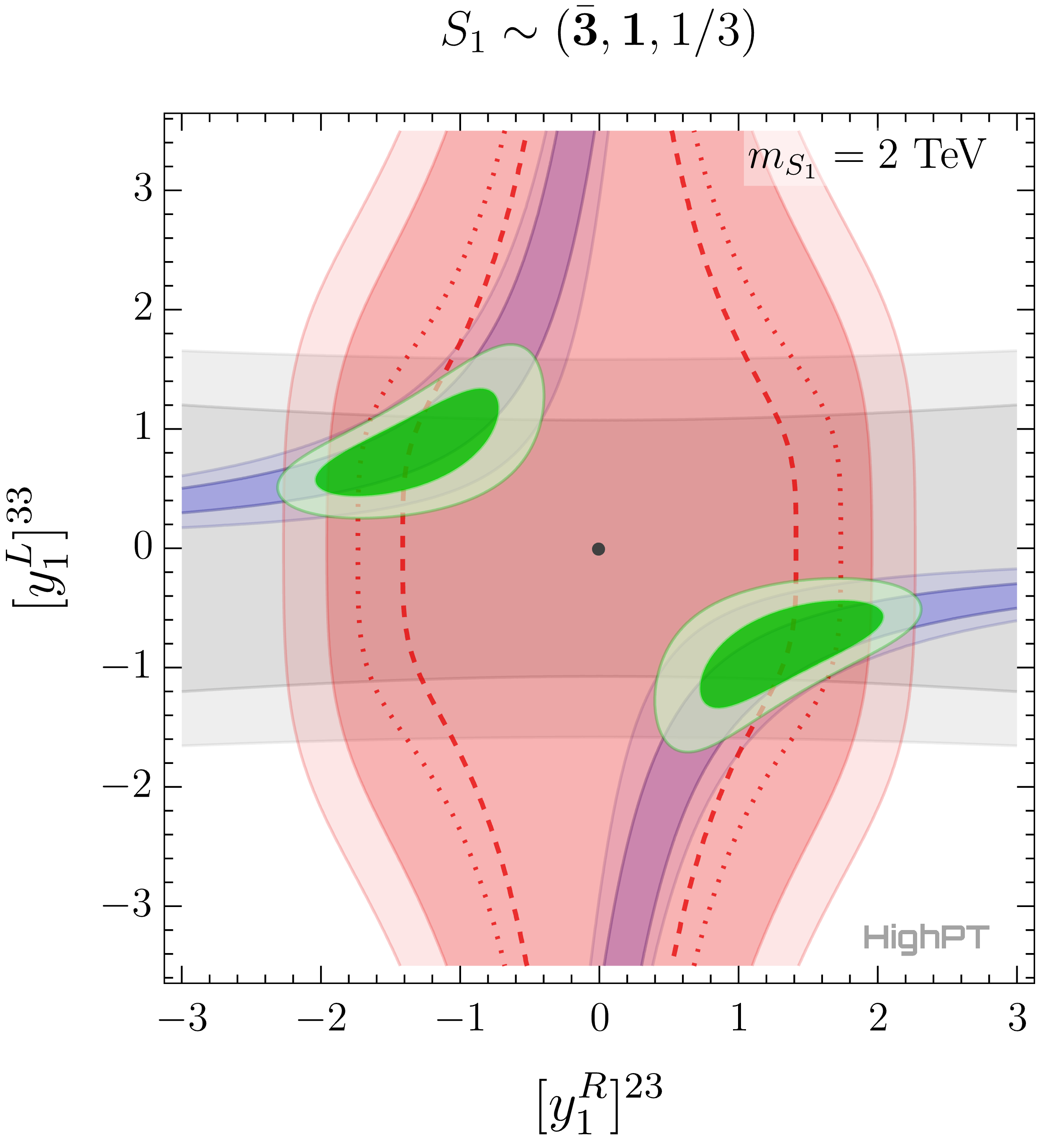}
    \caption{\sl}
    \end{subfigure}
    \\ \vspace{0.3cm}
    \begin{subfigure}[b]{0.49\textwidth}
    \centering
    \includegraphics[width=\textwidth]{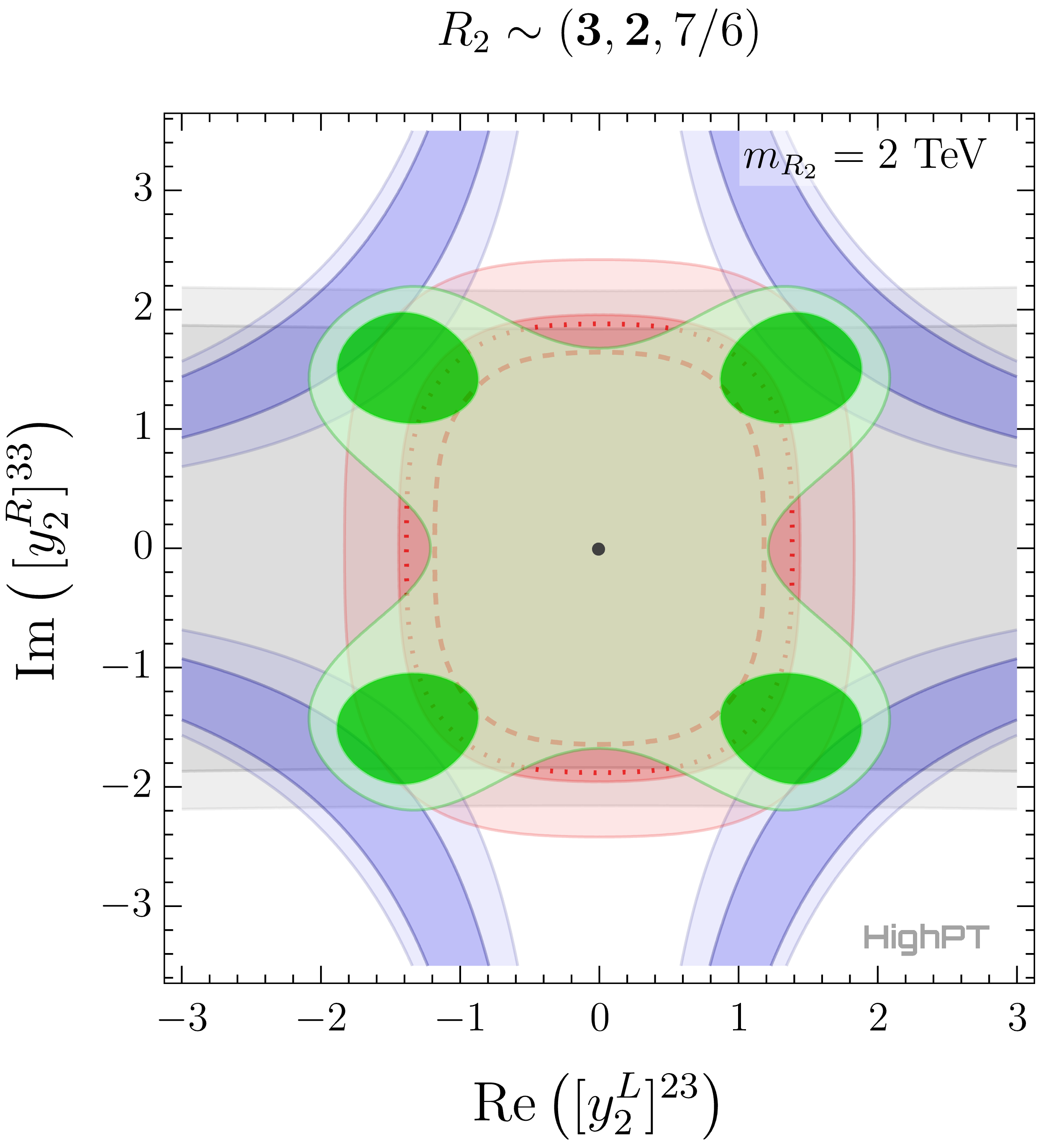}
    \caption{\sl}
    \end{subfigure}
    \caption{\sl\small Bounds on the leptoquark couplings from low-energy (blue), electroweak pole (gray) and high-$p_T$ LHC (red) observables. The combined fit is shown in green. For every bound we show the $1\sigma$ and $2\sigma$ regions. The dashed (dotted) red line indicates the projection of the $1\sigma$ ($2\sigma$) region for an integrated luminosity of 3 ab$^{-1}$. The leptoquark masses are set to $\Lambda = 2$ TeV. The minimum values for the combined $\chi^2$ in our fits are given by $\chi^2_{\text{min}} = 26.4 \,(a),\, 30.0 \,(b),\, 38.6 \,(c)$.}
    \label{fig:LQcombined}
\end{figure}
%%%%%%%%%%%%%%%%%%
The last scenario we consider is $\smash{\mathcal{C}_{lequ}^{(1)}=4\,\mathcal{C}_{lequ}^{(3)}}$, which is predicted by the $R_2$ leptoquark. The corresponding constraints are shown in Fig.~\ref{fig:SMEFTcombined}(c) for the flavor indices entering the $b\to c\tau\nu$ transition, with the same color code as before. This scenario is peculiar since purely real effective coefficients would induce contributions to $R_D$ and $R_{D^\ast}$ with different signs, which are incompatible with current data~\cite{Sakaki:2013bfa,Angelescu:2018tyl,Becirevic:2018uab,Becirevic:2018afm,Becirevic:2022tsj,Crivellin:2022mff}. In other words, an imaginary part of the scalar/tensor coefficients is needed to simultaneously explain the deviations observed in $R_D$ and $R_{D^\ast}$, as shown in Fig.~\ref{fig:SMEFTcombined}. Electroweak and Higgs constraints are not shown in this plot since they turn out to be weak in comparison to flavor bounds at the EFT level. LHC constraints are dominated by $pp\to\tau\nu$ and they appear to probe a small portion of the favored flavor-region. However, this conclusion should be taken with caution since the propagation effects of the leptoquark have a non-negligible effects in the LHC observables, as will be shown in the following.

\subsubsection*{Concrete models}

From the EFT examples discussed above, it is clear that the Drell-Yan tails provide complementary information to low-energy observables, being particularly useful to single out the viable solutions of the $R_{D^{(\ast)}}$ anomalies. However, there are limitations of the EFT approach that must be kept in mind. First of all, there can be non-negligible corrections to the EFT description of the LHC observables if the EFT cutoff $\Lambda$ is not sufficiently larger than the partonic center-of-mass energy. Moreover, there are correlations among low- and high-energy observables that are only manifest within the concrete models.

The constraints on the concrete models are shown in Fig.~\ref{fig:LQcombined} for the leptoquarks $U_1$ (upper left), $S_1$ (upper right) and $R_2$ (bottom), with the leptoquark masses fixed to $2$~TeV, in agreement with current constraints from leptoquark pair-production at the LHC~\cite{Angelescu:2021lln}. For each scenario, a minimal set of two Yukawa couplings has been chosen to induce the SMEFT operators needed to explain $R_{D^{(\ast)}}$ in Fig.~\ref{fig:SMEFTcombined}. The leptoquark Lagrangians are defined in Table~\ref{tab:mediators}, with their tree-level matching to the SMEFT given in Table~\ref{tab:LQmatching}.

From Fig.~\ref{fig:LQcombined} we see that the three models are viable explanations of $R_{D^{(\ast)}}$ and we confirm that there is a complementarity of the low- and high-energy constraints. The high-$p_T$ constraints turn out to be slightly relaxed in all cases in comparison to the EFT computation, due to the propagation effects of the leptoquarks,
%%%%%%%%%%%%%%%
\begin{align}
    \dfrac{1}{(t-m^2)^2} \simeq \dfrac{1}{m^4} \left(1+ \dfrac{2t}{m^2}+\dots\right)\,,\qquad\quad t\in(-s,0)\,,
\end{align}
%%%%%%%%%%%%%%%
where $m$ denotes the leptoquark mass and we assume without loss of generality that the leptoquark is exchanged in the $t$-channel. The first power-correction on $t/m^2$ comes with a relative negative sign which reduces the cross-section estimated with the EFT~\cite{Iguro:2020keo,Jaffredo:2021ymt}. 

Going from the EFT description to concrete models also allows us to obtain additional constraints arising from the correlation of the different SMEFT operators. An example is given by the electroweak constraints for the $R_2$ model in Fig.~\ref{fig:LQcombined}(c), which are not present for the minimal set of EFT operators contributing to the charged currents in Fig.~\ref{fig:SMEFTcombined}(c). This correlation is also the reason why LHC constraints seem to be weak in Fig.~\ref{fig:SMEFTcombined}(b), but become relevant for the full $S_1$ models in Fig.~\ref{fig:LQcombined}(b).

%%%%%%%%%%%%%%%%%%%%%%%%%%%%%%%%%%%%%%%%%%%%%%%%%%%%%%%%%%%%%%%%%%%%%%%%%%
\section{Summary and Outlook}
\label{sec:outlook}
%%%%%%%%%%%%%%%%%%%%%%%%%%%%%%%%%%%%%%%%%%%%%%%%%%%%%%%%%%%%%%%%%%%%%%%%%%

In this paper, we have explored New Physics effects in semileptonic transition using the high-energy tails of the monolepton and dilepton searches at the LHC. We introduced a general parametrization of the  Drell-Yan scattering amplitudes in terms of form-factors describing generic New Physics at tree level, such as EFT contributions or new (bosonic) propagating degrees of freedom with $\cO(1~{\rm TeV})$ masses. For the SMEFT, this allowed us to systematically include the leading $\cO(1/\Lambda^{4})$ corrections to the $pp\to \ell\ell$ and $pp\to \ell\nu$ differential cross-sections. These corrections come from the New Physics squared contributions generated by the $d=6$ quark/lepton dipoles and semileptonic four-fermion operators, and from the SM-interfering terms generated by the dimension-$8$ operators with four fermions and two derivatives. 

We provided, for the first time, the complete high-$p_T$ Drell-Yan likelihood for the full set of relevant $d\leq 8$ SMEFT operators with arbitrary flavor indices. This was achieved by recasting the most recent run-II data-sets by ATLAS and CMS in the monolepton ($e\nu$, $\mu\nu$, $\tau\nu$) and dilepton ($ee$, $\mu\mu$, $\tau\tau$, $e\mu$, $e\tau$, $\mu\tau$) production channels as shown in Table~\ref{tab:lhc-searches}. These results are compiled in the {\tt Mathematica} package \HighPT, developed as a tool to facilitate phenomenological studies of New Physics flavor models at high-$p_T$~\cite{<HighPT:_A_Tool_for_High-pT_Collider_Studies_Beyond_the_Standard_Model><Allwicher;Lukas><Faroughy;Darius><Jaffredo;Florentin><Sumensari;Olcyr><Wilsch;Felix><2022>}. Furthermore, we derived single-parameter limits on the Wilson coefficients for the dipole and semileptonic four-fermion operators with any possible flavor combination. We also extracted two-parameter limits for specific vector operators, where we analyzed the effects of a non-diagonal CKM matrix with different quark-flavor alignments. These results are intended to be the initial steps towards a global analysis of the semileptonic sector of the SMEFT including all three fermion generations, where the goal is to combine LHC data from high-$p_T$ tails with complementary experimental probes such as electroweak precision tests and flavor observables~\cite{Allwicher:2022mcg}.

With the aim of providing a more complete EFT analysis, we also looked into the limitations of the SMEFT description of the Drell-Yan tails at the LHC. For different quark flavors, we assessed for several four-fermion operators the impact of a quadratic truncation of the EFT expansion at $\cO(1/\Lambda^4)$ and confronted it to the linear truncation at $\cO(1/\Lambda^2)$. We found in most instances that the New Physics squared corrections are too large to be neglected, and in the case of heavy-flavor initial quarks these corrections completely drive the limits even when disposing of the last experimental bins of the differential distributions. We explicitly checked this feature using the jack-knife and clipping procedures for specific examples \cite{10.2307/2334280,1604.06444}. Furthermore, following previous work in the literature \cite{Boughezal:2021tih,Kim:2022amu}, we estimated the relative contributions at $\cO(1/\Lambda^4)$ between dimension-$6$ and (the often neglected) dimension-$8$ operators. In this case we explicitly showed that the precision of the measurements in the tails of the dilepton and monoleptons distributions (for all three lepton generations) are only sensitive to dimension-$8$ effects if: (i) the initial states are first generation valence quarks and (ii) the dimension-$8$ Wilson coefficients are uncorrelated from the dimension-$6$ ones. This last requirement, however, may only be possible in fairly complicated ultraviolet setups.    
 
In addition to the SMEFT, we also computed the LHC likelihoods for all possible leptoquark mediators contributing non-resonantly to Drell-Yan production via $t/u$-channel exchange. These can also be found in \HighPT in full generality for each leptoquark state. We derived single-parameter constraints on each leptoquark with arbitrary couplings, and two-parameter limits to illustrate the interplay between bounds from lepton flavor conserving and LFV Drell-Yan production modes. Finally, as a highlight of our results, we considered the example of LFU tests in $B$-meson decays based on the $b\to c\ell\nu$ transition, showing that the LHC limits are complementary to low-energy data and to electroweak precision measurements, for the main scenarios aiming to accommodate the observed discrepancies in low-energy data. This analysis has been performed within the SMEFT, but also within viable leptoquark scenarios, including the leptoquark propagation effects in the LHC observables which are more accurate than naively recasting the EFT results.

Our study leaves room for a few improvements needed to fully exploit the potential of LHC data to constrain flavor-physics scenarios. These include the incorporation of electroweak and QCD corrections to the LHC observables for the SMEFT~\cite{Degrande:2020evl,Dawson:2021ofa} and leptoquark models~\cite{Haisch:2022lkt}, and the joint determination of quark PDFs and the BSM couplings~\cite{Greljo:2021kvv}, which would increase the accuracy of our constraints. The measurement of double differential-distributions is also a direction to be explored, as it could provide a useful handle to increase the sensitivity of Drell-Yan searches to New Physics effects~\cite{Panico:2021vav}. These improvements and directions are foreseen in the future, in addition to a complete model-independent combination of our Drell-Yan constraints with electroweak and low-energy flavor data~\cite{Allwicher:2022mcg}.

\section*{Acknowledgements}
We are grateful to Gino Isidori for encouraging us to pursue this project. We also thank Adam Falkowski for discussions regarding~Ref.~\cite{Breso-Pla:2021qoe}. This project has received support from the European Union’s Horizon 2020 research and innovation programme under the Marie Skłodowska-Curie grant agreement No~860881-HIDDeN. 
L.A., D.A.F. and F.W. received funding from the European Research Council (ERC) under the European Union’s Horizon 2020 research and innovation programme under grant agreement 833280 (FLAY), and by the Swiss National Science Foundation (SNF) under contract 200020-204428.

\appendix

\section{Validation of LHC Recast}
\label{app:recast-quality}
The constraints presented in this work are based on recasts of the ATLAS and CMS searches listed in Table~\ref{tab:lhc-searches}. To validate our recasts with the experimental searches, we simulated the Drell-Yan process for the New Physics models considered using our simulation pipeline discussed in Sec.~\ref{sec:recasts}.
We provide the information on the quality of our recast in Table~\ref{tab:recast_quality}. The second column should be interpreted as the acceptance times efficiency $(\mathcal{A}\times\epsilon)$ obtained in our recasts divided by values provided in the CMS and ATLAS papers for the specific benchmark New Physics models listed in the third column. We find an overall good agreement, except for the $e\tau$, $\mu\tau$ and $\tau\tau$ channels in which deviations are up to $\mathcal{O}(1)$. Note, in particular, that for these channels the main limitation of our recasts is the $\tau$-tagging efficiencies which are flat in $p_T$ and~$\eta$. This could be improved in the future by going beyond the default settings of Delphes.

\begin{table}[t]
    \renewcommand{\arraystretch}{1.3}
    \centering
    \begin{tabular}{ccccc}
        Search & Experiment & Ref. &$\dfrac{\left.\mathcal{A}\times\epsilon\right|_\mathrm{recast}}{\left.\mathcal{A}\times\epsilon\right|_\mathrm{search}}$ & Models 
        \\[0.3cm]\hline\hline
        $pp\to\tau\tau$ & ATLAS & \cite{ATLAS:2020zms} & $33\%-57\%$ & $H$ (0.2, 0.3, 0.4, 0.6, 1.0, 1.5, 2.0 and 2.5~TeV) 
        \\
        $pp\to\mu\mu$ & CMS & \cite{CMS:2021ctt} & $93\%-96\%$ & $Z^\prime$ (0.4, 0.6, 1.0, 1.5, 2.0 and 2.5~TeV) 
        \\
        $pp\to ee$ & CMS & \cite{CMS:2021ctt} & $58\%-69\%$ & $Z^\prime$ (0.4, 0.6, 1.0, 1.5, 2.0 and 2.5~TeV) 
        \\ \hline
        $pp\to\tau\nu$ & ATLAS & \cite{ATLAS:2021bjk} &  $93\%-167\%$ & $W'$ ($1, 2, 3, 4$ and $5$ TeV)
        \\
        $pp\to\mu\nu$ & ATLAS & \cite{ATLAS:2019lsy} & $127\%-145\%$ & $W'$ ($2$ and $7$ TeV)
        \\
        $pp\to e\nu$ & ATLAS & \cite{ATLAS:2019lsy} & $87\%-100\%$ & $W'$ ($2$ and $7$ TeV)
        \\ \hline
        $pp\to \tau\mu$ & CMS & \cite{CMS:2022fsw} & $180\%$ & $Z'$ ($1.6$ TeV) 
        \\
        $pp\to \tau e$ & CMS & \cite{CMS:2022fsw} & $150\%$ & $Z'$ ($1.6$ TeV) 
        \\
        $pp\to \mu e$ & CMS & \cite{CMS:2022fsw} & $97\%$ & $Z'$ ($1.6$ TeV)\\  \hline\hline
    \end{tabular}
    \caption{\sl \small Validation of our simulation of the New Physics signal used as a benchmark in the experimental analysis. In the fourth column, we compare against the \textit{acceptance $\times$ efficiency} $(\mathcal{A}\times\epsilon)$ provided by the experimental collaborations.}
    \label{tab:recast_quality}
\end{table}

% % % % % % % % % % %

\section{Notation and Conventions}
\label{sec:conventions}

We consider the same notation of Ref.~\cite{Jenkins:2013zja,Jenkins:2013wua,Alonso:2013hga} for the operators in the Warsaw basis. Quark and lepton doublets are denoted by $q$ and $l$, while up and down quarks and lepton singlets are denoted by $u$, $d$ and $e$, respectively. Our convention for the covariant derivative is given by
%%%%%%%%%%%%%%%%
\begin{equation}
D_\mu=\partial_\mu + i g^\prime \, Y B_\mu + i g\, \frac{\tau^I}{2} W^I_\mu + i g_3\, T^A G_\mu^A \,,
\end{equation}
%%%%%%%%%%%%%%%%
where $T^A=\lambda^A/2$ are the $SU(3)_c$ generators, $\tau^I$ are the Pauli matrices and $Y$ denotes the hypercharge. The Yukawa couplings are defined in the flavour basis as
%%%%%%%%%%%%%%%%
\begin{equation}
-\mathcal{L}_{\mathrm{yuk}} = H^\dagger\, \bar{d}\, Y_d\, q + \widetilde{H}^\dagger \,\bar{u}\, Y_u\, q + H^\dagger \, \bar{e}\, Y_e\, l+\mathrm{h.c.}\,,
\end{equation}
%%%%%%%%%%%%%%%%

\noindent where $Y_f$ (with $f\in \lbrace u,d,e \rbrace$) denote the Yukawa matrices and flavor indices have been omitted, $H$ corresponds to the SM Higgs doublet with the conjugate field defined as $\widetilde H\equiv \epsilon H^*$ and the $SU(2)$ anti-symmetric tensor is defined as $\epsilon \equiv i\tau^2$.
We work in the basis where $Y_\ell$ and $Y_d$ are diagonal matrices, while 
$Y_u$ contains the CKM matrix, $V \equiv V_{\mathrm{CKM}}$\,.\\

\subsection{SMEFT conventions}
\label{sec:SMEFTconventions}
For Hermitian semileptonic operators, we define the Hermitian conjugate $\cC^\dagger$ of a Wilson coefficients $\cC$ that can be a two-tensor or four-tensor in quark/lepton flavor space as:
\begin{align}
    [\,\cC^\dagger\,]_{\alpha\beta}&\equiv[\,\cC^*\,]_{\beta\alpha}\,,\\[0.3em]
    [\,\cC^\dagger\,]_{ij}&\equiv[\,\cC^*\,]_{ji}\,,\\[0.3em]
    [\,\cC^\dagger\,]_{\alpha\beta ij}&\equiv[\,\cC^*\,]_{\beta\alpha ji}\,.
\end{align}
These coefficients have a redundancy under the flavor index swappings $\alpha\!\leftrightarrow\! \beta$ and/or $i\!\leftrightarrow\!j$. One can remove this redundancy by adopting the following convention: for Hermitian Wilson coefficients with four flavor-indices, we fix the lepton indices to $\alpha\le\beta$, which also determines the ordering of the quark-flavor indices of the semileptonic four fermion operators if $\alpha < \beta$. For the case $\alpha=\beta$ we adopt the ordering $i\le j$. For operators with two indices, we also adopt the convention $\alpha \leq \beta$ and $i \leq j$.~\footnote{This choice agrees with the conventions adapted by \cmd{WCxf}~\cite{Aebischer:2017ugx} with the exception of the $\smash{\C{qe}{ij\alpha\beta}}$ operator in the original Warsaw basis that we dub~{$\big{[}C_{eq}\big{]}_{\alpha\beta ij}$\,.}
}

In order to keep expressions compact, it is useful to introduce the following notation for the signed sums of Wilson coefficients associated to operators with the same field content, but different gauge/Lorentz structure:
\begin{align}\label{eq:signed_sum_WC}
    \cC^{\,(i\,\pm\, j\,\pm\,k\,\pm\,\ldots)}\ &\equiv\ \cC^{(i)}\pm\cC^{(j)}\pm \cC^{(k)}\pm \ldots
\end{align}

\subsection{Form-factor rotations from the weak basis to the mass basis}
\label{sec:FFrotations}
In this Appendix, we provide the rotations between the mass and weak basis for the form-factors. Our expressions are given in terms of the matrices $V_u$ and $V_d$, which are the left-handed rotations to the mass basis for up-type and down-type quarks, respectively. The CKM matrix can then be expressed as $V = V_u^\dagger V_d$. These rotations read:
\begin{align}
    \mathcal{F}_{V}^{XL,\,ud} &\to 
    V_u \cF_{V}^{XL,\,ud} V_d^\dagger \,, &
    \mathcal{F}_{I \neq V}^{XL,\,ud} &\to
    \cF_{I \neq V}^{XL,\,ud} V_d^\dagger \,, &
    \mathcal{F}_{I \neq V}^{XR,\,ud} &\to
    V_u \cF_{I \neq V}^{XR,\,ud} \,, \nonumber\\[0.35em]
    \mathcal{F}_{V}^{XL,\,uu} &\to
    V_u \cF_{V}^{XL,\,uu} V_u^\dagger \,, &
    \mathcal{F}_{I \neq V}^{XL,\,uu} &\to
    \cF_{I \neq V}^{XL,\,uu} V_u^\dagger \,, & & \\[0.35em]
    \mathcal{F}_{V}^{XL,\,dd} &\to
    V_d \cF_{V}^{XL,\,dd} V_d^\dagger \,, &
    \mathcal{F}_{I \neq V}^{XL,\,dd} &\to
    \cF_{I \neq V}^{XL,\,dd} V_d^\dagger \,. & &\nonumber
\end{align}
The matching relations of these form-factors to the SMEFT and to models with concrete mediators are provided in the weak basis in Appendix~\ref{sec:FF_SMEFT} and \ref{sec:FF_concrete_UVmodel}, respectively. 
The rotation to the down-aligned basis can be obtained from the above by setting $V_u = V^\dagger$ and $V_d = \mathbb{1}$, while the up-aligned one with $V_u =\mathbb{1}$ and $V_d = V$.
The rotations for all the remaining form-factors can be obtained via:
\begin{align}
    \left[\cF_{V}^{XY,\,ud}\right] &= \left[\cF_{V}^{XY,\,du}\right]^\dagger \,,  
    &
    \\[0.35em] 
    \left[\cF_{I=S,T}^{LL,\,ud}\right] &= \left[\cF_{I=S,T}^{RR,\,du}\right]^\dagger \,, 
    &
    \left[\cF_{I=S,T}^{RR,\,ud}\right] &= \left[\cF_{I=S,T}^{LL,\,du}\right]^\dagger \,, 
    \\[0.35em]
    \left[\cF_{S}^{RL,\,ud}\right] &= \left[\cF_{S}^{LR,\,du}\right]^\dagger \,, 
    &
    \left[\cF_{S}^{LR,\,ud}\right] &= \left[\cF_{S}^{RL,\,du}\right]^\dagger \,, 
    \\[0.35em]
    \left[\cF_{D_\ell}^{RX,\,ud}\right] &= -\left[\cF_{D_\ell}^{LX,\,du}\right]^\dagger \,,
    &
    \left[\cF_{D_\ell}^{LX,\,ud}\right] &= -\left[\cF_{D_\ell}^{RX,\,du}\right]^\dagger \,,
    \\[0.35em]
    \left[\cF_{D_q}^{XL,\,ud}\right] &= -\left[\cF_{D_q}^{XR,\,du}\right]^\dagger \,,
    &
    \left[\cF_{D_q}^{XR,\,ud}\right] &= -\left[\cF_{D_q}^{XL,\,du}\right]^\dagger \,,
    \\[0.35em]
    \left[\cF_{I = S,T}^{LL,\,qq}\right] &= \left[\cF_{I = S,T}^{RR,\,qq}\right]^\dagger \,, 
    &
    \left[\cF_{S}^{RL,\,qq}\right] &= \left[\cF_{S}^{LR,\,qq}\right]^\dagger \,, 
    \\[0.35em]
    \left[\cF_{D_\ell}^{RX,\,qq}\right] &= -\left[\cF_{D_\ell}^{LX,\,qq}\right]^\dagger \,,
    &
    \left[\cF_{D_q}^{XL,\,qq}\right] &= -\left[\cF_{D_q}^{XR,\,qq}\right]^\dagger \,.
\end{align}

\section{Form-factors in the SMEFT}
\label{sec:FF_SMEFT}
%----------------------------------------------------
\subsection{Scalar and tensor form-factors}
\label{sec:SMEFT_match_scalar_tensor}

In this Appendix, we provide the full matching between the form-factor coefficients and the SMEFT Wilson coefficients. Notice that contributions coming from the redefinition of input parameters are not included in the matching conditions below.

%---------------------------------------------------
\subsubsection*{Neutral currents}
The matching to the SMEFT Wilson coefficients depends on whether the process is up-quark or down-quark initiated, $\bar u_i u_j\to \ell_\alpha^-\ell_\beta^+$ and $\bar d_i d_j\to \ell_\alpha^-\ell_\beta^+$. It is given by:
\begin{align}
    \cF^{RR,\, uu}_{S\,(0,0)}&= -\frac{v^2}{\Lambda^2}\,\cC_{lequ}^{(1)} \,,
    &
    \cF^{RL,\,dd}_{S\,(0,0)}&= \frac{v^2}{\Lambda^2}\,\cC_{ledq} \,,
    \\[0.35em]
    \cF^{RR,\, uu}_{T\,(0,0)}&= -\frac{v^2}{\Lambda^2}\,\cC_{lequ}^{(3)} \,.
    &\nonumber
\end{align}

\subsubsection*{Charged currents}
For the monolepton processes $\bar u_i d_j\to \ell^-_\alpha \nu_\beta$ (and their conjugates), the matching reads:
\begin{align}
    \cF^{RR,\,du}_{S\,(0,0)}&= \frac{v^2}{\Lambda^2}\,{\cC_{lequ}^{(1)}} \,,
    &
    \cF^{RL,\,du}_{S\,(0,0)}&= \frac{v^2}{\Lambda^2}\,{\cC_{ledq}} \,,
    \\[0.35em]
    \cF^{RR,\,du}_{T\,(0,0)}&= \frac{v^2}{\Lambda^2}\,{\cC_{lequ}^{(3)}} \,.
    &\nonumber
\end{align}
%

%----------------------------------------------------
\subsection{Vector form-factors}
%----------------------------------------------------
\label{sec:SMEFT_match_vector}

\subsubsection*{Neutral currents $\bar u_iu_j\to l_\alpha^-l^+_\beta$}

The matching of the coefficients $\cF^{XY}_{V\, (0,0)}$ to the SMEFT for up-quark initiated processes is given by:
\begin{align}
\begin{split}
\cF_{V\,(0,0)}^{LL,\,uu} & = \frac{v^2}{\Lambda^2}\,\cC_{lq}^{\,(1-3)}  + \frac{v^4}{2\Lambda^4}\,\cC_{l^2q^2H^2}^{\,(1+2-3-4)} + \frac{v^2 m_Z^2}{2\Lambda^4} \left[ g_l^L\,\cC_{q^2H^2D^3}^{\,(1-2-3+4)} + g_u^L \, \cC_{l^2H^2D^3}^{\,(1-2+3-4)} \right] ,\label{eq:FVLL00uu}\\[0.35em]
\cF_{V\,(0,0)}^{LR,\,uu} & = \frac{v^2}{\Lambda^2}\,\cC_{lu} + \frac{v^4}{2\Lambda^4}\,\cC_{l^2u^2H^2}^{\,(1+2)} + \frac{v^2 m_Z^2}{2\Lambda^4} \left[ g_l^L\,\cC_{u^2H^2D^3}^{\,(1-2)} + g_u^L \, \cC_{l^2H^2D^3}^{\,(1-2+3-4)} \right] \,,\\[0.35em]
\cF_{V\,(0,0)}^{RL,\,uu} & = \frac{v^2}{\Lambda^2}\,\cC_{qe} + \frac{v^4}{2\Lambda^4}\,\cC_{q^2e^2H^2}^{\,(1-2)} + \frac{v^2 m_Z^2}{2\Lambda^4} \left[ g_l^R\,\cC_{q^2H^2D^3}^{\,(1-2-3+4)} + g_u^L \,\cC_{e^2H^2D^3}^{\,(1-2)} \right] \,,\\[0.35em]
\cF_{V\,(0,0)}^{RR,\,uu} & = \frac{v^2}{\Lambda^2}\,\cC_{eu} + \frac{v^4}{2\Lambda^4}\,\cC_{e^2u^2H^2} + \frac{v^2 m_Z^2}{2\Lambda^4} \left[ g_l^R\,\cC_{u^2H^2D^3}^{\,(1-2)} + g_u^R \,\cC_{e^2H^2D^3}^{\,(1-2)}  \right] \,.
\end{split}
\end{align} 
The higher-order coefficients $\cF_{V\,(1,0)}^{XY}$ and $\cF_{V\,(0,1)}^{XY}$ are generated in the SMEFT at $d=8$ from momentum-dependent contact operators in the class {\bf$\psi^4D^2$}. These read:
\begin{align}
\cF_{V\,(1,0)}^{LL,\,uu} & = \frac{v^4}{\Lambda^4} \,\cC_{l^2q^2D^2}^{\,(1+2-3-4)}\ \,, 
&
\cF_{V\,(0,1)}^{LL,\,uu}   &=   2\frac{v^4}{\Lambda^4}\, \cC_{l^2q^2D^2}^{\,(2-4)}\,, \label{eq:FVLL11uu}
\\[0.35em]
\cF_{V\,(1,0)}^{LR,\,uu} & = \frac{v^4}{\Lambda^4} \,\cC_{l^2u^2D^2}^{\,(1+2)}\ \,, 
&
\cF_{V\,(0,1)}^{LR,\,uu}   &=   2\frac{v^4}{\Lambda^4}\, \cC_{l^2u^2D^2}^{\,(2)}\,, 
\\[0.35em]
\cF_{V\,(1,0)}^{RL,\,uu} & = \frac{v^4}{\Lambda^4} \,\cC_{q^2e^2D^2}^{\,(1+2)}\ \,, 
&
\cF_{V\,(0,1)}^{RL,\,uu}   &=   2\frac{v^4}{\Lambda^4}\, \cC_{q^2e^2D^2}^{\,(2)}\,, 
\\[0.35em]
\cF_{V\,(1,0)}^{RR,\,uu} & = \frac{v^4}{\Lambda^4} \,\cC_{e^2u^2D^2}^{\,(1+2)}\ \,, 
& 
\cF_{V\,(0,1)}^{RR,\,uu}   &=   2\frac{v^4}{\Lambda^4}\, \cC_{e^2u^2D^2}^{\,(2)}\,. 
\end{align}
The matching of the pole residues to the SMEFT is given by:
\begin{align}
\begin{split}
\delta \cS_{(Z)}^{LL,\,uu}
= &-2 \frac{m_Z^2}{\Lambda^2}\,\left[g_{l}^L\,\cC_{Hq}^{\,(1-3)} + g_{u}^L\,\cC_{Hl}^{\,(1+3)}\right] 
+ \frac{v^2 m_Z^2}{\Lambda^4}\cC_{Hl}^{\,(1+3)}\cC_{Hq}^{\,(1-3)} \\
&-\frac{v^2 m_Z^2}{\Lambda^4} \left[ g_l^L \left(\cC_{q^2H^4D}^{\,(1)} -2\cC_{q^2H^4D}^{\,(2)}\right) + g_u^L \left( \cC_{l^2H^4D}^{\,(1)} +2 \cC_{l^2H^4D}^{\,(2)} \right) \right] \\ 
&+ \frac{m_Z^4}{2\Lambda^4} \left[ g_l^L\,\cC_{q^2H^2D^3}^{\,(1-2-3+4)} + g_u^L \, \cC_{l^2H^2D^3}^{\,(1-2+3-4)} \right]  \,,\\[0.35em]
\delta \cS_{(Z)}^{LR,\,uu}
= &-2\frac{m_Z^2}{\Lambda^2}\,\left[g_{l}^L\,\cC_{Hu} + g_{u}^R\,\cC_{Hl}^{\,(1+3)}\right] + \frac{v^2 m_Z^2}{\Lambda^4} \cC_{Hl}^{\,(1+3)}\cC_{Hu} \\
&-\frac{v^2 m_Z^2}{\Lambda^4} \left[g_l^L \cC_{u^2H^4D} + g_u^R \left( \cC_{l^2H^4D}^{\,(1)} +2 \cC_{l^2H^4D}^{\,(2)} \right)  \right] \\ 
&+ \frac{m_Z^4}{2\Lambda^4} \left[ g_l^L\,\cC_{u^2H^2D^3}^{\,(1-2)} + g_u^R \, \cC_{l^2H^2D^3}^{\,(1-2+3-4)} \right]   \,,\\[0.35em]
\delta \cS_{(Z)}^{RL,\,uu}
= &-2 \frac{m_Z^2}{\Lambda^2}\,\left[g_{l}^R\,\cC_{Hq}^{\,(1-3)} + g_{u}^L\,\cC_{He}\right] + \frac{v^2 m_Z^2}{\Lambda^4}\,\cC_{He}\,\cC_{Hq}^{\,(1-3)} \\
&-\frac{v^2 m_Z^2}{\Lambda^4} \left[ g_l^R \left(\cC_{q^2H^4D}^{\,(1)} -2\cC_{q^2H^4D}^{\,(2)}\right)  + g_u^L \cC_{e^2H^4D}  \right] \\
&+ \frac{m_Z^4}{2\Lambda^4} \left[ g_l^R\,\cC_{q^2H^2D^3}^{\,(1-2-3+4)} + g_u^L \,\cC_{e^2H^2D^3}^{\,(1-2)} \right] \,,\\[0.35em]
\delta \cS_{(Z)}^{RR,\,uu}
= &-2\frac{m_Z^2}{\Lambda^2}\,\left[g_{l}^R\,\cC_{Hu} + g_{u}^R\,\cC_{He}\right] + \frac{v^2 m_Z^2}{\Lambda^4} \left[\cC_{He}\,\cC_{Hu} - g_L^R \cC_{u^2H^4D} - g_u^R \cC_{e^2H^4D}  \right] \\
&+ \frac{m_Z^4}{2\Lambda^4} \left[ g_l^R\,\cC_{u^2H^2D^3}^{\,(1-2)} + g_u^R \,\cC_{e^2H^2D^3}^{\,(1-2)}  \right] \,.
\end{split}
\end{align}

\subsubsection*{Neutral currents $\bar d_id_j\to \ell_\alpha^-\ell^+_\beta$}

For down-quark initiated processes the matching for the leading coefficient $\cF^{XY}_{V\, (0,0)}$ is given by:
\begin{align}
\cF_{V\,(0,0)}^{LL,\,dd} &=\frac{v^2}{\Lambda^2}\,\cC_{lq}^{\,(1+3)}+\frac{v^4}{2\Lambda^4}\, \cC_{l^2q^2H^2}^{\,(1+2+3+4)} + \frac{v^2 m_Z^2}{2\Lambda^4} \left[ g_l^L\,\cC_{q^2H^2D^3}^{\,(1-2+3-4)\,\dagger} + g_d^L \, \cC_{l^2H^2D^3}^{\,(1-2+3-4)} \right] ,\label{eq:FVLL00dd}\\[0.35em]
\cF_{V\,(0,0)}^{LR,\,dd} & = \frac{v^2}{\Lambda^2}\,\cC_{ld} + \frac{v^4}{2\Lambda^4}\,\cC_{l^2d^2H^2}^{\,(1+2)} + \frac{v^2 m_Z^2}{2\Lambda^4} \left[ g_l^L\,\cC_{d^2H^2D^3}^{\,(1-2)\,\dagger} + g_d^{R} \, \cC_{l^2H^2D^3}^{\,(1-2+3-4)} \right] \,,\\[0.35em]
\cF_{V\,(0,0)}^{RL,\,dd} & = \frac{v^2}{\Lambda^2}\,\cC_{qe} + \frac{v^4}{2\Lambda^4}\,\cC_{q^2e^2H^2}^{\,(1+2)} + \frac{v^2 m_Z^2}{2\Lambda^4} \left[ g_l^R\,\cC_{q^2H^2D^3}^{\,(1-2+3-4)\,\dagger} + g_d^L \,\cC_{e^2H^2D^3}^{\,(1-2)} \right] \,,\\[0.35em]
\cF_{V\,(0,0)}^{RR,\,dd} & = \frac{v^2}{\Lambda^2}\,\cC_{ed} + \frac{v^4}{2\Lambda^4}\,\cC_{e^2d^2H^2} + \frac{v^2 m_Z^2}{2\Lambda^4} \left[ g_l^R\,\cC_{d^2H^2D^3}^{\,(1-2)\,\dagger} + g_d^R \,\cC_{e^2H^2D^3}^{\,(1-2)}  \right] \,.
\end{align} 
The higher order coefficients $\cF_{V\,(1,0)}^{XY}$ and $\cF_{V\,(0,1)}^{XY}$ read:
\begin{align}
\cF_{V\,(1,0)}^{LL,\,dd} & = \frac{v^4}{\Lambda^4} \,\cC_{l^2q^2D^2}^{\,(1+2+3+4)}\ \,, 
& 
\cF_{V\,(0,1)}^{LL,\,dd}   &=   2\frac{v^4}{\Lambda^4}\, \cC_{l^2q^2D^2}^{\,(2+4)}\,, \label{VLLd}
\\[0.35em]
\cF_{V\,(1,0)}^{LR,\,dd} & = \frac{v^4}{\Lambda^4} \,\cC_{l^2d^2D^2}^{\,(1+2)}\ \,, 
&
\cF_{V\,(0,1)}^{LR,\,dd}   &=   2\frac{v^4}{\Lambda^4}\, \cC_{l^2d^2D^2}^{\,(2)}\,, \label{VLRd}
\\[0.35em]
\cF_{V\,(1,0)}^{RL,\,dd} & = \frac{v^4}{\Lambda^4} \,\cC_{q^2e^2D^2}^{\,(1+2)}\ \,, 
&
\cF_{V\,(0,1)}^{RL,\,dd}   &=   2\frac{v^4}{\Lambda^4}\, \cC_{q^2e^2D^2}^{\,(2)}\,, \label{VRLd}
\\[0.35em]
\cF_{V\,(1,0)}^{RR,\,dd} & = \frac{v^4}{\Lambda^4} \,\cC_{e^2d^2D^2}^{\,(1+2)}\ \,, 
& 
\cF_{V\,(0,1)}^{RR,\,dd}   &=   2\frac{v^4}{\Lambda^4}\, \cC_{e^2d^2D^2}^{\,(2)}\,, \label{VRRd}
\end{align}
and the pole residues are given by:
\begin{align}
\begin{split}
\delta \cS_{(Z)}^{LL,\,dd}
= &-2 \frac{m_Z^2}{\Lambda^2}\,\left[g_{l}^L\,\cC_{Hq}^{\,(1+3)} + g_{d}^L\,\cC_{Hl}^{\,(1+3)}\right] 
+ \frac{v^2 m_Z^2}{\Lambda^4}\cC_{Hl}^{\,(1+3)}\cC_{Hq}^{\,(1+3)} \\
&-\frac{v^2 m_Z^2}{\Lambda^4} \left[ g_l^L \left(\cC_{q^2H^4D}^{\,(1)} +2\cC_{q^2H^4D}^{\,(2)}\right) + g_d^L \left( \cC_{l^2H^4D}^{\,(1)} +2 \cC_{l^2H^4D}^{\,(2)} \right) \right] \\ 
&+ \frac{m_Z^4}{2\Lambda^4} \left[ g_l^L\,\cC_{q^2H^2D^3}^{\,(1-2+3-4)} + g_d^L \, \cC_{l^2H^2D^3}^{\,(1-2+3-4)} \right]  \,,\\[0.35em]
\delta \cS_{(Z)}^{LR,\,dd}
= &-2\frac{m_Z^2}{\Lambda^2}\,\left[g_{l}^L\,\cC_{Hd} + g_{d}^R\,\cC_{Hl}^{\,(1+3)}\right] + \frac{v^2 m_Z^2}{\Lambda^4} \cC_{Hl}^{\,(1+3)}\cC_{H{d}} \\
&-\frac{v^2 m_Z^2}{\Lambda^4} \left[g_l^L \cC_{d^2H^4D} + g_d^R \left( \cC_{l^2H^4D}^{\,(1)} +2 \cC_{l^2H^4D}^{\,(2)} \right)  \right] \\ 
&+ \frac{m_Z^4}{2\Lambda^4} \left[ g_l^L\,\cC_{d^2H^2D^3}^{\,(1-2)} + g_d^{R} \, \cC_{l^2H^2D^3}^{\,(1-2+3-4)} \right]   \,,\\[0.35em]
\delta \cS_{(Z)}^{RL,\,dd}
= &-2 \frac{m_Z^2}{\Lambda^2}\,\left[g_{l}^R\,\cC_{Hq}^{\,(1+3)} + g_{d}^L\,\cC_{He}\right] + \frac{v^2 m_Z^2}{\Lambda^4}\,\cC_{He}\,\cC_{Hq}^{\,(1+3)} \\
&-\frac{v^2 m_Z^2}{\Lambda^4} \left[ g_l^R \left(\cC_{q^2H^4D}^{\,(1)} +2\cC_{q^2H^4D}^{\,(2)}\right)  + g_d^L \cC_{e^2H^4D}  \right] \\
&+ \frac{m_Z^4}{2\Lambda^4} \left[ g_l^R\,\cC_{q^2H^2D^3}^{\,(1-2+3-4)} + g_d^L \,\cC_{e^2H^2D^3}^{\,(1-2)} \right] \,,\\[0.35em]
\delta \cS_{(Z)}^{RR,\,dd}
&= -2\frac{m_Z^2}{\Lambda^2}\,\left[g_{l}^R\,\cC_{Hd}+ g_{d}^R\,\cC_{He}\right] + \frac{v^2 m_Z^2}{\Lambda^4}\left[\cC_{He}\,\cC_{Hd} - g_L^R \cC_{d^2H^4D} - g_d^R \cC_{e^2H^4D} \right] \\
&+ \frac{m_Z^4}{2\Lambda^4} \left[ g_l^R\,\cC_{d^2H^2D^3}^{\,(1-2)} + g_d^R \,\cC_{e^2H^2D^3}^{\,(1-2)}  \right]   \,.
\end{split}
\end{align}

\subsubsection*{Charged-currents $\bar u_id_j\to \ell_\alpha^-\bar\nu_\beta$} 

The matching of the leading form-factor coefficients $\cF_{V\,(0,0)}^{LL(LR)}$ is given by
\begin{align}
\nonumber
\left[\cF_{V\,(0,0)}^{LL,\,ud}\right]_{\alpha\beta ij} & =\ 2 \frac{v^2}{\Lambda^2}\,\left[\cC_{l q}^{\,(3)}\right]_{\alpha\beta ij}+\frac{v^4}{\Lambda^4}\,\left(\left[\cC_{l^2q^2H^2}^{(3)}\right]_{\alpha\beta ij}+i(1-\delta_{ij})\left[\cC_{l^2q^2H^2}^{(5)}\right]_{\alpha\beta i j}\right)+ \label{eq:FVLL00ud}\\
&- \frac{g^2}{2}\frac{v^4}{2\Lambda^4} \left[\left( \cC_{l^2H^2D^3}^{\,(3)} - \cC_{l^2H^2D^3}^{\,(4)\,\dagger} \right)\mathbb{1}_{q} + \left( \cC_{q^2H^2D^3}^{\,(3)\,\dagger} - \cC_{q^2H^2D^3}^{\,(4)} \right)\mathbb{1}_{l}  \right]_{\alpha\beta ij}\,.
\end{align}
Notice that the $d=8$ operator $\smash{\cO_{l^2q^2H^2}^{(5)}=\epsilon^{I\!J\!K} (\bar l \gamma^\mu \tau^I l) (\bar q \gamma_\mu \tau^J q) (H^\dag \tau^K H)}$ only contributes to flavor violating processes and therefore does not enter into neutral currents, but does affect charged currents like e.g. $u\bar s\to \ell^\pm\nu$ at order $\cO(1/\Lambda^4)$. The effects of this operator are small because they only interfere with CKM suppressed transitions in the SM. For the higher order regular coefficients we obtain the following matching to the SMEFT:
\begin{align}  \label{eq:FVLL11ud}
\cF_{V\,(1,0)}^{LL,\,ud} &= 2\frac{v^4}{\Lambda^4}\,\cC^{\,(3+4)}_{l^2q^2D^2} \,,
&
\cF_{V\,(0,1)}^{LL,\,ud} &= 4\frac{v^4}{\Lambda^4}\,\cC^{\,(4)}_{l^2q^2D^2}\,.
\end{align}
The matching of the pole residues is given by:
\begin{align}
\nonumber
\delta \cS_{(W)}^{LL,\,ud}
= &\ \frac{g^2}{2} \frac{v^2}{\Lambda^2} \left[\cC_{Hl}^{\,(3)\,\dagger}\,\mathbb{1}_{q}+ \cC_{Hl}^{\,(3)}\, \mathbb{1}_{\ell} \right]
+ \frac{g^2}{2} \frac{v^4}{\Lambda^4}\, \cC_{Hl}^{(3)\,\dagger}\, \cC_{Hq}^{\,(3)} \\ \nonumber
&+ \frac{g^2}{2}\frac{v^4}{2\Lambda^4}\left[ \cC_{l^2H^4D}^{\,(2^\dagger-3^\dagger+4)} \mathbb{1}_{q} + \cC_{q^2H^4D}^{\,(2-3+4^\dagger)} \mathbb{1}_{l}  \right] \\
&- \frac{g^2}{2}\frac{v^2 m_W^2}{2\Lambda^4} \left[\left( \cC_{l^2H^2D^3}^{\,(3)} - \cC_{l^2H^2D^3}^{\,(4)\,\dagger} \right)\mathbb{1}_{q} + \left( \cC_{q^2H^2D^3}^{\,(3)\,\dagger} - \cC_{q^2H^2D^3}^{\,(4)} \right)\mathbb{1}_{l}  \right]  \,,
\\[0.35em] \nonumber
\delta \cS_{(W)}^{LR,\,ud} 
= & \frac{g^2}{4} \frac{v^2}{\Lambda^2}\, \cC_{Hud}\,\mathbb{1}_{l} \,,
\end{align}
and $\delta \cS_{(W)}^{LL,\,du}=\delta \cS_{(W)}^{LL,\,ud}$ and $\delta \cS_{(W)}^{LR,\,du}=\delta \cS_{(W)}^{LR,\,ud\,\dagger}$.

%----------------------------------------------------
\subsection{Dipole form-factors}
%----------------------------------------------------
\label{sec:SMEFT_match_dipole}

\subsubsection*{Neutral currents}

The matching conditions for the $Z$ boson and photon pole coefficients are give by:  
\begin{align}
\cS^{RR,qq}_{D_l\,(\gamma)} &= \cS^{RL,qq}_{D_l\,(\gamma)} = - \sqrt{2} e Q_q \frac{v^2}{\Lambda^2} \left( s_w \cC_{eW} - c_w \cC_{eB} \right) \mathbb{1}_q \,,
\\
\cS^{LR,qq}_{D_l\,(\gamma)} &= \cS^{LL,qq}_{D_l\,(\gamma)} = \sqrt{2} e Q_q \frac{v^2}{\Lambda^2} \left( s_w \cC_{eW}^\dagger - c_w \cC_{eB}^\dagger \right) \mathbb{1}_q \,,
\\
\cS^{RR,qq}_{D_l\,(Z)} &= \cS^{RL,qq}_{D_l\,(Z)} = - \sqrt{2} g_{q}^{R/L} \frac{v^2}{\Lambda^2} \left( c_w \cC_{eW} + s_w \cC_{eB} \right) \mathbb{1}_q \,,
\\
\cS^{LR,qq}_{D_l\,(Z)} &= \cS^{LL,qq}_{D_l\,(Z)} = \sqrt{2} g_{q}^{R/L} \frac{v^2}{\Lambda^2} \left( c_w \cC_{eW}^\dagger + s_w \cC_{eB}^\dagger \right) \mathbb{1}_q \,,
\end{align}
\begin{align}
\cS^{RR,dd}_{D_q\,(\gamma)} &= \cS^{LR,dd}_{D_q\,(\gamma)} = \sqrt{2} e Q_e \frac{v^2}{\Lambda^2} \left( s_w \cC_{dW} - c_w \cC_{dB} \right) \mathbb{1}_\ell \,,
\\
\cS^{RL,dd}_{D_q\,(\gamma)} &= \cS^{LL,dd}_{D_q\,(\gamma)} = - \sqrt{2} e Q_e \frac{v^2}{\Lambda^2} \left( s_w \cC_{dW}^\dagger - c_w \cC_{dB}^\dagger \right) \mathbb{1}_\ell \,,
\\
\cS^{RR,dd}_{D_q\,(Z)} &= \cS^{LR,dd}_{D_q\,(Z)} = \sqrt{2} g_{l}^{R/L} \frac{v^2}{\Lambda^2} \left( c_w \cC_{dW} + s_w \cC_{dB} \right) \mathbb{1}_\ell \,,
\\
\cS^{RL,dd}_{D_q\,(Z)} &= \cS^{LL,dd}_{D_q\,(Z)} = - \sqrt{2} g_{l}^{R/L} \frac{v^2}{\Lambda^2} \left( c_w \cC_{dW}^\dagger + s_w \cC_{dB}^\dagger \right) \mathbb{1}_\ell \,,
\end{align}
\begin{align}
\cS^{RR,uu}_{D_q\,(\gamma)} &= \cS^{LR,uu}_{D_q\,(\gamma)} = - \sqrt{2} e Q_e \frac{v^2}{\Lambda^2} \left( s_w \cC_{uW} + c_w \cC_{uB} \right) \mathbb{1}_\ell \,,
\\
\cS^{RL,uu}_{D_q\,(\gamma)} &= \cS^{LL,uu}_{D_q\,(\gamma)} = \sqrt{2} e Q_e \frac{v^2}{\Lambda^2} \left( s_w \cC_{uW}^\dagger + c_w \cC_{uB}^\dagger \right) \mathbb{1}_\ell \,,
\\
\cS^{RR,uu}_{D_q\,(Z)} &= \cS^{LR,uu}_{D_q\,(Z)} = - \sqrt{2} g_{l}^{R/L} \frac{v^2}{\Lambda^2} \left( c_w \cC_{uW} - s_w \cC_{uB} \right) \mathbb{1}_\ell \,,
\\
\cS^{RL,uu}_{D_q\,(Z)} &= \cS^{LL,uu}_{D_q\,(Z)} = \sqrt{2} g_{l}^{R/L} \frac{v^2}{\Lambda^2} \left( c_w \cC_{uW}^\dagger - s_w \cC_{uB}^\dagger \right) \mathbb{1}_\ell \,.
\end{align}

\subsubsection*{Charged currents}

The $W$ boson pole coefficients read:
\begin{align}
\cS^{RL,ud}_{D_l\,(W)} &= \sqrt{2} g \frac{v^2}{\Lambda^2} \cC_{eW} \mathbb{1}_q \,,
&
\cS^{LL,du}_{D_l\,(W)} &= - \sqrt{2} g \frac{v^2}{\Lambda^2} \cC_{eW}^\dagger \mathbb{1}_q \,,
\\
\cS^{LR,ud}_{D_q\,(W)} &= - \sqrt{2} g \frac{v^2}{\Lambda^2} \cC_{dW} \mathbb{1}_\ell \,,
&
\cS^{LL,du}_{D_q\,(W)} &= \sqrt{2} g \frac{v^2}{\Lambda^2} \cC_{dW}^\dagger \mathbb{1}_\ell \,,
\\
\cS^{LR,du}_{D_q\,(W)} &= - \sqrt{2} g \frac{v^2}{\Lambda^2} \cC_{uW} \mathbb{1}_\ell \,,
&
\cS^{LL,ud}_{D_q\,(W)} &= \sqrt{2} g \frac{v^2}{\Lambda^2} \cC_{uW}^\dagger \mathbb{1}_\ell \,.
\end{align}

\section{Form-factors in Concrete UV Models}
\label{sec:FF_concrete_UVmodel}

We now give the matching relations to the pole form-factors for the tree-level mediators collected in Table~\ref{tab:mediators}. The poles below are defined in terms of the mass and width of each mediator as $\Omega_i\equiv m^2_i-im_i\Gamma_i$.
\subsection{Scalar form-factors}

\subsubsection*{Neutral-currents $\bar u_iu_j\to \ell_\alpha^-\ell^+_\beta$\,}

\begin{align}
    \frac{1}{v^2}\left[\cF^{LL,\,uu}_{S,\,\text{Poles}}\right]_{\alpha\beta ij} &=
    -\frac{\big{[}\xi_H^\ell\big{]}_{\alpha\beta}\big{[}\xi_H^u\big{]}_{ij}}{\hat s - \Omega_{H}}
    +\frac{\big{[}\xi_A^\ell\big{]}_{\alpha\beta}\big{[}\xi_A^u\big{]}_{ij}}{\hat s - \Omega_{A}}
    -\frac{\frac{1}{2} {[y_2^L]_{i\beta}}{[y_2^R]_{j\alpha}^\ast}}{\hat t-\Omega_{R_2^{(5/3)}}}
    +\frac{\frac{1}{2} {[y_1^L]_{j\beta}}{[y_1^R]_{i\alpha}^\ast}}{\hat u-\Omega_{S_1}}
    \,,\\[0.35em]
    \frac{1}{v^2}\left[\cF^{LR,\,uu}_{S,\,\text{Poles}}\right]_{\alpha\beta ij} &= -\frac{\big{[}\xi_H^\ell\big{]}_{\alpha\beta}\big{[}\xi_H^u\big{]}_{ij}}{\hat s - \Omega_{H}}
    -\frac{\big{[}\xi_A^\ell\big{]}_{\alpha\beta}\big{[}\xi_A^u\big{]}_{ij}}{\hat s - \Omega_{A}}\,.
    \end{align}
    
\subsubsection*{Neutral-currents $\bar d_id_j\to \ell_\alpha^-\ell^+_\beta$\,} 

    \begin{align}
    \frac{1}{v^2}\left[\cF^{LL,\,dd}_{S,\,\text{Poles}}\right]_{\alpha\beta ij} &=-\frac{\big{[}\xi_H^\ell\big{]}_{\alpha\beta}\big{[}\xi_H^d\big{]}_{ij}}{\hat s - \Omega_{H}}
    +\frac{\big{[}\xi_A^\ell\big{]}_{\alpha\beta}\big{[}\xi_A^d\big{]}_{ij}}{\hat s - \Omega_{A}}\,,\\[0.35em]
    \frac{1}{v^2}\left[\cF^{LR,\,dd}_{S,\,\text{Poles}}\right]_{\alpha\beta ij} &=
    -\frac{\big{[}\xi_H^\ell\big{]}_{\alpha\beta}\big{[}\xi_H^d\big{]}_{ij}}{\hat s - \Omega_{H}}
    -\frac{\big{[}\xi_A^\ell\big{]}_{\alpha\beta}\big{[}\xi_A^d\big{]}_{ij}}{\hat s - \Omega_{A}}
    -\frac{2{[x_1^L]_{i\beta}}{[x_1^R]_{j\alpha}^\ast}}{\hat t-\Omega_{U_1}}
    +\frac{2{[x_2^R]_{i\alpha}^\ast}{{[x_2^L]}_{j\beta}}}{\hat u-\Omega_{V_2^{(4/3)}}}
    \,.
\end{align}

\subsubsection*{Charged-currents $\bar u_id_j\to \ell_\alpha^-\bar\nu_\beta$} 

\begin{align}
    \frac{1}{v^2}\left[\cF^{LL,\,ud}_{S,\,\text{Poles}}\right]_{\alpha\beta ij} &= 
    -\frac{\big{[}\xi_{H^+}^{\ell_R}\big{]}^*_{\beta\alpha}\big{[}\xi_{H^+}^{q_L}\big{]}_{ij}}{\hat s - \Omega_{H^-}}
    + \frac{\frac{1}{2}\,[y_2^L]_{i\beta}{[y_2^R]_{j\alpha}^\ast}}{\hat t-\Omega_{R_2^{(2/3)}}}
    - \frac{\frac{1}{2}\,{[y_1^L]_{j\beta}}{[y_1^R]_{i\alpha}^\ast}}{\hat u-\Omega_{S_1}}
    \,,\\[0.35em]
    \frac{1}{v^2}\left[\cF^{LR,\,ud}_{S,\,\text{Poles}}\right]_{\alpha\beta ij} &= 
    -\frac{\big{[}\xi_{H^+}^{\ell_R}\big{]}^*_{\beta\alpha}\big{[}\xi_{H^+}^{q_R}\big{]}_{ij}}{\hat s - \Omega_{H^-}}
    -\frac{2\,[x_1^L]^{i\beta}{[x_1^R]_{j\alpha}^\ast}}{\hat t-\Omega_{U_1}}
    +\frac{2\,[x_2^L]^{j\beta}{[x_2^R]_{i\alpha}^\ast}}{\hat u-\Omega_{V_2^{(1/3)}}}
    \,,\\[0.35em]
    \frac{1}{v^2}\left[\cF^{RL,\,ud}_{S,\,\text{Poles}}\right]_{\alpha\beta ij} &=
    -\frac{\big{[}\xi_{H^+}^{\ell_L}\big{]}^*_{\beta\alpha}\big{[}\xi_{H^+}^{q_L}\big{]}_{ij}}{\hat s - \Omega_{H^-}}
    -\frac{2\,{[x_1^L]_{j\alpha}^*}[\bar{x}_1^R]_{i\beta}}{\hat t-\Omega_{U_1}}
    +\frac{[\widetilde{x}_2^R]_{j\beta}{[\widetilde{x}_2^L]_{i\alpha}^*}}{\hat u-\Omega_{\widetilde{V}_2^{(1/3)}}}
    \,,\\[0.35em]
    \frac{1}{v^2}\left[\cF^{RR,\,ud}_{S,\,\text{Poles}}\right]_{\alpha\beta ij} &= 
    -\frac{\big{[}\xi_{H^+}^{\ell_L}\big{]}^*_{\beta\alpha}\big{[}\xi_{H^+}^{q_R}\big{]}_{ij}}{\hat s - \Omega_{H^-}}
    +\frac{\frac{1}{2}\,[\widetilde{y}_2^R]_{i\beta}{[\widetilde{y}_2^L]_{j\alpha}^\ast}}{\hat t-\Omega_{\widetilde{R}_2^{(2/3)}}}
    +\frac{\frac{1}{2}\,[\bar{y}_1^R]_{j\beta}{[{y}_1^L]_{i\alpha}^\ast}}{\hat u-\Omega_{S_1}}
    \,.
\end{align}

\subsection{Vector form-factors}

\subsubsection*{Neutral-currents $\bar u_iu_j\to \ell_\alpha^-\ell^+_\beta$\,}

\begin{align}
    \frac{1}{v^2}\left[\cF^{LL,\,uu}_{V,\,\text{Poles}}\right]_{\alpha\beta ij} &=
     \frac{{[g_1^l]_{\alpha\beta}}[g_1^q]_{ij}}{\hat s-\Omega_{Z'}}
    -\frac{{[g_3^l]_{\alpha\beta}}[g_3^q]_{ij}}{\hat s-\Omega_{W'}} 
    +\frac{2\,{{[x_3^L]}_{i\beta}}{[x_3^L]_{j\alpha}^\ast}}{\hat t-\Omega_{U_3}^{(5/3)}} 
    \nonumber\\&
    -\frac{\frac{1}{2}\,{[y_3^L]_{j\beta}}{[y_3^L]_{i\alpha}^\ast}}{\hat u-\Omega_{S_3^{(1/3)}}}
    -\frac{\frac{1}{2}\,{{[y_1^L]}_{j\beta}}{[y_1^L]_{i\alpha}^*}}{\hat u-\Omega_{S_1}}
    \,,\\[0.35em]
    \frac{1}{v^2}\left[\cF^{LR,\,uu}_{V,\,\text{Poles}}\right]_{\alpha\beta ij} &=
    \frac{{[g_1^l]_{\alpha\beta}}[g_1^u]_{ij}}{\hat s-\Omega_{Z'}}
    +\frac{\frac{1}{2}\,{[y_2^L]_{i\beta}}{[y_2^L]_{j\alpha}^*}}{\hat t-\Omega_{R_2^{(5/3)}}}
    +\frac{{[\widetilde{x}_2^L]_{i\beta}}{[\widetilde{x}_2^L]_{i\alpha}^*}}{\hat u-\Omega_{\widetilde{V}_2^{(1/3)}}}
    \,,\\[0.35em]
    \frac{1}{v^2}\left[\cF^{RL,\,uu}_{V,\,\text{Poles}}\right]_{\alpha\beta ij} &= 
    \frac{{[g_1^e]_{\alpha\beta}} [g_1^q]_{ij} }{\hat s-\Omega_{Z'}}
    +\frac{\frac{1}{2}\,{[y_2^R]_{i\beta}}{[y_2^R]_{j\alpha}^\ast}}{\hat t-\Omega_{R_2^{(5/3)}}}
    -\frac{{[x_2^R]_{j\beta}}{[x_2^R]_{i\alpha}^\ast}}{\hat u-\Omega_{V_2^{(1/3)}}}
    \,,\\[0.35em]
    \frac{1}{v^2}\left[\cF^{RR,\,uu}_{V,\,\text{Poles}}\right]_{\alpha\beta ij} &= 
    \frac{{[g_1^e]_{\alpha\beta}}[g_1^u]_{ij}}{\hat s-\Omega_{Z'}}
    +\frac{{[\widetilde{x}_1^R]}_{i\beta}[\widetilde{x}_1^R]_{j\alpha}^\ast}{\hat t-\Omega_{\widetilde{U}_1}}
    -\frac{\frac{1}{2}\,{[y_1^R]_{j\beta}}{[y_1^R]_{i\alpha}^\ast}}{\hat u-\Omega_{S_1}}
    \,.
\end{align}

\subsubsection*{Neutral currents $\bar d_id_j\to \ell_\alpha^-\ell^+_\beta$}

\begin{align}
    \frac{1}{v^2}\left[\cF^{LL,\,dd}_{V,\,\text{Poles}}\right]_{\alpha\beta ij} &=
     \frac{{[g_1^l]_{\alpha\beta}}[g_1^q]_{ij}}{\hat s-\Omega_{Z'}}
    +\frac{{[g_3^l]_{\alpha\beta}}[g_3^q]_{ij}}{\hat s-\Omega_{W'}} 
    +\frac{{[x_1^L]_{i\beta}}{[x_1^L]_{j\alpha}^\ast}}{\hat t-\Omega_{U_1}}
    \nonumber\\&
    +\frac{{[x_3^L]_{i\beta}}{[x_3^L]_{j\alpha}^\ast}}{\hat t-\Omega_{U_3^{(2/3)}}}
    -\frac{{[y_3^L]_{j\beta}}{[y_3^L]_{i\alpha}^\ast}}{\hat u-\Omega_{S_3^{(4/3)}}}\,,\\[0.35em]
    \frac{1}{v^2}\left[\cF^{LR,\,dd}_{V\,\text{Poles}}\right]_{\alpha\beta ij} &= 
    \frac{{[g_1^l]_{\alpha\beta}} [g_1^d]_{ij}}{\hat s-\Omega_{Z'}}
    +\frac{\frac{1}{2}\,{[\widetilde{y}_2^L]_{i\beta}}{[\widetilde{y}_2^L]_{j\alpha}^\ast}}{\hat t-\Omega_{\widetilde{R}_2^{(2/3)}}}
    -\frac{{[x_2^L]_{j\beta}}{[x_2^L]_{i\alpha}^\ast}}{\hat u-\Omega_{V_2^{(4/3)}}},\\[0.35em]
    \frac{1}{v^2}\left[\cF^{RL,\,dd}_{V,\,\text{Poles}}\right]_{\alpha\beta ij} &= 
    \frac{{[g_1^e]_{\alpha\beta}} [g_1^q]_{ij}}{\hat s-\Omega_{Z'}}
    +\frac{\frac{1}{2}{[y_2^R]_{i\beta}}{[y_2^R]_{j\alpha}^\ast}}{\hat t-\Omega_{R_2^{(2/3)}}}
    -\frac{{[x_2^R]_{j\beta}}{[x_2^R]_{i\alpha}^\ast}}{\hat u-\Omega_{V_2^{(4/3)}}}\,,\\[0.35em]
    \frac{1}{v^2}\left[\cF^{RR,\,dd}_{V,\,\text{Poles}}\right]_{\alpha\beta ij} &= 
    \frac{{[g_1^e]_{\alpha\beta}} [g_1^d]_{ij}}{\hat s-\Omega_{Z'}}
    +\frac{[{x}_1^R]_{i\beta}[{x}_1^R]_{j\alpha}^\ast}{\hat t-\Omega_{U_1}}
    +\frac{\frac{1}{2}\,{[\widetilde{y}_1^R]_{j\beta}}{[\widetilde{y}_1^R]_{i\alpha}^\ast}}{\hat u-\Omega_{\widetilde{S}_1}}\,.
\end{align}

\subsubsection*{Charged-currents $\bar u_id_j\to \ell_\alpha^-\bar\nu_\beta$}

\begin{align}
    \frac{1}{v^2}\left[\cF^{LL,\,ud}_{V,\,\text{Poles}}\right]_{\alpha\beta ij} &=
    \frac{2\,{[g_3^l]_{\alpha\beta}}[g_3^q]_{ij}}{\hat s-\Omega_{W'}}
    +\frac{[x_1^L]_{i\beta}{[x_1^L]_{j\alpha}^\ast}}{\hat t-\Omega_{U_1}}
    -\frac{[x_3^L]_{i\beta}{[x_3^L]_{j\alpha}^\ast}}{\hat t-\Omega_{U_3^{(2/3)}}}
    \nonumber\\&
    +\frac{\frac{1}{2}\,{[y_1^L]_{j\beta}}{[y_1^L]_{i\alpha}}^*}{\hat u-\Omega_{S_1}}
    -\frac{\frac{1}{2}\,{[y_3^L]_{j\beta}}{[y_3^L]_{i\alpha}}^*}{\hat u-\Omega_{S_3^{(1/3)}}},\\[0.35em]
    \frac{1}{v^2}\left[\cF^{LR,\,ud}_{V,\,\text{Poles}}\right]_{\alpha\beta ij} &= 0 \,,\\[0.35em]
    \frac{1}{v^2}\left[\cF^{RL,\,ud}_{V,\,\text{Poles}}\right]_{\alpha\beta ij} &= 0 \,,\\[0.35em]
    \frac{1}{v^2}\left[\cF^{RR,\,ud}_{V,\,\text{Poles}}\right]_{\alpha\beta ij} &=
    \frac{[\widetilde{g}_1^l]_{\alpha\beta}{[\widetilde{g}_1^q]_{ij}}}{\hat s-\Omega_{\widetilde{Z}}}
    +\frac{{[x_1^R]_{j\alpha}^\ast}{[\bar{x}_1^R]_{i\beta}}}{\hat t-\Omega_{U_1}}
    -\frac{\frac{1}{2}[\bar{y}_1^R]_{j\beta}{[y_1^R]_{i\alpha}^\ast}}{\hat u-\Omega_{S_1}}
    \,.
\end{align}

\subsection{Tensor form-factors}

\subsubsection*{Neutral-currents $\bar u_iu_j\to \ell_\alpha^-\ell^+_\beta$}

\begin{align}
    \frac{1}{v^2}\left[\cF^{LL,\,uu}_{T,\,\text{Poles}}\right]_{\alpha\beta ij} &=
    -\frac{\frac{1}{8}\,{[y_2^L]_{i\beta}}{[y_2^R]_{j\alpha}}^*}{\hat t-\Omega_{R_2^{(5/3)}}}
    -\frac{\frac{1}{8}\,{[y_1^L]_{j\beta}}{[y_1^R]_{i\alpha}}^*}{\hat u-\Omega_{S_1}}
    \,.
\end{align}

\subsubsection*{Charged-currents $\bar u_id_j\to \ell_\alpha^-\bar\nu_\beta$}
 
\begin{align}
    \frac{1}{v^2}\left[\cF^{LL,\,ud}_{T,\,\text{Poles}}\right]_{\alpha\beta ij} &=
      \frac{\frac{1}{8}\,[y_2^L]_{i\beta}{[y_2^R]_{j\alpha}^\ast}}{\hat t-\Omega_{R_2^{(2/3)}}}
    + \frac{\frac{1}{8}\,{[y_1^L]_{j\beta}}{[y_1^R]_{i\alpha}^\ast}}{\hat u-\Omega_{S_1}}\,,\\[0.35em]
    \frac{1}{v^2}\left[\cF^{RR,\,ud}_{T,\,\text{Poles}}\right]_{\alpha\beta ij} &=
    \frac{\frac{1}{8}\,[\widetilde{y}_2^R]_{i\beta}{[\widetilde{y}_2^L]_{j\alpha}^\ast}}{\hat  t-\Omega_{\widetilde{R}_2^{(2/3)}}}
    - \frac{\frac{1}{8}\,[\bar{y}_1^R]_{j\beta}{[{y}_1^L]_{i\alpha}^\ast}}{\hat u-\Omega_{S_1}} \,.
\end{align}

%----------------------------------------------------
\section{Form-factors in the $\nu$SMEFT}
%----------------------------------------------------
\label{sec:nuSMEFT}

The matching of the form-factors with dimension-6 operators involving light right-handed neutrinos read:
\begin{align}
\cF_{V\,(0,0)}^{RR,ud}&= \frac{v^2}{\Lambda^2}\,\cC_{eNud}\,, \\
\cF_{S\,(0,0)}^{RL,ud}&= \frac{v^2}{\Lambda^2}\,\cC_{lNuq}\,, \\
\cF_{S\,(0,0)}^{RR,ud}&= -\frac{v^2}{\Lambda^2}\,\cC_{lNqd}^{\,(1)}\,, \\
\cF_{T\,(0,0)}^{RR,ud}&= -\frac{v^2}{\Lambda^2}\,\cC_{lNqd}^{\,(3)}\,.
\end{align}

%================================
%================================
%================================
%================================
%================================
%================================
%================================

% - - - - - - - - - - - - - - - - - - - - - - - - - - - - - - - - - 
\section{Semileptonic SMEFT operators}
\label{sec:SMEFT-ops}

The SMEFT operators of dimension $d=6$ \cite{Grzadkowski:2010es} and $d=8$ \cite{Murphy:2020rsh} that are relevant to our study are defined in Table~\ref{tab:dim6_ops}, and Tables~\ref{tab:dim8_ops_1} and \ref{tab:dim8_ops_2}, respectively.

\begin{table}[p!]
  \begin{center}
\begin{tabular}{@{\hspace{1em}}c@{\hspace{2em}}c@{\hspace{2em}}c@{\hspace{2em}}c@{\hspace{1em}}}
  %  \hline\hline
    \bf $d=6$ &   \bf $\psi^4$   &\bf $pp\to\ell\ell$  &\bf $pp\to\ell\nu$ \\
\hline\hline
$\cO_{lq}^{(1)}$   & $(\bar l_\alpha \gamma^\mu l_\beta)(\bar q_i \gamma_\mu q_j)$     &  $\checkmark$  & --\\  
$\cO_{lq}^{(3)}$   & $(\bar l_\alpha \gamma^\mu\tau^I l_\beta)(\bar q_i \gamma_\mu\tau^I q_j)$  &  $\checkmark$   & $\checkmark$  \\ 
\hline
$\cO_{lu}$         & $(\bar l_\alpha \gamma^\mu l_\beta)(\bar u_i \gamma_\mu u_j)$     &  $\checkmark$   & --\\
$\cO_{ld}$         & $(\bar l_\alpha \gamma^\mu l_\beta)(\bar d_i \gamma_\mu d_j)$     & $\checkmark$   & --\\
\hline
$\cO_{eq}$         & $(\bar e_\alpha \gamma^\mu e_\beta)(\bar q_i \gamma_\mu q_j)$     &  $\checkmark$   & --\\
$\cO_{eu}$         & $(\bar e_\alpha \gamma^\mu e_\beta)(\bar u_i \gamma_\mu u_j)$     &  $\checkmark$   & --\\
$\cO_{ed}$         & $(\bar e_\alpha \gamma^\mu e_\beta)(\bar d_i \gamma_\mu d_j)$     &  $\checkmark$   & --\\
\hline
$\cO_{ledq}$ + h.c.       & $(\bar l_\alpha e_\beta)(\bar d_i q_j)$  &  $\checkmark$   & $\checkmark$ \\
$\cO_{lequ}^{(1)}$ + h.c. & $(\bar l_\alpha e_\beta)\varepsilon(\bar q_i u_j)$  & $\checkmark$   & $\checkmark$ \\
$\cO_{lequ}^{(3)}$ + h.c. & $(\bar l_\alpha \sigma^{\mu\nu} e_\beta)\varepsilon(\bar q_i \sigma_{\mu\nu} u_j)$  & $\checkmark$    &$\checkmark$   \\
    \hline\hline
    \end{tabular}
    
\vspace{1cm}

\begin{tabular}{@{\hspace{1em}}c@{\hspace{2em}}c@{\hspace{2em}}c@{\hspace{2em}}c@{\hspace{1em}}}
%\hline\hline
 \bf$d=6$ &   \bf$\psi^2H^2D$   &\bf $pp\to\ell\ell$  &\bf $pp\to\ell\nu$ \\
\hline\hline
$\cO_{Hl}^{(1)}$   & $(\bar l_\alpha \gamma^\mu l_\beta)(H^\dagger i\overleftrightarrow{D}_\mu H)$     &  $\checkmark$  & --\\  
$\cO_{Hl}^{(3)}$   & $(\bar l_\alpha \gamma^\mu\tau^I l_\beta)(H^\dagger i\overleftrightarrow{D}^I_\mu H)$  &  $\checkmark$   & $\checkmark$   \\
\hline
$\cO_{Hq}^{(1)}$   & $(\bar q_i \gamma^\mu q_j)(H^\dagger i\overleftrightarrow{D}_\mu H)$    &  $\checkmark$  & --\\  
$\cO_{Hq}^{(3)}$   & $(\bar q_i \gamma^\mu\tau^I q_j)(H^\dagger i\overleftrightarrow{D}^I_\mu H)$  &  $\checkmark$   & $\checkmark$   \\
\hline
$\cO_{He}$         & $(\bar e_\alpha \gamma^\mu e_\beta)(H^\dagger i\overleftrightarrow{D}_\mu H)$  &  $\checkmark$   & --  \\ 
$\cO_{Hu}$         & $(\bar u_i \gamma^\mu u_j)(H^\dagger i\overleftrightarrow{D}_\mu H)$  &  $\checkmark$   & --  \\
$\cO_{Hd}$         & $(\bar d_i \gamma^\mu d_j)(H^\dagger i\overleftrightarrow{D}_\mu H)$  &  $\checkmark$   & --  \\ 
\hline
$\cO_{Hud}$ + h.c.        & $(\bar u_i \gamma^\mu d_j)(\widetilde H^\dagger iD_\nu H)$  & --   & $\checkmark$  \\ 
\hline\hline
\end{tabular} 
 \vspace{1cm}

\begin{tabular}{@{\hspace{1em}}c@{\hspace{2em}}c@{\hspace{2em}}c@{\hspace{2em}}c@{\hspace{1em}}}
%\hline\hline
 \bf$d=6$ &   \bf$\psi^2XH$ + h.c.   &\bf $pp\to\ell\ell$  &\bf $pp\to\ell\nu$ \\
\hline\hline
$\cO_{eW}$   & $(\bar l_\alpha \sigma^{\mu\nu} e_\beta)\, \tau^I H W^I_{\mu\nu}$     &  $\checkmark$  &  $\checkmark$\\  
$\cO_{eB}$   & $(\bar l_\alpha \sigma^{\mu\nu} e_\beta)\, H B_{\mu\nu}$  &  $\checkmark$   & --   \\
\hline
$\cO_{uW}$   & $(\bar q_i \sigma^{\mu\nu} u_j)\, \tau^I \widetilde{H} W^I_{\mu\nu}$     &  $\checkmark$  &  $\checkmark$\\  
$\cO_{uB}$   & $(\bar q_i \sigma^{\mu\nu} u_j)\, \widetilde{H} B_{\mu\nu}$  &  $\checkmark$   & --  \\
\hline
$\cO_{dW}$   & $(\bar q_i \sigma^{\mu\nu} d_j)\, \tau^I H W^I_{\mu\nu}$     &  $\checkmark$  &  $\checkmark$\\  
$\cO_{dB}$   & $(\bar q_i \sigma^{\mu\nu}\,d_j) H B_{\mu\nu}$  &  $\checkmark$   & --   \\
\hline\hline
\end{tabular}    
  \caption{\sl\small SMEFT $d=6$ operators that contribute to the processes $pp\to\ell\ell$ and $pp\to\ell\nu$. Semileptonic operators ({\bf$\psi^4$}) are collected in the upper table, whereas Higgs-current ({\bf$\psi^2 H^2 D$}) and dipole operators ({\bf$\psi^2 X H$})  appear in the middle and bottom tables, respectively. We use the operators in the Warsaw basis~\cite{Grzadkowski:2010es}, where we renamed the operator $\cO_{qe}$ to $\cO_{eq}$ to conveniently have lepton- before quark-flavor indices.}
\label{tab:dim6_ops}
  \end{center}
\end{table}
% - - - -
\begin{table}[p!]
  \begin{center}
    \begin{tabular}{@{\hspace{1em}}c@{\hspace{2em}}c@{\hspace{2em}}c@{\hspace{1em}}c@{\hspace{2em}}}
   % \hline\hline
 \bf$d=8$ &   \bf$\psi^4H^2$   &\bf $pp\to\ell\ell$  &\bf $pp\to\ell\nu$ \\
\hline\hline
$\cO_{l^2q^2H^2}^{(1)}$  & $(\bar l_\alpha \gamma^\mu l_\beta) (\bar q_i \gamma_\mu q_j) (H^\dag H)$ & $\checkmark$  & -- \\
$\cO_{l^2q^2H^2}^{(2)}$  & $(\bar l_\alpha \gamma^\mu \tau^I l_\beta) (\bar q_i \gamma_\mu q_j) (H^\dag \tau^I H)$ &  $\checkmark$ & -- \\
$\cO_{l^2q^2H^2}^{(3)}$  & $(\bar l_\alpha \gamma^\mu \tau^I l_\beta) (\bar q_i \gamma_\mu \tau^I q_j) (H^\dag H)$ &  $\checkmark$   &  $\checkmark$ \\ 
$\cO_{l^2q^2H^2}^{(4)}$  & $(\bar l_\alpha \gamma^\mu l_\beta) (\bar q_i \gamma_\mu \tau^I q_j) (H^\dag \tau^I H)$ & $\checkmark$  & -- \\
$\cO_{l^2q^2H^2}^{(5)}$  & $\epsilon^{I\!J\!K} (\bar l_\alpha \gamma^\mu \tau^I l_\beta) (\bar q_i \gamma_\mu \tau^J q_j) (H^\dag \tau^K H)$ & --  &  $\checkmark$ \\ 
\hline
$\cO_{l^2u^2H^2}^{(1)}$  &  $(\bar l_\alpha \gamma^\mu l_\beta) (\bar u_i \gamma_\mu u_j) (H^\dag H)$  &  $\checkmark$   & -- \\
$\cO_{l^2u^2H^2}^{(2)}$  &  $(\bar l_\alpha \gamma^\mu \tau^I l_\beta) (\bar u_i \gamma_\mu u_j) (H^\dag \tau^I H)$  &  $\checkmark$  & -- \\
$\cO_{l^2d^2H^2}^{(1)}$  &  $(\bar l_\alpha \gamma^\mu l_\beta) (\bar d_i \gamma_\mu d_j) (H^\dag H)$  &  $\checkmark$   & -- \\
$\cO_{l^2d^2H^2}^{(2)}$  &  $(\bar l_\alpha \gamma^\mu \tau^I l_\beta) (\bar d_i \gamma_\mu d_j) (H^\dag \tau^I H)$  &  $\checkmark $  & -- \\
\hline
$\cO_{e^2q^2H^2}^{(1)}$  &  $(\bar e_\alpha \gamma_\mu e_\beta) (\bar q_i \gamma^\mu q_j) (H^\dag H)$  &  $\checkmark$   & -- \\
$\cO_{e^2q^2H^2}^{(2)}$  &  $(\bar e_\alpha \gamma_\mu e_\beta) (\bar q_i \gamma^\mu \tau^I q_j) (H^\dag \tau^I H)$  &  $\checkmark$   & -- \\
\hline
$\cO_{e^2u^2H^2}$  &  $(\bar e_\alpha \gamma^\mu e_\beta) (\bar u_i \gamma_\mu u_j) (H^\dag H)$ &  $\checkmark$   & -- \\
$\cO_{e^2d^2H^2}$  &   $(\bar e_\alpha \gamma^\mu e_\beta) (\bar d_i \gamma_\mu d_j) (H^\dag H)$  & $\checkmark$   & -- \\
\hline\hline
\end{tabular} 

\vspace{1cm}
  
\begin{tabular}{@{\hspace{1em}}c@{\hspace{2em}}c@{\hspace{2em}}c@{\hspace{2em}}c@{\hspace{1em}}}
%\hline\hline
\bf $d=8$ &\bf   $\psi^4D^2$  &\bf $pp\to\ell\ell$  &\bf $pp\to\ell\nu$ \\
\hline\hline
$\cO_{l^2q^2D^2}^{(1)}$  &  $D^\nu (\bar l_\alpha \gamma^\mu l_\beta) D_\nu (\bar q_i \gamma_\mu q_j)$   &  $\checkmark$  & -- \\
$\cO_{l^2q^2D^2}^{(2)}$  &  $(\bar l_\alpha \gamma^\mu \overleftrightarrow{D}^\nu l_\beta) (\bar q_i \gamma_\mu \overleftrightarrow{D}_\nu q_j)$  &  $\checkmark$   & -- \\
$\cO_{l^2q^2D^2}^{(3)}$  & \ \ \  $D^\nu (\bar l_\alpha \gamma^\mu \tau^I l_\beta) D_\nu (\bar q_i\gamma_\mu \tau^I q_j)$   &  $\checkmark$  & $\checkmark$ \\
$\cO_{l^2q^2D^2}^{(4)}$  &  $(\bar l_\alpha \gamma^\mu \overleftrightarrow{D}^{I\nu} l_\beta) (\bar q_i \gamma_\mu \overleftrightarrow{D}^I_\nu q_j)$ & $\checkmark$   & $\checkmark$ \\
\hline
$\cO_{l^2u^2D^2}^{(1)}$  &  $D^\nu (\bar l_\alpha \gamma^\mu l_\beta) D_\nu (\bar u_i \gamma_\mu u_j)$   & $\checkmark$   & -- \\
$\cO_{l^2u^2D^2}^{(2)}$  &  $(\bar l_\alpha \gamma^\mu \overleftrightarrow{D}^\nu l_\beta) (\bar u_i \gamma_\mu \overleftrightarrow{D}_\nu u_j)$  &  $\checkmark$  & -- \\
$\cO_{l^2d^2D^2}^{(1)}$  &  $D^\nu (\bar l_\alpha \gamma^\mu l_\beta) D_\nu (\bar d_i \gamma_\mu d_j)$   &  $\checkmark$   & -- \\
$\cO_{l^2d^2D^2}^{(2)}$  &  $(\bar l_\alpha \gamma^\mu \overleftrightarrow{D}^\nu l_\beta) (\bar d_i \gamma_\mu \overleftrightarrow{D}_\nu d_j)$  &  $\checkmark$  & -- \\
\hline 
$\cO_{e^2q^2D^2}^{(1)}$  &  $D_\nu (\bar e_\alpha \gamma_\mu e_\beta) D^\nu (\bar q_i \gamma^\mu q_j)$   &  $\checkmark$  & -- \\
$\cO_{e^2q^2D^2}^{(2)}$  &  $(\bar e_\alpha \gamma_\mu \overleftrightarrow{D}_\nu e_\beta) (\bar q_i \gamma^\mu \overleftrightarrow{D}^\nu q_j)$  &  $\checkmark$   & -- \\
\hline
$\cO_{e^2u^2D^2}^{(1)}$  &  $D^\nu (\bar e_\alpha \gamma^\mu e_\beta) D_\nu (\bar u_i \gamma_\mu u_j)$ &  $\checkmark$   & -- \\
$\cO_{e^2u^2D^2}^{(2)}$  &  $(\bar e_\alpha \gamma^\mu \overleftrightarrow{D}^\nu e_\beta) (\bar u_i \gamma_\mu \overleftrightarrow{D}_\nu u_j)$&  $\checkmark$  & -- \\
$\cO_{e^2d^2D^2}^{(1)}$  &  $D^\nu (\bar e_\alpha \gamma^\mu e_\beta) D_\nu (\bar d_i \gamma_\mu d_j)$  & $\checkmark$   & -- \\
$\cO_{e^2d^2D^2}^{(2)}$  &  $(\bar e_\alpha \gamma^\mu \overleftrightarrow{D}^\nu e_\beta) (\bar d_i \gamma_\mu \overleftrightarrow{D}_\nu d_j)$&  $\checkmark$   & -- \\
    \hline\hline
    \end{tabular} 
  \caption{\sl\small SMEFT $d=8$ four-fermion operators that contribute to the processes $pp\to\ell\ell$ and $pp\to\ell\nu$. We use the basis defined in Ref.~\cite{Murphy:2020rsh}, where we relabeled the operators $\smash{\cO_{q^2 e^2 H^2}^{(k)}}$ and $\smash{\cO_{q^2 e^2 D^2}^{(k)}}$ to $\smash{\cO_{e^2 q^2 H^2}^{(k)}}$ and $\smash{\cO_{e^2 q^2 D^2}^{(k)}}$, respectively, to conveniently have lepton- before quark-flavor indices.}
  \label{tab:dim8_ops_1}
  \end{center}
\end{table}
% % - - - -

% % - - - - - - - - - - - - - - - - - - - - - - - - - - - - - - - - - - 
\begin{table}[t!]
  \begin{center}
    \begin{tabular}{@{\hspace{1em}}c@{\hspace{1em}}c@{\hspace{1em}}c@{\hspace{1em}}c@{\hspace{1em}}}
    %\hline\hline
    \bf $d=8$ &\bf   $\psi^2H^4D$   &\bf $pp\to\ell\ell$  &\bf $pp\to\ell\nu$ \\
\hline\hline
$\cO_{l^2H^4D}^{(1)}$  &  $ i (\bar l_\alpha \gamma^{\mu} l_\beta) (H^{\dag} \overleftrightarrow{D}_{\mu} H) (H^{\dag} H)$   &  $\checkmark$  & --  \\
$\cO_{l^2H^4D}^{(2)}$  &  $ i (\bar l_\alpha \gamma^{\mu} \tau^I l_\beta) [(H^{\dag} \overleftrightarrow{D}_{\mu}^I H) (H^{\dag} H) + (H^{\dag} \overleftrightarrow{D}_{\mu} H) (H^{\dag} \tau^I H)]$   & $\checkmark$   &  $\checkmark$ \\
$\cO_{l^2H^4D}^{(3)}$  &  $ \epsilon^{IJK} (\bar l_\alpha \gamma^{\mu} \tau^I l_\beta) (H^{\dag} \overleftrightarrow{D}_{\mu}^J H) (H^{\dag} \tau^K H)$   & --  &  $\checkmark$\\
$\cO_{l^2H^4D}^{(4)}$  &  $ \epsilon^{IJK} (\bar l_\alpha \gamma^{\mu} \tau^I l_\beta) (H^{\dag} \tau^J H) (D_{\mu} H)^{\dag} \tau^K H$    &  --  & $\checkmark$\\
\hline
$\cO_{q^2H^4D}^{(1)}$  &  $ i (\bar q_i \gamma^{\mu} q_j) (H^{\dag} \overleftrightarrow{D}_{\mu} H) (H^{\dag} H)$  & $\checkmark$ & --\\
$\cO_{q^2H^4D}^{(2)}$  &  $ i (\bar q_i \gamma^{\mu} \tau^I q_j) [(H^{\dag} \overleftrightarrow{D}_{\mu}^I H) (H^{\dag} H) + (H^{\dag} \overleftrightarrow{D}_{\mu} H) (H^{\dag} \tau^I H)]$ & $\checkmark$ &  $\checkmark$\\
$\cO_{q^2H^4D}^{(3)}$  &  $ i \epsilon^{IJK} (\bar q_i \gamma^{\mu} \tau^I q_j) (H^{\dag} \overleftrightarrow{D}_{\mu}^J H) (H^{\dag} \tau^K H)$  & -- &  $\checkmark$\\
$\cO_{q^2H^4D}^{(4)}$  &  $ \epsilon^{IJK} (\bar q_i \gamma^{\mu} \tau^I q_j) (H^{\dag} \tau^J H) (D_{\mu} H)^{\dag} \tau^K H$  &  --  &  $\checkmark$\\
\hline
$\cO_{e^2H^4D}$  &  $ i (\bar e_\alpha \gamma^{\mu} e_\beta) (H^{\dag} \overleftrightarrow{D}_{\mu} H) (H^{\dag} H)$    &  $\checkmark$ & --  \\
$\cO_{u^2H^4D}$  &  $ i (\bar u_i \gamma^{\mu} u_j) (H^{\dag} \overleftrightarrow{D}_{\mu} H) (H^{\dag} H)$  &  $\checkmark$  & --  \\
$\cO_{d^2H^4D}$  &  $ i (\bar d_i \gamma^{\mu} d_j) (H^{\dag} \overleftrightarrow{D}_{\mu} H) (H^{\dag} H)$  &  $\checkmark$ & --  \\
    \hline\hline
    \end{tabular}
    
    \vspace{1cm}

    \begin{tabular}{@{\hspace{1em}}c@{\hspace{1em}}c@{\hspace{1em}}c@{\hspace{1em}}c@{\hspace{1em}}}
    %\hline\hline
    \bf $d=8$ &\bf   $\psi^2H^2D^3$   &\bf $pp\to\ell\ell$  &\bf $pp\to\ell\nu$ \\
\hline\hline
$\cO_{l^2H^2D^3}^{(1)}$  &  $ i (\bar l_\alpha \gamma^{\mu} D^\nu l_\beta) \,(D_{(\mu}D_{\nu)} H)^{\dag} H$  &  $\checkmark$  & --  \\
$\cO_{l^2H^2D^3}^{(2)}$  &  $ i (\bar l_\alpha \gamma^{\mu} D^\nu l_\beta) \,H^{\dag} (D_{(\mu}D_{\nu)}  H)$  &  $\checkmark$  & --  \\
$\cO_{l^2H^2D^3}^{(3)}$  &  $ i (\bar l_\alpha \gamma^{\mu}\tau^I D^\nu l_\beta)\, (D_{(\mu}D_{\nu)} H)^{\dag}\tau^I H)$  &  $\checkmark$  & $\checkmark$  \\
$\cO_{l^2H^2D^3}^{(4)}$  &  $ i (\bar l_\alpha \gamma^{\mu}\tau^I D^\nu l_\beta) \, H^{\dag} \tau^I (D_{(\mu}D_{\nu)} H)$  &  $\checkmark$  & $\checkmark$  \\
\hline
$\cO_{e^2H^2D^3}^{(1)}$  &  $ i (\bar e_\alpha \gamma^{\mu} D^\nu e_\beta) \,(D_{(\mu}D_{\nu)} H)^{\dag} H)$  &  $\checkmark$  & --  \\
$\cO_{e^2H^2D^3}^{(2)}$  &  $ i (\bar e_\alpha \gamma^{\mu} D^\nu e_\beta) \,H^{\dag} (D_{(\mu}D_{\nu)}  H)$  &  $\checkmark$  & --  \\
\hline
$\cO_{q^2H^2D^3}^{(1)}$  &  $ i (\bar q_i \gamma^{\mu} D^\nu q_j) \,(D_{(\mu}D_{\nu)} H)^{\dag} H$  &  $\checkmark$  & --  \\
$\cO_{q^2H^2D^3}^{(2)}$  &  $ i (\bar q_i \gamma^{\mu} D^\nu q_j) \,H^{\dag} (D_{(\mu}D_{\nu)}  H)$  &  $\checkmark$  & --  \\
$\cO_{q^2H^2D^3}^{(3)}$  &  $ i (\bar q_i \gamma^{\mu}\tau^I D^\nu q_j)\, (D_{(\mu}D_{\nu)} H)^{\dag}\tau^I H$  &  $\checkmark$  & $\checkmark$  \\
$\cO_{q^2H^2D^3}^{(4)}$  &  $ i (\bar q_i \gamma^{\mu}\tau^I D^\nu q_j) \, H^{\dag} \tau^I (D_{(\mu}D_{\nu)} H)$  &  $\checkmark$  & $\checkmark$  \\
\hline
$\cO_{u^2H^2D^3}^{(1)}$  &  $ i (\bar u_i \gamma^{\mu} D^\nu u_j) \, (D_{(\mu}D_{\nu)} H)^{\dag} H$  &  $\checkmark$  & --  \\
$\cO_{u^2H^2D^3}^{(2)}$  &  $ i (\bar u_i \gamma^{\mu} D^\nu u_j) \,H^{\dag} (D_{(\mu}D_{\nu)}  H)$  &  $\checkmark$  & --  \\
\hline
$\cO_{d^2H^2D^3}^{(1)}$  &  $ i (\bar d_i \gamma^{\mu} D^\nu d_j) \,(D_{(\mu}D_{\nu)} H)^{\dag} H$  &  $\checkmark$  & --  \\
$\cO_{d^2H^2D^3}^{(2)}$  &  $ i (\bar d_i \gamma^{\mu} D^\nu d_j) \,H^{\dag} (D_{(\mu}D_{\nu)}  H)$  &  $\checkmark$  & --  \\
    \hline\hline
    \end{tabular} 
  \caption{\sl\small SMEFT $d=8$ two-fermion operators, in the basis of Ref.~\cite{Murphy:2020rsh}, that contribute to the processes $pp\to\ell\ell$ and $pp\to\ell\nu$.}
  \label{tab:dim8_ops_2}
  \end{center}
\end{table}
% - - - -

% - - - - - - - - - - - - - - - - - - - - - - - - - - - - - - - - - 
\section{LHC Limits on SMEFT operators}
\label{sec:SMEFT-limits}

\begin{figure}[b!]
    \centering
    \resizebox{0.85\textwidth}{!}{
    \begin{tabular}{c c}
        \includegraphics[]{figures/lq1_33.pdf}   & \includegraphics[]{figures/lq3_33.pdf}  \\
        \includegraphics[]{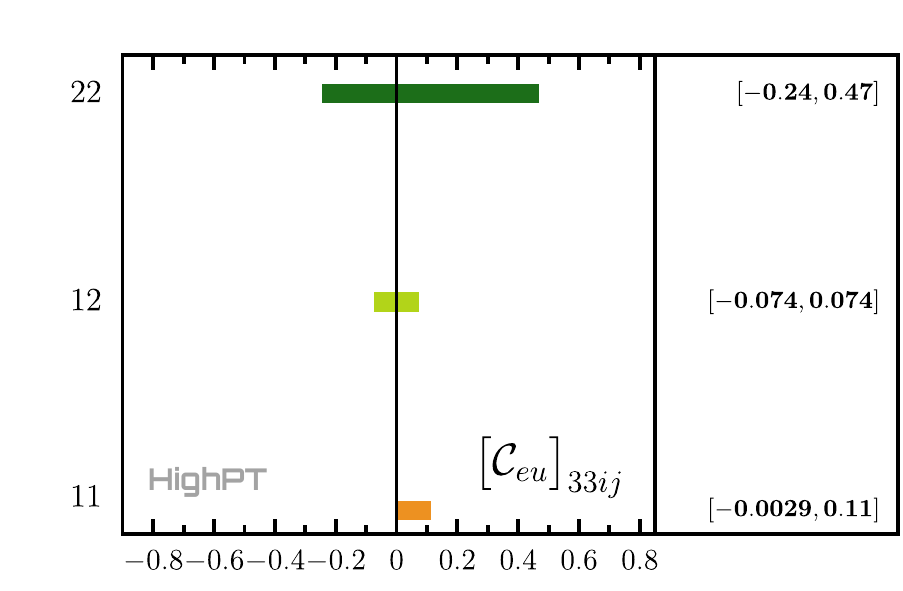}    & \includegraphics[]{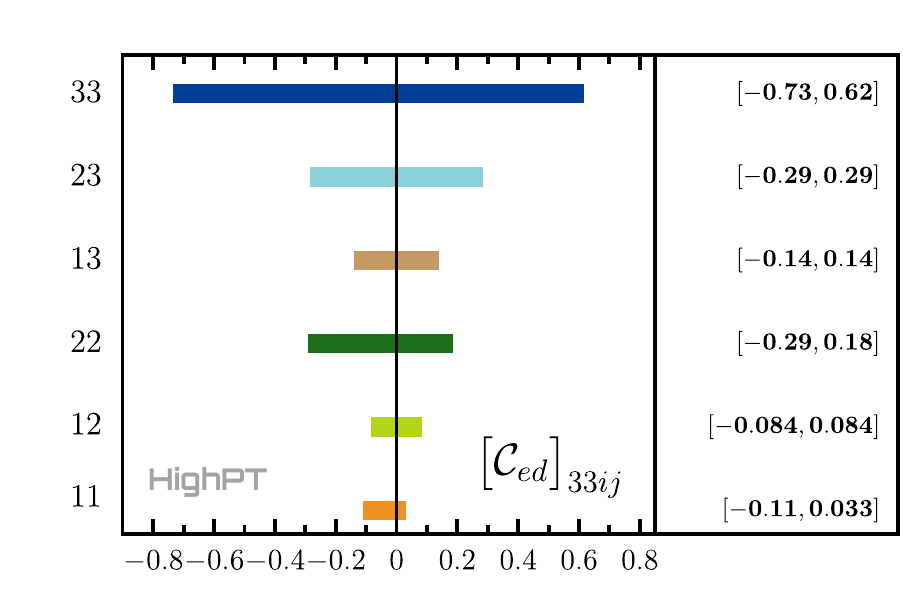}   \\
        \includegraphics[]{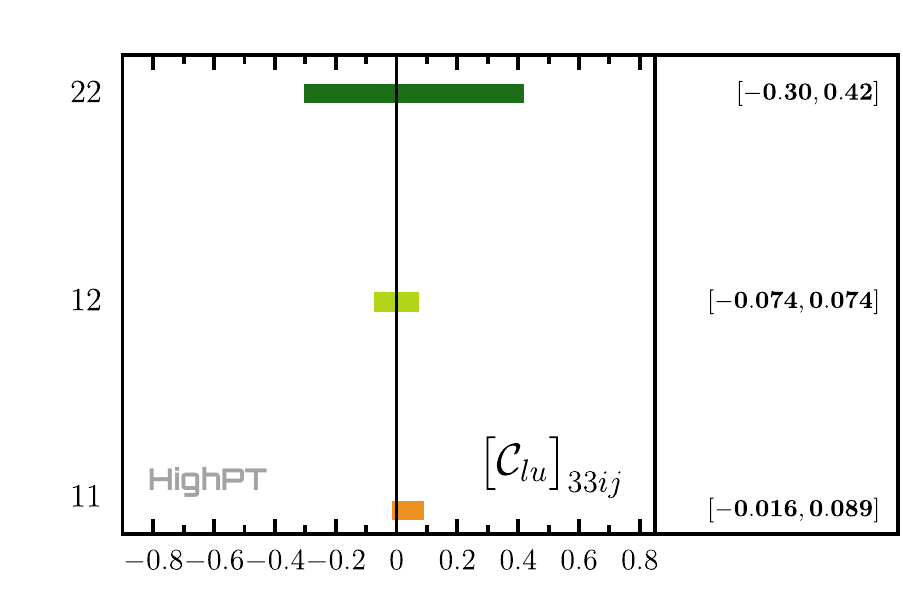}    & \includegraphics[]{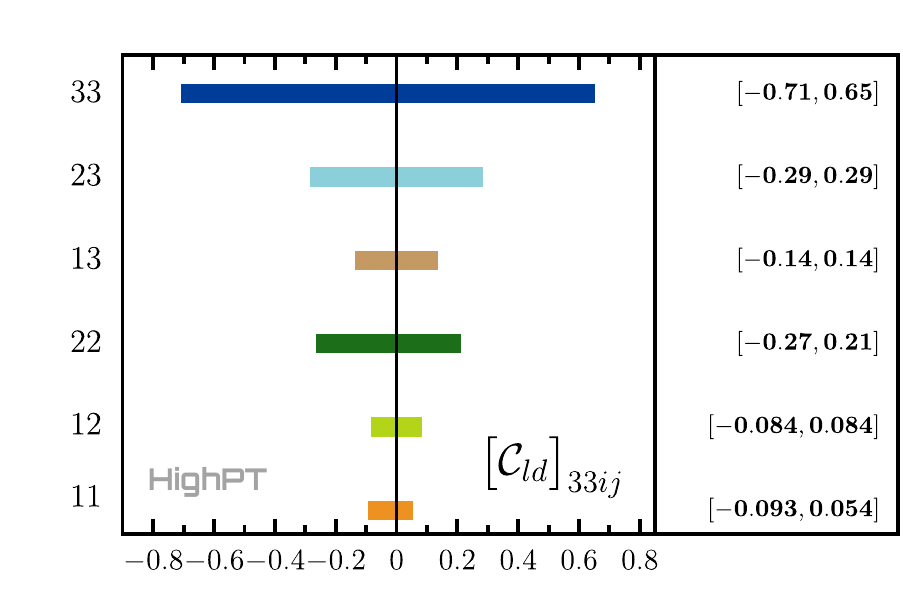}   \\
        \includegraphics[]{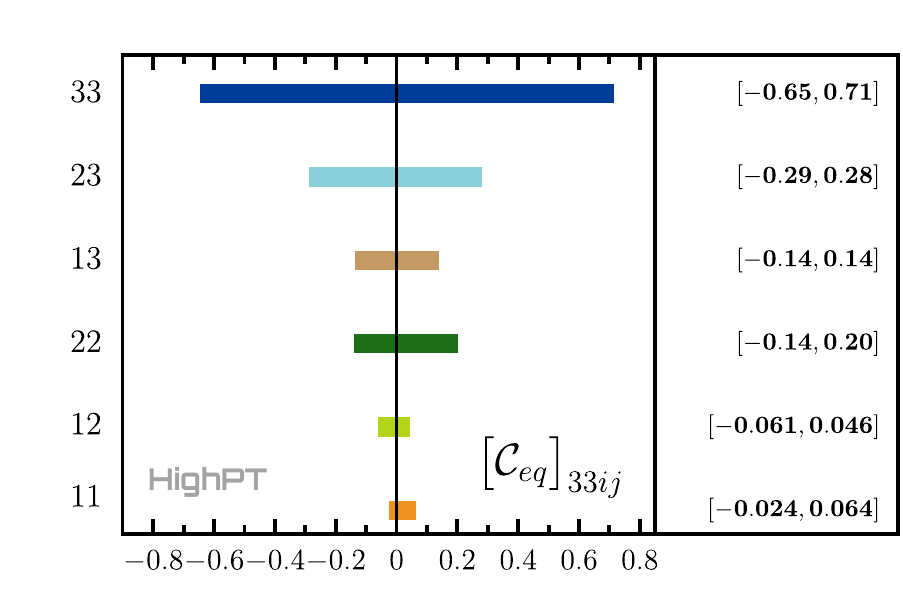}    & \includegraphics[]{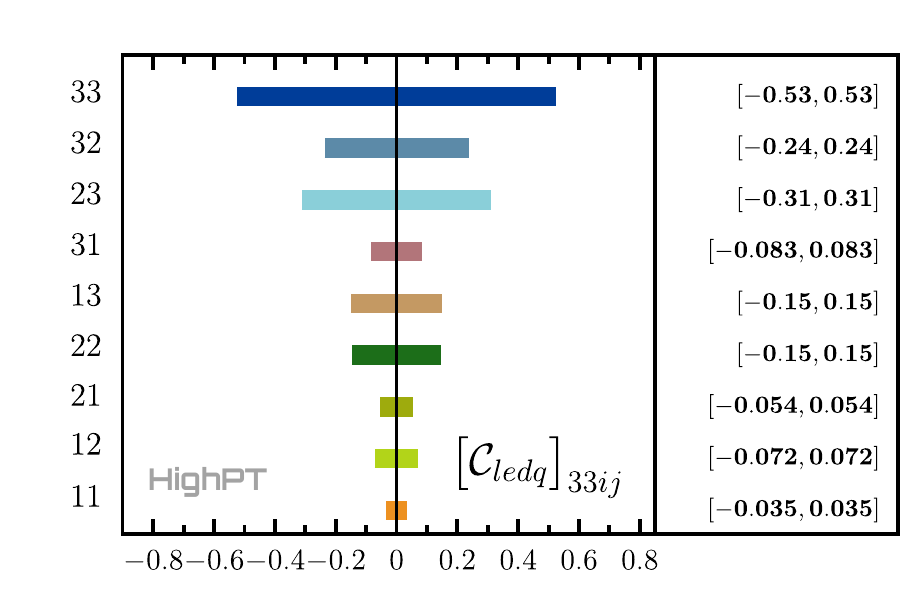} \\
        \includegraphics[]{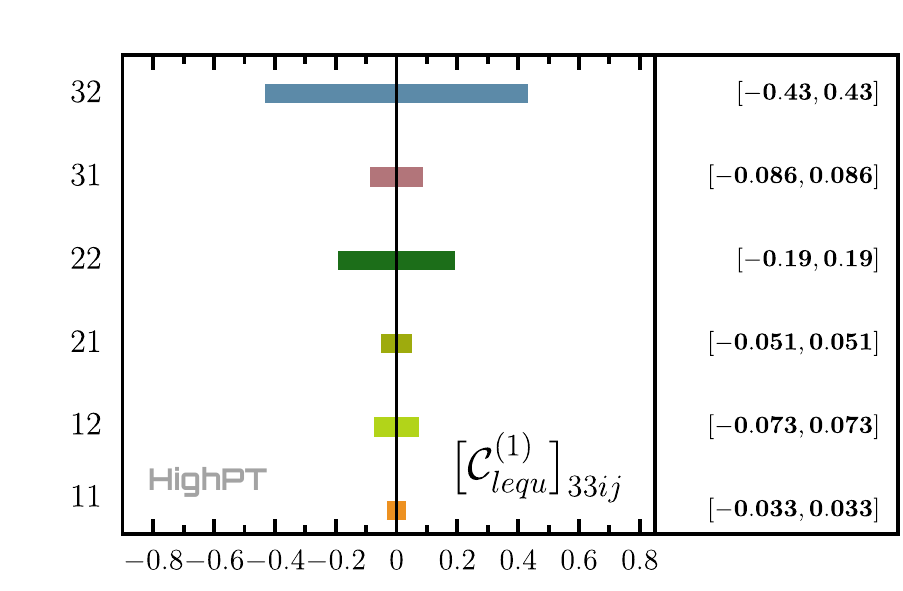} & \includegraphics[]{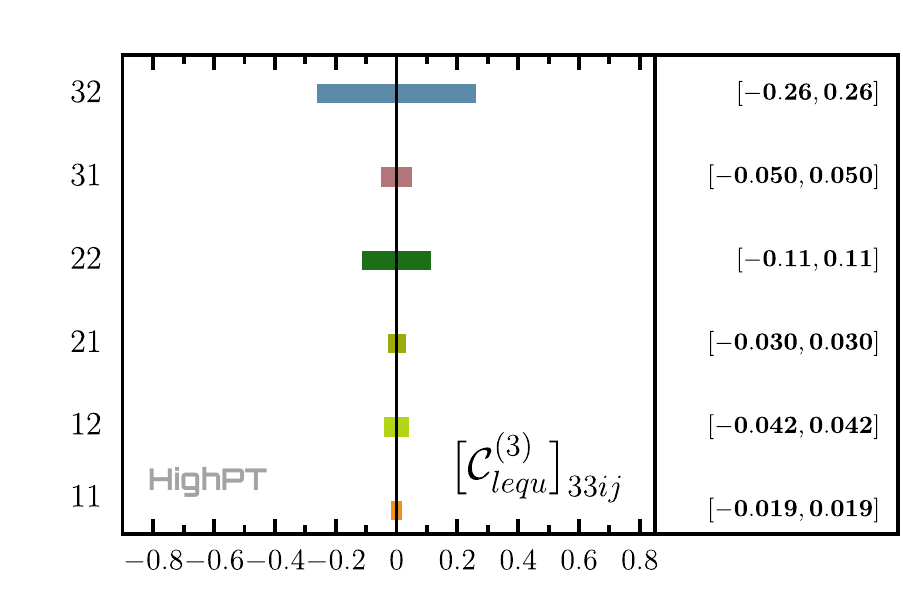}
    \end{tabular}
    }
    \caption{\sl\small LHC constraints on semileptonic $d=6$ Wilson coefficients with $\tau\tau$ flavor indices to $95\%$~CL accuracy, where a single coefficient is turned on at a time. Quark-flavor indices are denoted by~$ij$ and are specified on the left hand-side of each plot. All coefficients are assumed to be real and contributions to the cross-section up to and including $\cO(1/\Lambda^{4})$ are considered. The New Physics scale is chosen as~$\Lambda=1\,\mathrm{TeV}$. }
    \label{fig:single-WC-limits-tau-tau}
\end{figure}

\begin{figure}[b!]
    \centering
    \resizebox{0.9\textwidth}{!}{
    \begin{tabular}{c c}
        \includegraphics[]{figures/lq1_22.pdf}   & \includegraphics[]{figures/lq3_22.pdf}  \\
        \includegraphics[]{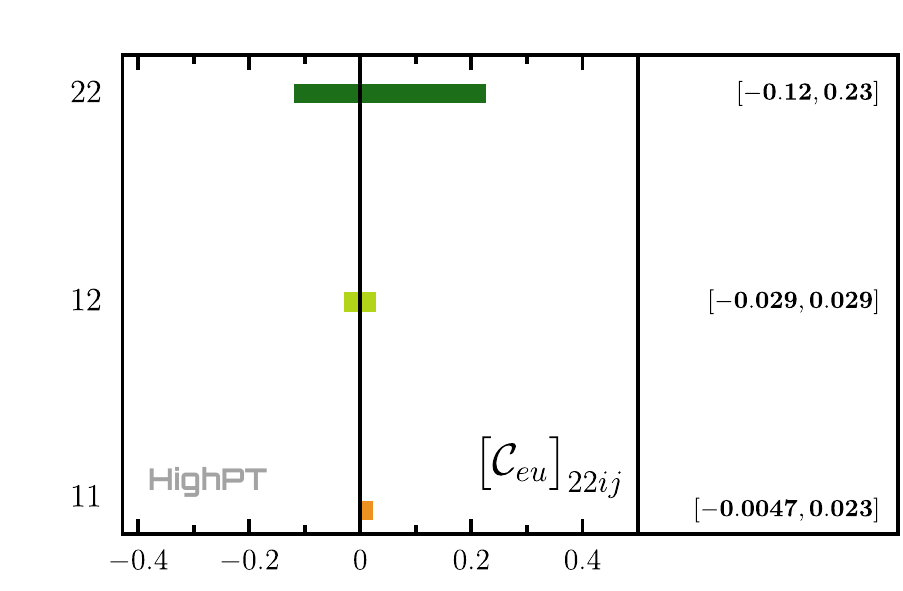}    & \includegraphics[]{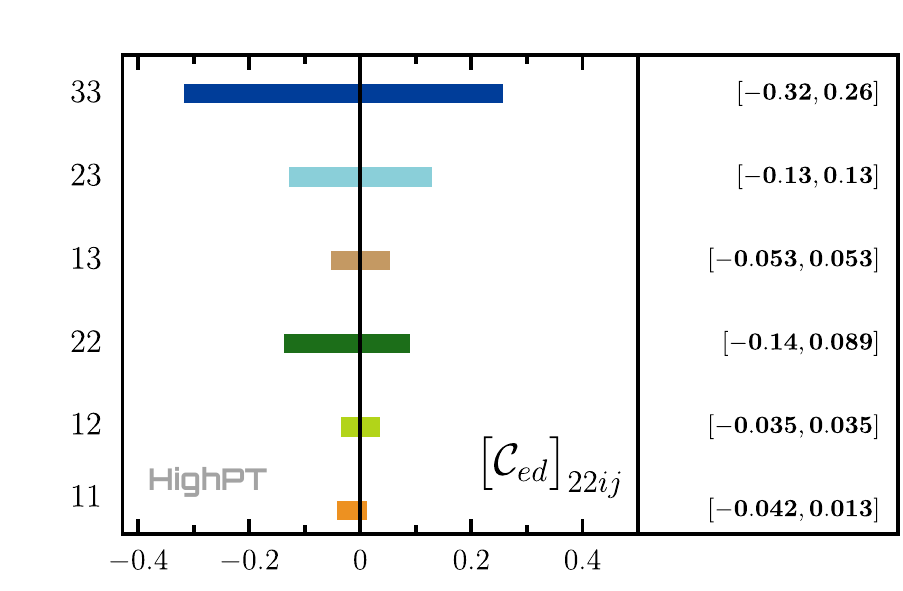}   \\
        \includegraphics[]{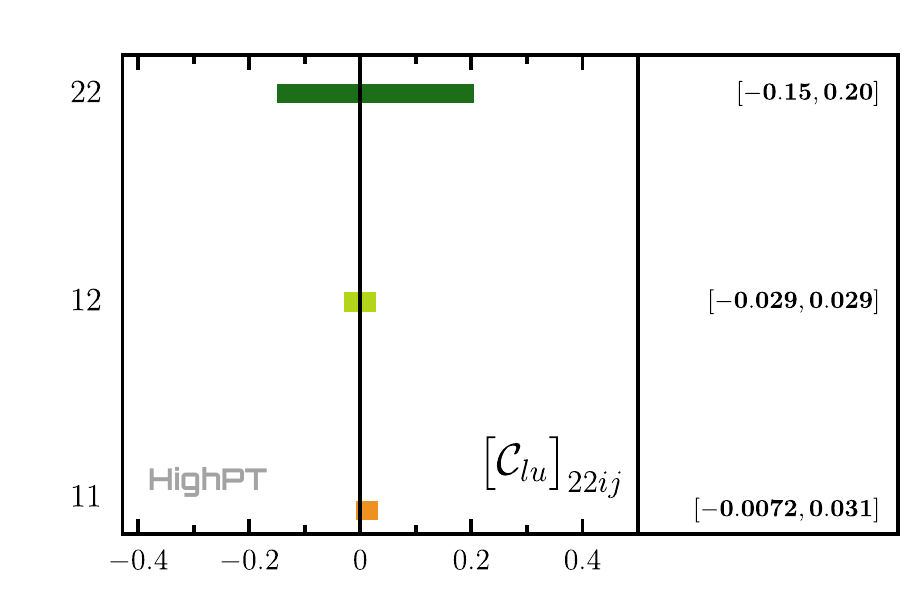}    & \includegraphics[]{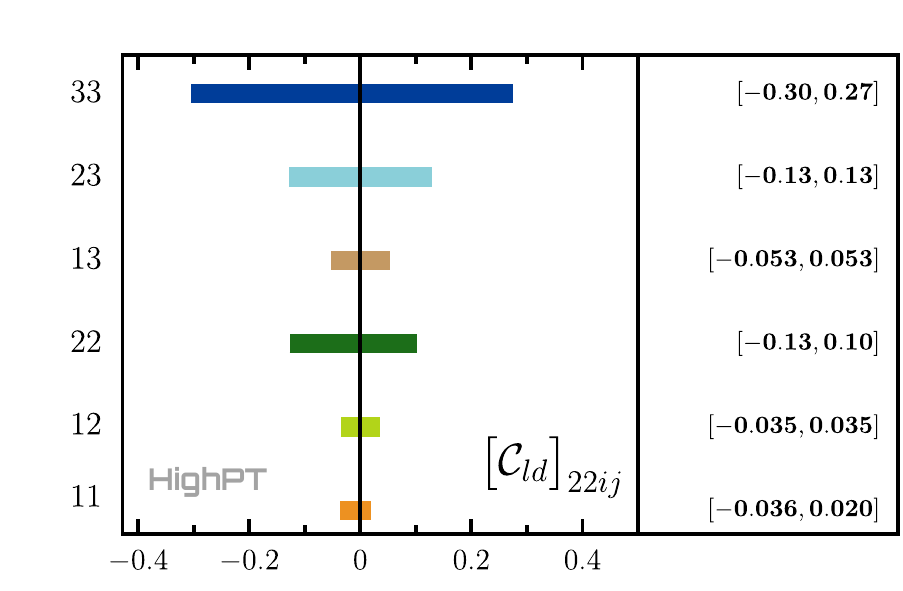}   \\
        \includegraphics[]{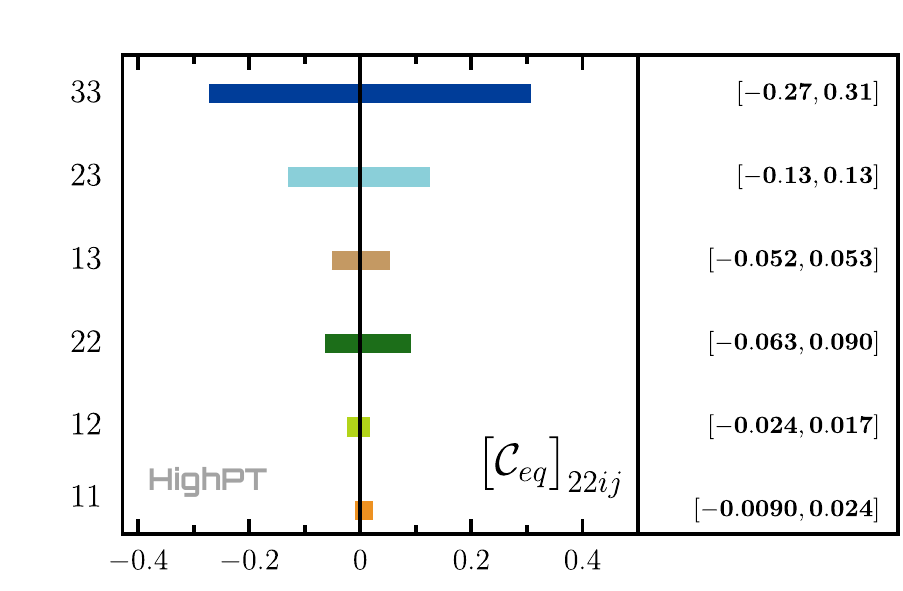}    & \includegraphics[]{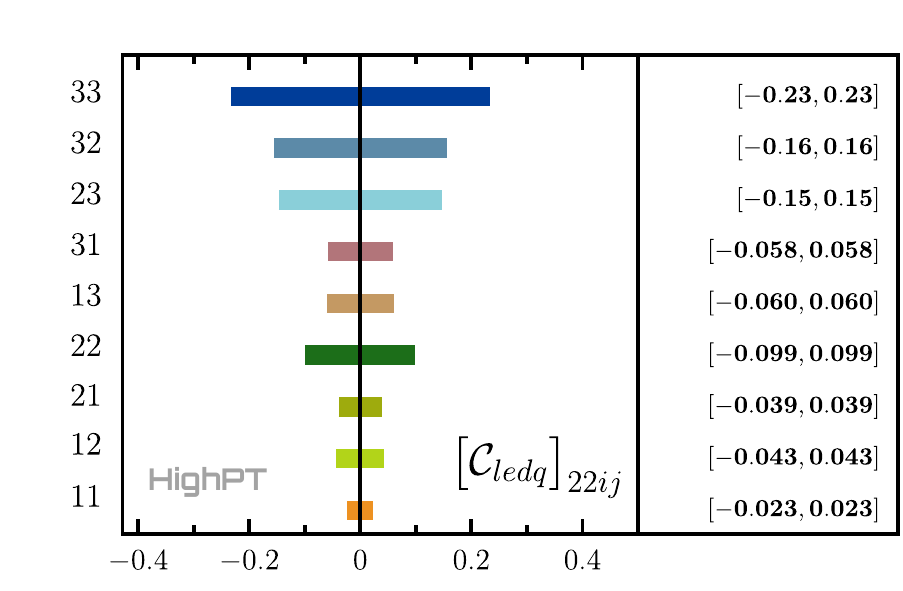} \\
        \includegraphics[]{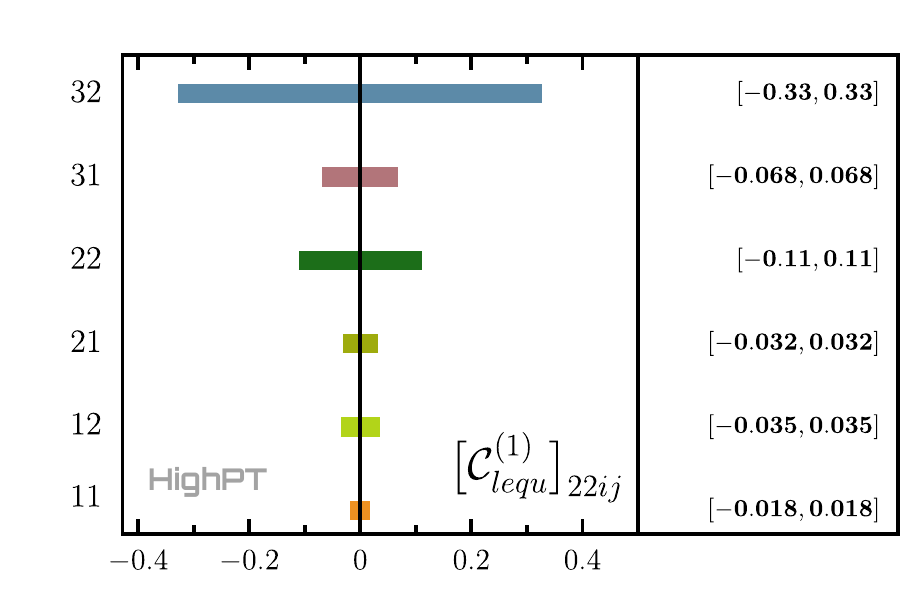} & \includegraphics[]{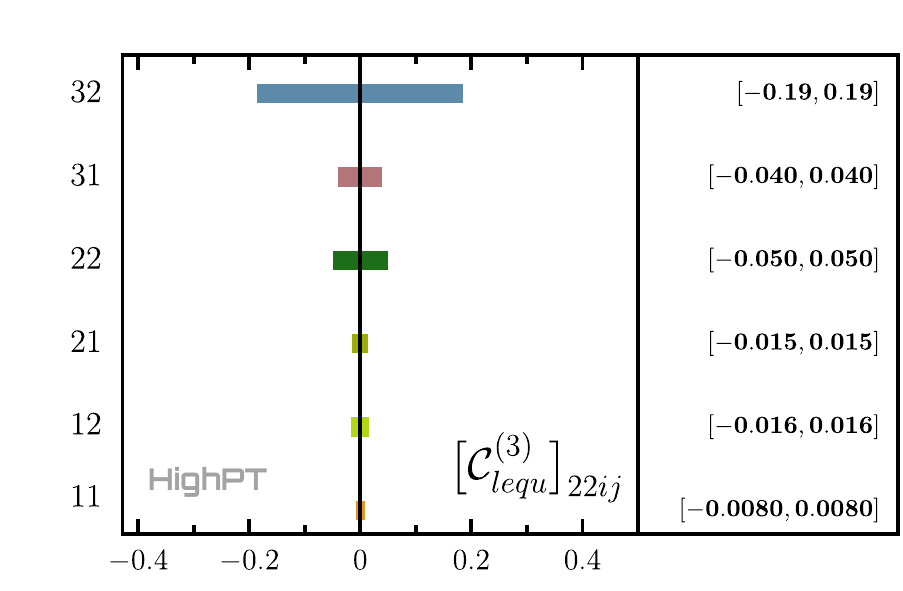}
    \end{tabular}
    }
    \caption{\sl\small LHC constraints on semileptonic $d=6$ Wilson coefficients with $\mu\mu$ flavor indices to $95\%$~CL accuracy, where a single coefficient is turned on at a time.  See caption of Fig.~\ref{fig:single-WC-limits-tau-tau}.}
    \label{fig:single-WC-limits-mu-mu}
\end{figure}

\begin{figure}[t!]
    \centering
    \resizebox{0.9\textwidth}{!}{
    \begin{tabular}{c c}
        \includegraphics[]{figures/lq1_11.pdf}   & \includegraphics[]{figures/lq3_11.pdf}  \\
        \includegraphics[]{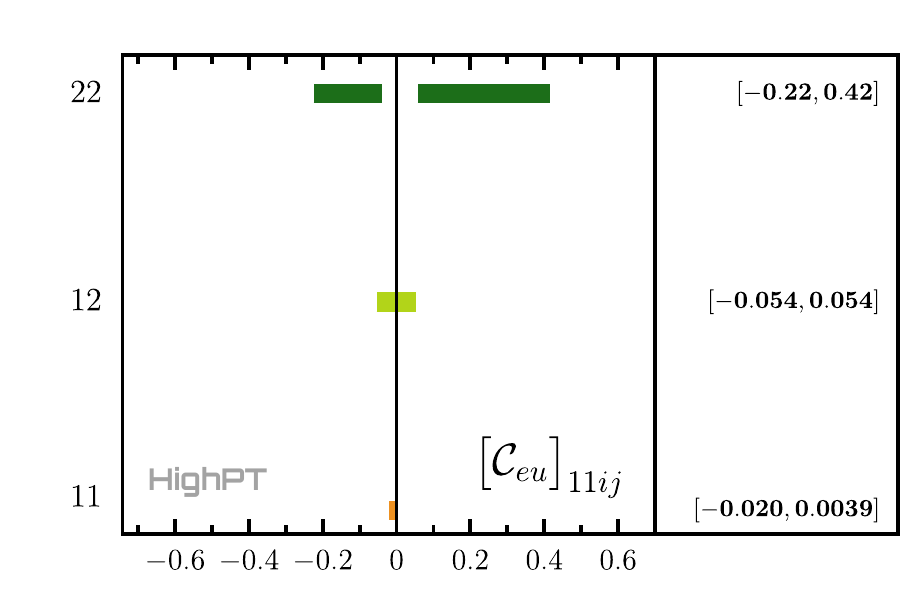}    & \includegraphics[]{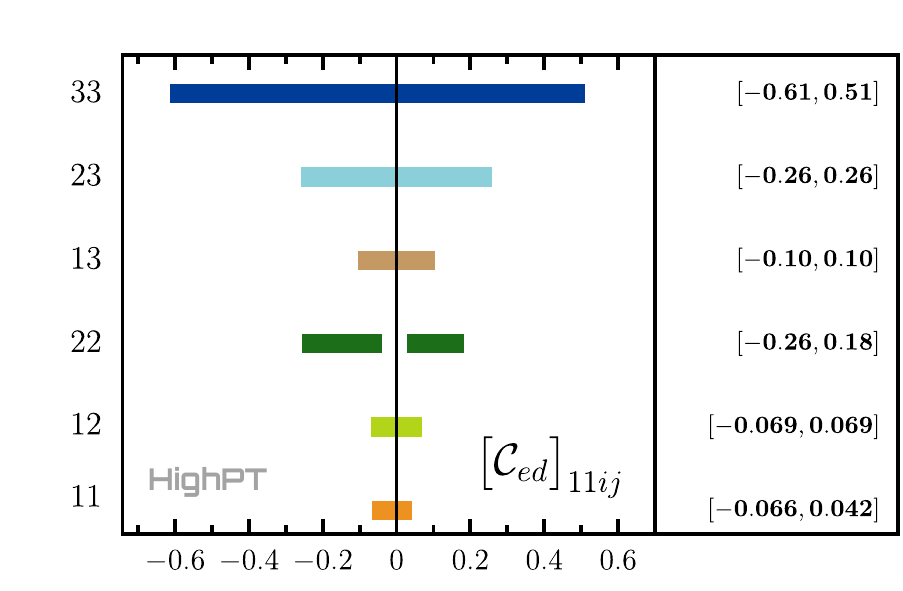}   \\
        \includegraphics[]{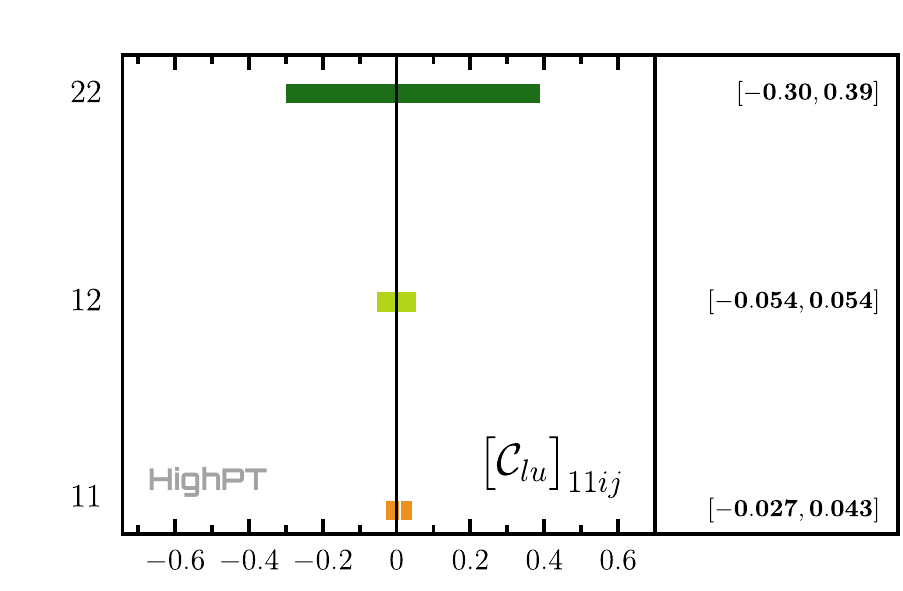}    & \includegraphics[]{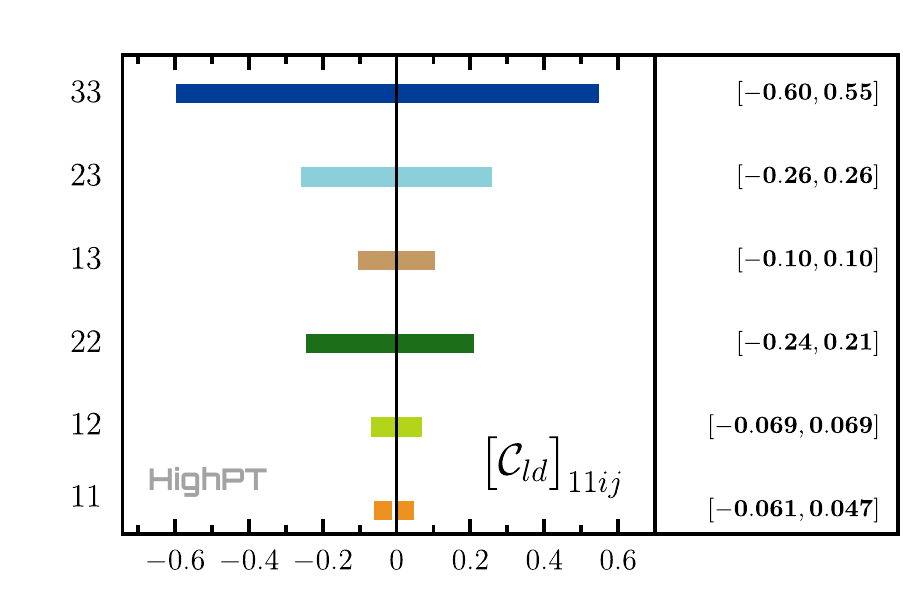}   \\
        \includegraphics[]{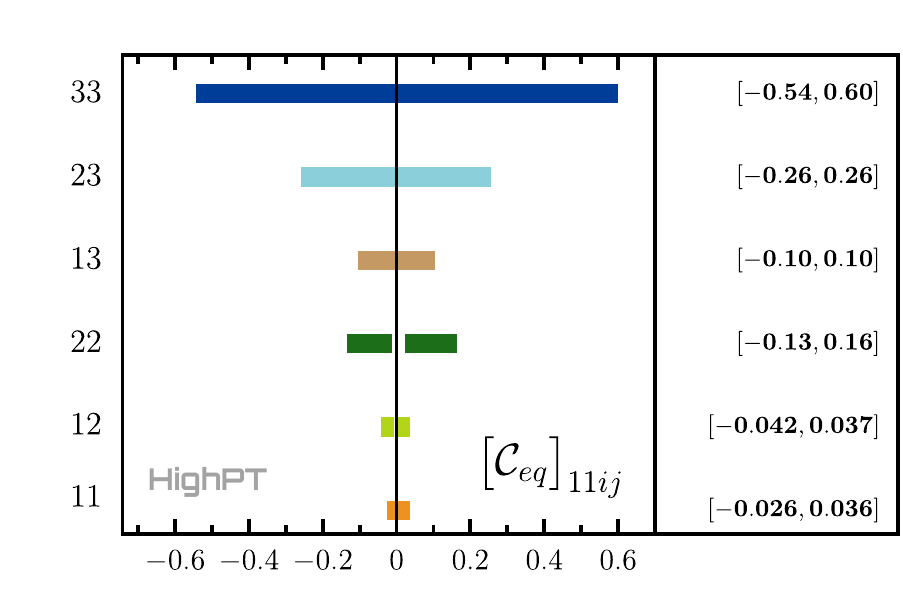}    & \includegraphics[]{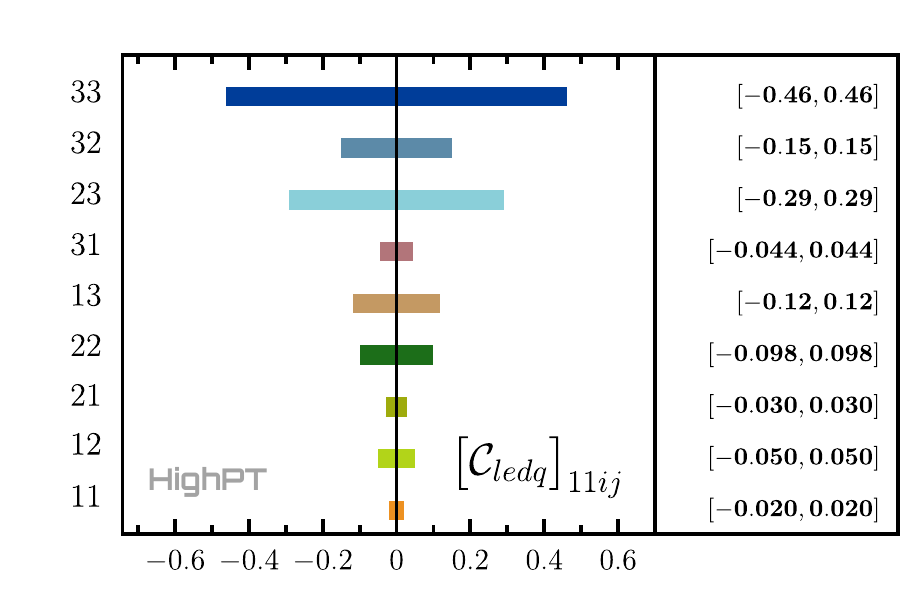} \\
        \includegraphics[]{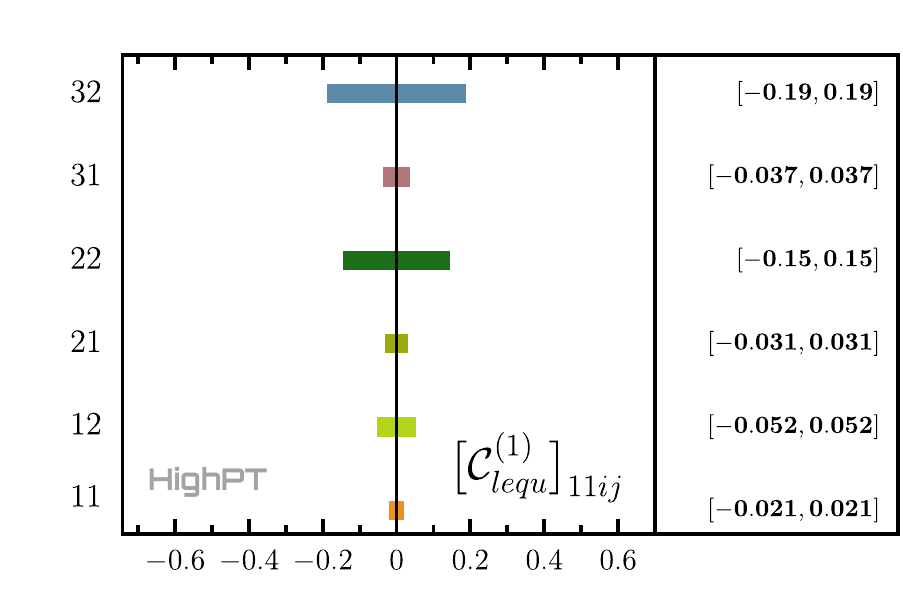} & \includegraphics[]{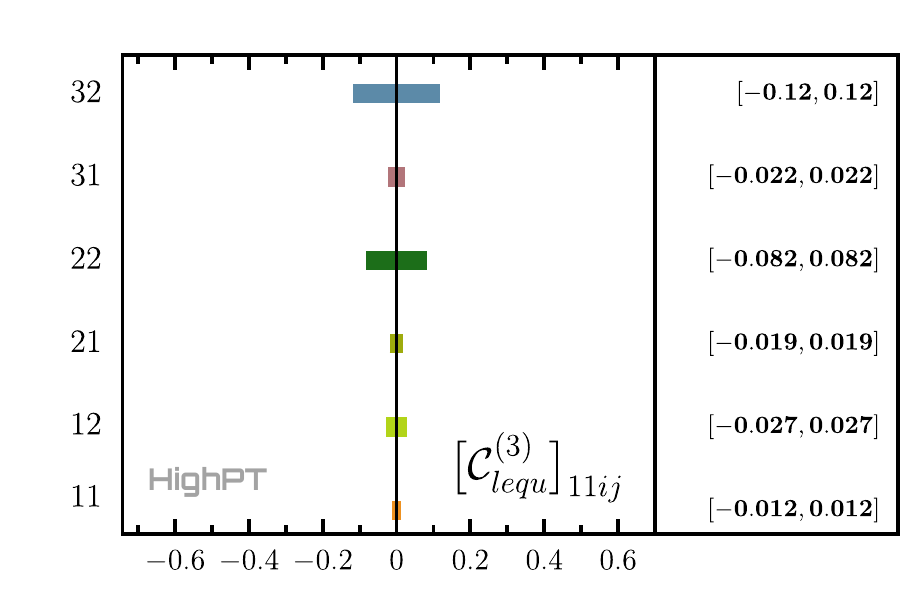}
    \end{tabular}
    }
    \caption{\sl\small LHC constraints on semileptonic $d=6$ Wilson coefficients with $ee$ flavor indices to $95\%$~CL accuracy, where a single coefficient is turned on at a time. See caption of Fig.~\ref{fig:single-WC-limits-tau-tau}. The largest deviation from the SM observed is~$\sim 2 \sigma$.}
    \label{fig:single-WC-limits-e-e}
\end{figure}

\begin{figure}[p!]
    \centering
    \resizebox{\textwidth}{!}{
    \begin{tabular}{c c c}
        \includegraphics[]{figures/lq1_23.pdf}   & \includegraphics[]{figures/lq3_23.pdf}   &
        \includegraphics[]{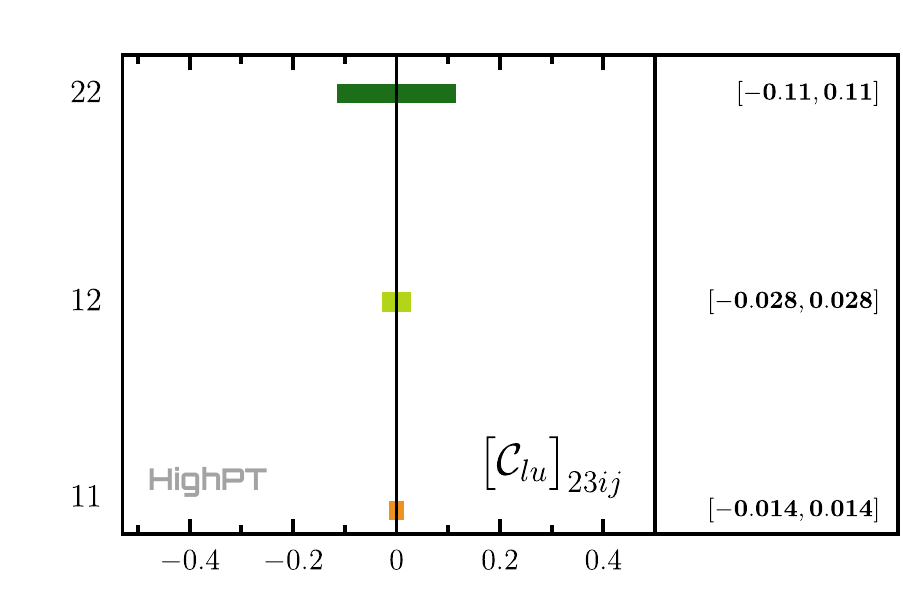}   \\ \includegraphics[]{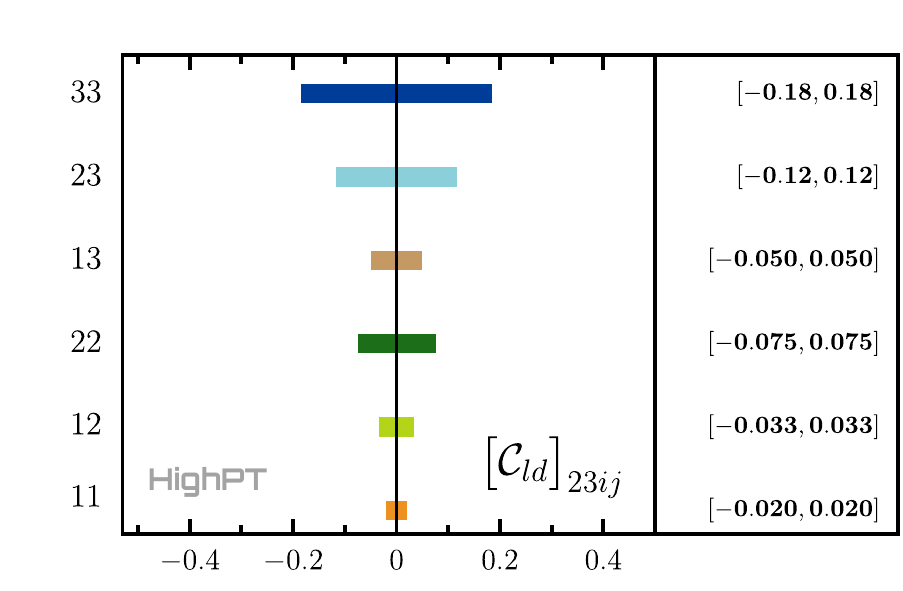}    &
        \includegraphics[]{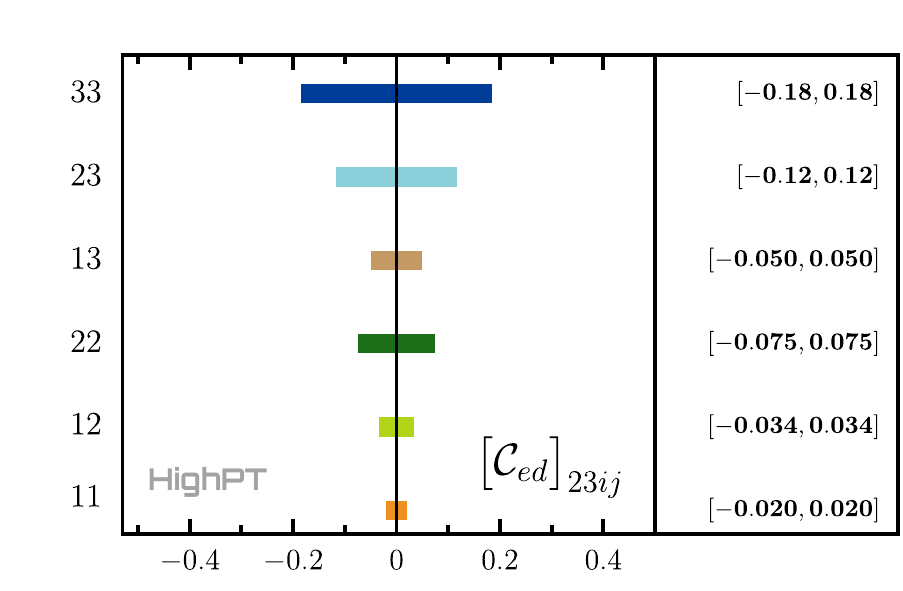}    & \includegraphics[]{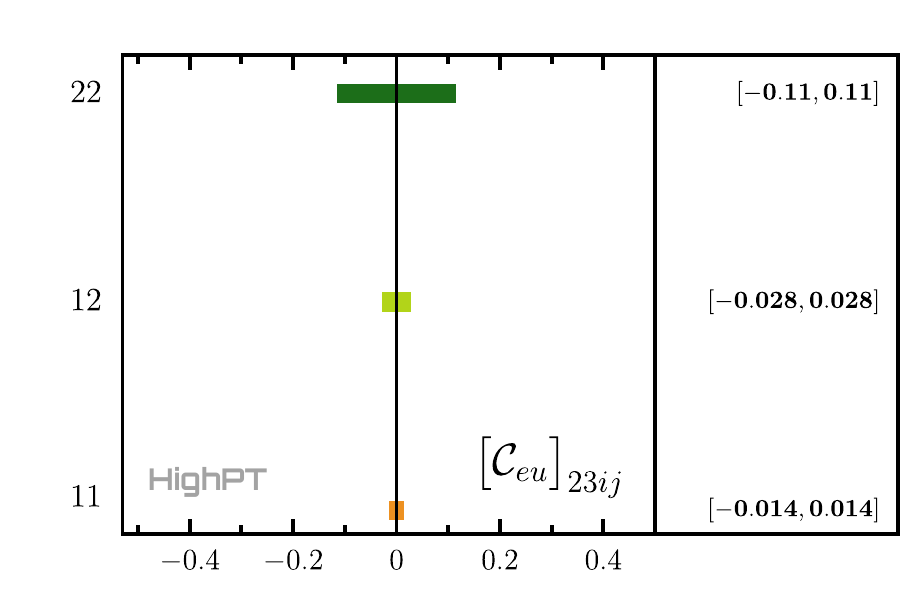}   \\
        \includegraphics[]{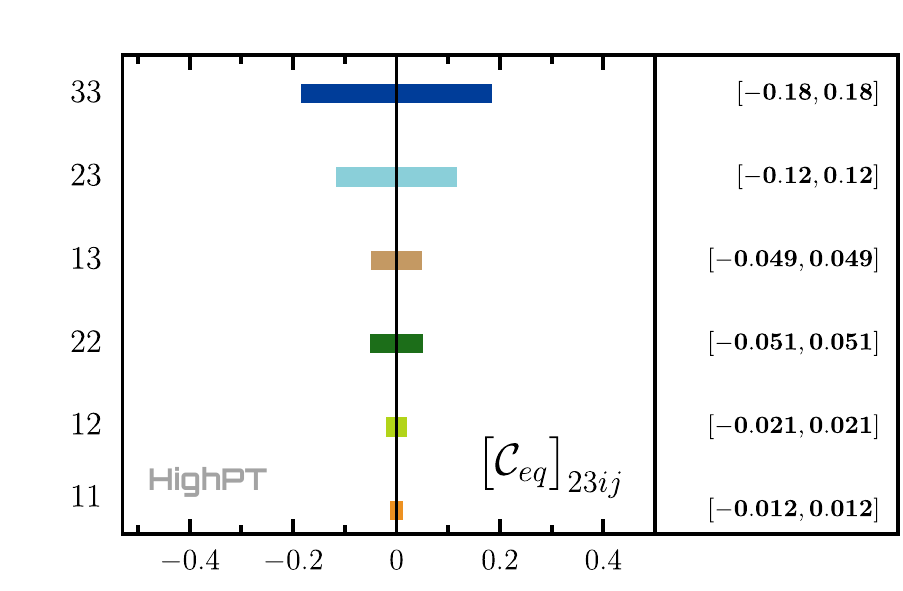}    &
        \includegraphics[]{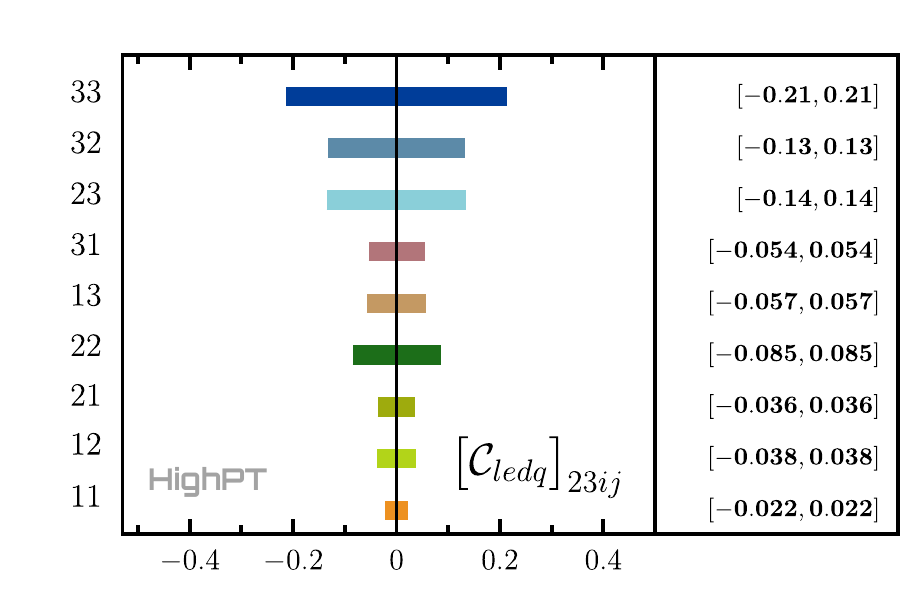}  &
        \includegraphics[]{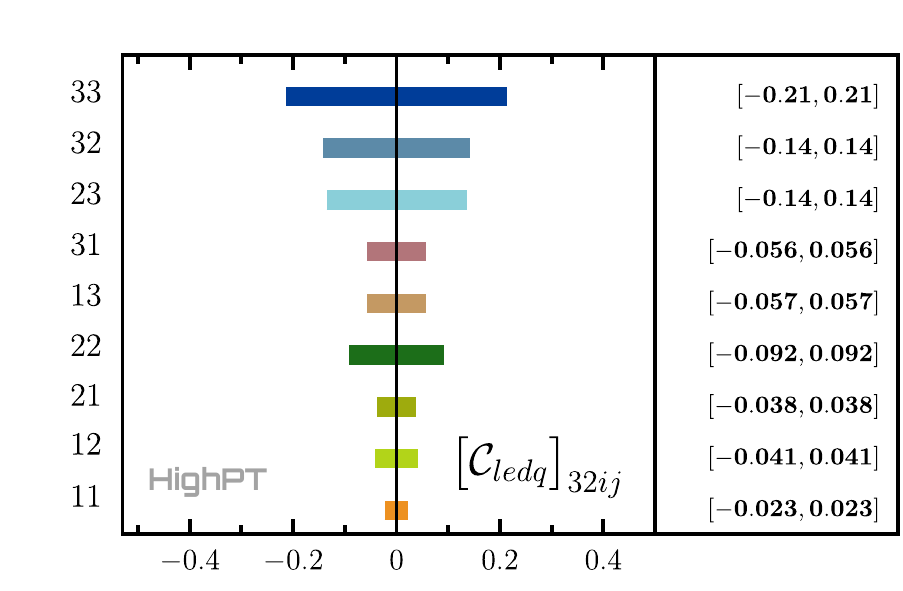} \\
        \includegraphics[]{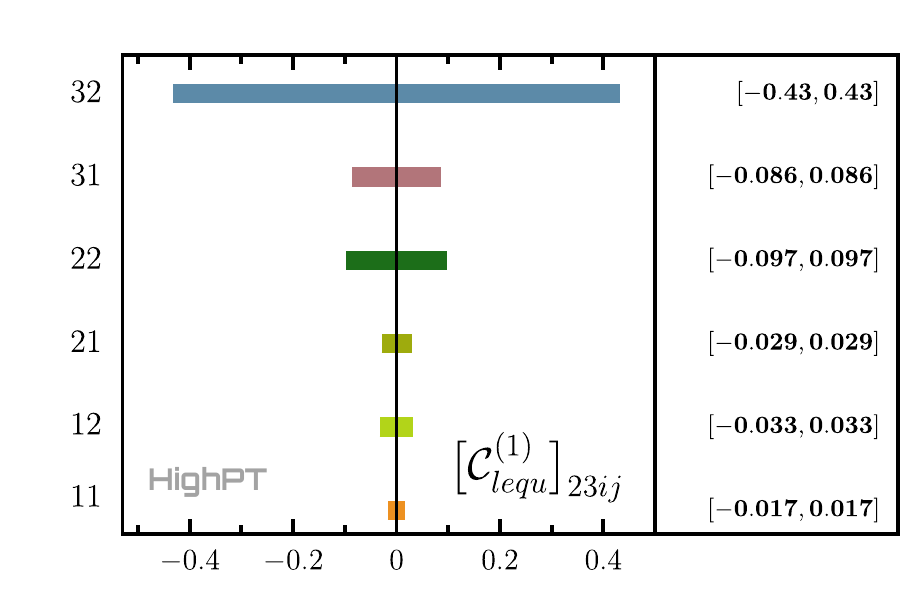} & \includegraphics[]{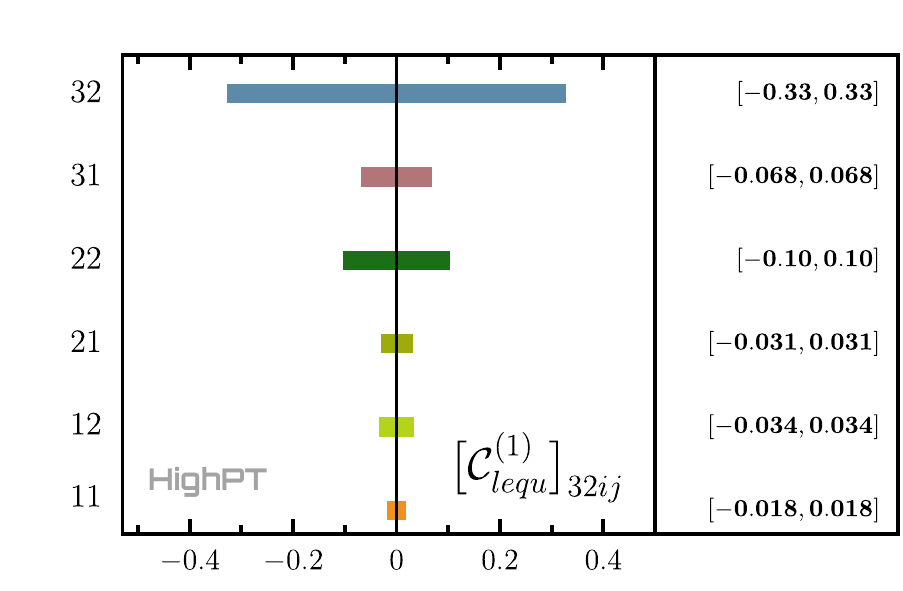} & \\
        \includegraphics[]{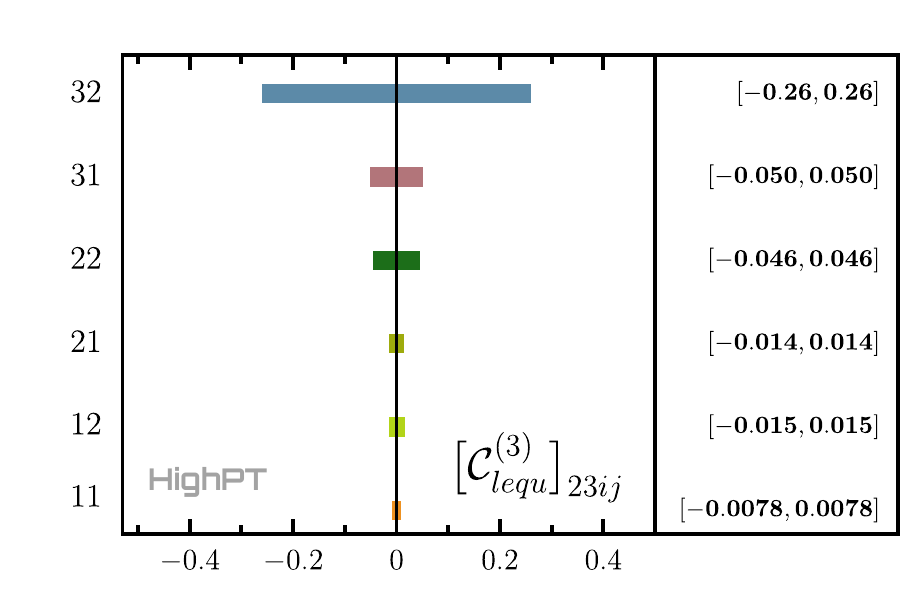} & \includegraphics[]{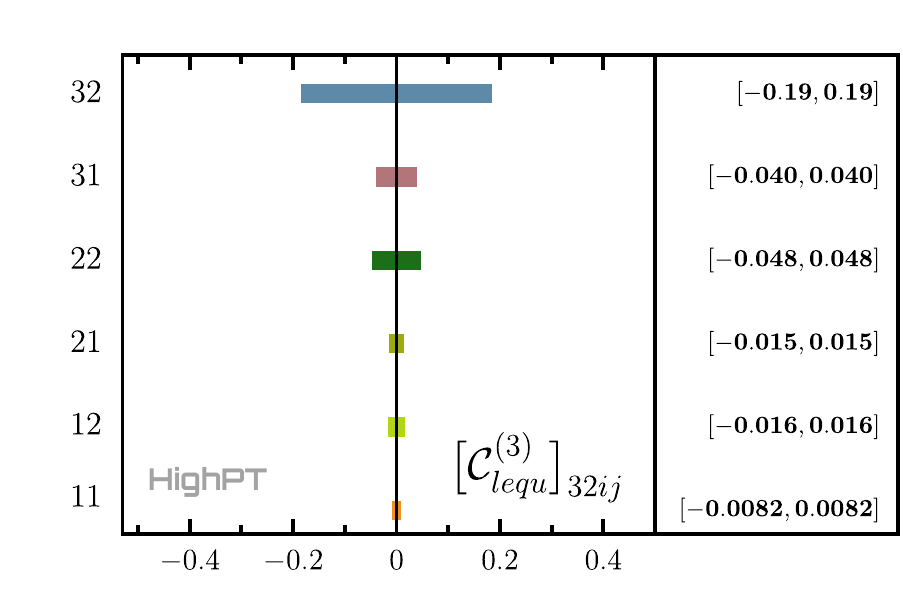} &
    \end{tabular}
    }
    \caption{\sl\small LHC constraints on $\psi^4$ semileptonic $d=6$ Wilson coefficients with $\mu\tau$ flavor indices, where a single coefficient is turned on at a time. See caption of Fig.~\ref{fig:single-WC-limits-tau-tau}. }
    \label{fig:single-WC-limits-tau-mu}
\end{figure}

\begin{figure}[p!]
    \centering
    \resizebox{1\textwidth}{!}{
    \begin{tabular}{c c c}
        \includegraphics[]{figures/lq1_13.pdf}   & \includegraphics[]{figures/lq3_13.pdf}   &
        \includegraphics[]{figures/lu_23.pdf}   \\ \includegraphics[]{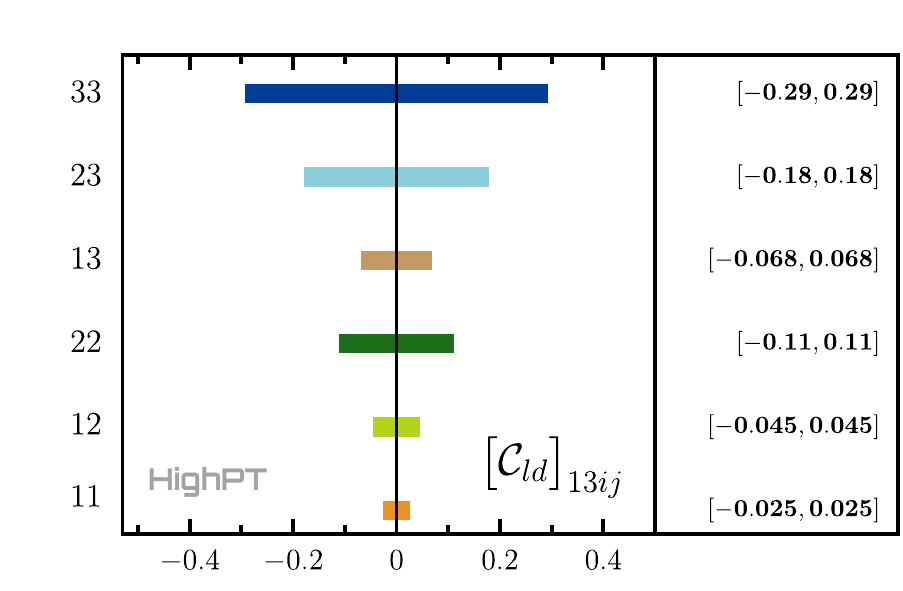}    &
        \includegraphics[]{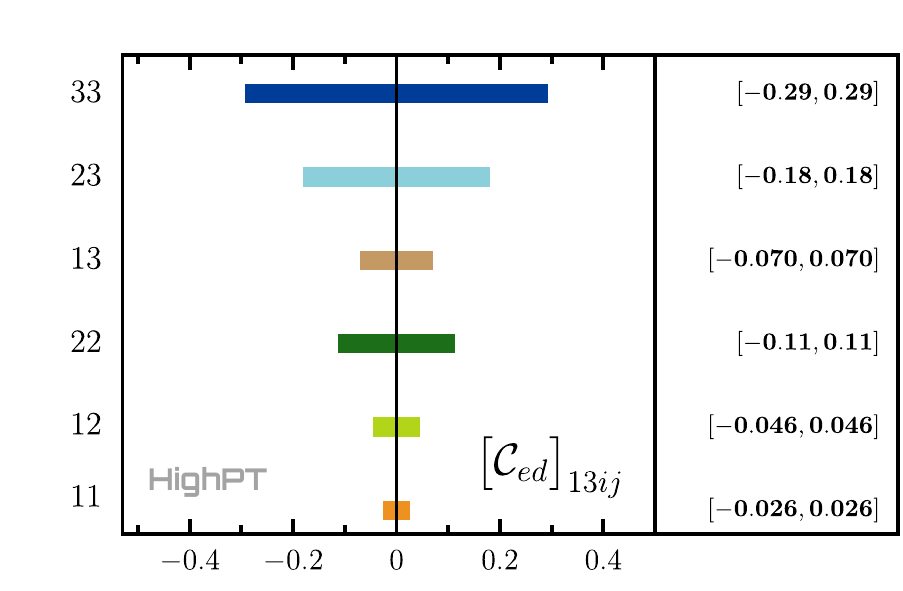}    & \includegraphics[]{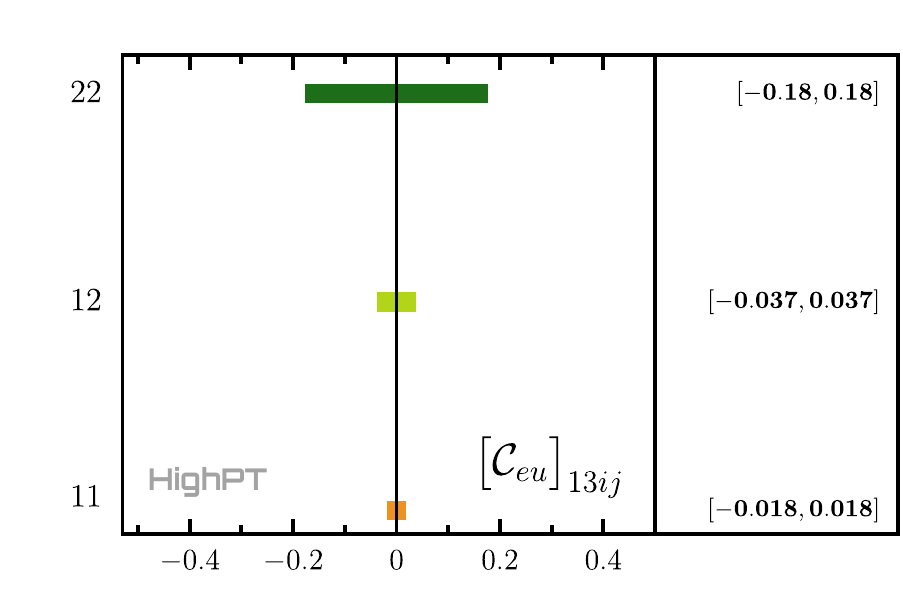}   \\
        \includegraphics[]{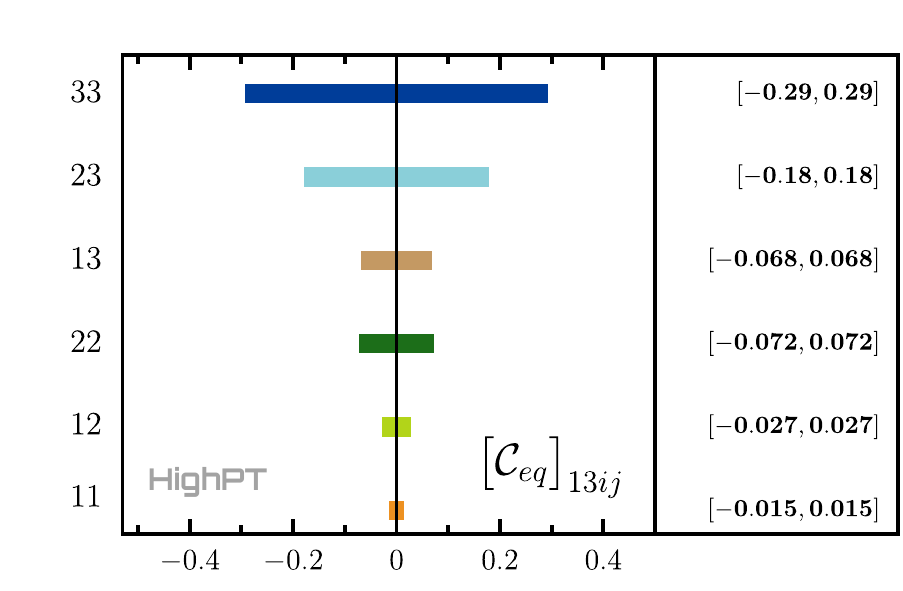}    &
        \includegraphics[]{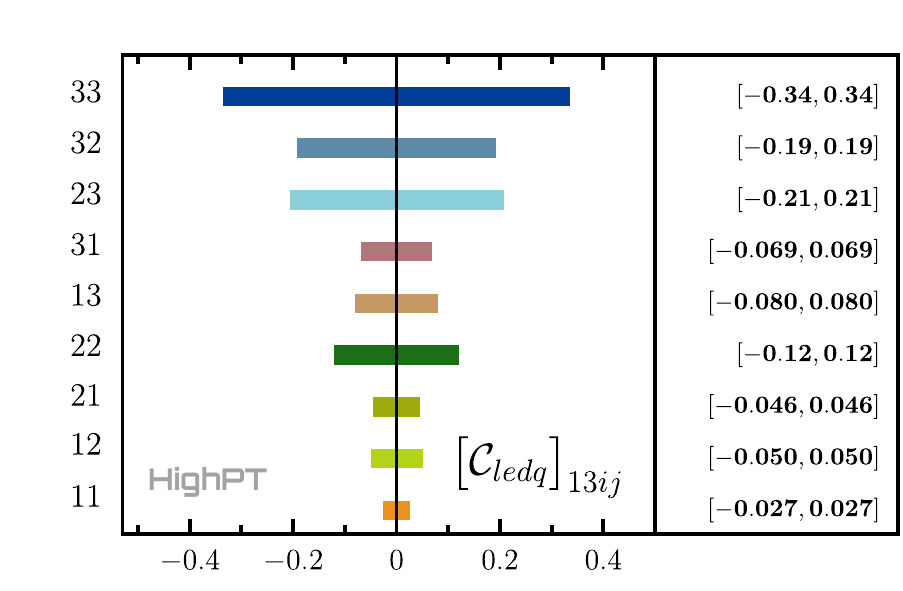}  &
        \includegraphics[]{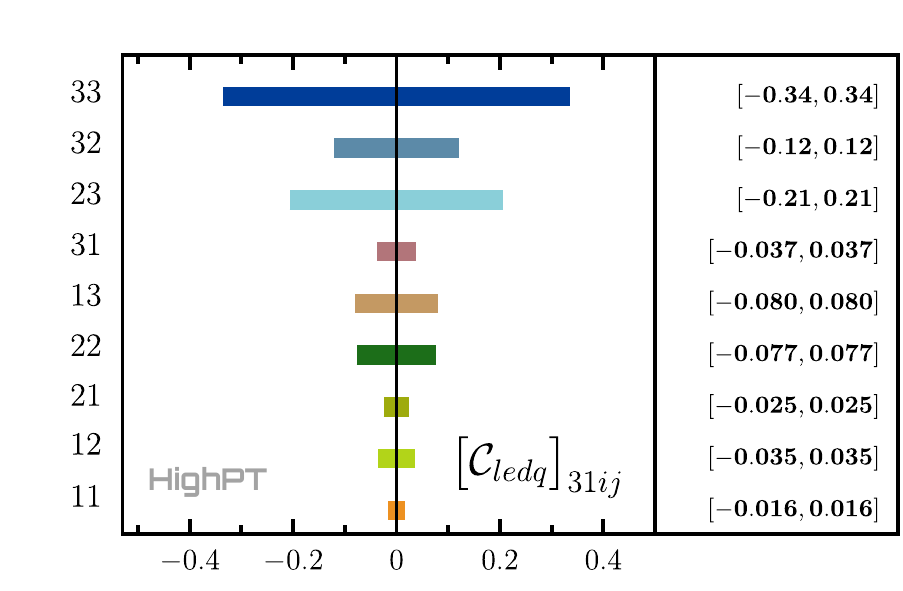} \\
        \includegraphics[]{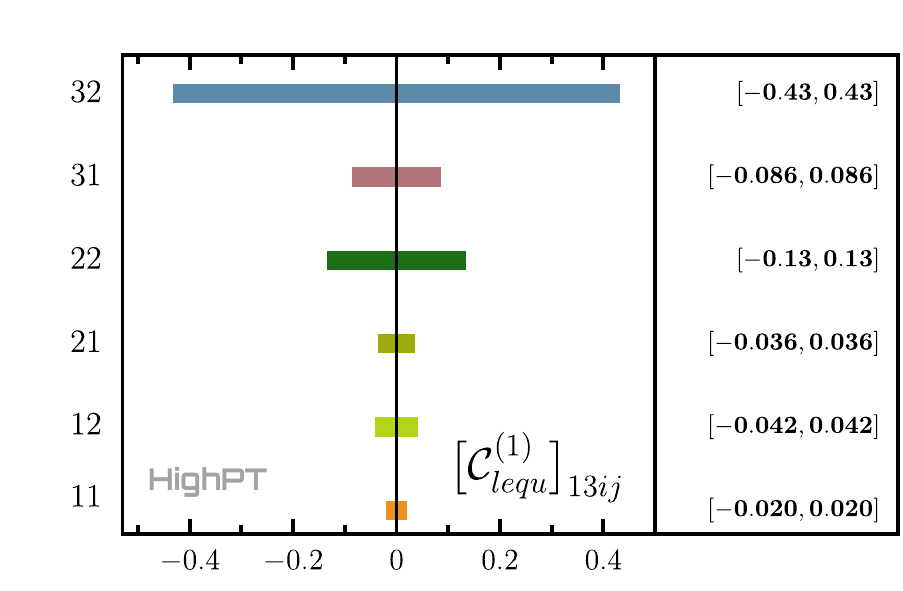} & \includegraphics[]{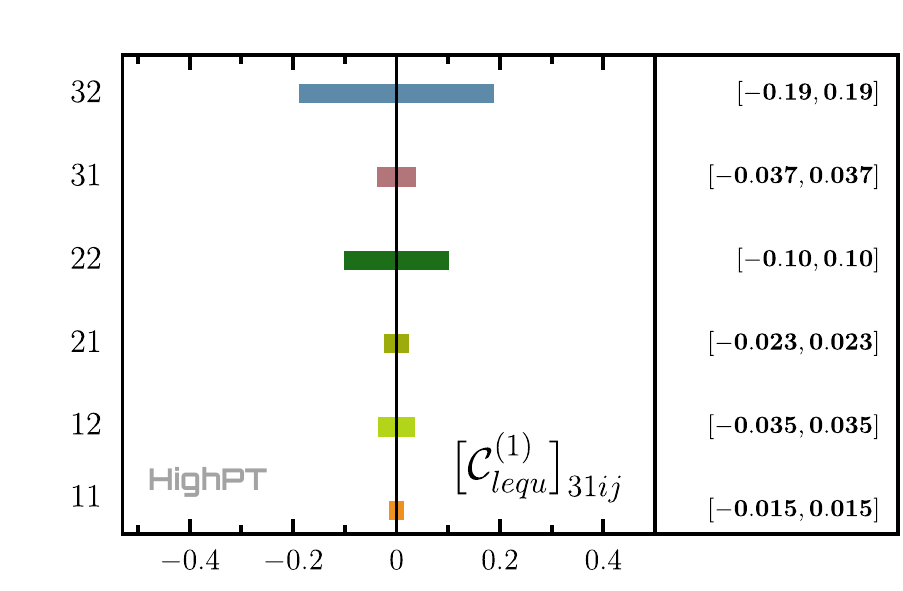} & \\
        \includegraphics[]{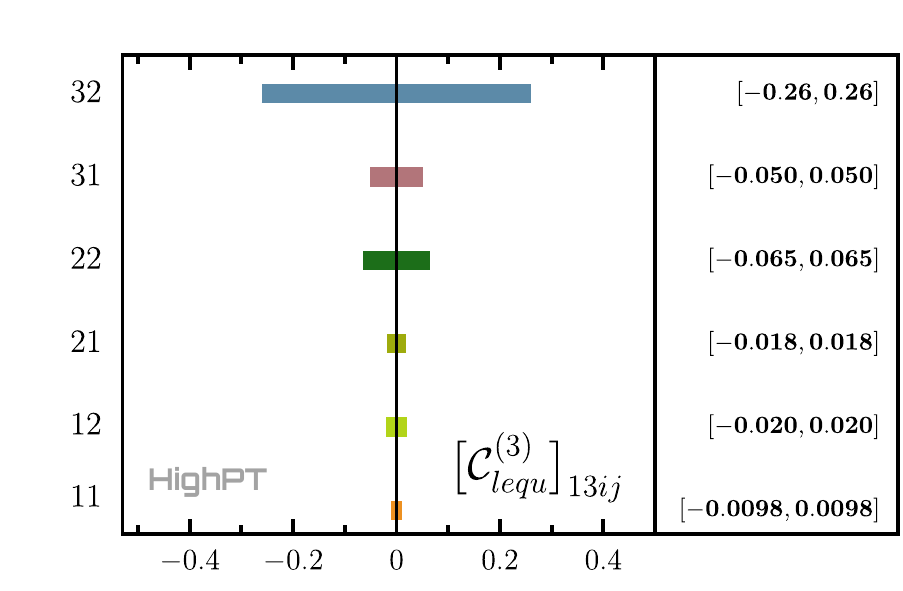} & \includegraphics[]{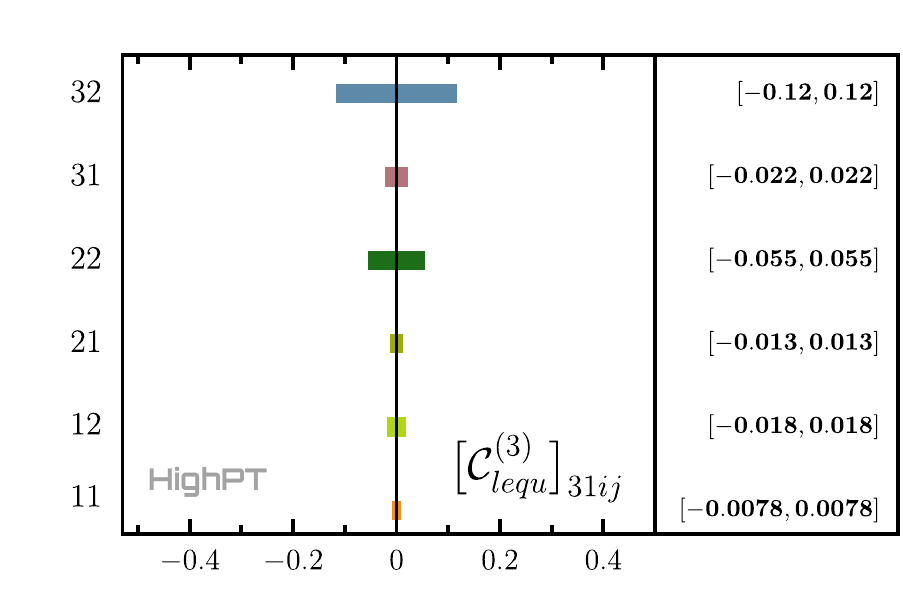} &
    \end{tabular}
    }
    \caption{\sl\small LHC constraints on $\psi^4$ semileptonic $d=6$ Wilson coefficients with $e\tau$ flavor indices, where a single coefficient is turned on at a time. See caption of Fig.~\ref{fig:single-WC-limits-tau-tau}.}
    \label{fig:single-WC-limits-tau-e}
\end{figure}

\begin{figure}[p!]
    \centering
    \resizebox{1\textwidth}{!}{
    \begin{tabular}{c c c}
        \includegraphics[]{figures/lq1_12.pdf}   & \includegraphics[]{figures/lq3_12.pdf}   &
        \includegraphics[]{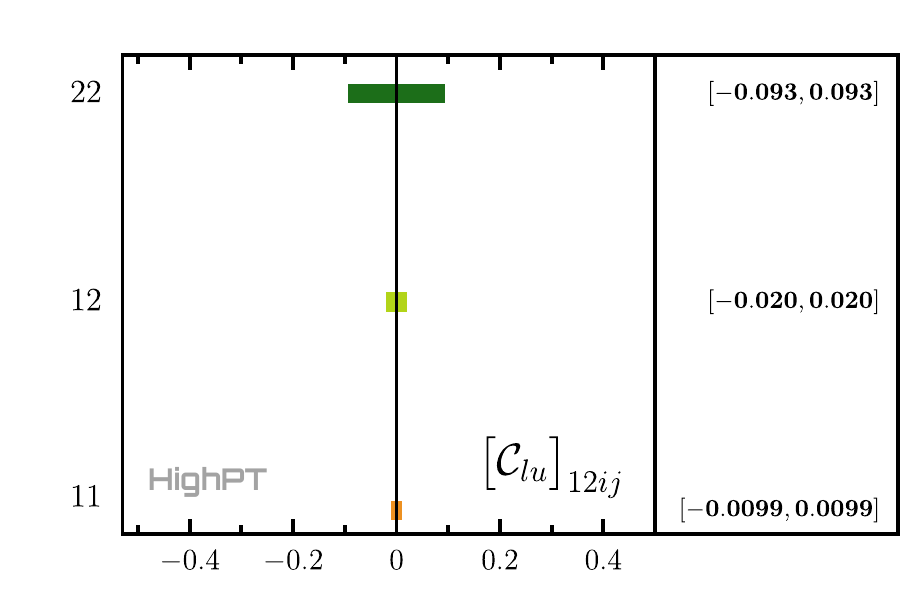}   \\ \includegraphics[]{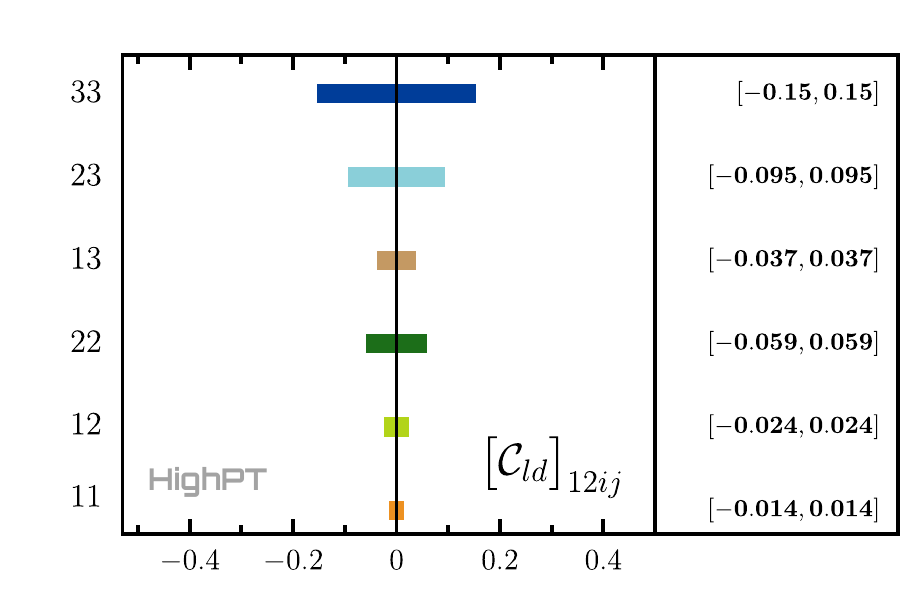}    &
        \includegraphics[]{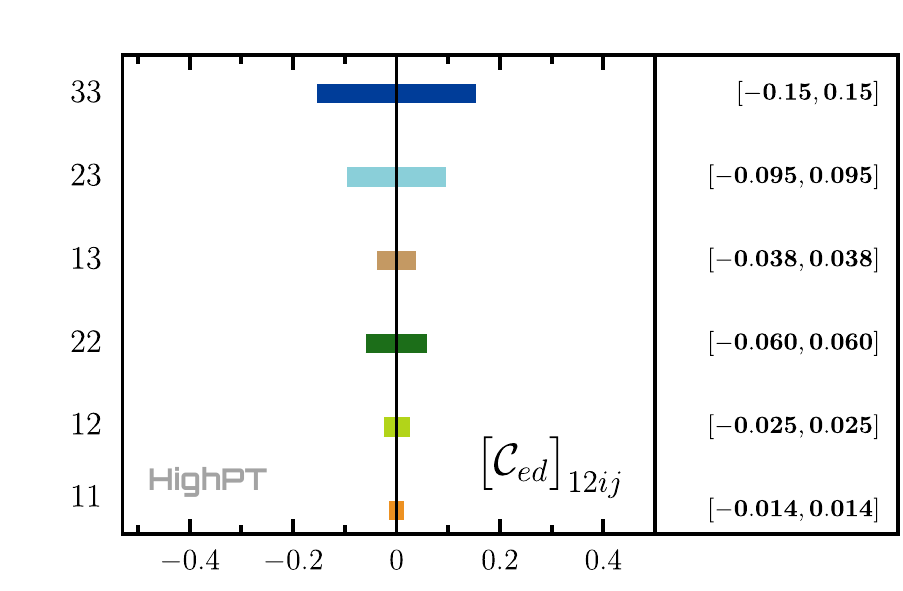}    & \includegraphics[]{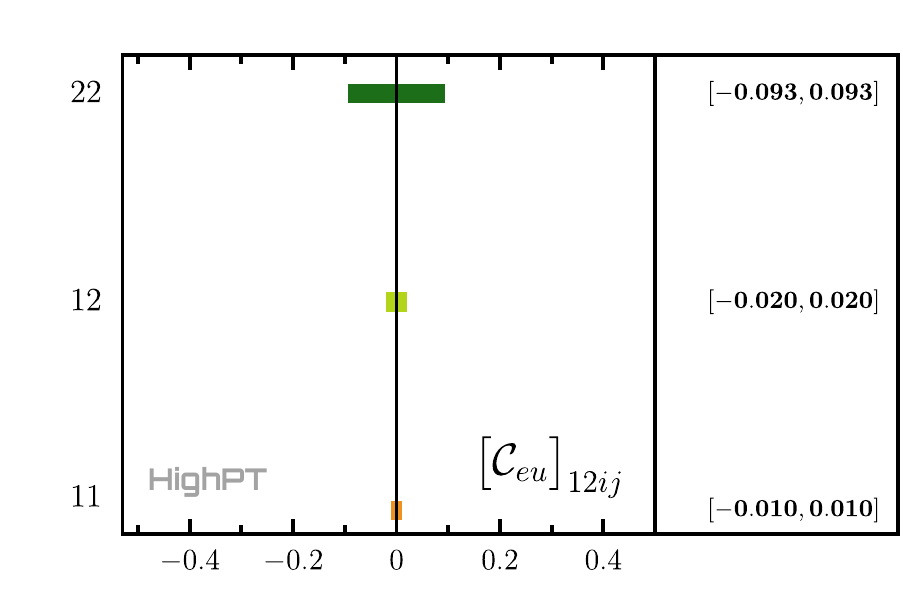}   \\
        \includegraphics[]{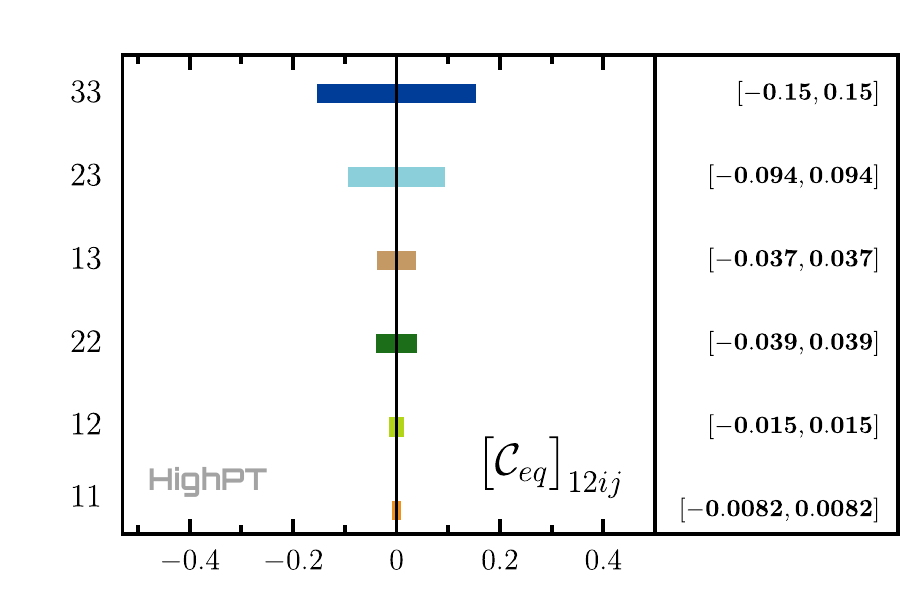}    &
        \includegraphics[]{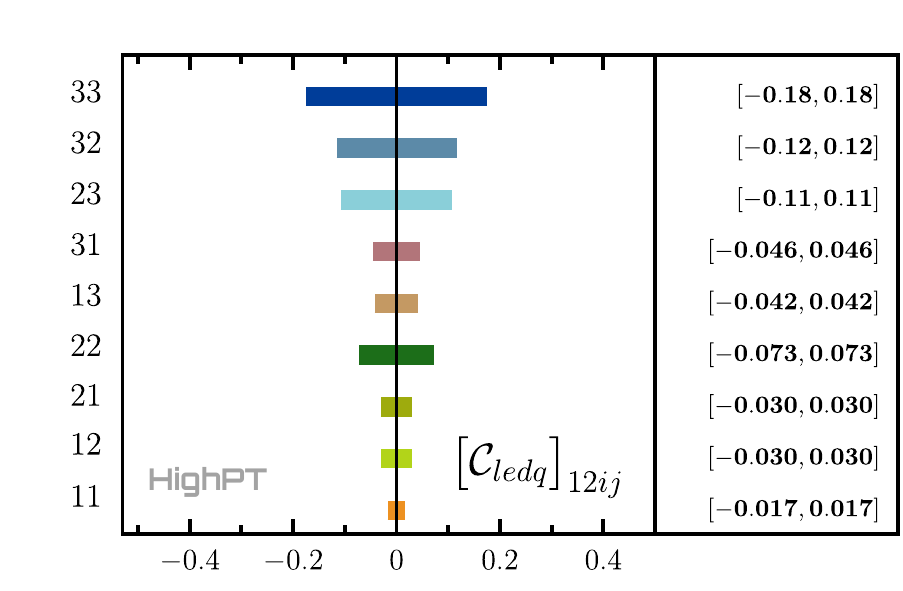}  &
        \includegraphics[]{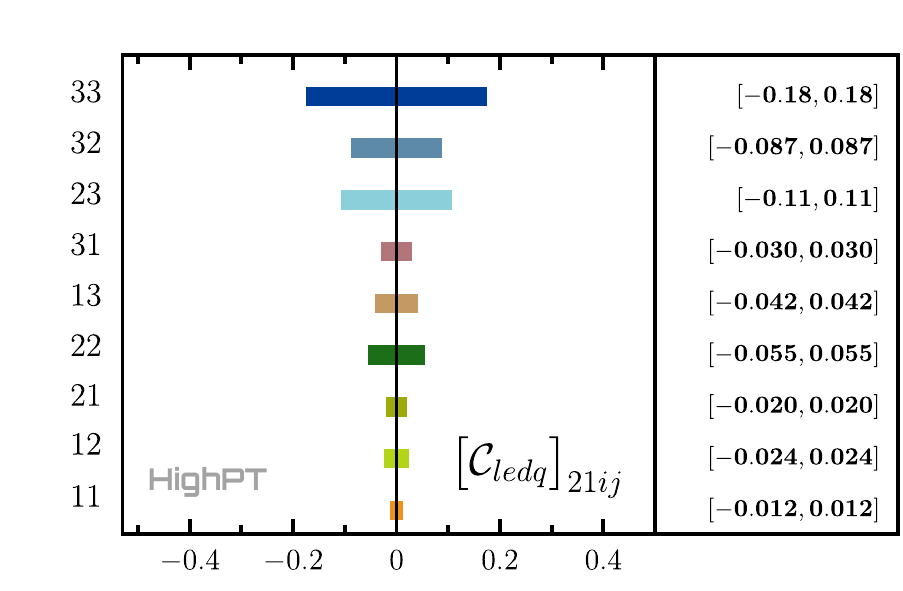} \\
        \includegraphics[]{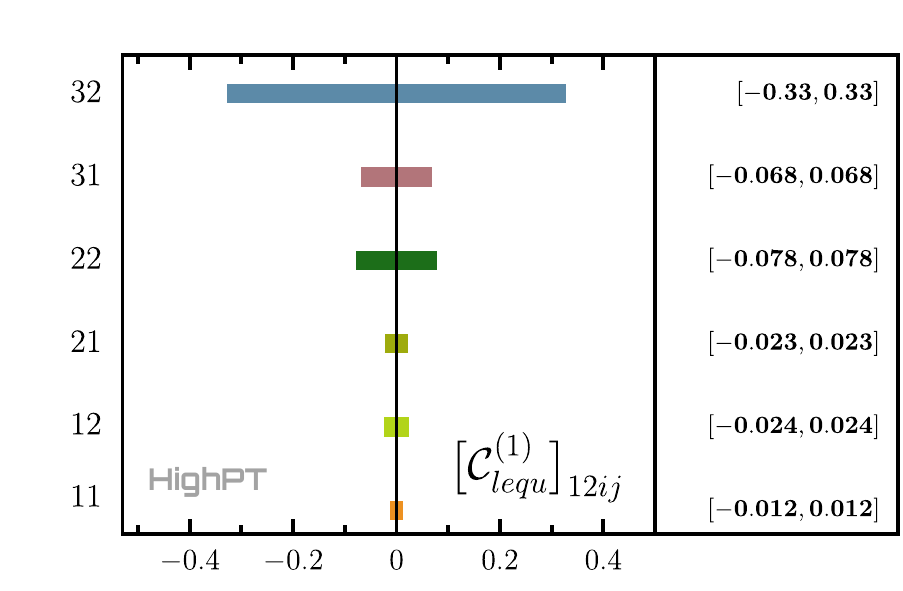} & \includegraphics[]{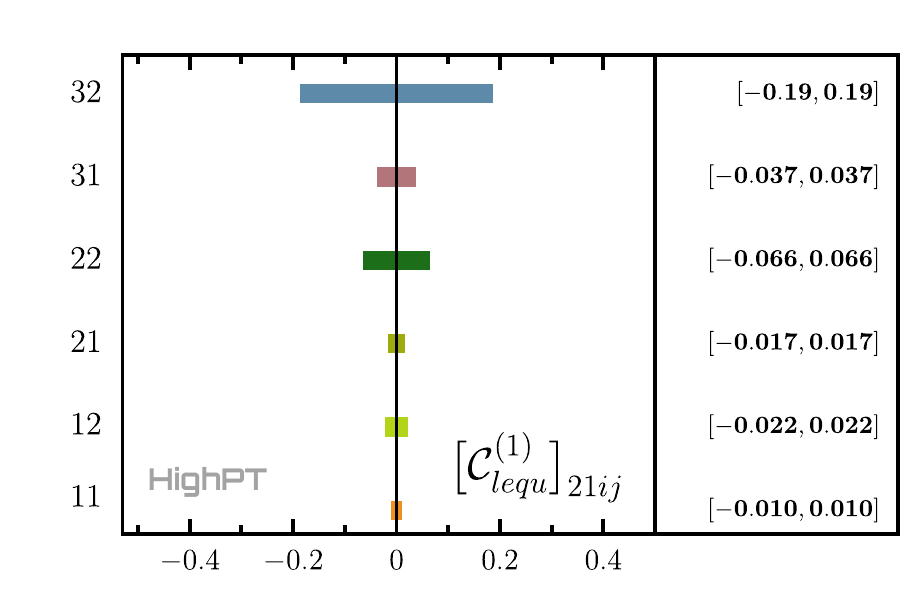} & \\
        \includegraphics[]{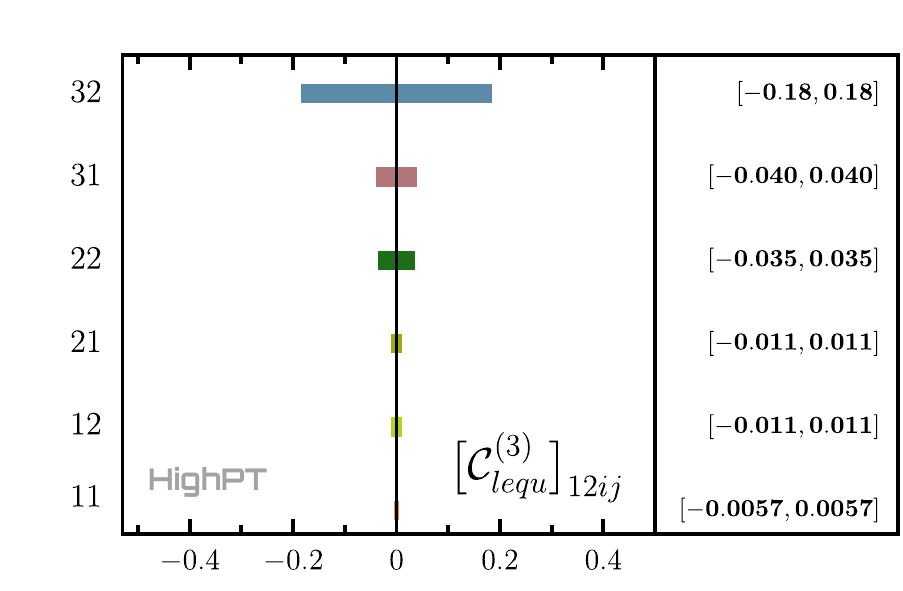} & \includegraphics[]{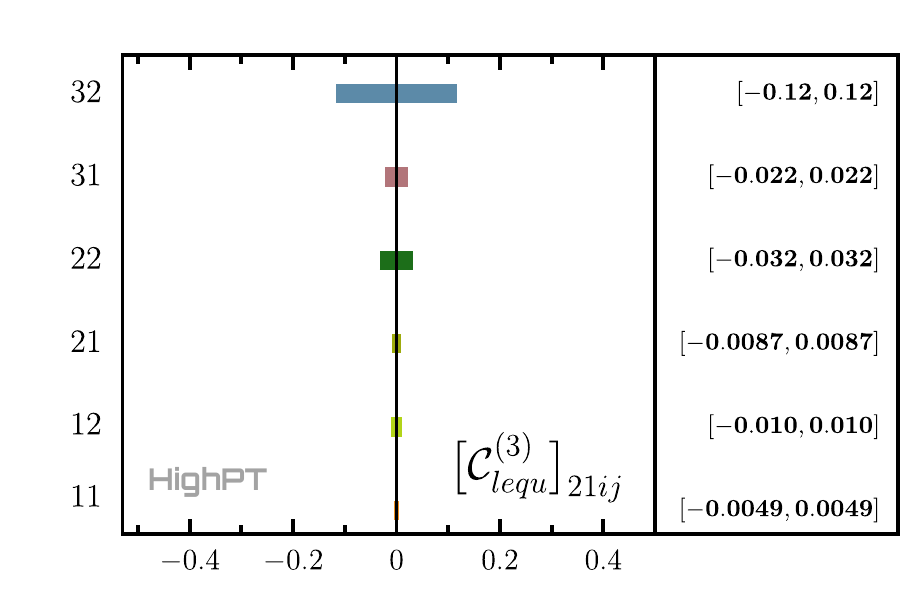} &
    \end{tabular}
    }
    \caption{\sl\small LHC constraints on semileptonic $d=6$ Wilson coefficients with $e\mu$ flavor indices, where a single coefficient is turned on at a time. See caption of Fig.~\ref{fig:single-WC-limits-tau-tau}.}
    \label{fig:single-WC-limits-mu-e}
\end{figure}

We report in this Appendix the high-$p_T$ limits derived on $d=6$ semileptonic operators, with all possible flavor indices, assuming a single coefficient at a time. These results are reported at $95\%$~CL for the $\tau\tau$, $\mu\mu$ and $ee$ channels in Figs.~\ref{fig:single-WC-limits-tau-tau}--\ref{fig:single-WC-limits-e-e}, and for the $\tau\mu$, $\tau e$ and $\mu e$ ones in Figs.~\ref{fig:single-WC-limits-tau-mu}--\ref{fig:single-WC-limits-mu-e}. Similar limits for the $d=6$ quark- and lepton-dipole operators are reported in Fig.~\ref{fig:single-WC-limits-dipoles}.

%%%%%%%%%%%%%%%%%%%
%%%%%%%%%%%%%%%%%%%

%%%%%%%%%%%%%%%%%%%
%%%%%%%%%%%%%%%%%%%

%%%%%%%%%%%%%%%%%%%
%%%%%%%%%%%%%%%%%%%

%%%%%%%%%%%%%%%%%%%
%%%%%%%%%%%%%%%%%%%

%%%%%%%%%%%%%%%%%%%
%%%%%%%%%%%%%%%%%%%

%%%%%%%%%%%%%%%%%%%
%%%%%%%%%%%%%%%%%%%

% - - - - Bibliography - - - - %
\clearpage
{
\bibliographystyle{JHEP}
\footnotesize
\bibliography{main}
}

\end{document}